\documentclass[12pt,a4paper,twoside]{report}
\usepackage{graphicx}
\begin{document}
\bibliographystyle{amsplain}
\pagestyle{myheadings}
\begin{titlepage}
\begin{center}
\vspace{0.8cm}
\begin{large}
{\bf JAGIELLONIAN UNIVERSITY}\\
{\bf INSTITUTE OF PHYSICS}\\
\end{large}
\vspace{2cm}
\begin{center}
\begin{Huge}
{\bf Proton induced spallation }\\
{\bf reactions in the energy range }\\
{\bf 0.1 - 10 GeV }\\
\end{Huge}
\end{center}
\vspace{1.5cm}
\begin{Large}
{\bf Anna Kowalczyk}
\end{Large}
\vspace{2cm}
\begin{figure}[!h]
\begin{center}
\includegraphics{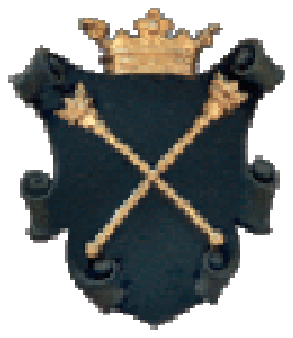}
\label{logo}
\end{center}
\end{figure}
\vspace{1cm}
\begin{center}
\begin{large}
{\bf A doctoral dissertation prepared at the Institute of Nuclear Physics 
of the Jagiellonian University, submitted to the Faculty of Physics, 
Astronomy and Applied Computer Science at the Jagiellonian University, 
conferred by Dr hab. Zbigniew Rudy}
\end{large}
\end{center}
\vspace{0.8cm}
\begin{large}
{\bf Cracow 2007}\\
\end{large}
\end{center}
\end{titlepage}

\newpage
\titlepage
\hbox{}
\thispagestyle{empty}
\newpage
\titlepage
\tableofcontents

\chapter{Introduction}
\markboth{ }{Chapter 1. Introduction}
\hspace*{0.6cm} 
Several possible scenarios of proton - nucleus reaction are considered 
nowadays.
According to one of the scenarios, incoming proton deposits energy into target 
nucleus. It knocks out a few nucleons and leaves excited residual nucleus.
Then, nucleons and various fragments are emitted from the excited residuum.
 This scenario is called {\sl spallation}. It is also possible, that the 
residuum splits up slowly ({\sl fissions}) into two parts, which then emit 
particles. This scenario is known as emission from {\sl fission} fragments.
But, it could be also that all fragments appear simultaneously. This 
would have features of a phase transition in nuclear matter and is called 
{\sl fragmentation}. These reaction scenarios are based on experimental 
observations of different final states. Generally, observation of 
one heavy nucleus (in respect to the mass of initial target), a small   
number of light fragments and numerous individual nucleons indicates 
spallation. Detection of a large number of intermediate size fragments 
indicates fragmentation. 
Nevertheless, spallation and fragmentation are correlated. Their 
differentiation is not clear, difficult and still under discussions. 
Previous studies indicate, that spallation is the most probable scenario of 
proton - nucleus reaction, what will be shown also in this dissertation. 
The main idea of this work concerns theoretical study of proton 
induced spallation reactions in wide range of incident energy and mass of 
target nuclei; fission and fragmentation are not discussed in details.

\begin{figure}[!ht]
\begin{center}
\includegraphics[height=12cm, width=12cm, angle=0]{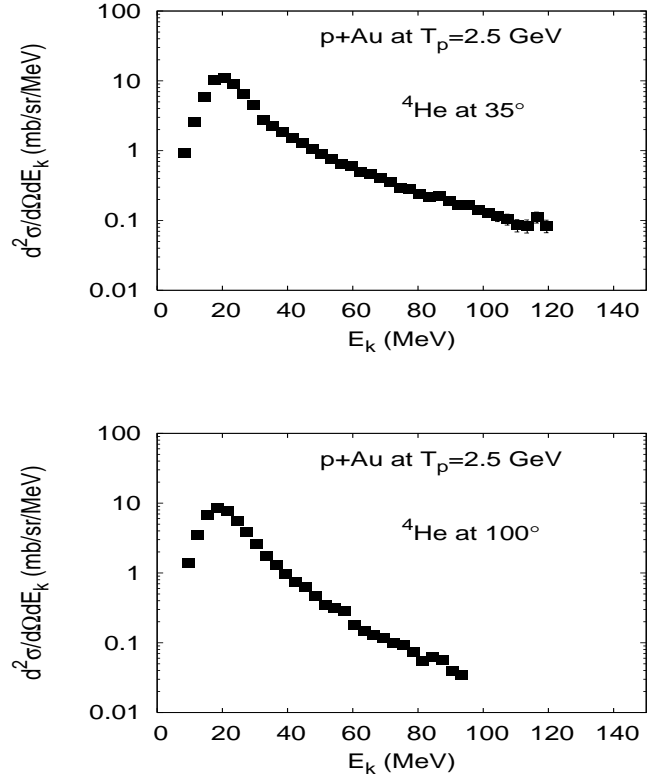}
\caption{{\sl Experimental observations of an example 2.5 GeV proton induced 
spallation reaction on Au target; inclusive alpha particles spectra measured 
by PISA at COSY \cite{Buba07}}}
\label{fig:exp_obs}
\end{center}
\end{figure}

The following definition of spallation process can be found in Nuclear Physics
Academic press:
{\sl "Spallation - a type of nuclear reaction in which the high-energy of 
incident particles causes the nucleus to eject more than tree
particles, thus changing both its mass number and its atomic number."}  
So, the term {\sl spallation} means a kind of nuclear reactions, where 
 hadron with high kinetic energy (100 MeV up to several GeV) interacts with 
 a target. First, this term was connected with observation of residuum of 
reaction corresponding to losses of mass of target nucleus from few up to 
several dozen nucleons. 
Nowadays, it means mechanism, in which high energy light particle causes 
production of numerous secondary particles from target nucleus, leaving cold 
residuum of spallation. As a result of such process also various Intermediate 
Mass Fragments (IMF), i.e. fragments with masses in range $4 < M_{IMF} < 20$, 
are observed.\\ 
From historical point of view, the possibility of heating a nucleus via 
bombarding by neutrons was suggested first time in 1936 by N. Bohr 
\cite{Bohr36}.
Studies of similar reactions were possible due to development of accelerator 
technics. It was in the end of fourties, when accelerators could provide 
projectiles with energies higher than 100 MeV \cite{Cunn47}. Experimentally, 
two - component spectra of emitted particles are observed: 
anisotropic high energy part, which dominates in forward angles 
(i.e. the high energy tail 
decreases at backward angles, as it is seen in the example Fig. 
\ref{fig:exp_obs}) and isotropic, low energy part.
These general features of spallation process are established experimentally.
A theoretical picture of an incident particle colliding successively with 
several nucleons inside target nucleus, losing a large 
fraction of its energy was proposed by Serber in 1947 \cite{Serb47}. Before, 
in 1937 Weisskopf considered possibility of emission of neutron from 
excited target nucleus \cite{Weis37}. 
In the end of fifties, Metropolis \cite{Metr58} 
and Dostrovsky \cite{Dostr58} (who used the ideas of Serber and Weisskopf)   
suggested description of spallation as two step process 
involving energy deposition and subsequent evaporation. They formulated and 
performed first Monte Carlo calculations of the reactions. Such treatment of 
spallation reactions is used from that time up to now.
\begin{figure}[!ht]
\begin{center}
\includegraphics[height=8cm, width=12cm, angle=0]{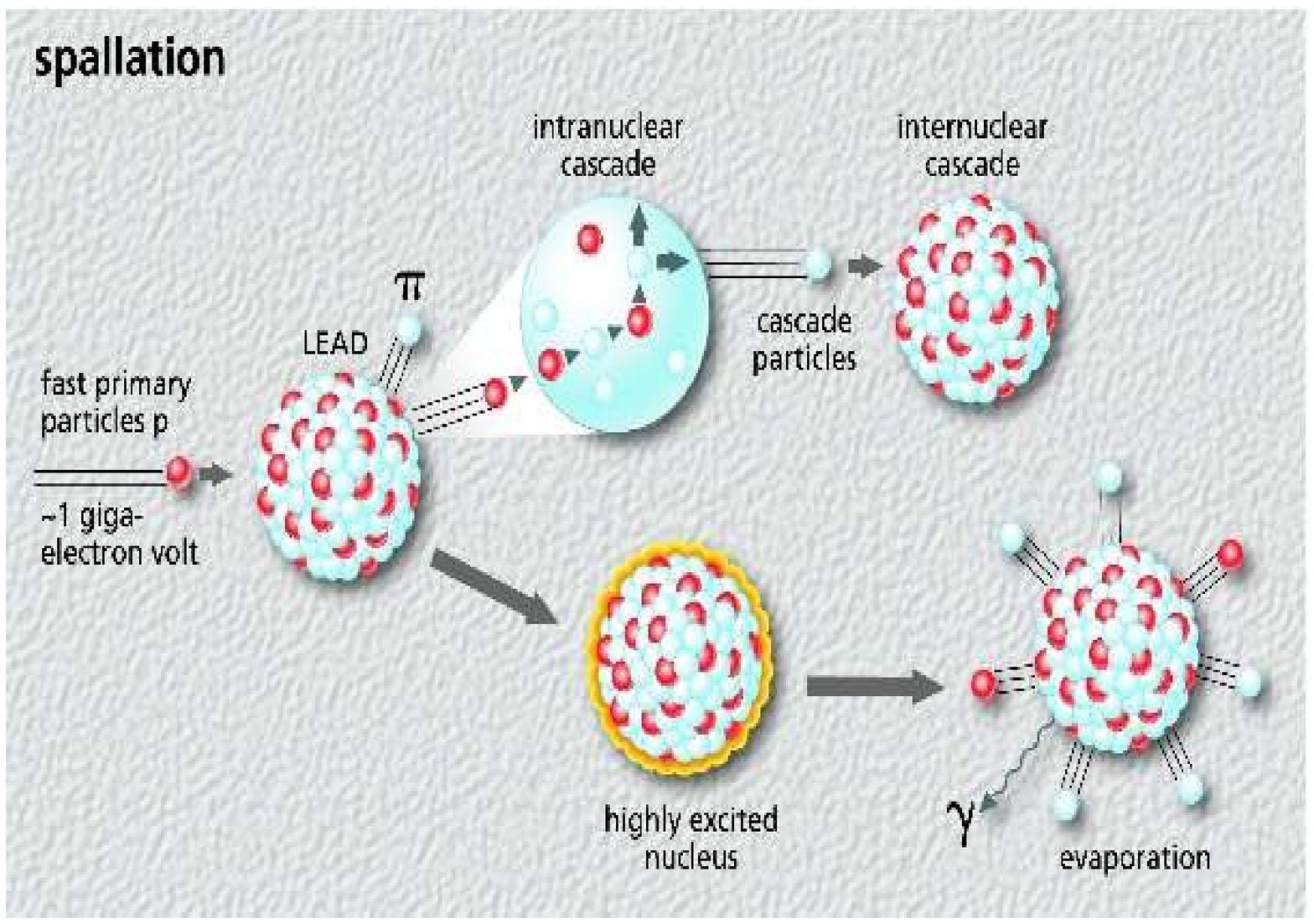}
\caption{{\sl The Spallation Process}}
\label{fig:spall_rys}
\end{center}
\end{figure}
In more details, the first, so-called fast stage of the spallation is highly 
non-equilibrated process. High energy proton causes an intra-nuclear 
cascade on a time scale $\sim 10^{-22}$s. The incident projectile goes 
through the target nucleus and deposits a significant amount of excitation 
energy and angular momentum, while ejecting only a few high energy nucleons 
 and, with a minor yield, pions and light ions. 
The result of the first stage is excited residual nucleus in thermodynamical 
equilibrium (totally or partly equilibrated), with excitation energy $\sim$ 
a few MeV/nucleon.\\
In case of thick target, i.e. system of several nuclei, the 
ejectiles, as secondary projectiles can cause so-called inter-nuclear 
cascade, placing individual nuclei into excited states, 
as illustrated in Fig. \ref{fig:spall_rys}.\\ 
The second, so-called slow stage of the spallation, consists in deexcitation of 
the residuum by evaporation of particles.
The isotropic emission (in the system of nucleus) of nucleons 
(mainly neutrons), light and heavy ions (d, t, He, Li, Be, B, ..., $\gamma$) 
takes place on a time scale $\sim 10^{-18}$ -  $10^{-16}$s.

From many years spallation reactions of medium and high energy protons with 
atomic nuclei are still of interest for many reasons. First of all, 
because knowledge of the reaction mechanism is still not complete. 
This is interesting both from theoretical 
and experimental point of view. Experimental data of double differential cross 
sections of emitted particles in the reactions are necessary for testing, 
validation and developing of theoretical models. 
It means, experimentally measured cross sections for exclusive elementary 
reactions (e.g. NN, N$\pi$, ...) are implemented in theoretical models. Then, 
results of calculations are compared with results of inclusive measurements. 
It is reasonable to study the reaction mechanism on the base of proton - 
nucleus rather than nucleus - nucleus collisions, where all processes start 
to be much more complicated (e.g. presence of distortions due to collective 
processes like compression, deformation, high spin \cite{Gold96}). 
Moreover, proton - nucleus reactions are important and indispensable also for 
experiments of nucleus - nucleus collisions (e.g. HADES \cite{Przy07}, 
CHIMERA \cite{Aiel95}). Results of proton - nucleus reactions facilitate 
extraction and interpretation of results of nucleus - nucleus reactions.       
Other reasons concern very broad range of applications (e.g. in medicine 
(radiation therapy), cosmology, accelerator technology). \\ 
Relatively huge number of produced neutrons suggested the idea of using 
spallation reactions as neutron sources. Nowadays, neutron beams are produced 
 in nuclear reactors. Reactors dedicated for such 
production generate also a lot of heat; about 190 MeV of energy is dissipated 
for single produced neutron. 
In accelerator based sources, neutrons are produced 
in a spallation process, with only about 30 MeV of energy dissipated for 
one generated neutron. During last decade several spallation sources 
(IPNS \cite{ICANS13}, ISIS \cite{ICANS13}, LANSCE \cite{ICANS11}, 
SINQ \cite{ICANS14}) became operational.\\
Spallation reactions are very important in accelerator technology (e.g. 
activation of detectors, radiation protection). 
The reactions are used for energy amplification, also for production of  
energy from nuclear waste and furthermore, transmutation of long - lived 
radioactive nuclei of nuclear waste to stable or short - lived, in order to 
avoid their long term storing \cite{Bowm92}.\\
\begin{figure}[!ht]
\begin{center}
\includegraphics[height=8cm, width=12cm, angle=0]{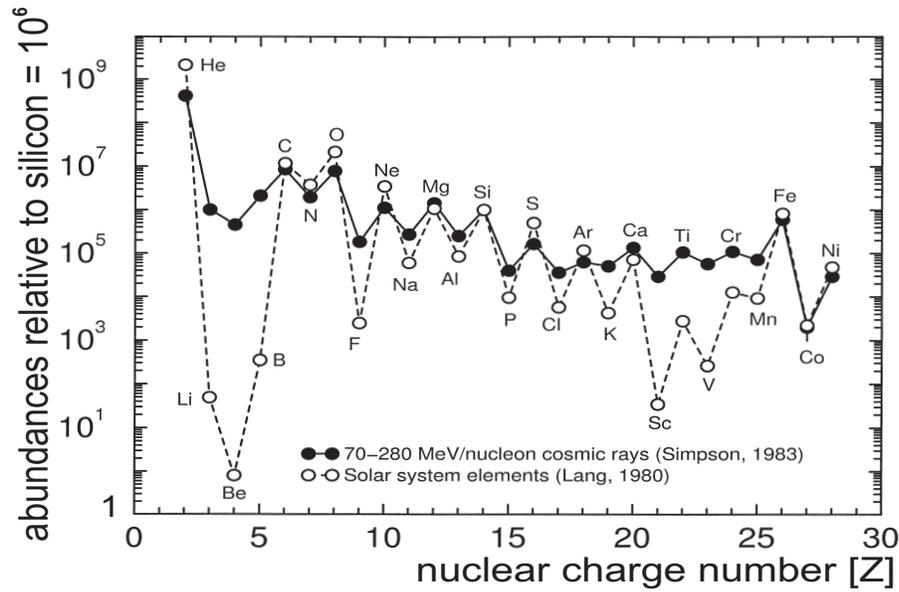}
\caption{{\sl The relative abundances of elements in cosmic rays measured at 
Earth and in the solar system, taken from \cite{Mash00}}}
\label{fig:astro_asp}
\end{center}
\end{figure}
Astrophysical models have to include spallation processes. If one compares  
abundances of cosmic rays and solar system elements, it is seen that Li, Be 
and B in cosmic rays are enriched by more than 6 orders of magnitude, as 
shown in Fig. \ref{fig:astro_asp}.
They were evidently produced in spallation reactions of hydrogen nuclei 
(which consist about 87 $\%$ of cosmic rays) with heavy elements (produced due 
to stars explosions). For more informations see \cite{Silb90, Mash00}.\\ 

Theoretical predictions of the process are important in each of mentioned 
above cases.
Several models have been constructed in order to describe the spallation 
process. First stage of the reaction is described by a class of microscopic 
models, e.g. \cite{Cugn97, Niit89, Aich91}. 
For the second stage statistical models are used, e.g. \cite{Gavr93, Furi00}. 
Nevertheless, the theoretical investigations are rather fragmentary and 
still not satisfactory. Global qualitative and quantitative informations of the
 spallation reactions are needed. 

Description of global average properties of proton induced spallation reactions
 in wide range of projectile energy and mass of target nuclei is the main  
subject of this work. This is investigated within a transport model based on 
Boltzmann - Uehling - Uhlenbeck (BUU) equation supplemented by a statistical 
evaporation model. The used transport model has been specially developed in 
the frame of this work, in order to enable description of such reactions. 

This work is organized as follows.  
In Chapter 2, present knowledge of the reaction dynamics and review of various 
theoretical models is given. Chapter 3 includes description of  
the Hadron String Dynamics (HSD) approach of the first stage of the spallation  
reaction. In Chapter 4 and Chapter 5, results concerning properties of the first
 stage of the reaction are presented and discussed. Chapter 6 concerns pions, 
which could be produced only in the first stage of reaction. In Chapter 7, 
statistical evaporation models for the second stage are recalled. 
Chapter 8 contains bulk model predictions of both stages of 
proton - nucleus reactions.  
In Chapter 9 and Chapter 10, the models results and comparison with available 
experimental data and with results of other models are presented, respectively. 
Finally, in Chapter 11, a summary and conclusions are given.\\


\chapter{Present knowledge of the reaction dynamics - basic theoretical models}
\label{chapt:theo_models}
\markboth{ }{Chapter 2. Basic theoretical models}

Our understanding of physical phenomena is expressed as modelling. 
At present, the broadest platform for such modelling is quantum mechanics 
approach. Unfortunately, many - body systems are usually an extreme challenge 
for existing methods of quantum mechanics. One has to rely on rather simple, 
much more straight formed concepts. In this Chapter, such general basic 
concepts will be shown as ingredients of typical models of nuclear reaction.
  
Several microscopic models have been constructed in order to describe the first
stage of proton - nucleus reaction. 
All of them have the same basis, they describe the reaction as a cascade of 
nucleon - nucleon collisions, but employing different assumptions. 
The main difference concerns implemented potential of nucleon - nucleus 
interaction.
One can distinguish the simplest models, which neglect features of the mean 
field dynamics and employ constant static potential, like a class of 
Intra - Nuclear Cascade (INC) models. Other, more sophisticated approaches 
comprise dynamically changing field and minimal fluctuations obtained due to 
use of test particle method, i.e. models based on Boltzmann - Uehling - 
Uhlenbeck (BUU) transport equation. 
There are also models, which include real fluctuations and particles 
correlations, employing two- and three- body potentials, e.g. Quantum 
Molecular Dynamics models.
The main ideas of the different types of nuclear reaction approaches are 
described below.

In this Chapter, devoted for model-like approach of investigation 
of nucleon - nucleus reactions, also so-called percolation model is presented. 
This rather basic model describes fragment mass distributions very well 
(Section \ref{sec:perc_mod}).

\section{Intranuclear Cascade model}

Intranuclear cascade (INC) model is used as a base in many existing codes for 
the first stage calculations.
Description of the nucleon - nucleus reaction in terms of binary nucleon - 
nucleon collisions inside nucleus is the basic assumption of the model.
In principle, the single particles approach of the INC is justified as long as 
the de Broglie wavelength $\lambda$ of the cascade particles is smaller than 
the average internucleon distance in the nucleus ($\approx$ 1.3 fm).
This indicates the low energy limit of the model 
(e.g. for projectile kinetic energy $T_{p}=100$ MeV $\Rightarrow$ $\lambda=2.7$
 fm, for $T_{p}=1000$ MeV $\Rightarrow$ $\lambda=0.7$ fm).
The INC calculations follow the history of individual nucleons that becomes 
involved in the nucleon - nucleon collisions in a semi - classical 
manner. It means, the momenta and coordinates (trajectories) of the particles 
are treated classically. The only quantum mechanical concept incorporated in 
the model is the Pauli principle.\\
The first code of INC has been created by Bertini \cite{Bert63}, in 1963.
Later, the conception was used also in other codes, e.g. by Yariv in his ISABEL 
code \cite{Yari79}. In the 80's and 90's, the next versions of INC model was 
developed by Cugnon et al. \cite{Cugn97}.\\
The main features of the standard INC approach are the following.\\   
The initial positions of target nucleons are chosen randomly in a sphere of 
radius $R=1.12 \cdot A_{T}^{1/3}$ fm, where: $A_{T}$ is the mass number of 
target nucleus.
Momenta of the nucleons are generated inside a Fermi sphere of radius 
$p_{F}=270$ MeV/c. 
Neutrons and protons are distinguished according to their isospin.
All nucleons are positioned in a fixed and constant, attractive potential well 
of $V_{0}=40$ MeV depth, inside the nuclear target volume.
The depth value is taken a bit higher than the Fermi energy 
($E_{F} \approx 38$ MeV), so that the target is stable during the reaction. 
The idea of fixed average potential is based on a relatively low number of 
particles emitted during the INC stage of reaction, which disturbes the mean 
field only slightly. 
The incident particle (of incident energy $T_{p}$) is provided with an impact 
parameter $b$, chosen randomly on a disc of radius $R$. It is positioned at 
the surface of nucleus, in the potential $V_{0}$. Its kinetic energy is equal 
to $T_{p}+V_{0}$.
Relativistic kinematics is used for description of the reaction (i.e. the total
 energy of a nucleon is connected with its momentum and mass by relation: 
$E^{2}=p^{2}+m^{2}$).
All nucleons are propagated in time; their momenta and positions 
are evolved in time as follows:
$\vec{r}(t + \delta t) = \vec{r}(t) + \frac{\vec{p}}{E} \delta t$, 
$\vec{p}(t + \delta t) = \vec{p}(t) - \nabla _{r} V_{0} \delta t$.\\
At time $t=0$, the incident nucleon is hitting the nuclear surface.
Next, all particles are moving along straight line trajectories, until two of 
them reach their minimum relative distance, or until one of them hits the 
nuclear surface.       
When a particle hits the nucleus surface from inside, two cases are considered.
 If the kinetic energy of the particle is lower than the $V_{0}$, it is 
reflected on the surface. If the kinetic energy is higher than the $V_{0}$, 
the particle is transmitted randomly with some probability, see Ref. 
\cite{Cugn97}. \\   
A collision takes place, if the minimum relative distance $d_{min}$ between 
 two particles fulfills the condition: 
$d_{min} \leq \sqrt{\frac{\sigma _{tot}(\sqrt{s})}{\pi}}$, where: $\sqrt{s}$ is
 the energy in the center of mass of the two particles, $\sigma _{tot}$ is the 
total collision cross section.
The two particles are scattered elastically or inelastically, according to the 
energy - momentum conservation. 
Inelastic collisions with high probability lead to the formation of Deltas 
($\Delta$'s). Mass of 
$\Delta$ is introduced with the Lorentzian distribution centered on the mean  
value equal to 1232 MeV, with the width $\Gamma$=110 MeV.\\
The following reactions are considered in the model:\\
$NN \Leftrightarrow NN$, $NN \Leftrightarrow N \Delta$, $N \Delta 
\Leftrightarrow N \Delta$, $\Delta \Delta \Leftrightarrow \Delta \Delta$, 
$\Delta \Leftrightarrow N \pi$. \\
All used cross sections and angular distributions are based on available 
experimental data \cite{Cugn97}.\\
Additional condition restricting a collision is Pauli blocking. Collisions and 
$\Delta$ - decays are avoided, when the presumed final states are already 
occupied. The final states phase - space occupation probability ($f_{i}$ and 
$f_{j}$, where $i$ and $j$ denote two particles predicted to be created in the 
final state) is evaluated by counting
 particles of the same kind, inside a reference volume in phase - space. The 
collision or decay is realized, when: $P_{ij}=(1-f_{i})(1-f_{j})$ is larger 
than a random number chosen between 0 and 1.\\   
The interaction process is stopped at time $t=t_{stop}$, determined by the 
average behaviour of some quantities (e.g. an excitation energy of the nucleus,
 see Ref. \cite{Cugn97}).
At the end of the cascade, all remaining $\Delta$'s are forced to decay. 

\section{Quantum Molecular Dynamics model}
\markboth{2.2 Quantum Molecular Dynamics model}{Chapter 2. Basic theoretical models}
Quantum Molecular Dynamics (QMD) \cite{Aich91, Niit95} model has been 
developed mainly in order to investigate fragments formation during 
proton - nucleus or nucleus - nucleus collisions. 
The model, 
as a N-body theory, describes the time evolution of correlations between 
particles,    
what is essential in consideration of the fragments formation.\\ 
In the approach, 
nucleons are spread out in phase - space with a Gaussian distribution. 
The coordinates and momenta of nucleons are designated simultaneously. 
The wave function of each nucleon ($i$) is assumed to have the following 
Gaussian form:
\begin{equation}
\psi _{i} (\vec{r}, \vec{p}, t)=\frac{1}{(2 \pi L)^{3/4}} e^{- \frac{(\vec{r}-\vec{r}_{i0})^{2}}{4L}} e^{i \vec{p} \vec{r}}
\end{equation}        
where: the Gaussian width - $L = 1.08 fm^{2}$ - corresponds to a root mean 
square radius of the nucleon of 1.8 fm, $\vec{r}_{i0}$ are the coordinates 
of the centers of the Gaussian wave packets.\\
The normalized Gaussian function represents one nucleon.\\
In the model, Szilard-Wigner densities are applied. Those give a semi-classical
 approximation, depending simultaneously on coordinates and momenta. The 
Szilard-Wigner density is defined by the following expression, 
constructed of a wave function: $\psi (x_{1}, ..., x_{n})$ \cite{Wign32}: 
\begin{eqnarray}
P(x_{1}, ..., x_{n}; p_{1}, ..., p_{n})=(\frac{1}{\hbar \pi})^{n} \int _{-\infty}^{\infty} ... \int _{-\infty}^{\infty} dy_{1} ... dy_{n} \psi (x_{1}+y_{1}, ..., x_{n}+y_{n})^{*} \nonumber \\
 \psi (x_{1}-y_{1}, ..., x_{n}-y_{n}) e^{\frac{2i}{\hbar}(p_{1}y_{1}+...+p_{n}y_{n})} 
\end{eqnarray}
This is called the probability function of the simultaneous values of 
coordinates: $x_{1}, ..., x_{n}$ and momenta: $p_{1}, ..., p_{n}$.   
It has the following properties. Integrated with respect to 
the $p$, it gives probabilities for the different values of the coordinates:
\begin{equation}
\int P(x_{1}, ..., x_{n}; p_{1}, ..., p_{n})dp = |\psi (x_{1}, ..., x_{n})|^{2}.
\end{equation}
Also, integrated with respect to the $x$, it gives quantum mechanical 
probabilities for the momenta $p_{1}, ..., p_{n}$:
\begin{equation}
\int P(x_{1}, ..., x_{n}; p_{1}, ..., p_{n})dx = |\int _{-\infty}^{\infty} ... \int _{-\infty}^{\infty} \psi (x_{1}, ..., x_{n}) e^{\frac{-i}{\hbar}(p_{1}x_{1}+...+p_{n}x_{n})}dx_{1}...dx_{n}|^{2}.
\end{equation}
Based on upper definition, the Szilard-Wigner representation of considered 
 in the model system is given by:
\begin{equation}
f(\vec{R}, \vec{p}, t)=\sum_{i} \frac{1}{(\pi \hbar)^{3}} exp(-\frac{(\vec{R}-\vec{R_{i}}(t))^{2}}{2L} -\frac{2L \cdot (\vec{p}-\vec{p_{i}}(t))^{2}}{\hbar ^{2}})    
\end{equation}
The boundary distributions, i.e. the densities in coordinate and momentum space
 are given by:
\begin{equation}
n(\vec{R}, t)=\int f(\vec{R}, \vec{p}, t)d^{3}p=\sum_{i} \frac{1}{(2 \pi L)^{3/2}} exp(-\frac{(\vec{R}-\vec{r_{i0}}(t))^{2}}{2L}),
\end{equation}
\begin{equation}
g(\vec{p}, t)=\int f(\vec{R}, \vec{p}, t)d^{3}R.
\end{equation}
 In order to construct the initial system,   
the centers of the Gaussians (i.e. nucleons) are chosen randomly in coordinate 
and momentum space, in the following way. 
First, the positions ($\vec{r_{i}}$) of nucleons are determined in a sphere of 
the radius $R=1.14 \cdot A^{1/3}$. The numbers are chosen randomly, rejecting 
those which would position the centers of two nucleons closer than 
$r_{min}=1.5$ fm. In the next step, the local density ($\rho (\vec{r_{i}})$), 
at the centers of all nucleons, generated by all the other nucleons is 
determined. Then, the local Fermi momentum ($p_{F}^{l}$) is calculated: 
$p_{F}^{l}=\hbar (\frac{3}{2}\pi ^{2} \rho (\vec{r_{i}}))^{1/3}$.
Finally, the momenta ($\vec{p_{i}}$) of all nucleons are chosen randomly, 
between zero and the local Fermi momentum ($p_{F}^{l}$).
Then, all random numbers, which position two nucleons in phase - space closer 
than: $(\vec{r_{i}}-\vec{r_{j}})^{2}(\vec{p_{i}}-\vec{p_{j}})^{2}=d_{min}$ are 
rejected, and must be chosen again. 
The initialization process lasts long time. Typically, only 1 of 
50 000 initializations is accepted under the criteria.  
But the finally accepted configurations are quite stable, usually no nucleon 
escapes from the nucleus in 300 fm/c, see \cite{Aich91}. 

After successful initialization, the nuclei (in case of nucleus - nucleus 
collisions) are boosted towards each other with the proper center of mass 
velocity.\\
During propagation, only positions ($r_{i}$) and momenta ($p_{i}$) of nucleons 
($i$) are changed, the width of the wave function is kept fixed. 
The mean values $(r_{i0}, p_{i0})$ are evaluated at time under the influence of
 two- and three- body interactions (important for preserving the correlations 
and fluctuations among nucleons), according to the classical Newtonian 
equation of motion: 
\begin{eqnarray}
\dot{r}_{i0} = \frac{\partial H}{\partial p_{i0}}, \nonumber \\
\dot{p}_{i0} = - \frac{\partial H}{\partial r_{i0}}, 
\end{eqnarray}   
with the Hamiltonian $H = T + U$, where: $T$ is total kinetic energy and $U$ is
 total potential energy of all nucleons. \\
Usually, the differential equations are solved using an Eulerian integration 
routine with a fixed time step $\Delta t$, where the momentum is evaluated at 
time points halfway between the times of the position determinations:    
\begin{eqnarray}
p_{i0}(n+1)=p_{i0}(n)- \nabla _{r} U_{i}(n+\frac{1}{2})\Delta t, \nonumber \\
r_{i0}(n+\frac{1}{2})=r_{i0}(n-\frac{1}{2})+\frac{p_{i0}(n)}{(p_{i0}(n)^{2}+m_{i}^{2})^{1/2}} \Delta t + \nabla _{p} U_{i}(n)\Delta t.
\end{eqnarray}

Assumed in the model total interaction is composed of a short range 
interactions between nucleons ($V^{loc} \approx \delta (r_{1}-r_{2}) + 
\delta (r_{1}-r_{2}) \delta (r_{1}-r_{3})$), a long range Yukawa interaction 
($V^{Yuk} \approx \frac{e^{-|r_{1}-r_{2}|/ \nu}}{|r_{1}-r_{2}|/ \nu}$, 
$\nu = 0.8$ fm) and a charge Coulomb interaction ($V^{Coul} \approx 
\frac{Z_{1} \cdot Z_{2}}{|r_{1}-r_{2}|}$, $Z_{1,2}$ stands for charges).\\
Potential acting on each particle is given by the expectation value of the two 
and three body interaction: 
\begin{equation}
U_{i}(t) = U_{i}^{(2)}(t) + U_{i}^{(3)}(t)
\end{equation}              
where the two body potential $U_{i}^{(2)}$ is given by local, Yukawa and 
Coulomb interaction terms, while the three body potential $U_{i}^{(3)}$ 
includes only the local interactions term, see Ref. \cite{Aich91}.

Two nucleons can collide if they come closer than $r=\sqrt{\sigma / \pi}$, 
where $\sigma$ is a total nucleon - nucleon cross section. 
Additionally, Pauli principle is taken into account.
In the model, the measured free nucleon - nucleon scattering cross section is 
used. However, the effective cross section is smaller because of the Pauli 
blocking of the final state. It means, whenever a collision has occurred, the 
phase - space around the final states of the scattering partners is checked. 
It is calculated, which percentage, $P_{1}$ and $P_{2}$, of the final phase - 
space for each of the two scattering partners, respectively, is already 
occupied by other nucleons. Then, the collision is blocked with a probability 
$P_{block} = P_{1} P_{2}$, or allowed with the probability $1 - P_{block}$. 
If a collision is blocked, the momenta of scattering partners are kept with 
values, which they had before scattering. \\     
The scattering angles of the single nucleon - nucleon collisions are chosen 
randomly, with the probability distribution known from the experimental 
nucleon - nucleon scattering \cite{Cugn81}.\\
Inelastic collisions lead to the formation of Deltas, which can be reabsorbed 
by the inverse reactions. \\
It is assumed in the QMD approach, that only these beam energies are accepted, 
at which no more than $84\%$ of all collisions are blocked. Therefore, the low 
energy limit of the model is kinetic energy $T_{lab}=20$ MeV/nucleon 
\cite{Aich91}.

\section{Percolation model}
\label{sec:perc_mod}
\markboth{2.3 Percolation model}{Chapter 2. Basic theoretical models}

Percolation model has been introduced in the eighties as minimum information 
approach, based on purely topological and statistical concepts. It was in order
 to describe fragment size distributions, as an outcome of nuclear 
fragmentation process within the simplest possible physical framework 
\cite{Rich01}.\\
However the model is flexible enough to allow for the inclusion of different 
physical mechanisms.\\
In general, one distinguishes between bond and site percolation model.

Site percolation corresponds to the case where each site is either 
occupied by one particle or is empty. One fixes a probability $p \in [0,1]$, 
and for each site generates a random number $\xi$, taken from a uniform 
distribution in the interval $[0,1]$. If $\xi \leq p$ the site is said to be 
occupied, if $\xi > p$ it is empty. Checking all sites, one obtains an 
ensemble of occupied and empty sites. Sets of occupied sites are called 
clusters. Finally, space is topologically covered with clusters and empty 
space.

Bond percolation corresponds to the case where each site is occupied, but 
neighbouring sites being bound or not to each other. The number of 
neighbouring sites is fixed by the bonds, which depend on the geometric 
structure of space occupation. A similar procedure, as in case of site 
percolation, works also in this case. For a fixed bond probability $p$ one 
considers a pair of neighbouring sites and generates a random number $\xi \in 
[0,1]$. If $\xi \leq p$ the sites are linked by a bond, if $\xi > p$ they are 
not. Checking all possible bonds between neighbouring sites one obtains 
clusters made of connected particles. There are clusters of different sizes, 
appearing with a given multiplicity, which depends on $p$. 

The first applications of percolation concepts were proposed by W. Bauer and 
collaborators \cite{Baue85, Baue86}, Campi and Desbois \cite{Camp85}.
While Campi and Desbois used a site percolation, Bauer with his group have 
proposed a model based on bond percolation theory. Nowadays, the bond percolation has been used as a base for models of fragmentation more often than the site one. Below, an outline of model proposed by Bauer et al. is presented.  \\
As any percolation model, it is based on two crucial ingredients: a description of the distribution of a set of points (i.e. nucleons) in a space and a criterion for deciding whether two given points are connected.
The target nucleons are represented by points occupying uniformly an approximately spherical volume on a simple cubic three-dimensional lattice in coordinate space.
The lattice spacing $d$ is computed from the normal nuclear density:
$d = \frac{1}{\rho_{0}^{1/3}} \approx 1.8$ $fm$, where: $\rho_{0} = 0.16$  $nucleons/fm^{3}$.
The number of points is equal to the number of target nucleons and is conserved during the calculation, therefore the conservation law of mass in the calculated fragmentation process is fulfilled.
Initially, each nucleon is connected by bonds (representing the short-ranged nuclear interactions) to its maximum six nearest neighbors, depending on its location in the target.
Then, a point-like proton, with an interaction radius $r$ collides the target. 
Because it is assumed that the motion of the nucleons in the target is 
neglected, the projectile sees a frozen image of the target (it is feasible, as
 typical Fermi motion speed of nucleon is significantly lower than speed of 
incoming proton). For a given impact parameter $b$, the proton removes from the lattice nucleons occupying a cylindrical channel with the radius $r$ along his straight path in the target. It is typically 6-8 nucleons.
All of remained nucleons, called spectators, are still connected via bonds.
These bonds are then broken with a probability $p$, which is a percolation parameter. The parameter $p$ should be related to some physical input.
For example, it is reasonable to assume that $p$ is a linear function of kinetic energy of the projectile, and also an increasing function of the excitation energy ($E^{*}$) of the spectators: $p = E^{*} / E_{B}$, where: $E_{B}$ is the nuclear matter binding energy per nucleon (16 MeV) \cite{Baue88}. \\
The breaking probability has to be also dependent on the impact parameter $b$ of the proton. Bauer et al. have used following dependence: $p(b) = \frac{p_{0}}{1+exp[(b-R)/a]}$, where: $p_{0} = p(b=0)$, $R$ is a radius of the target nucleus, $a = 1.0$ $fm$ is a diffuseness parameter \cite{Baue86, Baue88}.\\
For given parameter $b$, the breaking probability is assumed to be uniform for all bonds, independent of their position on the lattice.\\
Using the breaking probability $p$ as an input parameter, a Monte-Carlo algorithm decides for each bond individually whether it is broken or not, as follows.
For a given $p(b)$, a random number $\xi _{ijk} \in [0,1]$ is generated for each bond $B_{ijk}$, where: the indices $ijk$ correspond to the spatial location of the center of the bond on the lattice. If $\xi _{ijk} > p$, the bond $B_{ijk}$ is unbroken, if $\xi _{ijk} \leq p$, the bond is broken.
Then a cluster search algorithm \cite{Baue86} is used to find out which 
nucleons are still connected by bonds i.e. form clusters. 
Taking into account all impact parameters, inclusive mass and multiplicity distributions can be obtained. That can be compared to experimental results.\\
It is surprising that using only one free parameter and simple geometrical 
considerations, this model is able to reproduce experimental mass yield curves 
with a good accuracy, in particular, the power law behavior 
($ \sim A_{F}^{- \lambda}$) at small masses and the U-shape distribution of the 
whole mass range. \\
\indent Bauer's group have used such model also to study the possibility of observing a phase transition of nuclear matter in collisions of high energy protons ($>10$ GeV) with heavy targets. Since inclusive fragment mass distributions follow a power law behavior: $\sigma (A_{F}) \approx A_{F}^{- \lambda}$, for $A_{F} < A_{T}/3$, ($ \lambda = 2.6$ is independent of target mass for heavy targets) \cite{Hirs84} similar to the mass yield distribution of droplets condensing at the critical point in a van der Waals gas ( $ \sigma (m) \approx m^{- \tau}$, with the critical exponent $\tau = \frac{7}{3}$ ), they suggested that nuclear multifragmentation proceeds via a liquid-gas phase transition of nuclear matter.\\
Bauer et al. have accented the importance of their result for experimental study of the phase transition, that the critical events are not the ones with the highest multiplicities, but the ones with the highest value of standard deviation of mass distribution. 

In the percolation models, such as described above, it is assumed that nucleons
 are distributed uniformly in the sphere, and the total excitation energy is 
assumed to be uniformly distributed over the whole excited system of the 
spectators, what is equivalent to assumption that equilibrium is reached.
In this picture, the angular distribution of fragments should be forward peaked,
 because the momentum transfer from incident proton to the nucleus is in average
 in forward direction. In reality, as the incident energy increases, the mass 
fragments angular distribution grows from forward to sideward ($E_{p} \geq 10$ 
GeV) \cite{Mura02} or backward ($E_{p} \geq 100$ GeV) \cite{Fort80} peaked in 
laboratory frame, what contradicts to the picture of the fragmentation from an 
equilibrated system, and indicates that the nucleons may not be uniformly 
distributed spatially and the excitation may depend on the position inside the 
nucleus.

For the understanding of this sideward emission (for which not satisfactory 
explanation has been given so far) Hirata et al. \cite{Hira01} investigated the
 non-equilibrium dynamical effects, such as non-spherical nuclear formation. 
They formulated Non-Equilibrium Percolation (NEP) model, which they use in a 
combined framework with a transport model \cite{Hira01}. The main differences 
between equilibrium percolation and the NEP model are following. The initial 
conditions of percolation, instead of putting nucleons on sites with some 
assumed occupation probability, is taken from the results of the dynamical 
transport model calculations. The bond breaking probability is assumed to be 
dependent on the position and momenta of the nucleons. It is calculated by 
considering excitation energy, distance and momentum difference between the 
nucleon pairs, instead of giving a common breaking probability for the bonds 
connecting nearest neighbor sites. 

Analysing fragmentation process with the NEP model, considering calculated 
effects, Hirata et al. have found following mechanism of sideward enhanced 
fragments emission. Based on fragments formation point distributions, both in 
case of central and peripheral collisions, the fragments are formed mainly near 
the surface of the nuclei. It is due to fact, that along the incident proton path, nucleons collide with the leading proton or secondary cascade particles. Since they have large kinetic energies, they increase the bond breaking probability. As a result the fragment formation is suppressed along the incident proton path. Fragments are formed mainly in the cold region around the hot zone. Their formation points are distributed non-spherically, in a doughnut shaped region. Hirata et al. noticed that this effect alone does not generate any anisotropy in angular distributions.\\
Based on analysis of fragments energy distributions, they found that the energy
 distribution is effected by Coulomb repulsion. 
Calculated results after the Coulomb expansion well reproduce the qualitative behavior of experimental data \cite{Hira01}.
This is Coulomb repulsion between formed fragments that pushes and accelerates them sideways of the doughnut region. It means, the Coulomb repulsion modifyies the angular distribution from forward peaked to sideward peaked. 

Bond percolation model with non-equilibrium effects, have been investigated also by Yamaguchi and Ohnishi \cite{Yama04}. 
They introduced to the model isospin dependence, by assuming that neutron-neutron and proton-proton bonds are always broken, while neutron-proton bonds make the nucleus bound. Additionally, by comparing calculated fragments energy spectra with experimental data, they have got an agreement, when taken density of fragmenting nuclei is not equal the normal nuclear density $\rho_{0}$, but of around $ \sim \rho_{0}/3$. It means, target nucleus, after being heated by incident proton, would expand up to rather low, mechanically unstable density.\\
They have also considered, that the incident proton heats up either cylindrical or conic shaped region around its path in the target. They have found, that in 
case of cylindrical heated region, formed fragments are pushed, by Coulomb 
repulsion, more strongly in sideward directions. If a conic shaped region is 
heated up, fragments are pushed in rather backward directions.

\chapter{Specific models for fast stage of proton-nucleus collision}
\label{chapt:HSD}
\markboth{ }{Chapter 3. Specific models}

In the frame of this work, dynamical analysis of fast stage of proton - nucleus
 reactions are performed within transport approaches: 
Boltzmann-Uehling-Uhlenbeck (BUU) \cite{Niit89, Bert88} and Hadron 
String Dynamics (HSD) \cite{WCass, Geis98} models. The models have been 
specially developed in order to enable description of considered here 
reactions.\\

The BUU model is used to calculate proton - nucleus reactions in 
projectile kinetic energy range only up to about 2.5 GeV. 
That is because included in the 
model processes go mainly through single resonances excitations (i.e. $\Delta$,
 $N$(1440), $N$(1535)), what is correct in this energy range. While an incident
 energy increases, the density of produced resonances also increases. In this 
case, a possible proper description of processes requires taking into 
consideration hadron - hadron reactions on the level of elementary quark - 
quark interactions. This can be done by employment of a string model 
(e.g. FRITIOF model \cite{Ande93}), where during inelastic collision, two 
interacting hadrons are excited due to longitudinal energy-momentum transfer 
\cite{Ande93}.  
The formed excitation, so-called string, represents a prehadronic stage. It is 
characterized by the incoming quarks and a tubelike colour force field 
\cite{Perk00} spanned in between. The string is then allowed to decay into 
final state hadrons, with conservation of the four-momentum, according to e.g. 
Lund string fragmentation model \cite{Ande83}. \\
The HSD approach includes the FRITIOF scheme of string dynamics and the Lund 
model of hadron production through string fragmentation. It is employed, in 
particular, for incident energies higher than projectile energy 2.5 GeV. For 
projectile energies lower than about 2.5 GeV, the BUU code is used in the 
frame of the HSD model. \\
Below, a description of the approaches is given. \\ 

Both the models are based on transport equation. 

\section{Transport equation}
\markboth{3.1 Transport equation}{Chapter 3. Specific models}

Historically, the transport equation originate from classical Boltzmann 
equation for one-body phase-space distribution function $f(\vec{r},\vec{v},t)$ 
defined such that\\ $\int f(\vec{r},\vec{v},t)\, d^{3}\upsilon\, d^{3}r$ is the
 number of particles at time $t$ positioned in element volume $d^{3}r$ around 
$\vec{r}$, which have velocities in volume element of velocity space 
$d^{3}\upsilon$ around $\vec{v}$. \\ 
Let's consider particles, in which an external force $F$ with mass $m$ acts and
 assume initially that no collisions take place between the particles. 
In time $t+\delta t$ the velocity $\vec{v}$ of each particle will change to 
$\vec{v}+(\frac{\vec{F}}{m})\delta t$ and its position $\vec{r}$ will change 
to $\vec{r}+\vec{v}\delta t$. Thus the number of particles 
$f(\vec{r},\vec{v},t)\, d^{3}\upsilon\, d^{3}r$ is equal to the number of 
particles $f(\vec{r}+\vec{v}\delta t\,,\,\vec{v}+(\frac{\vec{F}}{m})\delta t\,,
\,t+\delta t) \,d^{3}r'd^{3}\upsilon'$,
what is explained by the Liouville theorem:
\begin{quote}
\emph{The volume of phase-space element is constant, if movement of all 
particles inside is consistent with canonical Hamilton equation of motion.}
\end{quote}
and written as: 
\begin{equation}
f(\vec{r}+\vec{v}\delta t\,,\,\vec{v}+(\frac{\vec{F}}{m})\delta t\,,\,t+\delta 
t) - f(\vec{r},\vec{v},t) = 0
\label{eq:f_nocoll}
\end{equation}

If collisions occur between the particles, an additional element, i.e. 
collision term is needed. This gives the following equation describing 
evolution of the distribution function:
\begin{equation}
f(\vec{r}+\vec{v}\delta t\,,\,\vec{v}+(\frac{\vec{F}}{m})\delta t\,,\,t+\delta 
t) - f(\vec{r},\vec{v},t) = (\partial f/\partial t)_{coll}\delta t  
\label{eq:f_coll}
\end{equation}

Letting $\delta t \rightarrow 0$ and expanding into the Taylor series gives the Boltzmann equation: 
\begin{equation}
((\partial/\partial t)+\nabla_{r}\cdot \vec{v}+(\frac{\vec{F}}{m})\nabla_{V})
f(\vec{r},\vec{v},t) = (\partial f/\partial t)_{coll}
\label{eq:Boltz_0}
\end{equation}
An apparent form of the collision term $(\partial f/\partial t)_{coll}$ can be 
found considering an element volume {\sl A} at time $t$, around position 
($\vec{r},\vec{v}$) and an element volume {\sl B} at time $t+\delta t$, around 
position ($\vec{r}+\vec{v}\delta t,\vec{v}+(\frac{\vec{F}}{m})\delta t$). 
These two element volumes are so similar, that letting $\delta t \rightarrow 0$,
 particles knocked out from {\sl A}, due to collisions, will not get into 
{\sl B}. Particles being outside {\sl A}, during time $\delta t$, will get into
 {\sl A}, and they will be inside {\sl B}. So, the number of particles inside 
{\sl B}, at time $t+\delta t$, at $\delta t \rightarrow 0$, is equal to the 
initial number of particles inside {\sl A}, at time $t$, and a magnitude of 
relative modification of number of particles due to collisions, during time 
$\delta t$.        
Therefore, a form of the collision term $(\partial f/\partial t)_{coll}$ can be 
calculated as a difference between the number of collision in a time range 
($t,t+\delta t$), when one of particles {\sl after} collision is situated in 
element volume $d^{3}r d^{3}\upsilon$ around position ($\vec{r},\vec{v}$), and 
the number of collision in a time range ($t,t+\delta t$), when one of particles
 {\sl before} collision is situated in the same element volume $d^{3}r 
d^{3}\upsilon$ around position ($\vec{r},\vec{v}$). 
It can be done by assuming that the density of particles is low enough, that 
only binary collisions need be considered. It is also assumed that the velocity
 of particle is uncorrelated with its position in the space. It means that in 
element volume $d^{3}r$ the number of particles pairs with velocities in volume
 elements of velocity space $d^{3}\upsilon_{1}$ around $\upsilon_{1}$ and 
$d^{3}\upsilon_{2}$ around $\upsilon_{2}$ is equal to: 
$[f(\vec{r},\vec{v_{1}},t)\, d^{3}r\, d^{3}\upsilon_{1}][f(\vec{r},\vec{v_{2}},
t)\, d^{3}r \,d^{3}\upsilon_{2}]$. \\

The number of binary collisions ($\vec{v_{1}}$, $\vec{v_{2}}$ $\rightarrow$ 
$\vec{v_{3}}$, $\vec{v_{4}}$) inside element $d^{3}r$, in time range 
$(t\,,\, t+\delta t)$ is equal to: $[f(\vec{r},\vec{v_{2}},t) 
d^{3}\upsilon_{2}] |\vec{v_{1}} - \vec{v_{2}}|\sigma(\Omega)d\Omega\delta t$,\\
 where: \\
$\vec{v_{1}}$ and $\vec{v_{2}}$ are the velocities of the two particles before 
collision,\\ 
$\vec{v_{3}}$ and $\vec{v_{4}}$ are their velocities after the collision, \\ 
$\sigma(\Omega)$ is the differential cross section for a reaction, in the 
centre of mass reference frame, \\
$\Omega$ is the solid angle the particles are scattered into (the angle between
 vectors $\vec{v_{1}}-\vec{v_{2}}$ i $\vec{v_{3}}-\vec{v_{4}}$),\\ 
$|\vec{v_{1}} - \vec{v_{2}}|$ is the magnitude of the particles relative 
velocity before the collision,\\
$[f(\vec{r},\vec{v_{2}},t) d^{3}\upsilon_{2}] |\vec{v_{1}} - \vec{v_{2}}|$ is 
the density of particles flux equal to the product of particles density and 
their velocity. \\
The total number of collisions, where one of the particle {\sl before} 
collision is situated inside element $d^{3}r d^{3}\upsilon_{1}$ around 
($\vec{r}, \vec{v_{1}}$) is obtained multiplying the number of binary 
collisions by number of particles with velocity $\vec{v_{1}}$, inside element 
$d^{3}r$ and integrating over all possible $\vec{v_{2}}$ and $\Omega$: \\
\begin{equation}
\int d^{3}\upsilon_{2}\int d\Omega \sigma(\Omega)f(\vec{r},\vec{v_{2}},t)|\vec{
v_{1}} - \vec{v_{2}}|[f(\vec{r},\vec{v_{1}},t)d^{3}r\,d^{3}\upsilon_{1}]\delta t
\label{eq:numcoll_bef}
\end{equation}
Taking into consideration the inverse binary collision: ($\vec{v_{3}}$, 
$\vec{v_{4}}$ $\rightarrow$ $\vec{v_{1}}$, $\vec{v_{2}}$), and using analogical
 method as above, the total number of collisions, where one of the particle 
{\sl after} collision is situated inside element $d^{3}r d^{3}\upsilon_{1}$ 
around ($\vec{r}, \vec{v_{1}}$) is obtain: 
\begin{equation}
\int d^{3}\upsilon_{4}\int d\Omega \sigma'(\Omega)f(\vec{r},\vec{v_{4}},t)|\vec{
v_{3}} - \vec{v_{4}}|[f(\vec{r},\vec{v_{3}},t)d^{3}r\,d^{3}\upsilon_{3}]\delta t
\label{eq:numcoll_af}
\end{equation}
Because collisions $\vec{v_{1}}$, $\vec{v_{2}}$ $\rightarrow$ $\vec{v_{3}}$, 
$\vec{v_{4}}$ and $\vec{v_{3}}$, $\vec{v_{4}}$ $\rightarrow$ $\vec{v_{1}}$, 
$\vec{v_{2}}$ are inverse collisions, so: $\sigma(\Omega) = \sigma'(\Omega)$. \\
From covservarion law for energy and momentum: $|\vec{v_{1}} - \vec{v_{2}}| = 
|\vec{v_{3}} - \vec{v_{4}}|$. \\
From Liouville theorem: $d^{3}\upsilon_{1}d^{3}\upsilon_{2} = d^{3}\upsilon_{3}
d^{3}\upsilon_{4}$. 
Subtracting equations (\ref{eq:numcoll_af}) and (\ref{eq:numcoll_bef}), and 
using above assumptions the collision term can be written as:
\begin{equation}
(\partial f/\partial t)_{coll}=\int d^{3}\upsilon_{2} d^{3}\upsilon_{3} \sigma(
\Omega) d\Omega |\vec{v_{1}} - \vec{v_{2}}|(f_{3}f_{4} - f_{1}f_{2})
\label{eq:coll_term}
\end{equation}
where: 
$f_{1}\;\equiv\;f(\vec{r},\vec{v_{1}},t)$, $f_{2}\;\equiv\;f(\vec{r},\vec{v_{2}}
,t)$,
 $f_{3}\;\equiv\;f(\vec{r},\vec{v_{3}},t)$, $f_{4}\;\equiv\;f(\vec{r},\vec{v_{4}
},t)$.\\

Joining equations (\ref{eq:Boltz_0}) and (\ref{eq:coll_term}) one obtains the 
classical Boltzmann equation:
\begin{eqnarray}
\{\frac{\partial}{\partial t} \: + \: (\frac{\vec{p}_{1}}{m_{1}} \: + \:
 \frac{\partial U(\vec{r},\vec{p}_{1},t)}{\partial \vec{p}_{1}}) \: \frac{
\partial}{\partial \vec{r}} \: - \: \frac{\partial U(\vec{r},\vec{p}_{1},t)}{
\partial \vec{r}} \,
 \frac{\partial}{\partial \vec{p}_{1}}\} f(\vec{r},\vec{p}_{1},t) =
\nonumber \\
 \frac{4}{(2 \pi)^{3}} \: \int \, d^{3} {p}_{2} \, d^{3} {p}_{3} \, d\Omega \; 
\sigma(\Omega) |\vec{v_{1}} - \vec{v_{2}}| \; \delta^{3}(\vec{p}_{1}+\vec{p}_{2}
-\vec{p}_{3}-\vec{p}_{4}) \cdot [f_{3}f_{4} - f_{1}f_{2}]
\label{eq:Boltz_tot}
\end{eqnarray}
where: \\ 
$\vec{v} = \vec{p}/m$,\\ 
$\vec{F} = - (\partial U(\vec{r})/\partial \vec{r})$, $U(\vec{r})$ is position 
dependent potential. \\

In 1933 Uehling and Uhlenbeck have developed the equation by adding the Pauli 
factors \cite{Uehl33}.\\
Due to Pauli blocking, a collision can occur only if in the final state, there 
are free quantum states. \\Probability of finding the free quantum state in a 
phase-space volume is equal to: $[1 - f(\vec{r},\vec{p},t)]$, what is 
responsible for fermion Pauli blocking.\\ The probability of two particles 
collision with momenta $\vec{p}_{1}$ and $\vec{p}_{2}$ is equal to: 
\begin{equation}
[1 - f(\vec{r},\vec{p}_{3},t)][1 - f(\vec{r},\vec{p}_{4},t)] \equiv \overline{f}
_{3} \overline{f}_{4} 
\label{eq:Pfac1}
\end{equation}

\noindent Probability of inverse occurrence is equal to:
\begin{equation}
[1 - f(\vec{r},\vec{p}_{1},t)][1 - f(\vec{r},\vec{p}_{2},t)] \equiv 
\overline{f}_{1} \overline{f}_{2}
\label{eq:Pfac2}
\end{equation}
Expression (\ref{eq:Pfac1}) and (\ref{eq:Pfac2}) are Pauli factors.\\

Including in (\ref{eq:Boltz_tot}) the Pauli factors gives:
\begin{eqnarray}
\{\frac{\partial}{\partial t} \: + \: (\frac{\vec{p}_{1}}{m_{1}} \: + \:
 \frac{\partial U(\vec{r},\vec{p}_{1},t)}{\partial \vec{p}_{1}}) \: \frac{
\partial}{\partial \vec{r}} \: - \: \frac{\partial U(\vec{r},\vec{p}_{1},t)}{
\partial \vec{r}} \,
 \frac{\partial}{\partial \vec{p}_{1}}\} f(\vec{r},\vec{p}_{1},t) = 
\nonumber \\
\frac{4}{(2 \pi)^{3}} \: \int \, d^{3} {p}_{2} \, d^{3} {p}_{3} \, d\Omega \; \sigma(\Omega) 
|\vec{v_{1}} - \vec{v_{2}}| \; \delta^{3}(\vec{p}_{1}+\vec{p}_{2}-\vec{p}_{3}-\vec{p}_{4}) \nonumber \\
\cdot [f_{3}f_{4}\overline{f}_{1}\overline{f}_{2} - f_{1}f_{2}\overline{f}_{3}\overline{f}_{4}]
\label{eq:buu}
\end{eqnarray}
The equation is named Boltzmann-Uehling-Uhlenbeck (BUU) equation.\\

\section{Boltzmann-Uehling-Uhlenbeck model}
\markboth{3.2 Boltzmann-Uehling-Uhlenbeck model}{Chapter 3. Specific models}

The theory based on the transport equation (\ref{eq:buu}) (it means the 
Boltzmann equation with a self-consistent potential field, and with a collision
 term that respects the Pauli principle) was used first time to nuclear 
collisions description by Bertsch, in 1984 \cite{Bert84}. \\
The BUU equation is solved numericaly, using Monte Carlo method, representing 
the one-body phase-space distribution by discretized test particles:
\begin{equation}
f(\vec{r},\vec{p},t) = \frac{1}{N} \sum_{i=1}^{N \cdot A(t)} \delta^{3}(\vec{r}
-\vec{r}_{i}(t)) \; \delta^{3}(\vec{p}-\vec{p}_{i}(t)) 
\end{equation}
where: \\
$N$ is a number of test particles, \\
$A(t)$ is a number of real particles at time $t$ \\

Likewise, one collision is replaced by parallel collisions.\\

All of the test particles give part to the density of nuclear matter not in 
single points, but they are smeared with Gauss distribution. This way, the 
effect of quantum smearing is included.\\ 
The density is calculated on the grid $r_{g}$:
\begin{equation}
\rho(\vec{r_{g}})=\frac{1}{N}\sum_{i=1}^{NA}\frac{1}{(2\pi\Delta^{2})^{3/2}}exp
(\frac{-(\vec{r_{g}}-\vec{r_{i}})^{2}}{2\Delta^{2}})
\end{equation}
where: $\Delta$ is the Gauss width parameter (taken usually equal 1).\\

The initial coordinates of particles of target nucleus have Wood-Saxon 
distribution form:
\begin{equation}
\rho(r) = \frac{\rho_{0}}{1+exp(\frac{r - R}{a})}
\label{eq:wood_saxon}
\end{equation}
where: \\
$a = 0.025 \cdot A^{1/3}+0.29$ $[fm]$ = 0.5 $[fm]$ \\
$R = 1.124 \cdot A^{1/3}$ $[fm]$  \\
$\rho_{0} = 0.168$ $[nukleon/fm^{3}]$. \\

Initially, the target nucleus is in the rest, the total momentum (i.e. the sum 
of momenta of all particles) is equal to 0, but the local momenta of particles 
are determined homogeneously on the Fermi sphere with radius $p_{F}(r)$:
\begin{equation}
p_{F}(r) = (\frac{3 \pi ^{2}}{2} \rho(r))^{1/3}
\end{equation}

The projectile is a single proton. The test particles replacing the proton are 
distributed homogeneously on a thin cylinder, with a radius equal the radius of
 target. Thanks to this approach, each test particle has different impact 
parameter, so the results of calculations are averaged over all impact 
parameters.

The solution of the transport equation is the single-particle phase-space 
distribution function, depending on time.\\
The collision is numericaly evolved by fixed time steps.\\
The test particles propagate between collisions according to the classical 
Hamilton equations of motion: \\
\begin{equation}
\dot{\vec{p}_{i}} = - \frac{\partial U(\vec{r}_{i},\vec{p}_{i},t)}{\partial 
\vec{r}_{i}} 
\end{equation}

\begin{equation}
\dot{\vec{r}_{i}} = \vec{p}_{i} / \sqrt{m^{2} + p^{2}} + \frac{\partial 
U(\vec{r}_{i},\vec{p}_{i},t)}{\partial \vec{p}_{i}}
\end{equation}
where: $U(\vec{r}_{i},\vec{p}_{i},t)$ is a mean field potential, dynamically 
changing, calculated as a function of local density:
\begin{equation}
U(\vec{r})=\frac{3}{4} t_{0} \rho(\vec{r})+\frac{7}{8} t_{3} \rho(\vec{r})^{4/3}+V_{0} \int d^{3}\vec{r'} \frac{exp(-\mu |\vec{r} - \vec{r'}|)}{\mu |\vec{r} - \vec{r'}|} \rho(\vec{r'}) + V_{Coul}
\end{equation} 
where: $t_{0}$=-1124 MeV$\cdot$fm$^{3}$, $t_{3}$=2037 MeV$\cdot$fm$^{4}$, 
$V_{0}$=-378 MeV, $\mu$=2.175 fm$^{-1}$, see Ref. \cite{Niit89}. \\

The momentum and coordinates of all particles taking part in reaction are 
calculated in the successive time steps.  
Using the values of momentum and coordinates, all other quantities (i.e. 
nucleon density, mean field potential) are calculated. \\

The nucleon - nucleon collision at a fixed time step is introduced as follows: 
when two nucleons come closer than the distance $b_{max} = \sqrt{\sigma ^{max} 
_{NN}/ \pi}$ (where $\sigma ^{max} _{NN}$ is the maximal cross section for 
nucleon - nucleon interaction in nuclear matter (30 mb \cite {Niit89})) they 
are made to scatter, but if the final state is Pauli blocked, this collision is
 canceled. \\

The BUU model describes the propagation and mutual interaction of nucleons, 
Delta's, $N^{*}$ - resonances, and also $\pi$ and $\eta$ - mesons. \\
In the model, the following reaction channels are included: \\
$NN$ $\leftrightarrow$ $NN$, \\
$NN$ $\leftrightarrow$ $NR$, \\
$NR$ $\leftrightarrow$ $NR$, \\  
$\Delta \Delta$ $\leftrightarrow$ $NR$, \\
$R$ $\leftrightarrow$ $N \pi$, \\
$N$(1535) $\leftrightarrow$ $N \eta$, \\
$NN$ $\leftrightarrow$ $NN \pi$, \\
$\pi N$ $\rightarrow$ $\pi N$, \\ 
$\pi N$ $\rightarrow$ $\pi \pi N$, \\
where: $N$ is a nucleon,  
$R$ stands for a resonance $\Delta$, $N$(1440) or $N$(1535).\\
Cross sections for the reactions, used in the model calculations, are 
parametrizations of the experimental cross sections taken from \cite{PDG94}.\\

All resonances are allowed to decay into two particles (besides the decays due 
to collisions with other particles, e.g. $NR \rightarrow NN$). 
The decay of a resonance is determined by its width $\Gamma(M)$. The decay 
probability $P$ is calculated in every step of time, according to the 
exponential decay law: $P=1-e^{-\Gamma(M)/(\hbar \gamma)\delta t}$, 
where: $\Gamma(M)$ is the energy dependent width of the resonance,  
$\gamma$ is a Lorentz factor related to the velocity of the resonance,  
$\delta t$ is a time step size of the calculations.\\ 
For the $\Delta$ decay the parametrization given by Koch et al. \cite{Koch84} 
is used. Details concerning higher resonances can be found in \cite{Teis97}.\\
In each step of time it is decided if the resonance may decay and to which 
final state it may go. If the final state is Pauli blocked, the resonance decay 
is rejected.     

\section{Hadron String Dynamics model} 
\markboth{3.3 Hadron String Dynamics model}{Chapter 3. Specific models}

The HSD model is based on the same transport equation (\ref{eq:buu}) as the 
BUU approach and solved also by use of the test particle method.\\ 
In the HSD model, propagation of the following real particles are included: 
baryons ($p$, $n$, $\Delta$, $N$(1440), $N$(1535), $\Lambda$, $\Sigma$, 
$\Sigma ^{*}$, $\Xi$, $\Omega$), the corresponding antibaryons and mesons 
($\pi$, $K$, $\eta$, $\eta'$, $\rho$, $\omega$, $\phi$, $K^{*}$, $a_{1}$).

\subsection{Cross sections}

The {\bf low-energy} {\bf baryon-baryon} and {\bf meson-baryon} collisions 
(i.e. with the invariant energies below "string threshold": $\sqrt{s} < 2.65$ 
GeV and $\sqrt{s} < 2.1$ GeV \cite{Geis98}, respectively) are described using 
the explicit cross section, as in the BUU code. 
The following parametrization of the experimental total and elastic $p+p$, 
$p+n$, $\pi^{+}p$, $\pi^{-}p$, $K^{+}p$, $K^{-}p$ cross sections, taken from 
\cite{PDG94}, is used: 
\begin{equation}
\sigma (p) = A + B \cdot p^{n} + C \cdot ln^{2}(p) + D \cdot ln(p) [mb],  
\label{eq:sig_buu}
\end{equation}
where: $p$ is a momentum of incident proton in laboratory frame, \\ 
$A$, $B$, $n$, $C$, $D$ are constants, with adequate values for different 
processes cross sections, see Ref. \cite{GeisPhD}.  \\
For reaction channels: $\pi N$ $\rightarrow$ $\pi N$ and $\pi N$ $\rightarrow$ 
$\pi \pi N$, to have them consistent with the experimental inelastic 
pion-proton cross section below the string threshold, instead of 
(\ref{eq:sig_buu}), the following parametrization is used \cite{GeisPhD}:
\begin{eqnarray}
\sigma_{tot} (p_{\pi}) = 25.0 + 16.0 \cdot p_{\pi}^{-1.4} [mb], \nonumber \\
\sigma_{el} (p_{\pi}) = 3.5 + 14.0 \cdot p_{\pi}^{-1.1} [mb], 
\label{eq:sig_pi}
\end{eqnarray}
where: $p_{\pi}$ is a momentum of incident pion (in GeV/c).\\
Additionally, the channels: \\
$\rho N$ $\rightarrow$ $N \pi \pi$, $\rho \Delta$ $\rightarrow$ $N \pi \pi$ and \\
$\omega N$ $\rightarrow$ $N \pi \pi \pi$, $\omega \Delta$ $\rightarrow$ $N \pi \pi \pi$, \\
are included with an energy independent cross section equals to 30 mb. \\
Because of very low number of produced hyperons in low-energy proton-nucleus 
collisions, the hyperon($\Lambda$, $\Sigma$) - nucleon interactions are 
neglected in the model. \\
   
Angular distribution of elastic collisions depends on energy \cite{GeisPhD}, 
therefore the following parametrization of the {\bf differential elastic 
nucleon - nucleon cross section}, taken from Cugnon et al. \cite{Cugn81}, 
is applied: 
\begin{equation}
\frac{d \sigma}{d \Omega} = e^{A(s)t}, 
\label{eq:sig_Cug}
\end{equation}
where: 
$s$ is the invariant energy of collision squared (in GeV),\\ 
$t$ is the four-momentum transfer squared,\\
$A(s) = 6 \frac{[3.65 (\sqrt{s} - 1.8776)]^{6}}{1+ [3.65 (\sqrt{s} - 1.8776)]
^{6}}$. \\

In the approach, the {\bf high-energy elastic} and {\bf total} {\bf baryon - 
baryon} and {\bf meson - baryon} collisions (i.e. with energies above the 
"string threshold"), are related to the measured cross sections by:\\
$\sigma_{el} ^{N \Delta} (\sqrt{s}) = 0.5 (\sigma_{el} ^{pp} (\sqrt{s}) +
\sigma_{el} ^{pn} (\sqrt{s}))$, \\
$\sigma_{el} ^{\rho N} (\sqrt{s}) = \sigma_{el} ^{\rho \Delta} (\sqrt{s}) = 
\sigma_{el} ^{\omega \Delta} (\sqrt{s}) = ... = 0.5 (\sigma_{el} ^{\pi ^{+}N} 
(\sqrt{s}) + \sigma_{el} ^{\pi ^{-}N} (\sqrt{s}))$, \\
$\sigma_{el} ^{KN} (\sqrt{s}) = \sigma_{el} ^{K \Delta} (\sqrt{s}) = ... = 
\sigma_{el} ^{K^{+} p} (\sqrt{s})$, \\
$\sigma_{el} ^{\overline{K} N} (\sqrt{s}) = \sigma_{el} ^{\overline{K}\Delta} 
(\sqrt{s}) = ... = \sigma_{el} ^{K^{-} p} (\sqrt{s})$, \\
where dots stand for other combination in the incoming channel, with 
$K = (K^{+}, K^{0}, K^{*+}, K^{*0})$ and $\overline{K} = (K^{-}, 
\overline{K}^{0}, K^{*-}, \overline{K}^{*0})$ . The same relations are applied 
for the total cross sections. \\
The {\bf high-energy inelastic} {\bf baryon - baryon} and {\bf meson - baryon} 
cross sections obtained by this procedure are equal to 30 mb and 20 mb, 
respectively. This corresponds to the typical geometrical cross section 
(i.e. $\sigma_{inel} \approx \pi R^{2}$), so it should be reasonable input for 
the calculations. \\
Due to this fact, in order to include all baryon - baryon and meson - baryon 
high-energy inelastic cross sections, only final state rates must be specified. 
Therefore, the string model (i.e. FRITIOF model) is employed, which will be 
described hereafter in this work.\\

In the HSD model, the following {\bf meson - meson} reaction channels are 
included:\\
$\pi \pi$ $\leftrightarrow$ $\rho$,\\
$\pi \rho$ $\leftrightarrow$ $\phi$,\\
$\pi \rho$ $\leftrightarrow$ $a_{1}$, \\
$K \overline{K}$ $\leftrightarrow$ $\phi$,\\
$\pi K$ $\leftrightarrow$ $K^{*}$.\\
They are more probable at high-energy proton - nucleus collisions, but as 
they appear as secondary or higher order reactions, the average energy for such
 processes is rather low. In this case, the cross section within the 
Breit-Wigner parametrization is employed. Therefore, the reactions $a+b$ 
$\rightarrow$ $m_{R}$ $\rightarrow$ $c+d$, where: $a$, $b$, $c$, $d$ are the 
mesons in the initial and final state, respectively, and $m_{R}$ denotes the 
intermediate mesonic resonance ($ \rho$, $a_{1}$, $\phi$, $K^{*}$), are 
described by: 
\begin{equation}
\sigma(ab \rightarrow cd) = \frac{2J_{R}+1}{(2S_{a}+1)(2S_{b}+1)} \frac{4 \pi}
{p_{i}^{2}} \frac{s \Gamma_{R \rightarrow ab} \Gamma_{R \rightarrow cd}}
{(s-M_{R}^{2})^{2} + s\Gamma_{tot}^{2}}, 
\label{eq:sig_BW}
\end{equation}
where: \\ 
$S_{a}$ and $S_{b}$ are spins of the particles, 
$J_{R}$ is spin of resonance;\\
$\Gamma_{R \rightarrow ab}$ and $\Gamma_{R \rightarrow cd}$ are partial decay 
widths in the initial and final channels, \\
$M_{R}$ is the mass of the resonance, 
$\Gamma_{tot}$ is the total resonance width, \\
$p_{i}$ is the initial momentum in the resonance rest frame.\\   
The decay widths and the branching ratios for the mesonic channels are adopted 
from the nuclear data tables \cite{PDG94}, without introducing new parameters. 
\\ Additionally, strangeness production in meson - meson collisions is 
included, with an isospin averaged cross section \cite{Cass97}:
\begin{equation}
\sigma_{mm \rightarrow K \overline{K}}(s) = 2.7 (1-\frac{s_{0}}{s})^{0.76} [mb],
\label{eq:sig_strang}
\end{equation}
where: $s_{0} = 4 m_{K}^{2}$, $mm$ stands for all possible non-strange mesons 
in the incoming channel (e.g. $\pi \pi \rightarrow K \overline{K}$, 
$\pi \rho\rightarrow K \overline{K}$). \\

\subsection{String Model}

Quarks, as colour-charged particles cannot be found individually. They are 
confined in two- or three-quarks systems, i.e. colour neutral hadrons. 
The quarks, in a given hadron, exchange gluons. If one of the quarks is pulled 
away from the other quarks in the hadron, the colour force field, which 
consists of gluons holding the quarks together, stretches between this quark 
and its neighbers. 
Because of interaction between gluons \cite{Perk00}, the colour field lines 
are not spread out over all space, as the electromagnetic field lines do, but 
they are constrained to a thin tube-like region. While the quarks are pulled 
apart, more and more energy is added to the colour force field. So, in such a 
formation, the four-momentum can be accumulated. At some point, it is 
energetically possible for the field to break into new quarks. 
The four-momentum is conserved, because the energy of the colour force field is
 converted into the mass of the new quarks. Finally, the colour force field 
comes back to an unstretched state. \\
With this picture in mind, high-energy hadron - hadron interactions models, 
called string models \cite{Geis98} have been created, where the formation 
composed of quarks and the colour field between is called a string. \\     
\indent In the HSD approach, in order to describe the high-energy inelastic 
hadron - hadron collisions, FRITIOF model is applied \cite{Ande93}.
In the model, hadronic collision corresponds to large longitudinal and small 
transversal energy - momentum transfer. It means, to the stretching of 
longitudinally extended string-like colour force field along the beam 
direction, between constituent quarks of the incoming hadron. The created 
excitation, i.e. string is a dynamical object, which may decay into final state
 hadrons, according to the Lund fragmentation scheme \cite{Ande83}, 
implemented in the FRITIOF model.
The field between quarks is confined into a tube, called "flux tube", which is 
one-dimensional object. The uniform colour field contains constant amount of 
energy stored per unit length. The total energy ($E$) of the field is 
proportional to the length ($L$): $E = \kappa \cdot L$, where $\kappa \cong 1$ 
GeV/fm, is a string tension \cite{Ande83}. Due to longitudinal energy-momentum 
transfer, spread over some region, the colour separation occurs, i.e. as seen 
in CM string frame, there will be two extended parts of the string moving 
forward and backward, along the beam direction. The potential between quarks is
 linearly rising.
As result, the system breaks. Because of colour confinement there is never a 
single quark in isolation. After the string breaks, on its ends new quarks 
appear. The new $q \overline{q}$ pair is created from the available field 
energy. By that means, the energy of the initial string decrease (a part of the
 energy is used for $q \overline{q}$ pair production), but the breaking process
 is not finished yet. The quarks of the new created strings are also moving in 
opposite directions in the strings rest frames. Thereby the original system 
breaks into smaller and smaller pieces, until only physical hadrons remain, 
(i.e. baryons as bound systems of three quarks, antibaryons - three antiquarks 
and mesons as quark - antiquark systems).\\    
In the HSD approach, baryonic ($qq - q$) and mesonic ($q - \overline{q}$) 
strings are considered. \\
The implementation of the string model into the transport approach implies 
introduction of a time scale for the particle production processes. The time 
scale is given by a formation time $t_{f}$, which includes formation of a 
string, fragmentation of the string into small substrings due to 
$q \overline{q}$ and $q q \overline{q} \overline{q}$ production, and formation 
of physical hadrons.
The formation time should be also related to the spatial extension of the 
interacting hadrons. It means, it should be big enough so that quark - 
antiquark pair could reach a distance corresponding to a typical hadron radius,
 i.e. 0.6 - 0.8 fm \cite{GeisPhD}. 
In the HSD model, the formation time is a single fixed parameter for all 
hadrons, it is set to $t_{f}=0.8$ fm/c in the rest frame of the new produced 
particles \cite{Eheh96}. \\ The particles production proceeds as follows. \\ 
If a system contains originally e.g. $\overline{q}_{0}$ and $q_{0}$, moving in 
opposite directions with large energies, it breaks after some time into two 
parts by a production of a pair $q_{1}$ $\overline{q}_{1}$, at a space-time 
point ($x_{1}$, $t_{1}$). The new produced quarks also move in opposite 
directions. Two subsystems are created by $\overline{q}_{0}$ $q_{1}$ and 
$\overline{q}_{1}$ $q_{0}$. As a result, a colour force field between the new 
pair $q_{1}$ $\overline{q}_{1}$ vanishes. At a later time another pair 
$q_{2}$ $\overline{q}_{2}$ can be produced at ($x_{2}$, $t_{2}$), due to 
breaking e.g. $\overline{q}_{1}$ $q_{0}$ subsystem. Analogically, new 
subsystems $\overline{q}_{1}$ $q_{2}$ and $\overline{q}_{2}$ $q_{0}$ are 
created. A colour force field between the new quark pair $q_{2}$ 
$\overline{q}_{2}$ vanishes. Created subsystems either are hadrons or else 
will fragment further, until only hadrons remain. In the model, string 
fragmentation starts always in its center (in the CM frame of the string). 
All quark pairs production points are separated in a space-time.
A total momentum in the rest frame of a string is equal to zero.  
The energy and momentum in the production process is conserved. 
Masses of the finally produced hadrons are equal to the masses of physical 
hadrons. \\
In the HSD model, the production probability ($P$) of massive 
$s \overline{s}$ or $q q \overline{q} \overline{q}$ pairs is supressed in 
comparison to light quarks pairs production ($u \overline{u}$, 
$d \overline{d}$). Inserting the following constituent quark masses: 
$m_{u} = m_{d} = 0.3$ GeV and $m_{s} = 0.5$ GeV, one gets: 
$\frac{P(s \overline{s})}{P(u \overline{u})} = 0.3$ and 
$\frac{P(q q \overline{q} \overline{q})}{P(u \overline{u})} = 0.07$.\\
So, the suppression factors used in the model are: \\
$u$ : $d$ : $s$ : $uu$ = 1 : 1 : 0.3 : 0.07, see Ref. \cite{GeisPhD}.  \\
As a result, mainly mesons are produced. 

The model assumes that there is no final state interaction of the produced 
hadrons included in the model.\\
Because most of the strings, in a given space-time volume, fragment within a 
small time interval, the interaction of the string field spanned between the 
constituent quarks with other hadrons is not taken into account. But the 
secondary interactions of the quarks or diquarks inside the strings are 
considered in the approach. The following cross sections of such interactions 
are used \cite{GeisPhD}: \\
$\sigma (q - B) = \frac{1}{3} \sigma (B - B) \approx 10$ mb, \\
$\sigma (qq - B) = \frac{2}{3} \sigma (B - B) \approx 20$ mb, \\
$\sigma (qq - q) = \frac{2}{9} \sigma (B - B) \approx 6.6$ mb. \\ 
Because most of strings are stretched longitudinally, parallelly to each other,
 and radius of string is small, equal to about 0.2 - 0.3 fm \cite{GGBCM98}, 
there is no string - string interaction included in the HSD model.\\

The characterized above version of the HSD code has been used and developed
in order to describe proton induced spallation reactions in wide energy range,
and mass of target nuclei in the following aspects. First of all, to calculate 
properties of residual nuclei remaining after first stage of the reaction, 
forming an input for models of the second stage calculations.\\ 
Additionally, the code has been developed for description
of pions produced in proton - nucleus spallation reactions.
Moreover, a version of HSD code that allows for calculations of pion induced 
reactions has been prepared.\\
Each of the directions of the HSD code development will be presented and 
discussed in the next sections of this work.

\section{Further development of the model} 
\markboth{3.4 Further development of the model}{Chapter 3. Specific models}

As a result of the first stage of proton - nucleus reaction, apart from  
emitted particles (mainly nucleons and pions), a hot excited nucleus remained. 
Properties of the residual nucleus, i.e. mass ($A_{R}$), charge ($Z_{R}$), 
excitation energy ($E^{*}_{R}$), three - momentum ($\vec{p_{R}}$) and angular
 momentum ($L_{R}$) form an input for the second stage models. 
In order to calculate the properties, the original version of the HSD code has 
been modified.
Corrections ensuring energy and charge conservations have been made, 
which are important here, but played minor role in the original version. Also, 
corrections enabling calculations of various quantities in function of time, 
for very large times of propagation have been found.\\  
The properties are evaluated in the following way. \\
First, using that in the models calculations, four-momenta of all hadrons
 are propagated in time, the particles that have left the residual heavy 
fragment are identified. This can be done in two ways. 
In case of low-energy collisions, a sphere of observation, with radius equal 
to $R_{A}+2$ fm, where: $R_{A}$ denotes the radius of the target with mass 
number $A$, can be considered. All the particles inside the sphere are 
treated as belonging to residual nucleus, the particles outside - as emitted. 
In case of high-energy collisions, when target nucleus is moving faster 
during reaction, it is easier to consider a baryon density criterion. It 
means, particles, which are positioned in baryon density lower than 0.02 
nucleon/fm$^{3}$ 
are treated as emitted, the rest of the particles form a residual 
nucleus. At intermediate proton impact energy (0.1 - 2.0 GeV) both methods are 
equivalent. In the frame of this work, the second criterion is used. 
Nevertheless, both of the methods of classification are 
connected with some inaccuracy, concerning the very low energetical particles. 
Distributions of particles escaped from 
the sphere of observation characterize absence of very low energetical part. 
In contrary, distributions of particles classified on the base of the baryon 
density criterion characterize overabundance in the very low energetical part.
In this second case, it is because the density condition classifies incorrectly 
nucleons placed in the lowest density level of a residual nucleus, as emitted. 
Therefore, the density method needs to be completed by an additional condition
concerning the particles kinetic energy.
Particles in the nucleon density equal to 0.02 nucleon/fm$^{3}$ 
acquire momentum with a value from the range from zero to the local Fermi 
momentum, equal to about 130 MeV/c, what corresponds to about 10 MeV of kinetic 
energy. Compliance of that information gives a complete and correct condition, 
i.e. particles in nuclear density lower than 0.02 nucleon/fm$^{3}$ and with 
kinetic energy higher than 10 MeV are considered as emitted.\\  
Then, by exploring the conservation of total energy, mass number, momentum 
and angular momentum, for each parallel ensemble (see description of models),
 the average values of properties of the residual nuclei are calculated as 
function of time, according to formulas:         
\begin{eqnarray}
<E^{*}_{R}>(t)=E_{tot}-\sum^{N_{p}(t)}_{i=1}\sqrt{p_{i}^{2}+m_{i}^{2}}-M_{R}-E_{C}, 
\nonumber \\
<A_{R}>(t)=A_{T}+A_{P}-N_{p}(t), \nonumber \\
<\vec{p_{R}}>(t)=\vec{p_{tot}}-\sum^{N_{p}(t)}_{i=1} \vec{p_{i}}(t), \nonumber \\
<L>(t)=L_{tot}-\sum^{N_{p}(t)}_{i=1} \vec{r_{i}}(t) \times \vec{p_{i}}(t),
\label{eq:res_prop}
\end{eqnarray}
where: $N_{p}(t)$ denotes the number of emitted particles, 
$M_{R}$ is the mass of the residual nucleus, $A_{T}$ is a mass of original 
target, $A_{P}=1$ stands for incoming proton and $E_{C}$ is the the energy of
 Coulomb interaction between the emitted particles and the residual nucleus.

\section{Stopping time criteria for the first stage model calculations}
\markboth{3.5 Stopping time criteria}{Chapter 3. Specific models}

\begin{figure}[!ht]
\begin{center}
\includegraphics[height=6cm, width=8cm, angle=0]{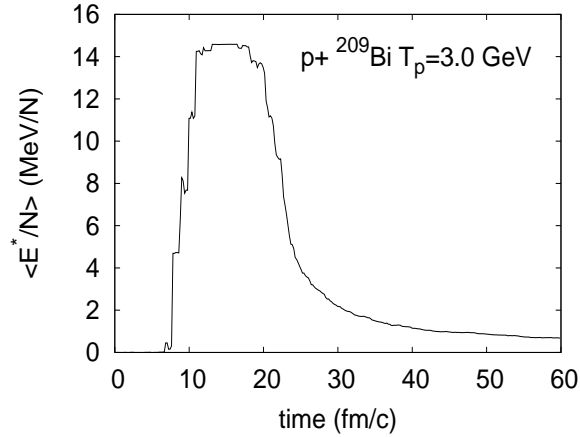}
\caption{{\sl Time variation of the average value of excitation energy per 
nucleon of the residual nucleus in p + Bi collision, at 3 GeV proton beam 
energy; results of the HSD model calculations}}
\label{fig:time_enexc}
\end{center}
\end{figure}

\begin{figure}[!ht]
\begin{center}
\includegraphics[height=6cm, width=8cm, angle=0]{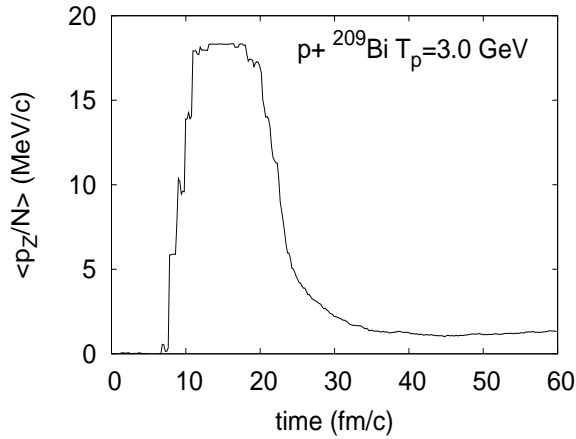}
\caption{{\sl Time variation of the average value of longitudinal momentum per 
nucleon of the residual nucleus in p + Bi collision, at 3 GeV proton beam 
energy; results of the HSD model calculations}}
\label{fig:time_pz}
\end{center}
\end{figure}

\begin{figure}[!ht]
\begin{center}
\includegraphics[height=6cm, width=8cm, angle=0]{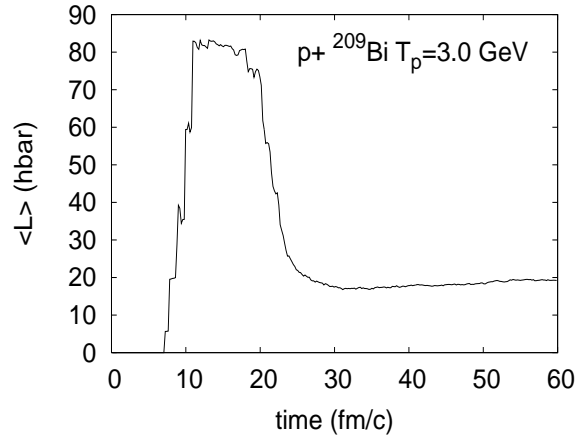}
\caption{{\sl Time variation of the average value of angular momentum of the 
residual nucleus in p + Bi collision, at 3 GeV proton beam energy;  
results of the HSD model calculations}}
\label{fig:time_angm}
\end{center}
\end{figure}

\begin{figure}[!ht]
\begin{center}
\includegraphics[height=6cm, width=8cm, angle=0]{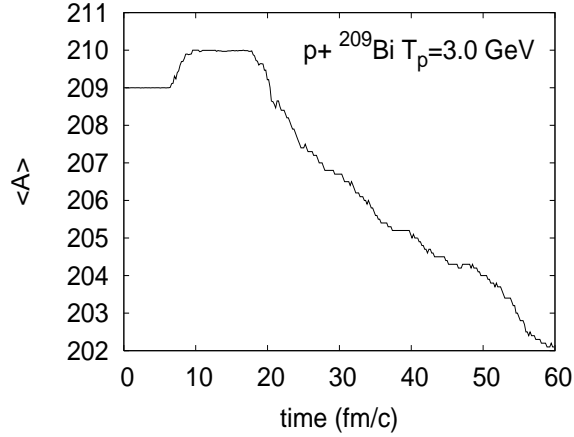}
\caption{{\sl Time variation of the average value of mass number of the 
residual nucleus in p + Bi collision, at 3 GeV proton beam energy; 
results of the HSD model calculations}}
\label{fig:time_mass}
\end{center}
\end{figure}

In order to describe both stages of the reaction, model of the first stage must 
be used together with a statistical model for the second stage of reaction. 
The outcome of the transport model calculations determines an input for 
the second stage model. Thus, it is important 
to assume a proper duration time of the first stage calculations.
In order to define the time at which the first stage calculations should be 
stopped, it must be verified, whether informations obtained from the 
transport models are sensitive to the time duration of the first stage of the 
reaction. For this purpose, time variation of the average values of four 
physical quantities: excitation energy per nucleon, momentum in beam direction 
per nucleon, angular momentum and mass 
number of the excited residual nucleus after the first stage, have been 
analysed. As example, the dependences for p + Bi collision, at 3.0 GeV 
proton beam energy are discussed below. 
The average values of presented quantities 
are evaluated according to the equations (\ref{eq:res_prop}). \\
First, let's look at the time evolution of the average value of excitation 
energy per nucleon of the residual nucleus shown in Fig. \ref{fig:time_enexc}.  
It is evident from the Figure, that it takes some time before proton 
approaches nucleus, i.e. about 10 fm/c (at the time {\sl zero}, projectile is 
placed in some distance from the target nucleus, ensuring that density 
distributions of projectile and target are not overlapping). 
Next, up to about 18 fm/c, the 
excitation energy gains some maximal value, which corresponds to the energy 
introduced into target nucleus by incoming proton. Then, the excitation energy 
drops very quickly in the range 20 - 30 fm/c of the duration time of the first 
stage of the reaction, whereas it is varying only a little at larger times. It 
is seen that the average value of excitation energy per nucleon starts to 
stabilize at time 35 - 40 fm/c. \\ 
Looking at Fig. \ref{fig:time_pz}, it is seen, that the average value of the 
momentum of the nucleus in the beam direction behaves very similar 
during first stage of the reaction as the average value of excitation energy.
Starting from about 10 fm/c, when proton reaches target nucleus, up to about 18 
fm/c, the longitudinal momentum per nucleon of the residual nucleus has a 
maximal value, adequate to the value of momentum introduced into target nucleus
 by incoming proton. Then, the average value drops quickly in the range 20 - 30
 fm/c of the duration time of the first stage of the reaction, because the 
momentum is carried out by nucleons escaping from the nucleus. At larger times 
the value varies only a little.  \\ 
Very similar behavior is observed for the average value of angular momentum of 
the residual nucleus during first stage of the reaction, presented in Fig. 
\ref{fig:time_angm}. 
At about 10 fm/c, incoming proton, having in 
average a non zero impact parameter, introduces some angular momentum into 
the nucleus. At around 18 fm/c and later the angular momentum lowers 
significantly due to nucleons escaping from the nucleus. This leakage of 
angular momentum stops at around 30 fm/c and the average value stabilizes. 
At later time, i.e. from about 40 fm/c, spurious slow increase of angular 
momentum is observed, which is unphysical and results from building-up 
inaccuracies of numerical calculations. That also indicates that 
the first stage calculations should be terminated at about 35 fm/c. \\
Looking at the time evolution of the average mass number of the residual 
nucleus, presented in Fig. \ref{fig:time_mass}, quite different behavior is 
observed.
It is seen, that at about 10 fm/c the projectile come into the target nucleus. 
As expected, the mass number is increased by one. But then, starting from 
about 18 fm/c, the mass number of the residual nucleus decreases monotonically 
with duration time of the first stage of reaction and does not stabilize at 
larger times, as it was in case of the other quantities. 
This indicates, that emission of particles in this model takes place all the 
time. Duration time of the cascade of the nucleon - nucleon collisions 
cannot be determined on the basis of behavior of the average mass number of 
the nucleus. Nevertheless, it accents the importance of assumption of a proper 
stopping time for the calculations, which has influence on a value of 
multiplicity of emitted particles. It has a negligible meaning for 
energy distributions of ejectiles, since, as a time evolution of excitation 
energy shows, from about 18 fm/c, particles with lower and lower energies are 
emitted.\\   
The behavior of excitation energy, longitudinal momentum and angular 
momentum as a function of time shows that choice of the stopping time is 
limited from two sides. The first stage calculations cannot be terminated too 
early, when interacting system is highly nonequilibrated. 
From the other side, also not too late, since for longer 
times numerical inaccuracies are increasing. 
Based on above dependences, it is concluded that the most reasonable duration 
time of transport models calculations of the first stage of proton induced 
reactions is equal to 35 fm/c. \\

\begin{figure}[!ht]
\vspace{-2cm}
\hspace{-2cm}
\begin{center}
\includegraphics[height=12cm, width=14cm, bbllx=0pt, bblly=245pt, bburx=594pt, bbury=842pt, clip=, angle=0]{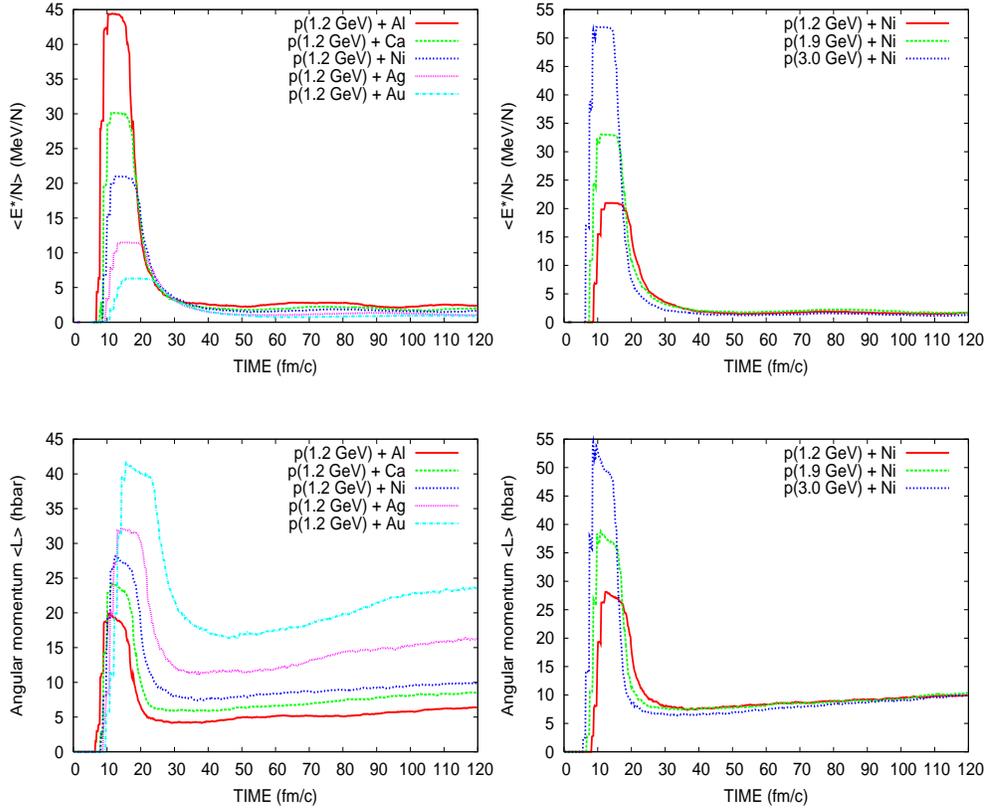}
\caption{{\sl Time variation of the average values of excitation energy per
nucleon and angular momentum of residual nuclei in proton induced reactions
on various targets, at different projectile energies; results of the HSD model 
calculations}}
\label{fig:fort_E_L_At_Tp}
\end{center}
\end{figure}

Behavior of the time dependences in case of reactions on other targets nuclei 
and other values of incident energy is displayed in Fig. 
\ref{fig:fort_E_L_At_Tp}. 
It is seen that stabilization of the average values in time depends both on 
incident energy and mass of target. The heavier target is used, the 
later stabilization occurs. Similarly, the lower projectile energy, the longer 
time of first stage calculations is needed in order to reach equilibrium. 
Unfortunately, it cannot be assumed one maximal time for all systems, because 
of building up of the numerical inaccuracies (see the time dependences of 
angular momentum), what starts at different time for different systems.  
Nevertheless, the deduced above stopping time equal to 35 fm/c will be treated 
in the frame of this work as an optimal time for each system. 
It is optimal for all used targets, since it has been 
established on the example of one of the heaviest targets. But taking into 
consideration dependence on incident energy, the following exceptions must be 
included. Calculations of first stage of reactions at projectile energies 
greater than or equal to 3.0 GeV, on all targets will be terminated after 
35 fm/c. At energies lower than 3.0 GeV, the HSD calculations will be stopped 
later, especially for heavier targets. \\     
  
\begin{figure}[!ht]
\begin{center}
\includegraphics[height=6cm, width=8cm, angle=0]{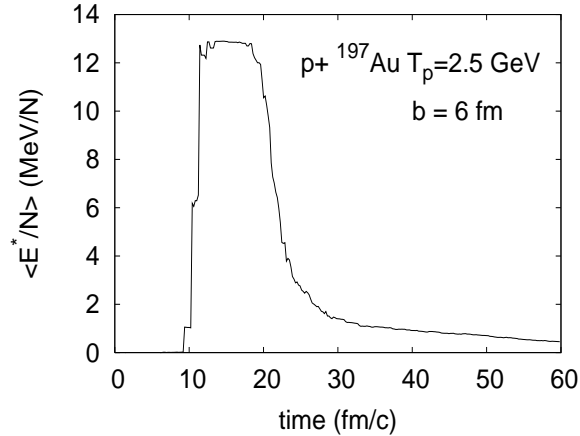}
\caption{{\sl Time variation of the average value of excitation energy per 
nucleon of the residual nucleus in p + Au collision at 2.5 GeV proton beam 
energy, the impact parameter b=6 fm; results of the HSD model calculations}}
\label{fig:time_enexc_b6}
\end{center}
\end{figure}

\begin{figure}[!ht]
\begin{center}
\includegraphics[height=6cm, width=8cm, angle=0]{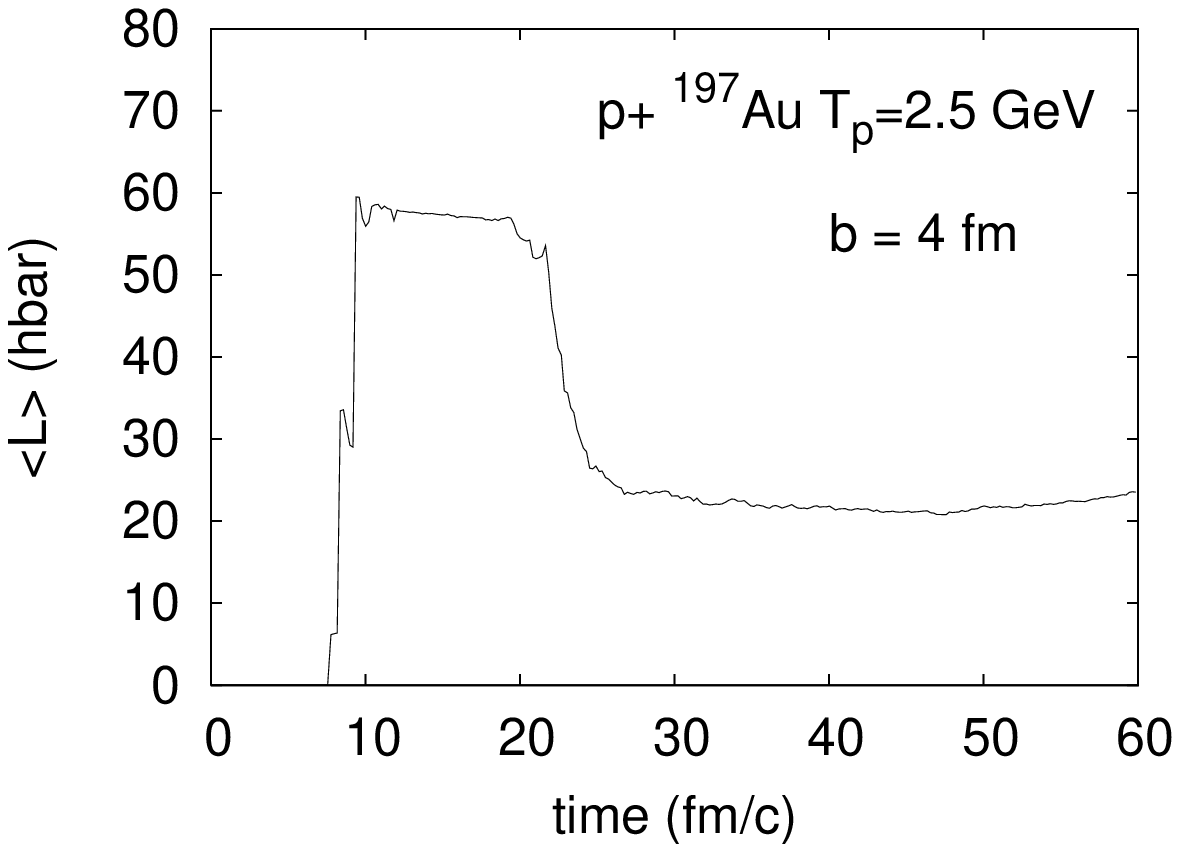}
\caption{{\sl Time variation of the average value of angular momentum of the 
residual nucleus in p + Au collision at 2.5 GeV proton beam energy, the impact 
parameter b=4 fm; results of the HSD model calculations}}
\label{fig:time_angm_b4}
\end{center}
\end{figure}

\begin{figure}[!ht]
\begin{center}
\includegraphics[height=6cm, width=8cm, angle=0]{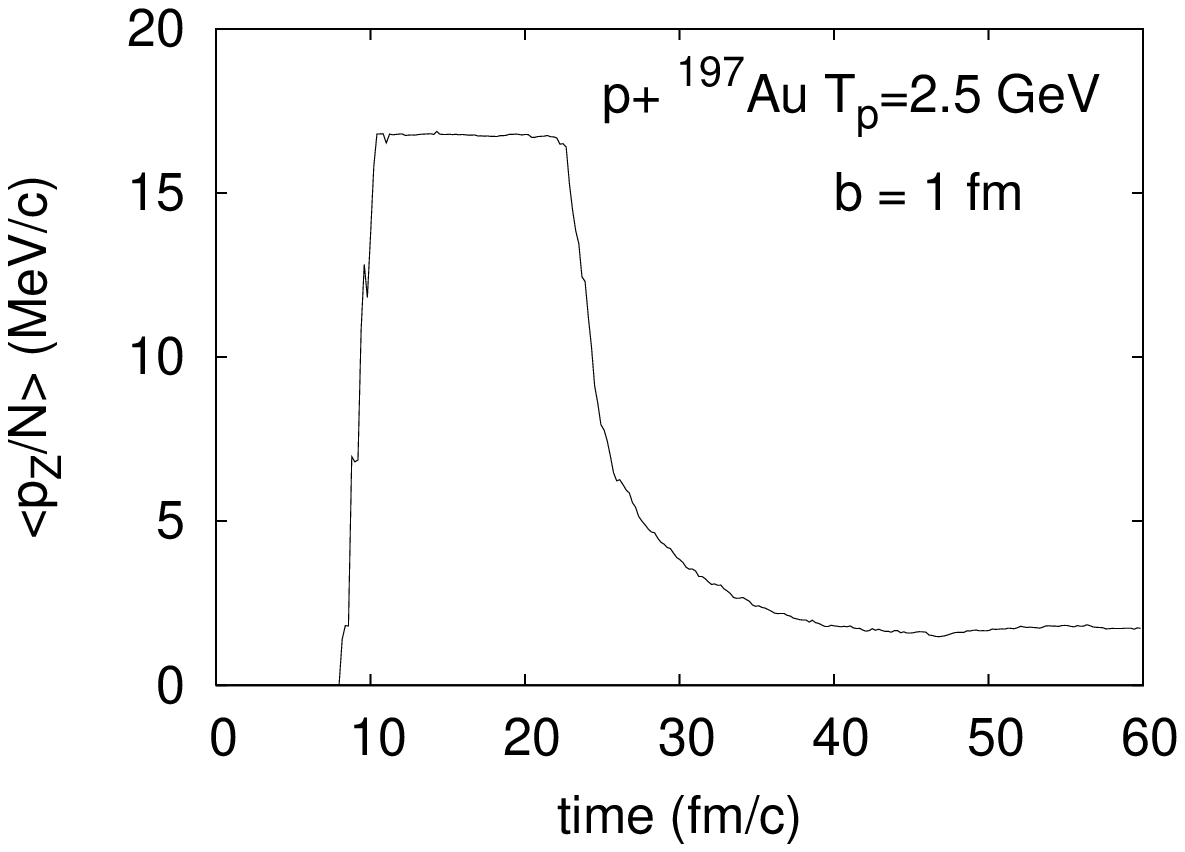}
\caption{{\sl Time variation of the average value of longitudimal momentum of 
the residual nucleus in p + Au collision at 2.5 GeV proton beam energy, the 
impact parameter b=1 fm; results of the HSD model calculations}}
\label{fig:time_pz_b0_b2}
\end{center}
\end{figure}

The above considerations are based on the results of calculations averaged over
 all impact parameters. For particular reaction p+Au at Tp=2.5 GeV, 
for specific impact parameters, e.g. b = 1 fm, 4 fm, 6 fm, the following 
dependences: time evolution of average excitation energy (presented in Fig. 
\ref{fig:time_enexc_b6}), angular momentum (Fig. \ref{fig:time_angm_b4}) and 
momentum in beam direction (Fig. \ref{fig:time_pz_b0_b2}) are obtained. 

The example dependences calculated for specific impact parameters agree with 
results obtained with averaging over all impact parameters; choice of time 
of calculations of first stage of the reaction ($\approx$ 35 - 45 fm/c) is 
satisfactory. 


\chapter{Bulk properties of the first stage of proton induced reactions}
\label{chapt:bulk_prop}
\markboth{ }{Chapter 4. Bulk properties of the first stage of reaction}

The following scenario of the first stage of proton - nucleus reactions takes 
place. High energy proton ($T_{p} > 100$ MeV) hits a target nucleus.
The intra-nuclear cascade starts to develop: the incoming proton, on its way,
collides with several target nucleons, transfers energy and momentum to them 
and may excite them into higher baryonic states. Depending on the position of 
the cascade nucleons inside the target, they either escape directly
from the nucleus or collide secondarily with a few other nucleons transfering 
further the energy and momentum. Some of the nucleons may leave the nucleus.
Development of the cascade can be seen by observing variations of spatial 
nucleon density with the reaction time. It indicates that proton induced 
reactions are low-invasive processes.

\begin{figure}[!ht]
\begin{center}
\includegraphics[height=17cm, width=14cm, angle=0]{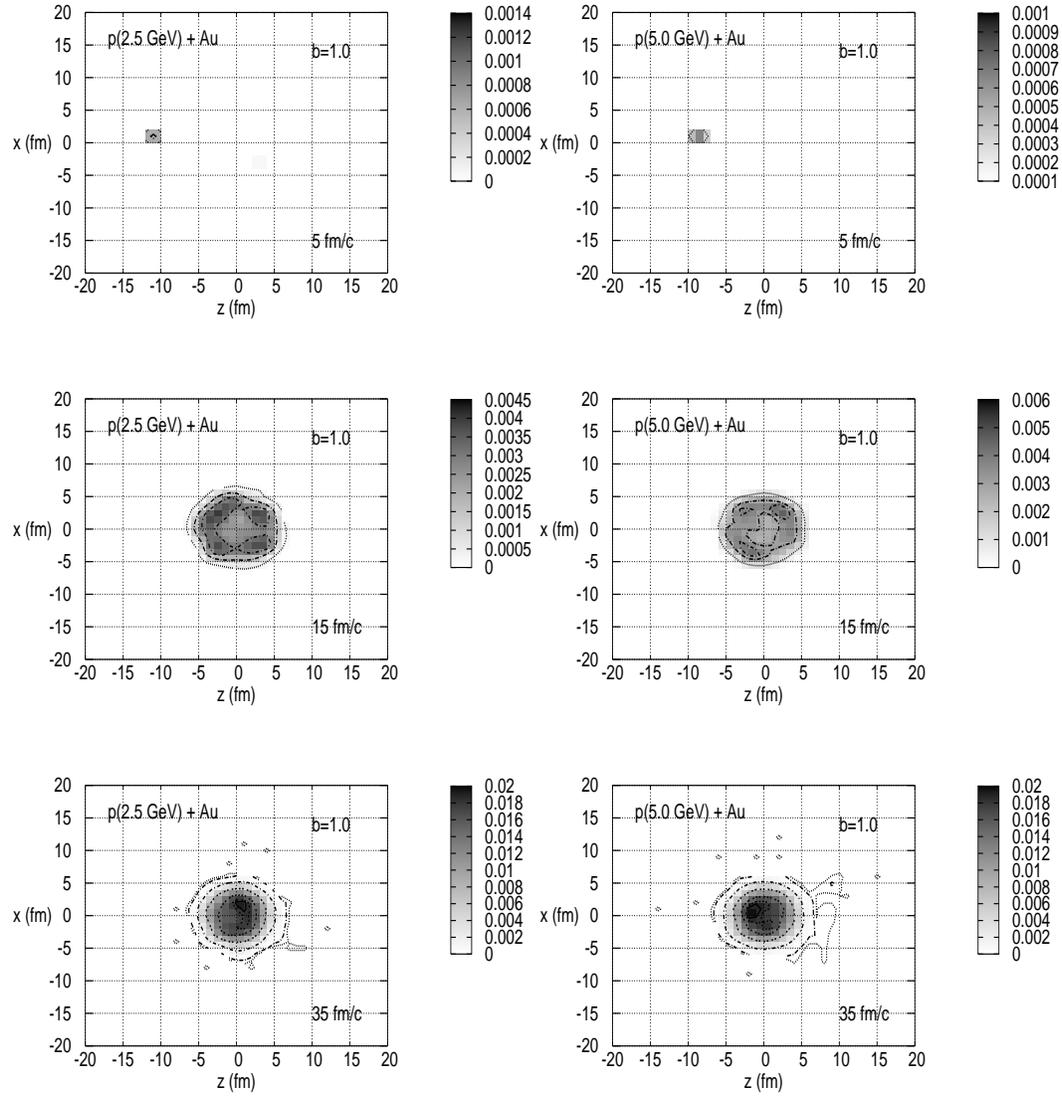}
\caption{{\sl Time evolution of nucleon density of nucleons with kinetic 
energies $E_{k}>50$ MeV, in central collisions of 2.5 GeV (left column)  
and 5.0 GeV (right column) proton with Au nucleus; results of the HSD model 
calculations (projections on xz plane)}}
\label{fig:17052_50_xz_pAu_2.5_5.0}
\end{center}
\end{figure}

\begin{figure}[!ht]
\begin{center}
\includegraphics[height=17cm, width=14cm, angle=0]{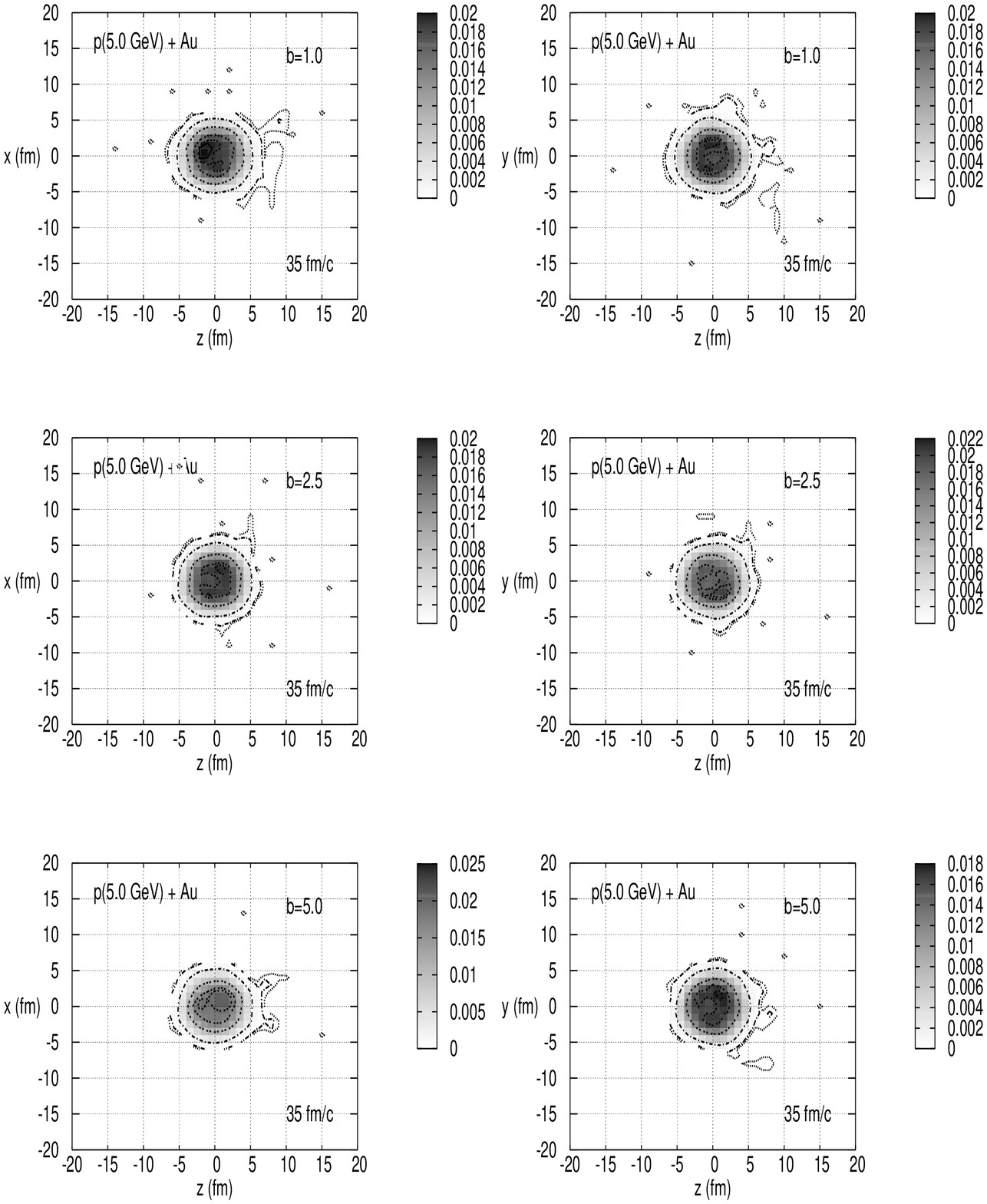}
\caption{{\sl Nucleon density of residual nuclei of nucleons with kinetic
energies $E_{k}>50$ MeV, in central (starting from the top, b=1fm), midcentral
 (b=2.5fm) and peripheral (b=5fm) collisions of 5.0 GeV proton with Au nucleus; 
results of the HSD model calculations (the density projections on xz (left 
column) and yz (right column) planes)}}
\label{fig:1705_50_xyz_pAu5.0}
\end{center}
\end{figure}

In Figure \ref{fig:17052_50_xz_pAu_2.5_5.0} the HSD simulations of time 
evolution of nucleon density in central collisions of 2.5 GeV and 5.0 GeV 
proton with Au nucleus are presented. The density of nucleons with kinetic 
energies $E_{k}>50$ MeV has been calculated, in order to see 
more clearly the development of the intra - nuclear cascade. 
Initially, before proton strikes the target nucleus, the maximal
kinetic energy of nucleons in the target is equal to the Fermi energy ($E_{F} 
\approx 35$ MeV). As a consequence of interaction nucleons get more energetical.
At the beginning of reaction, the incident proton is in distance of 15 fm 
from the center of target nucleus. Projectile enters the target from 
the left, $z$ is the axis in beam direction.
One can see in the Fig. \ref{fig:17052_50_xz_pAu_2.5_5.0}, that after 5 fm/c 
(1fm/c $\approx \frac{1}{3} \cdot 10^{-23}$ s), 
when proton has not entered the target yet, none of target nucleons has kinetic
 energy more than 50 MeV. 
Then, looking at the situation corresponding to density distribution sampled 
after 15 fm/c, one observes something like a wave going near the surface of 
target nucleus. Mainly nucleons from the most outer area of nucleus take part 
in the cascade. The center of target is not touched yet.
Situation after 35 fm/c, it means after first fast stage of the reaction 
is shown on the bottom plots of the Fig. \ref{fig:17052_50_xz_pAu_2.5_5.0}.
It is seen that the center of the nucleon density distribution is now  
occupied by the cascade nucleons. But the density is not spatially uniform. 
The central area corresponds to the maximum of the distribution. 
Levels of constant nucleon density (in units nucleon/fm$^{3}$), 
with values increasing to the center of the circular shaped distribution are 
clearly visible. It means that in the residual nuclei, the density distribution
 of shape like of initial target nucleus is reproduced (the density of initial 
nucleus have the Wood-Saxon distribution form (\ref{eq:wood_saxon})). 
Nevertheless, a small expansion of the nucleus is observed.
The level corresponding to the spatial density of value less than about 0.002 
nucleon/fm$^{3}$ is associated with the free nucleons knocked out of the 
target. \\ 
Distribution of the nucleon density of residual nuclei, in general, does not 
depend significantly on the centrality of proton - nucleus collision. This is 
illustrated in Fig. \ref{fig:1705_50_xyz_pAu5.0}, where  
two-dimensional projections of the nucleon density of residual nuclei, for 
different centralities of 5.0 GeV proton with Au nucleus collision are compared 
(i.e. for impact parameters of b=1.0, 2.5 and 5.0 fm).   
However, on the level corresponding to the lowest spatial density, associated 
with emitted nucleons, some differences are observed. They indicate that  
more nucleons are emitted in central than in peripheral collisions. 
\indent All the presented distributions of nucleon density show that in 
general, the incoming proton has caused only minor changes of density inside 
target nucleus. 

\begin{figure}[!ht]
\begin{center}
\includegraphics[height=9cm, width=12cm, angle=0]{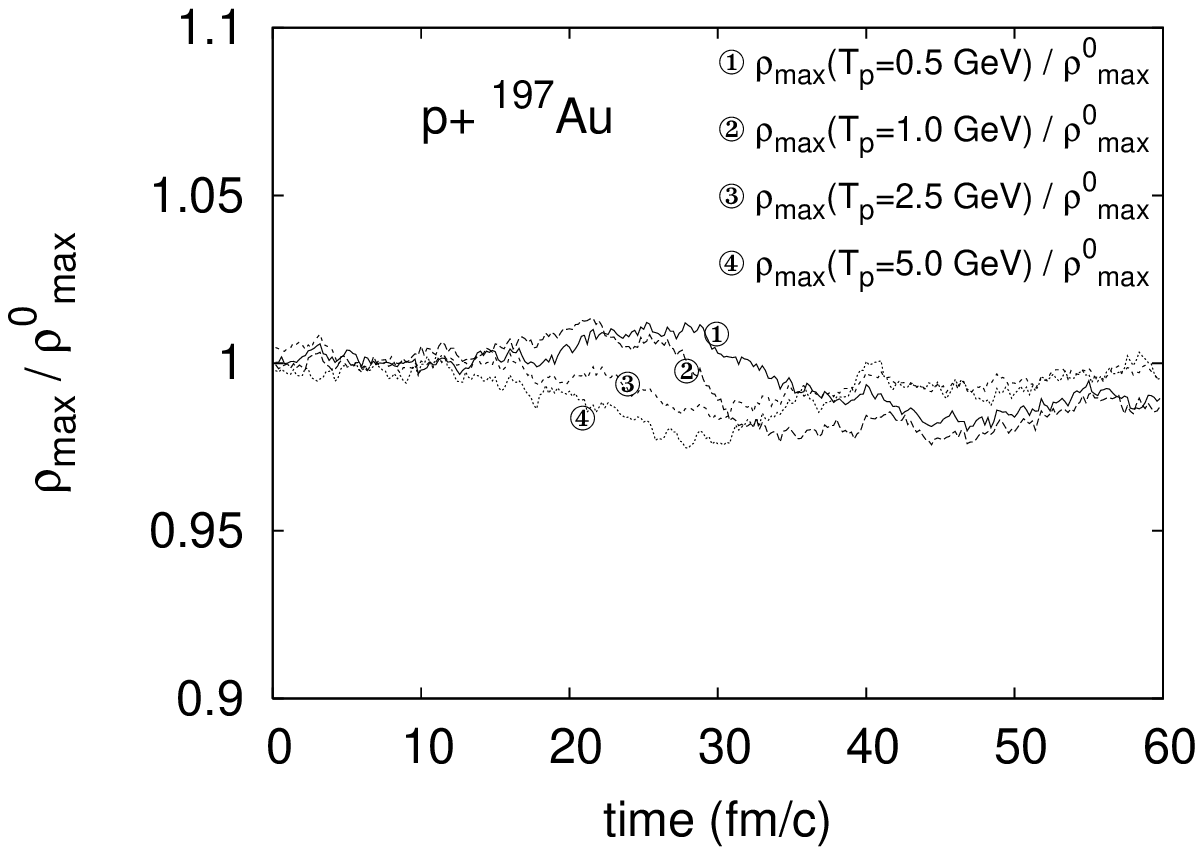}
\caption{{\sl Modifications of maximal nucleon density in the case of p+Au
collisions at 0.5 GeV, 1.0 GeV, 2.5 GeV and 5.0 GeV proton beam energy; 
results of the HSD model calculations}}
\label{fig:odn_DT_rhbmax}
\end{center}
\end{figure}

One of characteristics of created target nucleus is its maximal density. 
Modification of the density due to penetration of the nucleus by 
incoming proton is the characteristic feature of proton induced reactions.
Such modifications are presented qualitatively in Fig. \ref{fig:odn_DT_rhbmax},
 as time evolution of a ratio of maximal nuclear density in case of p+Au 
reaction at 0.5 GeV, 1.0 GeV, 2.5 GeV and 5.0 GeV of 
incident energy and the standard nuclear density $\rho ^{0}_{max}$.
It is evident from the Figure, that the incoming proton 
causes negligible modifications of nuclear density. Though, it is seen, that
 in case of low energy projectile the maximal density is first slightly 
increased and then it decreases. In case of higher energetical proton, first  
slight decrease and then an increase is observed. That could be explained by a 
fact that incoming low energy projectile almost stops inside target nucleus, 
causing increase of density, while higher energy projectile goes faster through 
the nucleus, pushing nucleons away. Then, the situation is changed 
respectively, due to acting of mean field potential. Nevertheless, 
the deviations of the presented ratios from the unity are of order of few 
percent, what proves that proton induced reactions are quite non-invasive 
processes.

\begin{figure}[!ht]
\begin{center}
\includegraphics[height=17cm, width=14cm, angle=0]{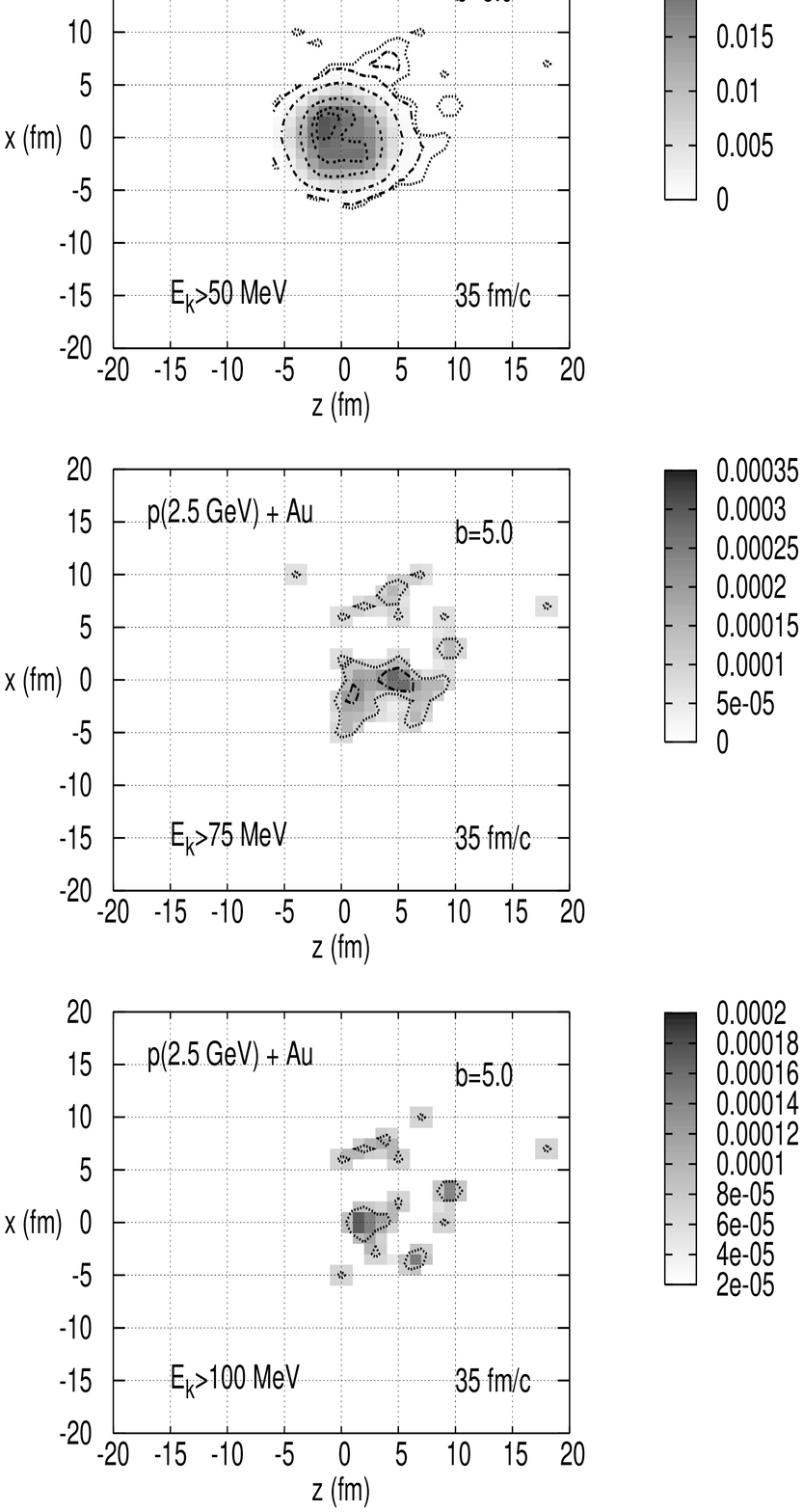}
\caption{{\sl Nucleon density of residual nuclei of nucleons with kinetic   
energies $E_{k}>50$ MeV (starting from the top), $E_{k} > 75$ MeV and 
$E_{k} > 100$ MeV, respectively, in peripheral (b=5fm) collisions of 2.5 GeV 
proton with Au nucleus; results of the HSD model calculations}}
\label{fig:1705_50_75_100_b.5.0_xz_pAu2.5}
\end{center}
\end{figure}

In order to check, how the kinetic energy is distributed inside the residual 
nuclei, the following test has been made. 
Based on p+Au reaction, at 2.5 GeV proton beam energy, nucleons,
which take part in the intra-nuclear cascade have been observed. It 
means, nucleons which gained a significant part of the energy of the 
projectile (i.e. of $E_{k} > 75$ MeV, $E_{k} > 100$ MeV). Most of the observed 
cases indicated homogeneous distribution of kinetic energy inside the residual 
nuclei. Partial heating up of the nuclei has been noticed in only about $1 \%$
 of the cases, in peripheral collisions. Illustration of the exceptional 
situation is shown in Fig. \ref{fig:1705_50_75_100_b.5.0_xz_pAu2.5}, where, 
starting from the top, the nucleon density of residual nuclei of nucleons with 
kinetic energies $E_{k} > 50$ MeV, $E_{k} > 75$ MeV and $E_{k} > 100$ MeV, 
respectively are plotted.  
Looking at the plot in the bottom of Fig. 
\ref{fig:1705_50_75_100_b.5.0_xz_pAu2.5} one can distinguish two differently 
excited parts. One part of nucleus is composed of nucleons with kinetic 
energies greater than 100 MeV, the second part - of nucleons with kinetic
energies lower than 100 MeV. The highly excited group of nucleons, it is about 
10 nucleons with an average total momentum equal to about 560(+/-80) MeV/c. 
The less excited part, it is till about 184 nucleons with an average total 
momentum equal to only about 190(+/-60) MeV/c. \\
Similar situation is observed, if looking at a bit lower energetical nucleons, 
the central plot of Fig. \ref{fig:1705_50_75_100_b.5.0_xz_pAu2.5}, nucleons 
with kinetic energies greater or lower than 75 MeV. 
In this case, the highly excited 
part consists of about 20 nucleons with an average total momentum equal to 
about 470(+/-90) MeV/c, and less excited part composed of 174 nucleons with an 
average total momentum equal to about 190(+/-60) MeV/c. \\
One can conclude, that the HSD simulations predict something 
like formation of two excited sources of evidently unequal masses. 
The smaller source consists of relatively few nucleons (up to $\sim$ 20) and is 
rather fast, $\beta \sim 0.025$ c. The larger source is built of $\sim$ 170 - 
180 nucleons and has velocity $\beta \sim 0.0012$ c. Similar observation has 
been drawn from phenomenological analysis of experimental data presented in 
\cite{Buba07}.


\chapter{Properties of residual nuclei after the first stage of proton - 
nucleus reactions}
\markboth{ }{Chapter 5. Properties of residual nuclei}

\begin{figure}[!htcb]
\vspace{-2cm}
\hspace{-2cm}
\includegraphics[height=20cm, width=18cm, bbllx=0pt, bblly=25pt, bburx=594pt, bbury=842pt, clip=, angle=0]{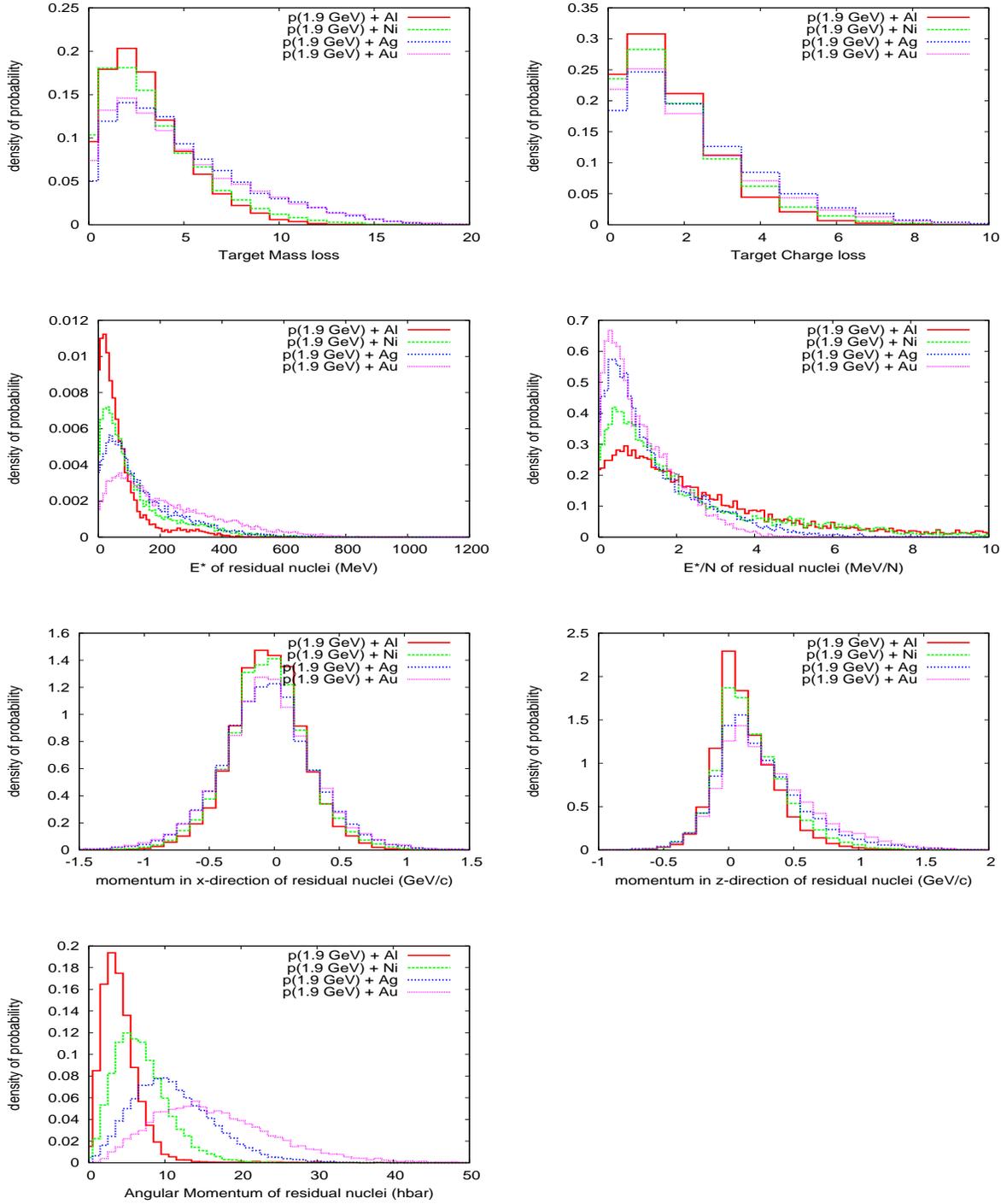}
\caption{{\sl One - dimensional distributions of properties of residual nuclei
after first stage of 1.9 GeV proton induced spallation reactions on 
$^{27}$Al, $^{58}$Ni, $^{107}$Ag and $^{197}$Au targets; results of the HSD 
model calculations}}
\label{fig:res_1.9}
\end{figure}

\begin{figure}[!htcb]
\vspace{-2cm}
\hspace{-2cm}
\includegraphics[height=20cm, width=18cm, bbllx=0pt, bblly=25pt, bburx=594pt, bbury=842pt, clip=, angle=0]{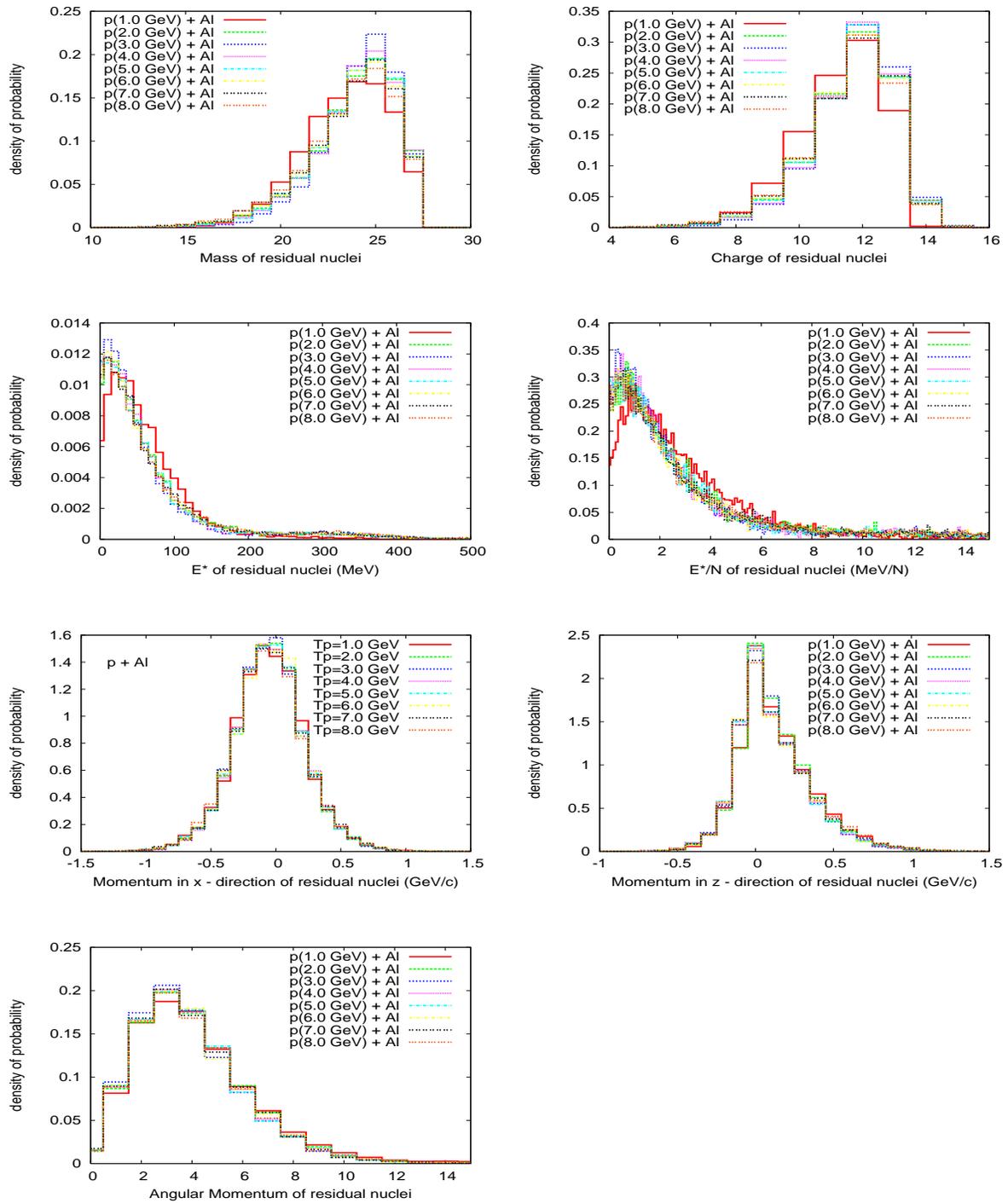}
\caption{{\sl One - dimensional distributions of properties of residual nuclei
after first stage of proton induced spallation reactions on 
$^{27}$Al target, at few values of projectile energy; results of the HSD model 
calculations}}
\label{fig:res_Al}
\end{figure}

As it is mentioned above in this work, as result of the first stage of proton 
- nucleus reaction, beside emitted particles, an excited nucleus 
remains. It differs from the initial target, in average, by only a few 
nucleons in mass number. Properties of the residual nucleus (i.e. mass 
($A_{R}$), charge ($Z_{R}$), excitation energy ($E^{*}_{R}$), three - momentum 
($\vec{p_{R}}$) and angular momentum ($L_{R}$)) are evaluated, in the frame of 
the HSD model, by exploring conservation laws, according to formulas 
\ref{eq:res_prop}. 
Results of calculations for reactions on various target nuclei, at different 
values of incident energy in range from 0.1 GeV to about 10 GeV, 
are discussed below.\\ 
Let's look first at one - dimensional distributions of the quantities evolving 
with projectile energy and mass of target. 
Histogrammed properties for exemplary reactions of 1.9 GeV proton on light 
($^{27}$Al), heavy ($^{197}$Au) and two intermediate mass ($^{58}$Ni and 
$^{107}$Ag) targets are displayed in Fig. \ref{fig:res_1.9}.   
Distributions for proton induced reaction on example Al target, at several 
values of proton beam energy are shown in Fig. \ref{fig:res_Al}.
The average values and standard deviations for the presented distributions are 
collected in Tables \ref{table: distr_1dim_1.9} and \ref{table: distr_1dim_Al}.

The Figures \ref{fig:res_1.9} and \ref{fig:res_Al} and values 
collected in the Tables \ref{table: distr_1dim_1.9} and 
\ref{table: distr_1dim_Al} indicate, that all of the distributions differ 
significantly with mass of target, but behave similarly for varied projectile 
energies. 

\begin{center}
\begin{table*}[tbp]
\caption{{\sl The average values and standard deviations for the one - 
dimensional distributions of properties of residual nuclei from 1.9 GeV proton 
induced reactions on several targets}}
\vspace{0.5cm}
\hspace{-1.0cm}
\begin{tabular}{|c||c|c|c|c|}
\hline
{\bf reaction} & {\bf p+Al} & {\bf p+Ni} & {\bf p+Ag} & {\bf p+Au} \\
\hline \hline
$<A>_{R}$ &23.96$\pm$2.25 &54.71$\pm$2.65 &102.24$\pm$3.55 &192.39$\pm$3.69 \\  
\hline 
$<Target$ $Mass$ $loss>$ &3.036$\pm$2.25 &3.29$\pm$2.65 &4.76$\pm$3.55 
&4.61$\pm$3.69 \\ 
\hline
$<Z>_{R}$ &11.64$\pm$1.42 &26.53$\pm$1.64 &45.05$\pm$1.98 &77.34$\pm$1.92 \\ 
\hline
$<Target$ $Charge$ $loss>$ &1.36$\pm$1.42 &1.47$\pm$1.64 &1.96$\pm$1.98 
&1.66$\pm$1.92 \\
\hline
$<E^{*}>_{R}$ $[$MeV$]$ &74.56$\pm$81.77 &122.84$\pm$120.41 &144.16$\pm$125.65 
&222.16$\pm$171.24 \\  
\hline
$<E^{*}/N>_{R}$ $[$MeV/N$]$ &3.27$\pm$3.63 &2.30$\pm$2.29 &1.43$\pm$1.27 
&1.16$\pm$0.91 \\ 
\hline
$<L>_{R}$ $[\hbar]$ &4.77$\pm$3.25 &7.41$\pm$3.59 &11.76$\pm$5.50 
&17.14$\pm$8.08 \\ 
\hline
$<p_{z}>_{R}$ $[$GeV/c$]$ &0.18$\pm$0.24 &0.22$\pm$0.27 &0.29$\pm$0.35 
&0.34$\pm$0.38 \\
\hline
$<p_{x}>_{R}$ $[$GeV/c$]$ &-0.0011$\pm$0.28 &-0.0014$\pm$0.31 
&-0.0058$\pm$0.36 &-0.0045$\pm$0.37 \\
\hline
\end{tabular}
\label{table: distr_1dim_1.9}
\end{table*}
\end{center}
\begin{center}
\begin{table*}[tbp]
\caption{{\sl The average values and standard deviations for the one - 
dimensional distributions of properties of residual nuclei from proton 
induced reactions on Al target, at several incident energies}}
\vspace{0.5cm}
\hspace{-1.0cm}
\begin{tabular}{|c||c|c|c|c|}
\hline
{\bf impact energy} &{\bf 1.0 GeV} &{\bf 2.0 GeV} &{\bf 3.0 GeV} 
&{\bf 4.0 GeV}\\
\hline \hline
$<A>_{R}$ &23.51$\pm$2.17 &24.032$\pm$2.20 &24.23$\pm$1.99 &24.12$\pm$2.085 \\ 
\hline 
$<Z>_{R}$ &11.28$\pm$1.35 &11.66$\pm$1.37 &11.78$\pm$1.31 &11.71$\pm$1.33 \\ 
\hline
$<E^{*}>_{R}$ $[$MeV$]$ &65.76$\pm$54.067 &73.46$\pm$84.27 &70.31$\pm$88.081 
&74.50$\pm$98.023 \\
\hline
$<E^{*}/N>_{R}$ $[$MeV/N$]$ &2.94$\pm$2.61 &3.22$\pm$3.75 &3.026$\pm$3.84  
&3.25$\pm$4.35 \\ 
\hline
$<L>_{R}$ $[\hbar]$ &4.75$\pm$2.65 &4.71$\pm$3.072 &4.85$\pm$4.48 
&4.99$\pm$5.23 \\ 
\hline
$<p_{z}>_{R}$ $[$GeV/c$]$ &0.18$\pm$0.24 &0.17$\pm$0.23 &0.16$\pm$0.24 
&0.16$\pm$0.25 \\
\hline
$<p_{x}>_{R}$ $[$GeV/c$]$ &-0.0021$\pm$0.27 &-0.00088$\pm$0.27 &-0.0030$\pm$0.27
 &0.0033$\pm$0.27 \\ 
\hline \hline \hline
{\bf impact energy} &{\bf 5.0 GeV} &{\bf 6.0 GeV} &{\bf 7.0 GeV} 
&{\bf 8.0 GeV}\\
\hline \hline
$<A>_{R}$ &24.027$\pm$2.19 &23.94$\pm$2.26 &23.86$\pm$2.35 &23.75$\pm$2.41 \\
\hline
$<Z>_{R}$ &11.66$\pm$1.39 &11.603$\pm$1.41 &11.58$\pm$1.46 &11.54$\pm$1.47 \\
\hline
$<E^{*}>_{R}$ $[$MeV$]$ &74.61$\pm$90.60 &76.48$\pm$98.98 &79.64$\pm$129.89 &
81.92$\pm$137.73 \\
\hline
$<E^{*}/N>_{R}$ $[$MeV/N$]$ &3.29$\pm$4.13 &3.41$\pm$4.55 &3.58$\pm$5.8 &3.71$\pm$6.29 \\
\hline
$<L>_{R}$ $[\hbar]$ &5.034$\pm$5.66 &5.48$\pm$8.25 &5.15$\pm$7.32 
&6.035$\pm$11.044 \\
\hline
$<p_{z}>_{R}$ $[$GeV/c$]$ &0.16$\pm$0.24 &0.16$\pm$0.25 &0.17$\pm$0.27 
&0.18$\pm$0.28 \\
\hline
$<p_{x}>_{R}$ $[$GeV/c$]$ &-0.000024$\pm$0.27 &0.0053$\pm$0.28 &0.0026$\pm$0.28
&-0.0051$\pm$0.29\\ 
\hline 
\end{tabular}
\label{table: distr_1dim_Al}
\end{table*}
\end{center}

\begin{figure}[!htcb]
\vspace{-2cm}
\hspace{-2cm}
\includegraphics[height=20cm, width=18cm, bbllx=0pt, bblly=25pt, bburx=594pt, bbury=842pt, clip=, angle=0]{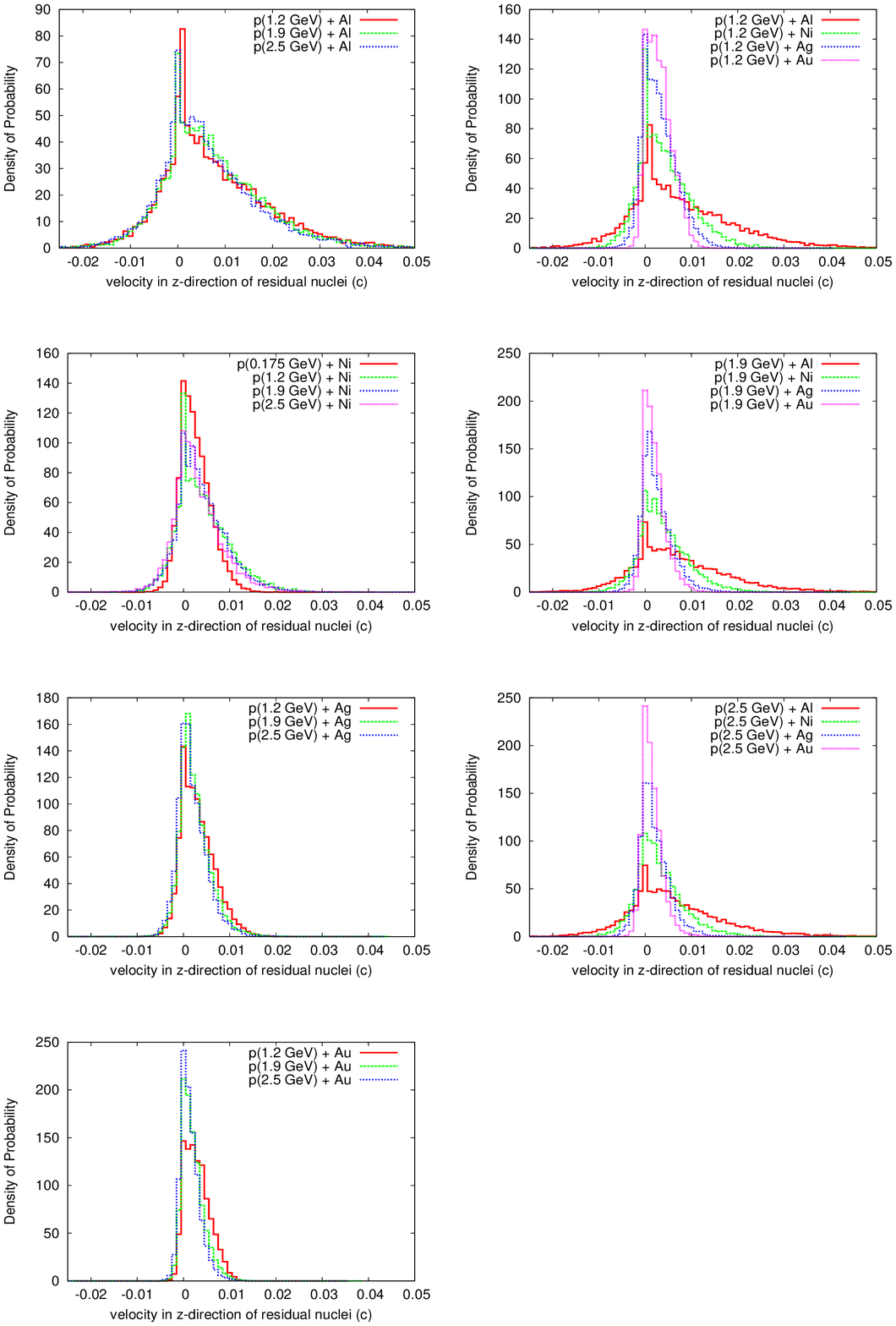}
\caption{{\sl One - dimensional distributions of velocity in beam direction of 
residual nuclei at the end of first stage of proton induced spallation 
reactions on various targets, at few example values of projectile energy; 
results of the HSD model calculations}}
\label{fig:vzres}
\end{figure}

Reason for these behaviors is connected with finite size of nuclei. 
Average values of excitation energy ($E^{*}_{R}$) and angular 
momentum ($L_{R}$) of residual nucleus and width of the distributions increase 
with mass of target nuclei, at fixed projectile energy. For a chosen target,  
 the distributions of $E^{*}_{R}$ very slightly depend on incident energy. 
Distributions of $L_{R}$ almost do not depend on incident energy. 
 This can be explained in an intuitive way by analyzing how the energy and 
momentum are deposited into a target nucleus. 
The $E^{*}_{R}$ left in the residual nucleus 
after the first stage of reaction is in average only a small fraction of the 
energy, which has been introduced into a target nucleus by incoming proton. 
Most of the energy is carried away by emitted particles (example energy 
balance is presented in Sec. \ref{sec:en_balance}). The number of 
ejected particles should be treated as a signature how many nucleons were 
involved in the intra - nuclear cascade. 
The more nucleon - nucleon collisions inside target nuclei, 
the larger part of the energy and angular momentum introduced by a projectile 
is deposited into the residual nucleus.  
E.g. due to a collision of the incident proton with a nucleon 
inside a target, a resonance, mainly Delta resonance could be excited. 
The Delta resonance has a short time of life, i.e. about 1 - 2 fm/c. 
It decays into nucleon and meson (mainly pion) or collides with other nucleon 
before the decay. As result, energy is accumulated inside target nucleus.   
The heavier target, the more collisions can occur, because the cascade of 
collisions can last for a longer time. 
For light targets, the excitation energy ($E^{*}_{R}$) is respectively small, 
because a number of collisions made by a projectile is low.     
In case of the excitation energy per nucleon ($E^{*}_{R}/N$) the situation is 
opposite, i.e. $E^{*}_{R}/N$ is a decreasing function of mass of target. 
This is just because in heavier targets, the amount of energy is distributed 
among more nucleons.
 
Similarly, distribution of angular momentum of residual nuclei almost does not 
depend on value of incident energy, but strongly depends on mass of target. 
The heavier target, the more angular momentum is deposited inside. This is 
because in the HSD model, quasi - classical approximation for calculations of 
angular momentum is used; at fixed impact energy, the 
maximal angular momentum of projectile is proportional to radius of target 
nuclei. 
Therefore, in case of heavier targets, more angular momentum is introduced by 
a projectile.   

Distributions of momentum in beam direction (z-component) and the momentum 
perpendicular to the beam direction (x-component) indicate that 
residual nuclei are moving according to initial direction of projectile. 
It is evidently seen (both from the Fig. \ref{fig:res_1.9} and Table 
\ref{table: distr_1dim_1.9}), that the average values of momentum in 
x-direction of residual nuclei are consistent with zero. 
\begin{center}
\begin{table*}[tbp]
\caption{{\sl The average values and standard deviations for distributions of 
velocity of residual nuclei ($<v_{z}>_{R}$ $[$c$]$) from proton induced 
reactions on several targets}}
\vspace{0.5cm}
\hspace{-1.0cm}
\begin{tabular}{|c||c|c|c|c|}
\hline
{\bf $<v_{z}>_{R}$ $[$c$]$} & {p+Al} & {p+Ni} & {p+Ag} & {p+Au} \\
\hline \hline
$T_{p}$=1.2 GeV &0.0082$\pm$0.011 &0.0046$\pm$0.0056 &0.0036$\pm$0.0036 & 
0.0035$\pm$0.0027 \\
\hline
$T_{p}$=1.9 GeV &0.0078$\pm$0.011 &0.0044$\pm$0.0054 &0.0029$\pm$0.0033 & 
0.0025$\pm$0.0024 \\
\hline
$T_{p}$=2.5 GeV &0.0069$\pm$0.010 &0.0036$\pm$0.0052 &0.0024$\pm$0.0032 & 
0.0019$\pm$0.0021 \\
\hline
\end{tabular}
\label{table: distr_vzres}
\end{table*}
\end{center}
Similar information about the reaction could be extracted from 
distributions of velocity in beam direction of residual nuclei. 
The distributions differ significantly with mass of target, but behave 
similarly for varied projectile energies, as shown in Fig. \ref{fig:vzres}. 
The average values and standard deviations of the values are presented in 
Table \ref{table: distr_vzres}. 
It is seen that the lighter target, the broader is the distribution. 
These are asymmetric, peaked in forward direction, with tails in 
backward direction. This indicates, that in average, the residual 
nuclei tend to move forward, according to the initial direction of projectile. 
          
\begin{figure}[!htcb]
\vspace{-2cm}
\hspace{-2cm}
\includegraphics[height=7cm, width=8cm, angle=0]{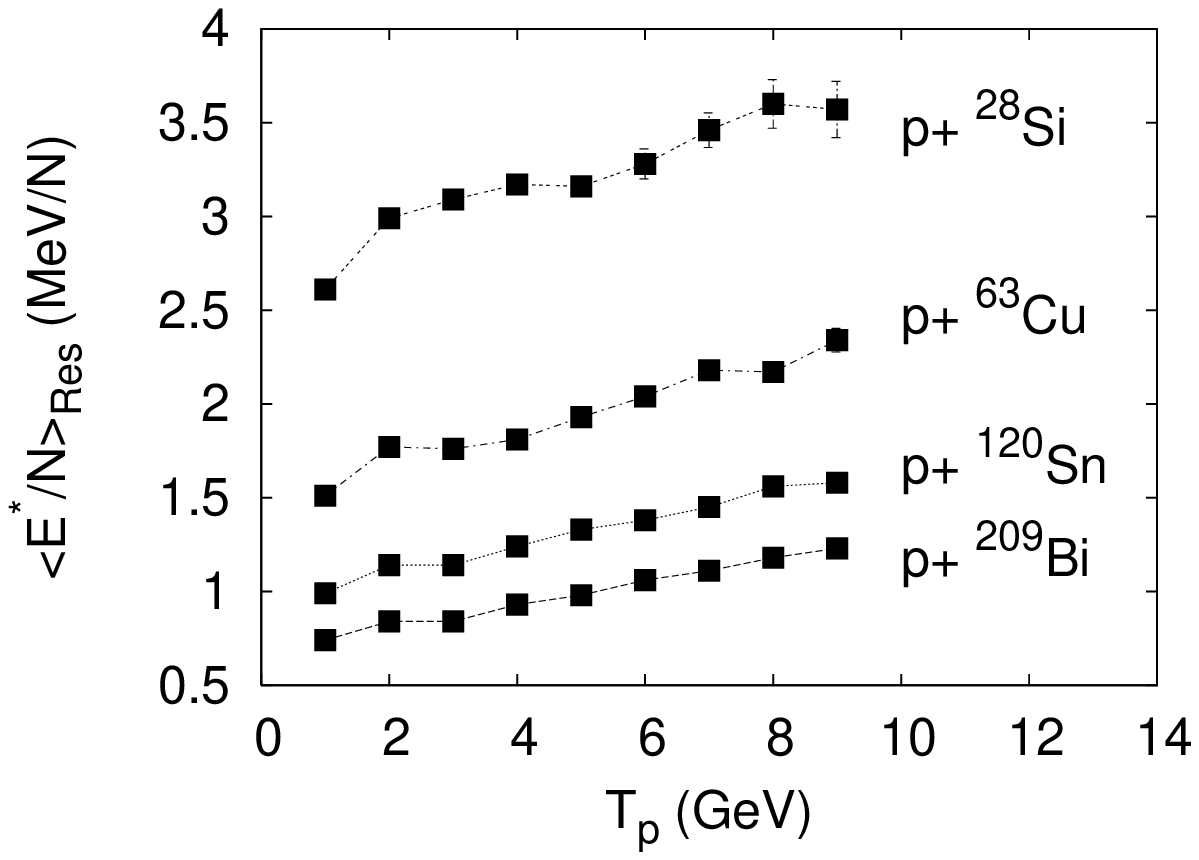}
\hspace{-0.6cm}
\includegraphics[height=7cm, width=8cm, angle=0]{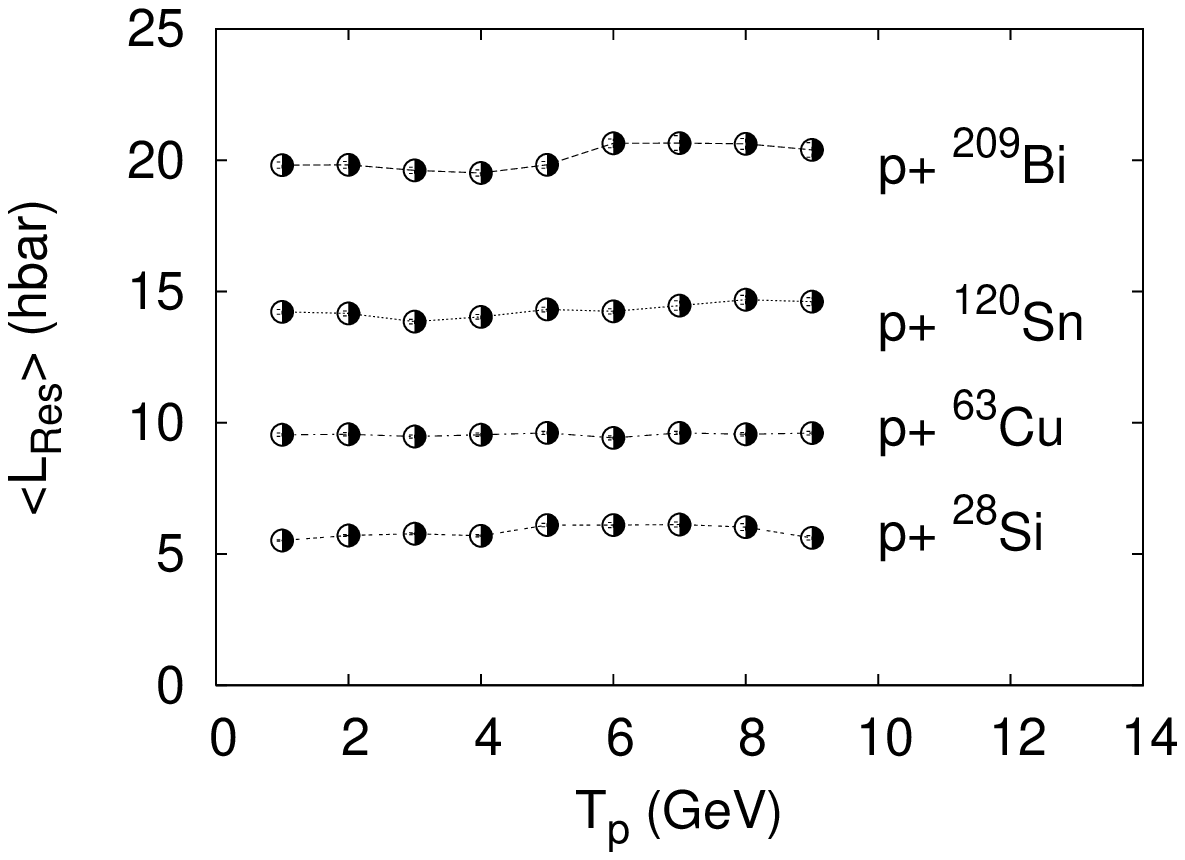}
\caption{{\sl Average values of excitation energy per nucleon (left) and 
angular momentum (right) of residual nuclei from proton induced reactions on 
Si, Cu, Sn and Bi target in function of incident energy; results of the HSD 
model calculations (error bars indicate values of standard
deviation of the average values, divided by a square root of number of events)}}
\label{fig:enexc_Lres}
\end{figure}

\begin{figure}[!htcb]
\vspace{-2cm}
\hspace{-2cm}
\includegraphics[height=6cm, width=8cm, angle=0]{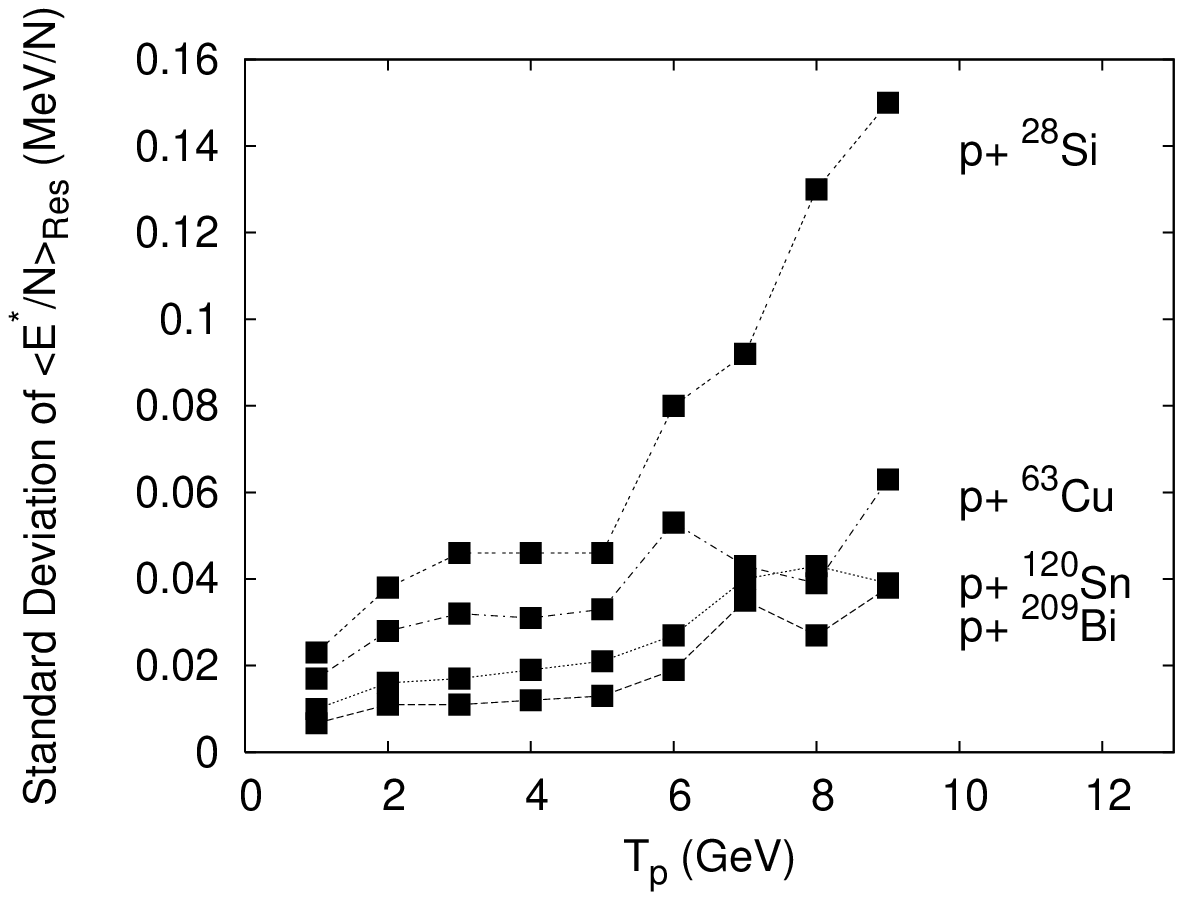}
\hspace{-0.6cm}
\includegraphics[height=6cm, width=8cm, angle=0]{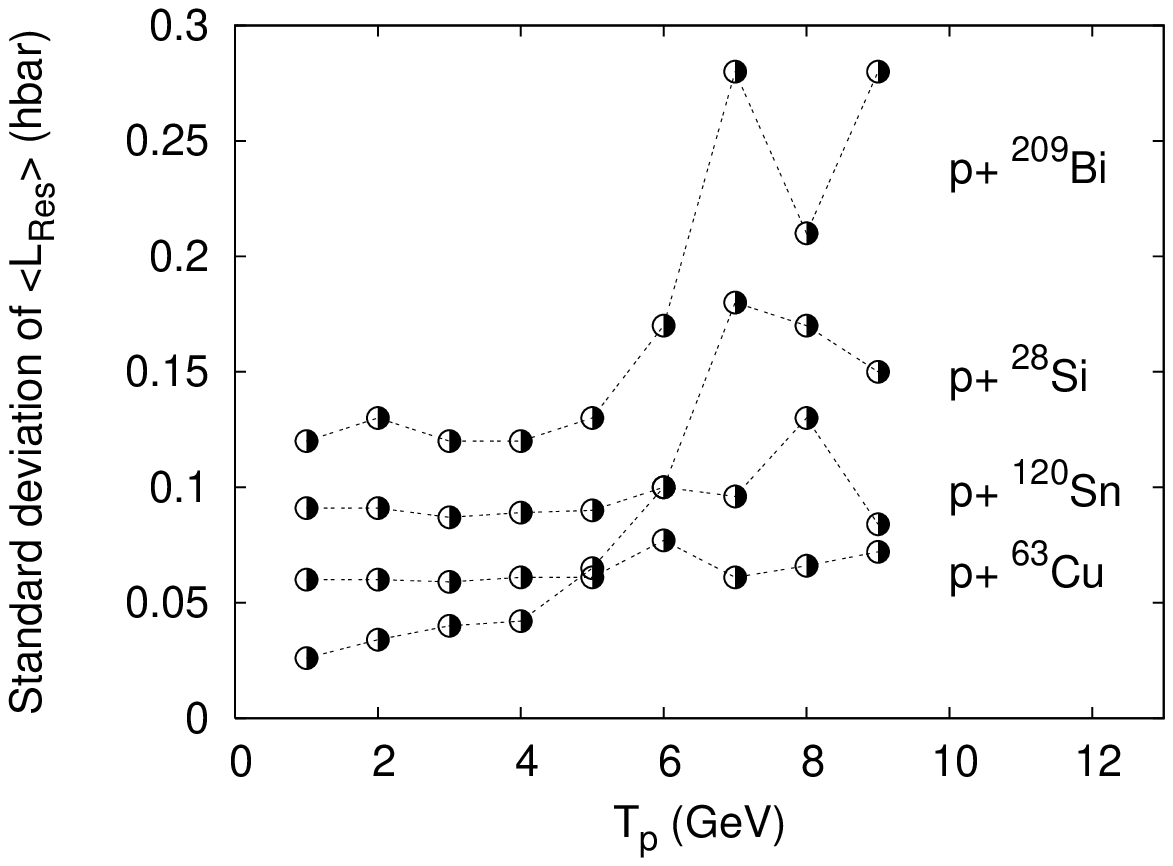}
\caption{{\sl Standard deviation of average values of excitation energy per 
nucleon (left) and of angular momentum (right) of residual nuclei from proton 
induced reactions on Si, Cu, Sn and Bi target, in function of incident energy; 
results of the HSD model calculations}}
\label{fig:stdev_enexc_Lres}
\end{figure}

In order to have a better view of the dependences, average values of the 
quantities and standard deviation of the average values, have been expressed as 
a function of incident energy and mass of target and presented in Figures 
\ref{fig:enexc_Lres} and \ref{fig:stdev_enexc_Lres}. 

\begin{figure}[!htcb]
\vspace{-2cm}
\hspace{-2cm}
\includegraphics[height=7cm, width=8cm, angle=0]{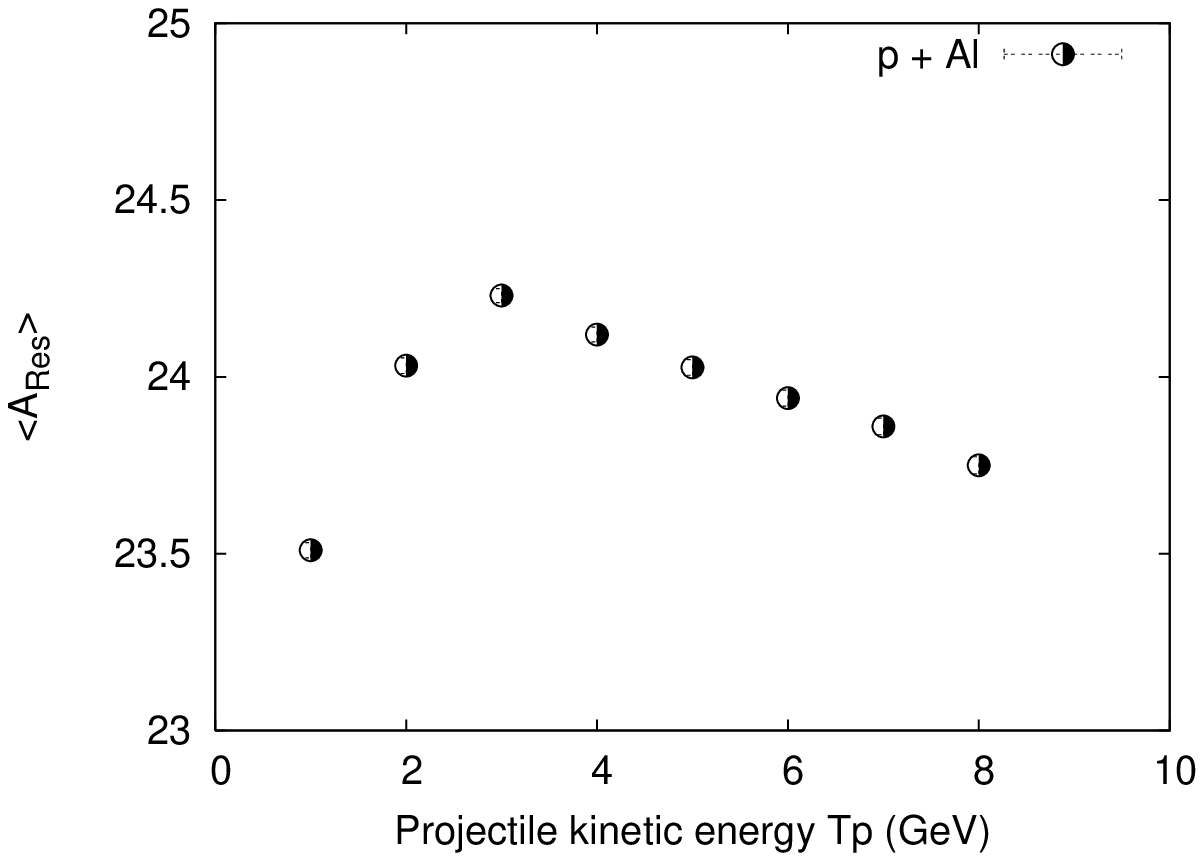}
\hspace{-0.6cm}
\includegraphics[height=7cm, width=8cm, angle=0]{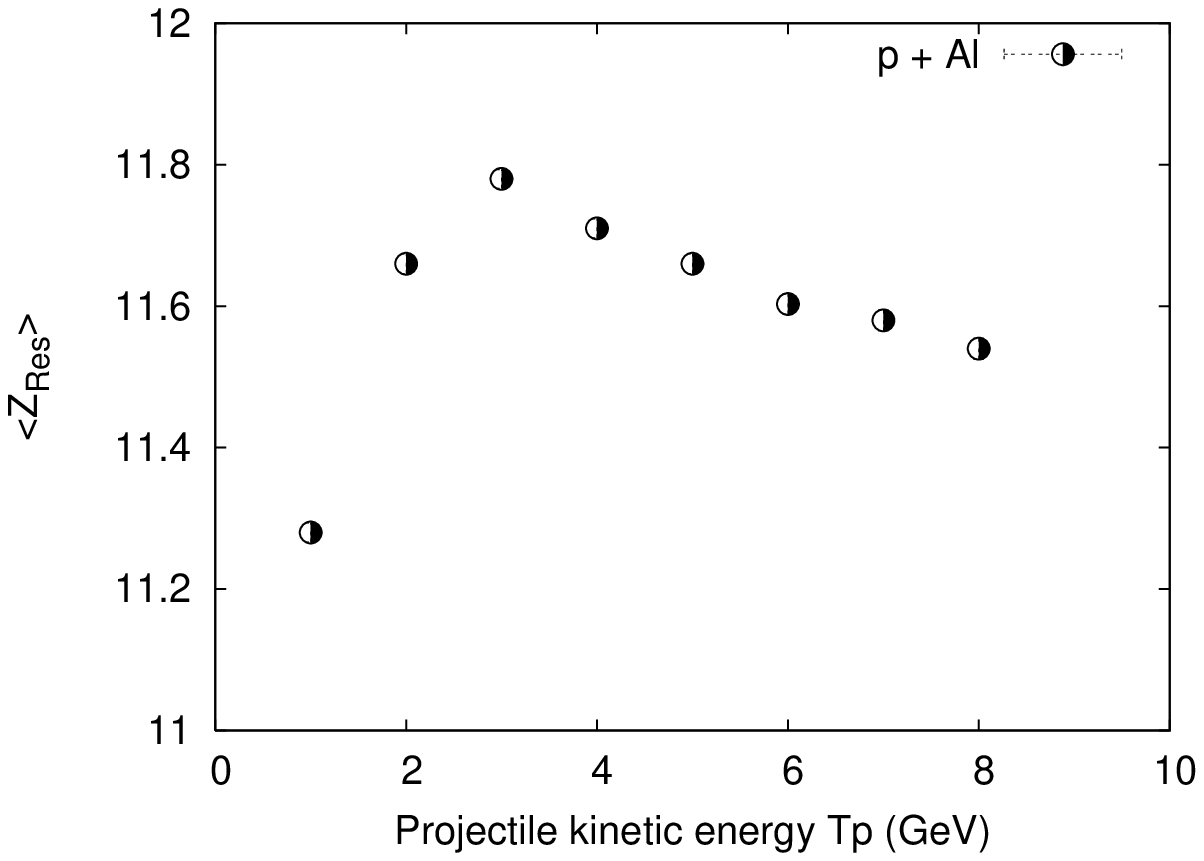}
\caption{{\sl Average values of mass number (left) and charge (right) of 
residual nuclei from proton induced reaction on Al target, in function of 
incident energy; results of the HSD model calculations (error bars indicate 
values of standard deviation of the average values, divided by a square root 
of number of events)}}
\label{fig:pAl_a_z}
\end{figure}

\begin{figure}[!htcb]
\vspace{-2cm}
\hspace{-2cm}
\includegraphics[height=7cm, width=8cm, angle=0]{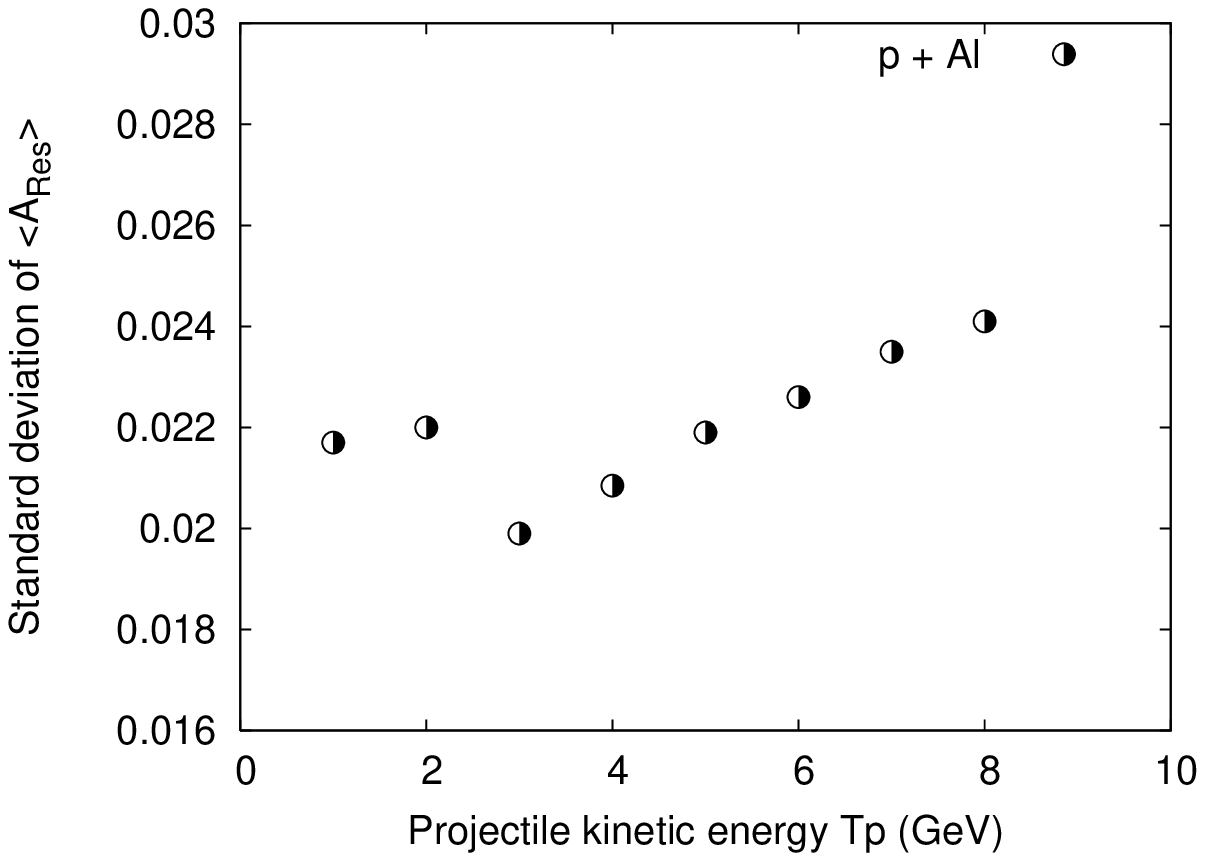}
\hspace{-0.6cm}
\includegraphics[height=7cm, width=8cm, angle=0]{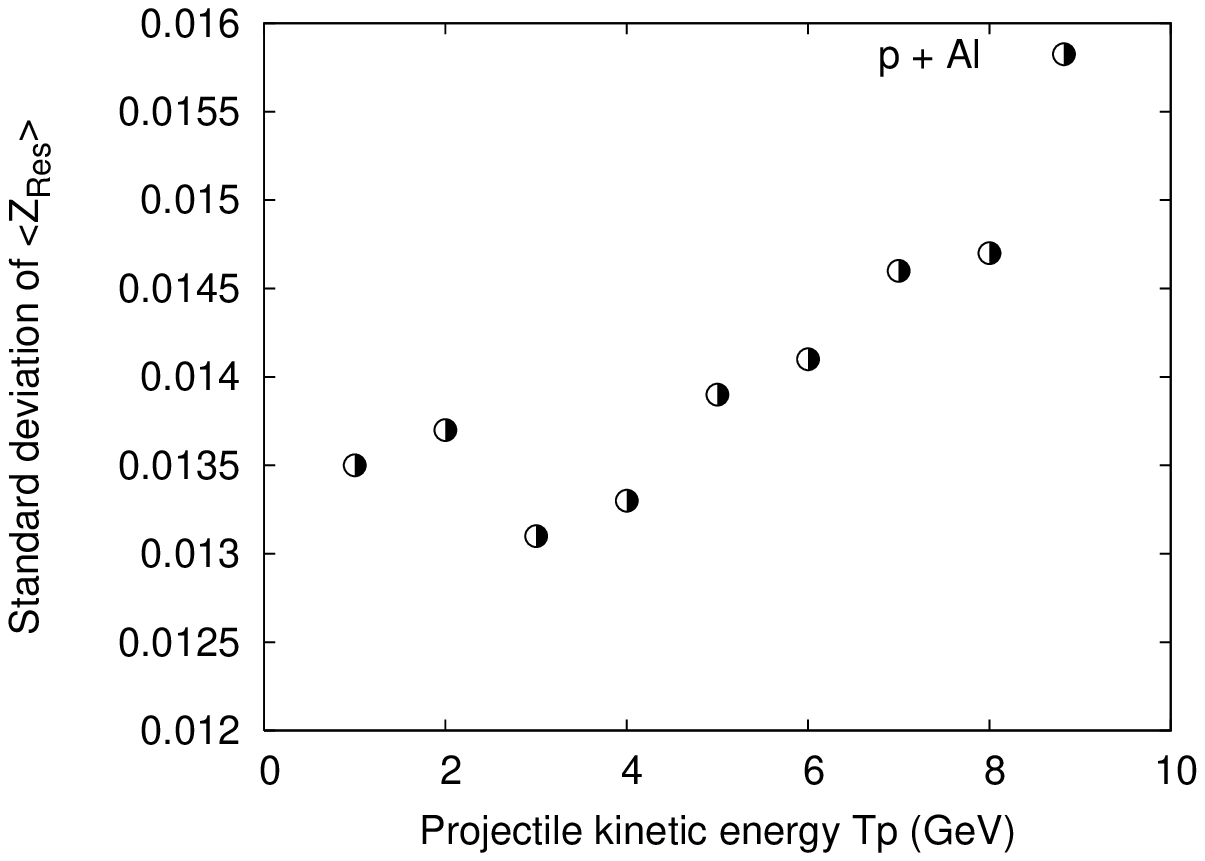}
\caption{{\sl Standard deviation of average values of mass number (left) and 
charge (right) of residual nuclei from proton induced reactions on Al target, in
 function of incident energy; results of the HSD model calculations}}
\label{fig:stdev_pAl_a_z}
\end{figure}

\begin{figure}[!htcb]
\vspace{-2cm}
\hspace{-2cm}
\begin{center}
\includegraphics[height=7cm, width=8cm, angle=0]{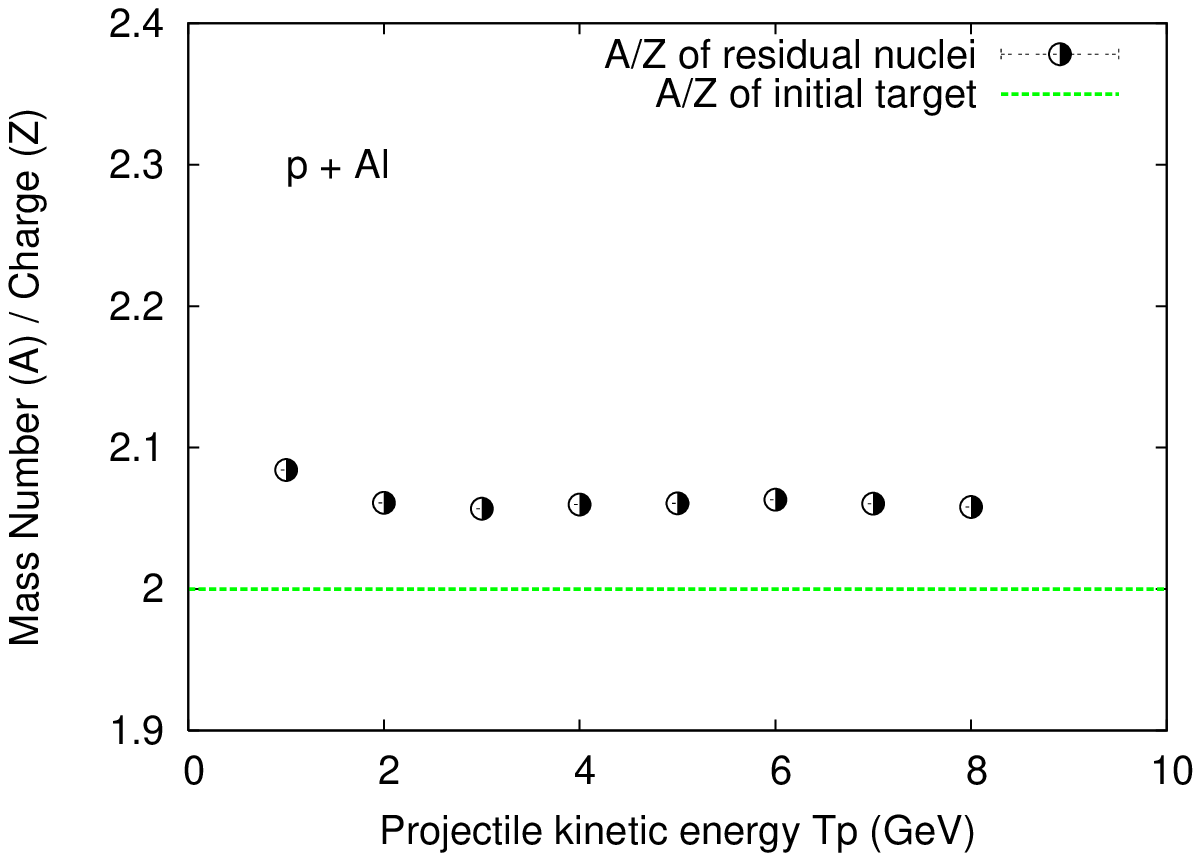}
\end{center}
\caption{{\sl Ratio of average values of mass number and charge of residual 
nuclei from proton induced reactions on Al target, in function of incident 
energy; results of the HSD model calculations}}
\label{fig:pAl_stos_az}
\end{figure}

Average values and standard deviations of mass number and charge of residual 
nuclei remaining after first stage of p+Al reaction, in function of projectile 
energy are presented in Figures \ref{fig:pAl_a_z} and \ref{fig:stdev_pAl_a_z}. 
It is interesting, that both the mass number and the charge, in average, 
increase up to 3.0 GeV of incident energy. Then decreasing is observed. 
The ratio of average values of the mass number and charge is constant as 
function of projectile energy, as shown in Fig. \ref{fig:pAl_stos_az}. 
Additionally, it exceeded the ratio for initial target nucleus.

\begin{figure}[!htcb]
\vspace{-2cm}
\hspace{-2cm}
\includegraphics[height=7cm, width=8cm, angle=0]{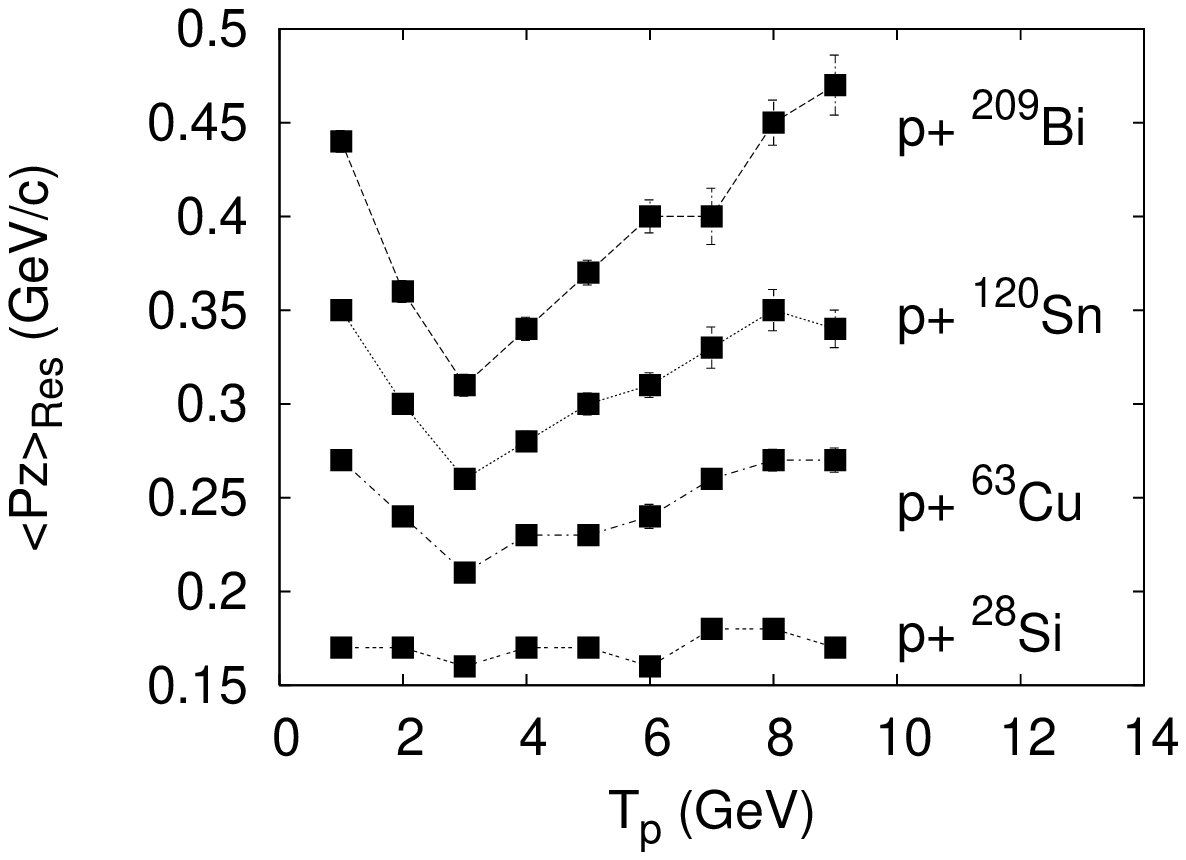}
\hspace{-0.6cm}
\includegraphics[height=7cm, width=8cm, angle=0]{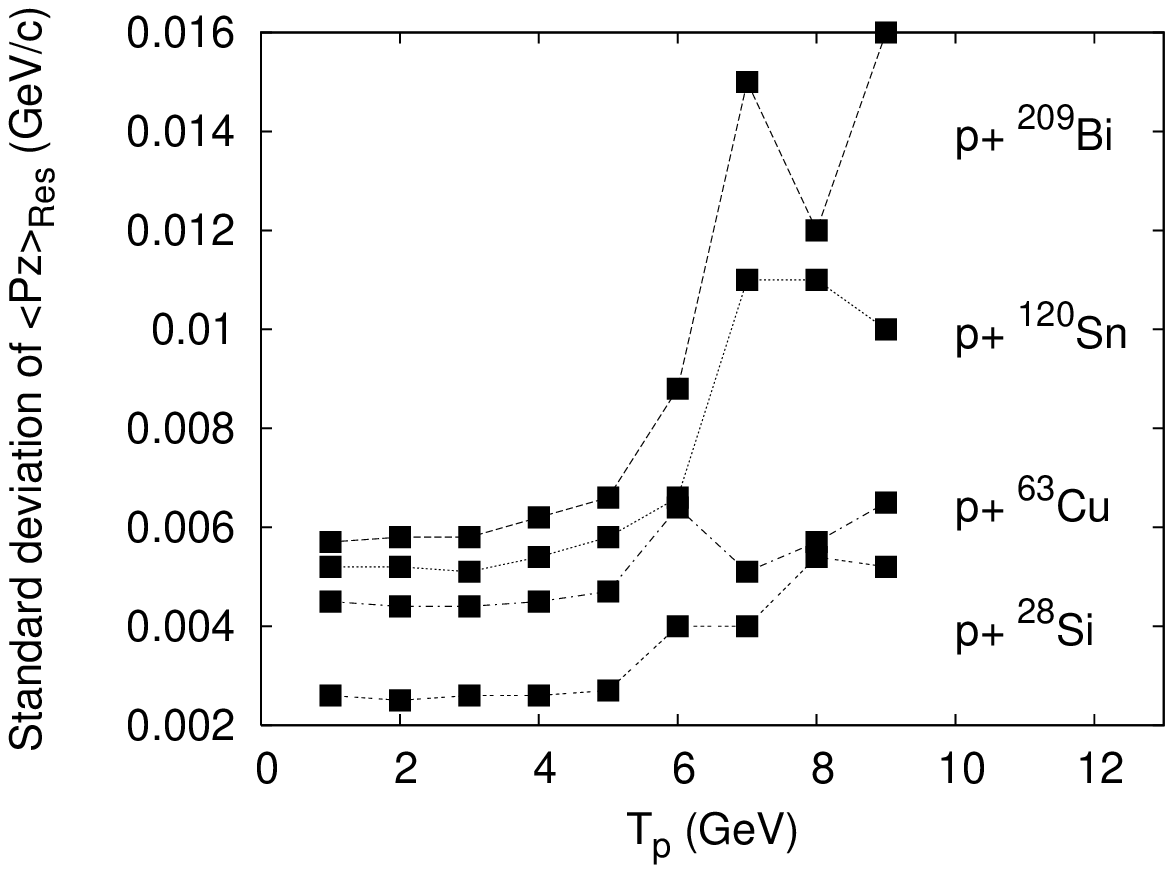}
\caption{{\sl Average values (left) and standard deviation (right) of the 
average values of momentum in beam direction of residual nuclei from proton 
induced reactions on several targets, in function of incident energy; 
results of the HSD model calculations}}
\label{fig:Pzres_stdev_Pzres}
\end{figure}

\begin{figure}[!htcb]
\vspace{-2cm}
\hspace{-2cm}
\includegraphics[height=7cm, width=8cm, angle=0]{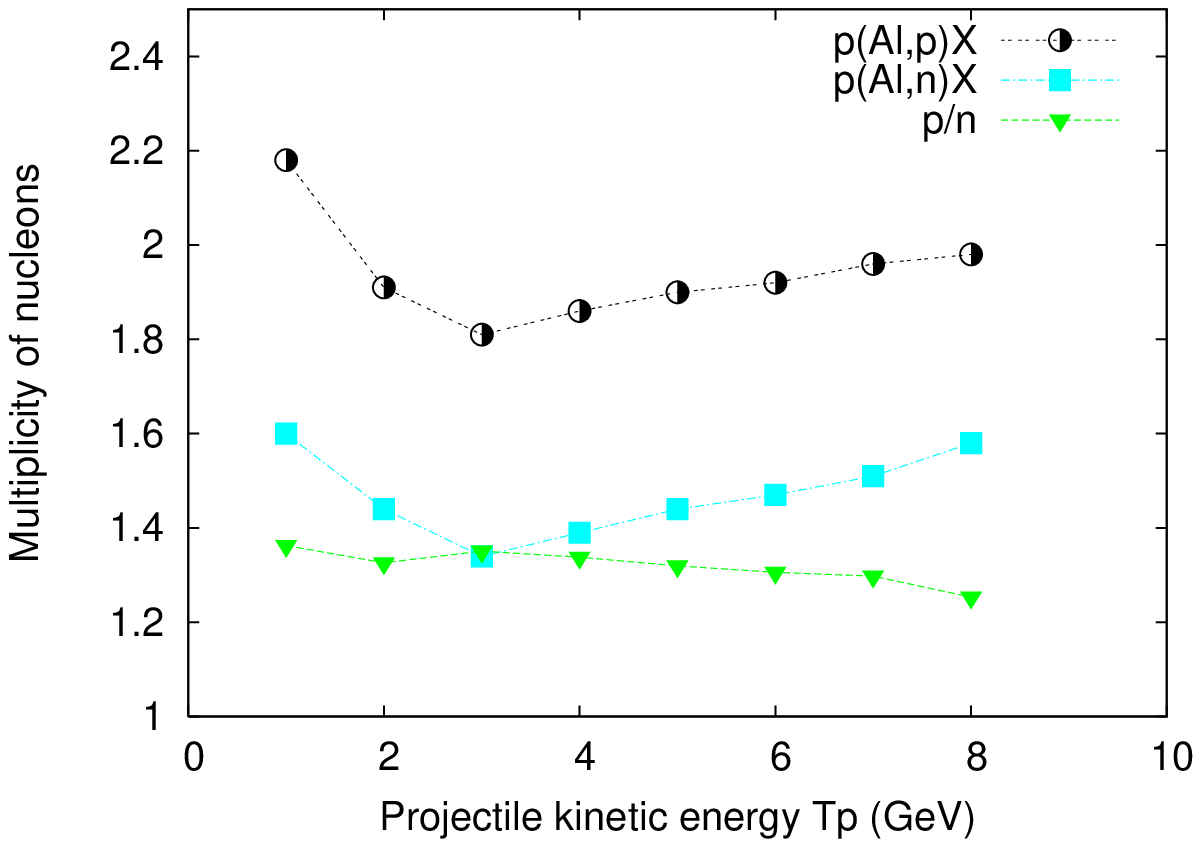}
\hspace{-0.6cm}
\includegraphics[height=7cm, width=8cm, angle=0]{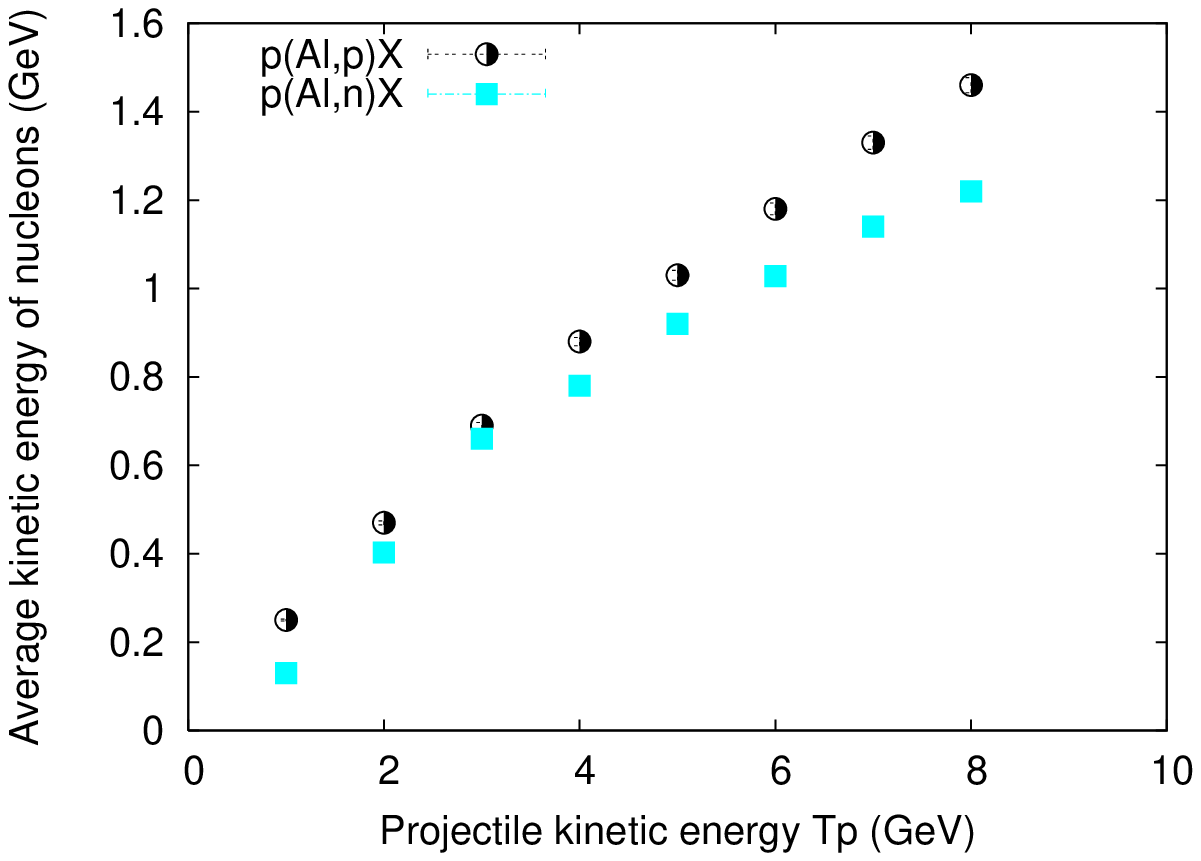}
\caption{{\sl Multiplicity (left) and average values of kinetic energy per 
individual nucleon (right) of nucleons emitted during first stage of proton 
induced reaction on Al target, as function of incident energy; 
results of the HSD model 
calculations (error bars indicate values of standard deviation of the average 
values, divided by a square root of number of events)}}
\label{fig:pAl_mult_ekin}
\end{figure}

Average values and standard deviations of momentum in beam direction of 
residual nuclei after first stage of proton induced reaction on several targets,
 at various incident energies are presented in Fig. \ref{fig:Pzres_stdev_Pzres}.
One can conclude, that the heavier the target, the more momentum in beam 
direction is deposited inside. What more, the maximum in mass dependence on 
projectile kinetic energy (at the incident energy equal to 3.0 GeV) for given 
target, see Fig. \ref{fig:pAl_a_z}, corresponds to the minimum of dependence of 
momentum in beam direction of residual nuclei in function of projectile 
energy.\\
Behavior of standard deviations as function of projectile kinetic 
energy is noteworthy. It is observed, that the higher incident energy, the 
broader are distributions of excitation energy, momentum, angular 
momentum, mass number, charge of residual nuclei.
This can be qualitatively explained taking into account results of simulations 
presented in the Fig. \ref{fig:pAl_mult_ekin} (i.e. dependence of multiplicity 
and energy of emitted nucleons during the first stage of the reactions on 
incident energy). 
The behavior of multiplicity of emitted nucleons as function of projectile 
energy perfectly agrees with the behavior of mass and charge of residual 
nuclei. It means, first, up to about 3.0 GeV of incident energy, number of 
emitted nucleons decreases, so that mass of residuum increases. With further 
increase of impact energy, more and more nucleons are emitted, so that mass of 
the residuum decreases. Obviously, average kinetic energy carried by emitted 
nucleons is a monotonically increasing function of projectile energy in the 
whole considered range of projectile energy, as shown in Fig. 
\ref{fig:pAl_mult_ekin}. Protons carried out in average more kinetic energy 
than neutrons, since more protons than neutrons are emitted during first stage 
of reaction. 
The p/n ratio is larger than unity, because of the extra proton 
correspondig to the incident particle (see Fig. \ref{fig:pAl_mult_ekin}). 
One can conclude that information of type of projectile is somehow remembered 
by the system during the first stage of reaction (this topic is undertaken also
 in Sec. \ref{sec:fission}). 

\begin{figure}[!htcb]
\vspace{-2cm}
\hspace{-2cm}
\includegraphics[height=7cm, width=8cm, angle=0]{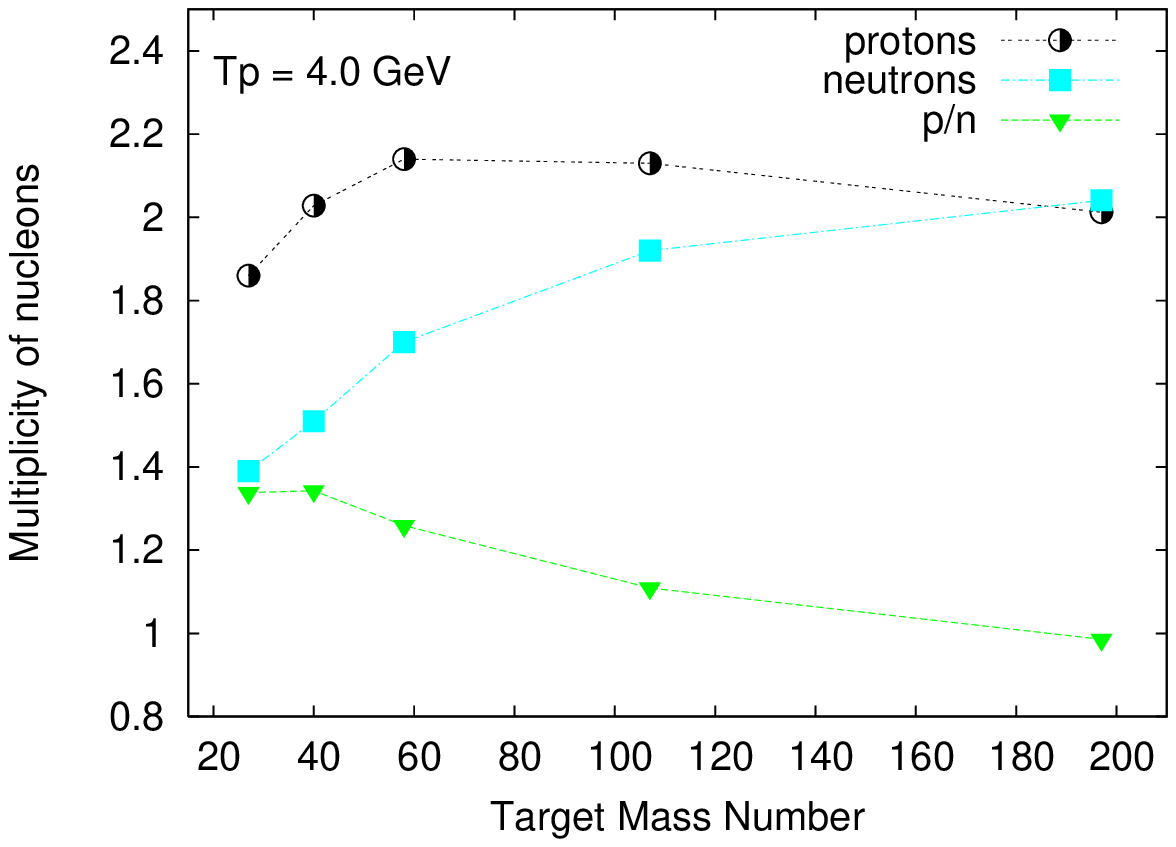}
\hspace{-0.6cm}
\includegraphics[height=7cm, width=8cm, angle=0]{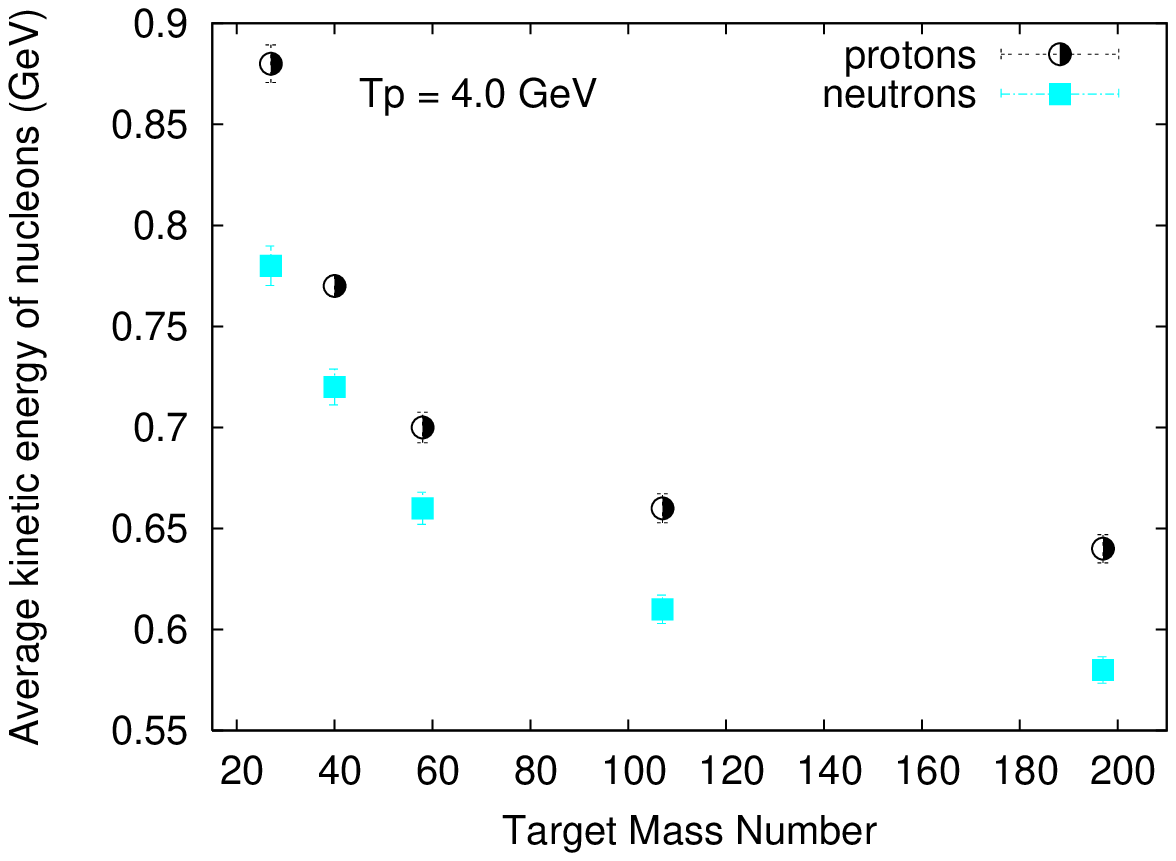}
\caption{{\sl Multiplicity (left) and average values of kinetic energy per 
individual nucleon (right) of nucleons emitted during first stage of 4.0 GeV 
proton induced reactions on a few targets, as function of target mass number; 
results of the HSD model calculations (error bars indicate values of standard
deviation of the average values, divided by a square root of number of events)}}
\label{fig:p4.0_mult_ekin}
\end{figure}

Looking at behaviors of average values of properties of residual nuclei, 
especially their momentum in beam direction, mass number and multiplicity of 
emitted nucleons during first stage of reaction as function
 of incident energy, it is seen that the average values first continuously 
decrease (or increase, respectively) with the incident energy, up to about 3 
GeV and then consistently, the opposite tendency is observed. The precise 
explanation of such behavior is difficult, it must be a result of a few 
simultaneous effects. Newertheless, considering elementary processes 
(implemented to the model, based on experimental data), one can suggest the 
following qualitative explanation. 
Such behaviors should be connected with contribution of different 
resonances excitations, which depends on the energy introduced by a projectile 
into the target nucleus. 
At low projectile energy (about 1.0 GeV) nucleons are emitted mainly due to 
elastic nucleon - nucleon scattering. More probable than at higher energies is 
also, that incoming proton stops inside target nuclei.  
At higher incident energies possibility of excitation of resonances (mainly, 
the most important here $\Delta$ resonance) increases. 
A bit less, but more energetic nucleons are emitted. In result, the average 
multiplicity of emitted nucleons decreases, 
so mass of residual nuclei increases and the average longitudinal momentum 
decreases.   
At higher incident energy (equal to about 3.0 GeV), from one side the 
probability of $\Delta$ resonance excitation becomes smaller, and from the 
other side probability of excitation of higher resonance is still too low. 
This corresponds to the minimum (or maximum) of the incident energy dependences
 of the considering quantities.  Then, with further increase of projectile 
energy, probability of excitation of other resonances increases. 
This causes, that more energy and momentum is deposited into residual 
nuclei. So, the average longitudinal momentum of residual nuclei increases. 
Simultaneously, more particles could be produced, since the ratio of inelastic 
to elastic collision increases. The multiplicity of 
emitted nucleons increases, so the average mass of remnants decreases.\\  
Behavior of multiplicity and average kinetic energy of emitted nucleons, 
as function of target mass number, at an example impact energy is shown in 
Fig. \ref{fig:p4.0_mult_ekin}.
One can see, that the heavier target, the smaller differences in 
number of emitted protons and neutrons. For the heaviest presented target, the 
p/n ratio is equal to the unity. One can conclude that in heavy targets, the 
information about a type of projectile is lost due to possible large number of
 mutual reactions. Total number of emitted nucleons increases with mass of 
target nucleus and stabilizes for heavier targets (with mass numbers greater 
than 60). 
In heavier targets, the cascade of intra-nuclear collisions can last longer, 
so that emitted nucleons carries in average less kinetic energy. The nucleon, 
on its way, going outside the nucleus, has to penetrate through nuclear skin of
 the nucleus, with high probability undergoes further collisions. As result, 
more particles with suppressed energies are emitted.    

\begin{figure}[!htcb]
\vspace{-3.5cm}
\hspace{-2.5cm}
\includegraphics[height=9cm, width=10cm, angle=0]{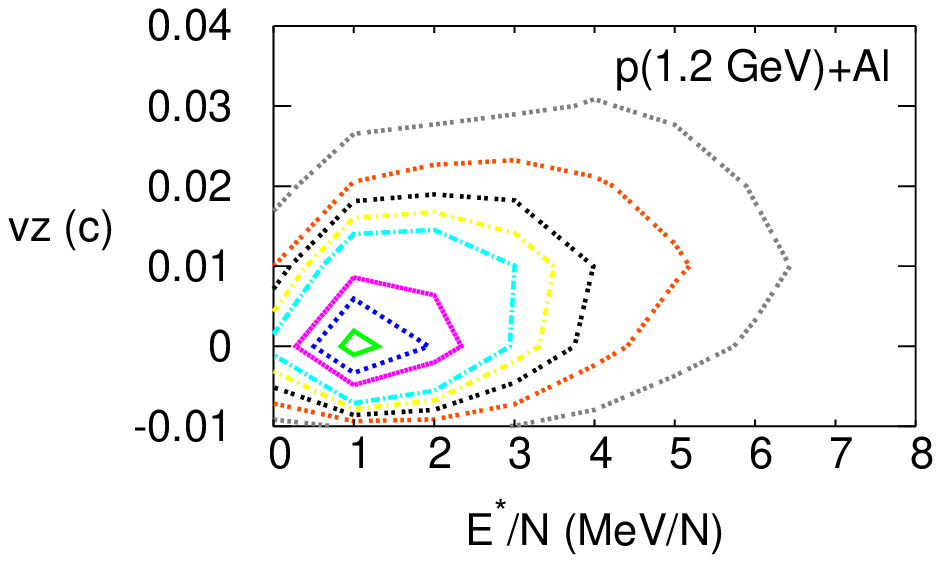}
\hspace{-2.5cm}
\includegraphics[height=9cm, width=10cm, angle=0]{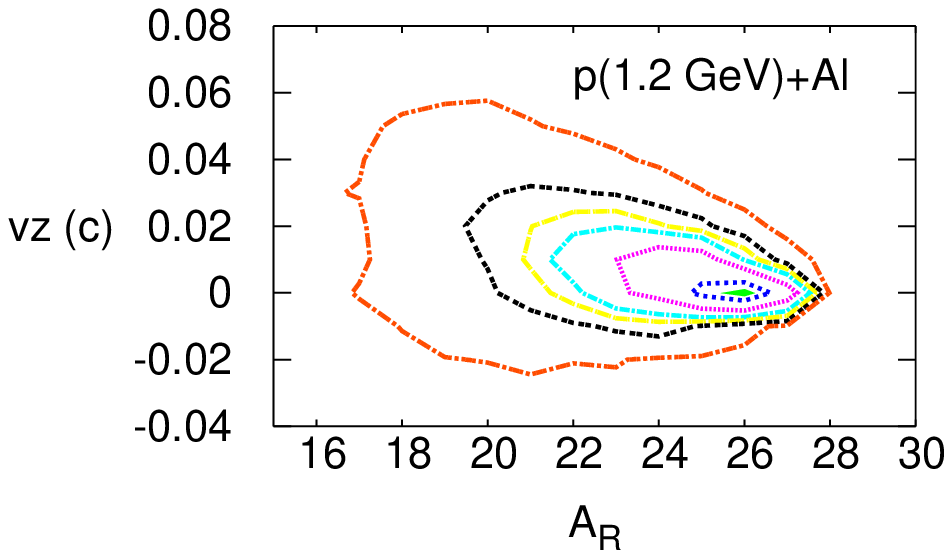}
\caption{{\sl Two - dimensional distributions of velocity in beam direction of 
residual nuclei and (a) excitation energy per nucleon of residual 
nucleus and (b) mass of residual nuclei, for exemplary reaction 
p+Al at 1.2 GeV of incident energy; results of the HSD model calculations 
(the most central contour line corresponds to a maximum of the distribution)}}
\label{fig:vz_ar_excnn}
\end{figure}

\begin{figure}[!htcb]
\vspace{-2cm}
\hspace{-2cm}
\includegraphics[height=6cm, width=8cm, angle=0]{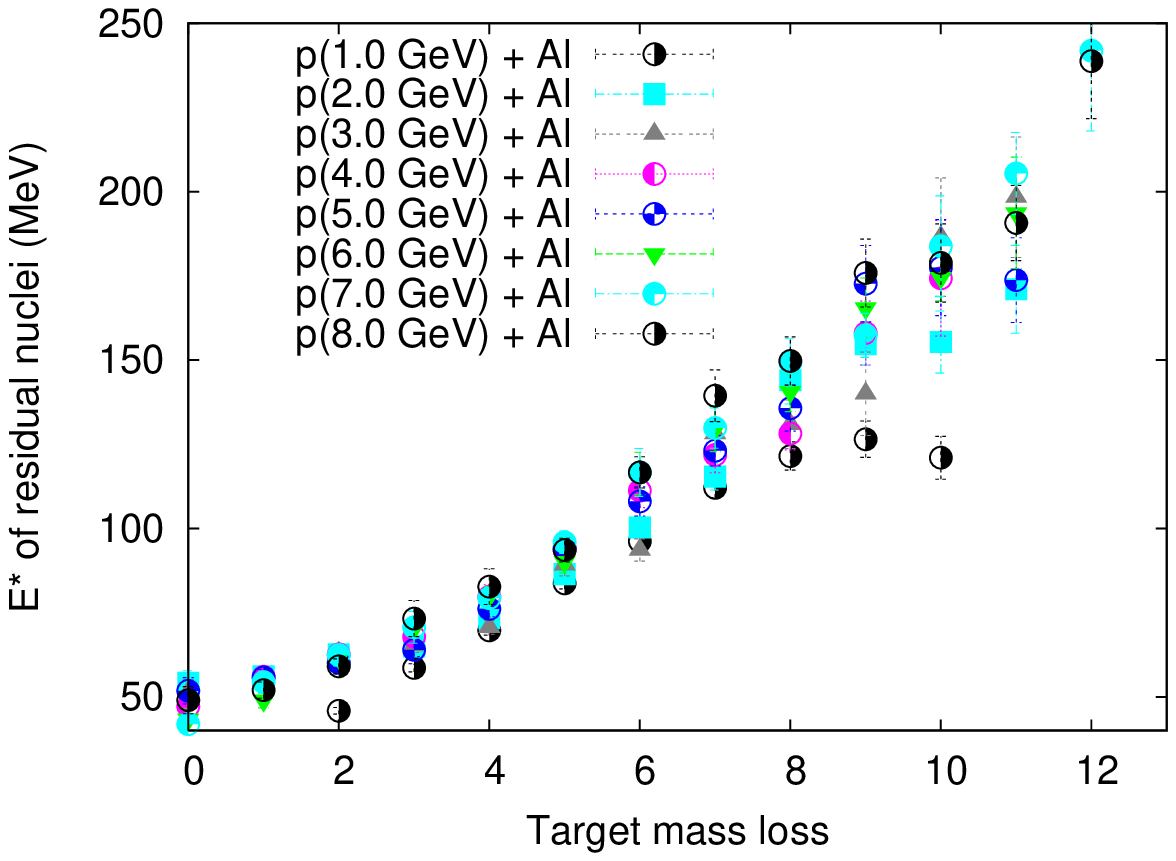}
\vspace{-0.1cm}
\includegraphics[height=6cm, width=8cm, angle=0]{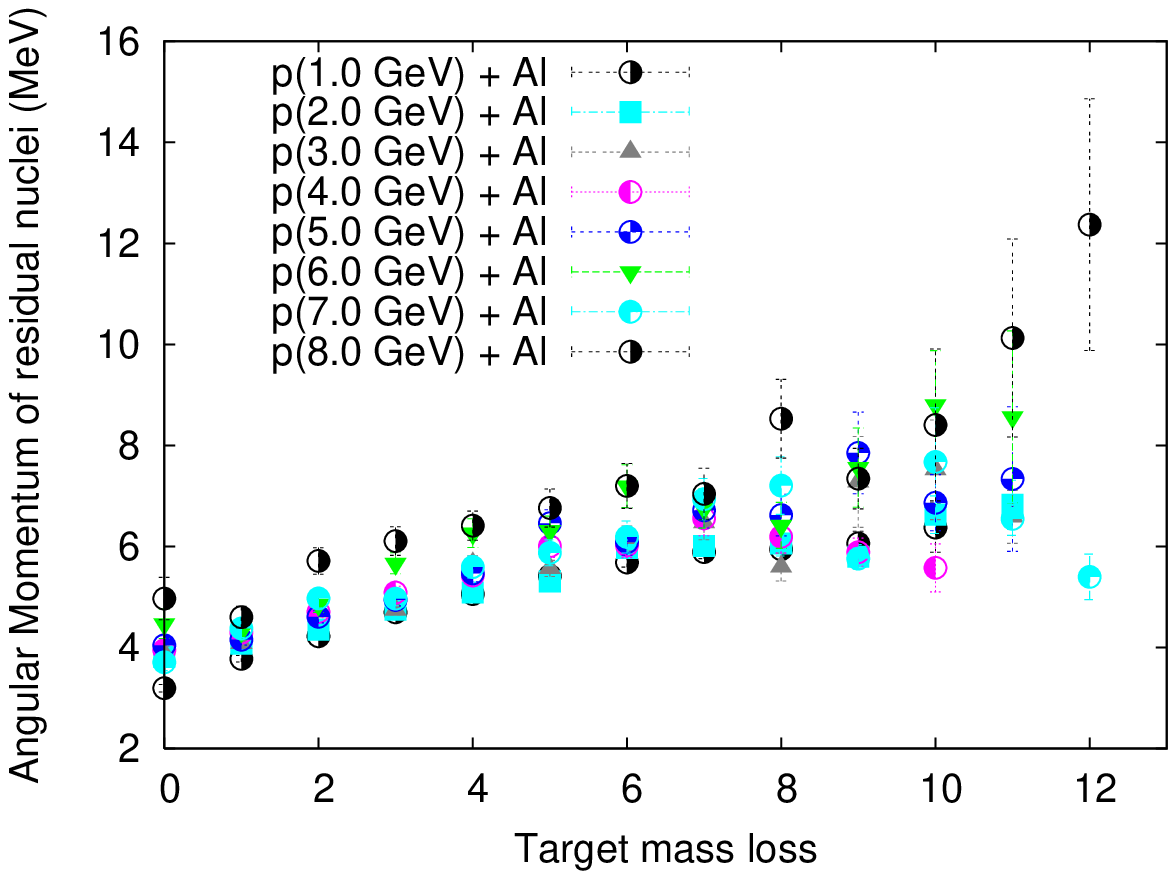}
\caption{{\sl Regression function of excitation energy (left) and angular 
momentum (right) of residual nuclei on target mass loss, for example p+Al 
reaction, at several values of incident energy; results of the HSD model 
calculations (error bars indicate values of standard deviation of average 
values of excitation energy and angular momentum, respectively, for adequate 
target mass loss, divided by a square root of number of events)}}
\label{fig:kr_da_exc_lres_Al}
\end{figure}

\begin{figure}[!htcb]
\vspace{-2cm}
\hspace{-2cm}
\includegraphics[height=6cm, width=8cm, angle=0]{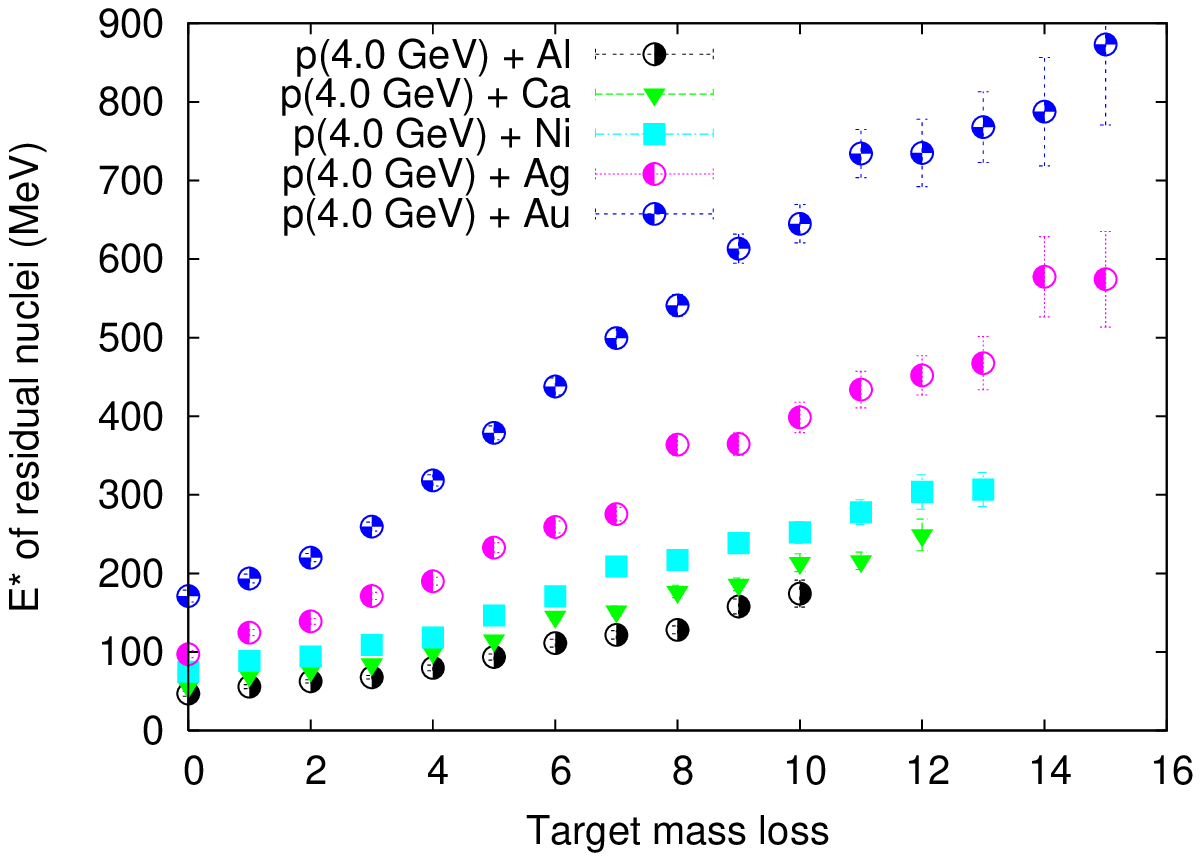}
\vspace{-0.1cm}
\includegraphics[height=6cm, width=8cm, angle=0]{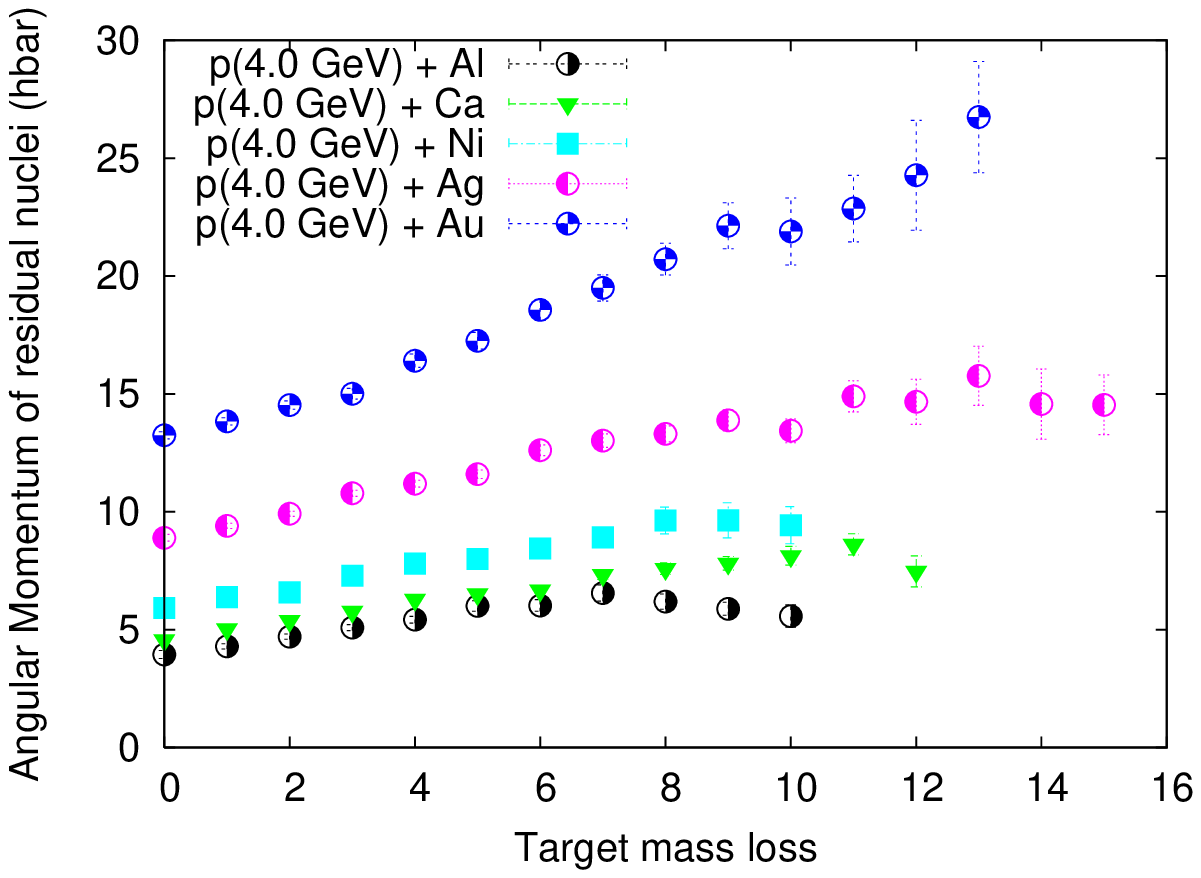}
\caption{{\sl Regression function of excitation energy (left) and angular 
momentum (right) of residual nuclei on target mass loss, 
for proton induced reactions on Al, Ca, Ni, Ag and Au nuclei, at an example 
value 4.0 GeV of incident energy; results of the HSD model calculations 
(error bars indicate values of standard deviation of average values of 
 excitation energy  and angular momentum for adequate target mass loss, 
divided by a square root of number of events)}}
\label{fig:kr_da_exc_lres_4.0}
\end{figure}

More informations about the excited nuclei is provided by two - dimensional 
correlation plots of their properties. Example dependences of excitation energy,
 mass of residual nuclei and their velocity in the beam direction are shown in 
Fig. \ref{fig:vz_ar_excnn} (velocity versus excitation energy per nucleon and 
velocity versus residual mass).   
It is meaningful, that velocity in beam direction of residual nuclei  
increases with increasing excitation energy deposited inside the residual 
nuclei. This indicates, that the energy is comprised also in onward movement of 
residual nuclei. 
Simultaneously, the velocity increases with decreasing mass of residual 
nuclei, what is equivalent to increase of number of ejected particles.  
One can conclude that also excitation energy of residual nuclei should 
increase with decrease of mass of the nuclei.
In Figures \ref{fig:kr_da_exc_lres_Al} and \ref{fig:kr_da_exc_lres_4.0}, 
regression function of excitation energy on target mass loss and  
regression function of angular momentum on target mass loss are presented    
(for residual nuclei formed in p+Al reaction, at a few impact energies and 
proton induced reaction on several targets at example 4.0 GeV incident energy, 
respectively). 
The dependences are roughly linear functions of the target mass loss.   

It is seen, that the heavier remnants (for the same initial target), the lower 
average excitation energy is deposited inside.
The average excitation energy increases with decreasing of the remnant mass.
Based on the Fig. \ref{fig:kr_da_exc_lres_4.0}, one can conclude that 
particularly the dependence on mass of target is quite pronounced. It is 
because, as it is already mentioned above, in heavier targets, where the 
cascade lasts longer, more energy can be accumulated.  
In the Figures \ref{fig:kr_da_exc_lres_Al} and \ref{fig:kr_da_exc_lres_4.0},
it is also seen, that in the cases of the lowest mass of residual nuclei, (i.e.
the target mass loss higher than about 7 nucleons), the increase of excitation 
energy and angular momentum is not only less pronounced, but even a slight 
decreases is observed. 
In such cases, i.e. more ejected particles, the intra - nuclear cascade is 
lasting for longer time, so that particles with rather moderated energies are 
emitted.

\begin{figure}[!htcb]
\vspace{-2cm}
\hspace{-1cm}
\includegraphics[height=19cm, width=18cm, bbllx=0pt, bblly=80pt, bburx=594pt, bbury=842pt, clip=, angle=0]{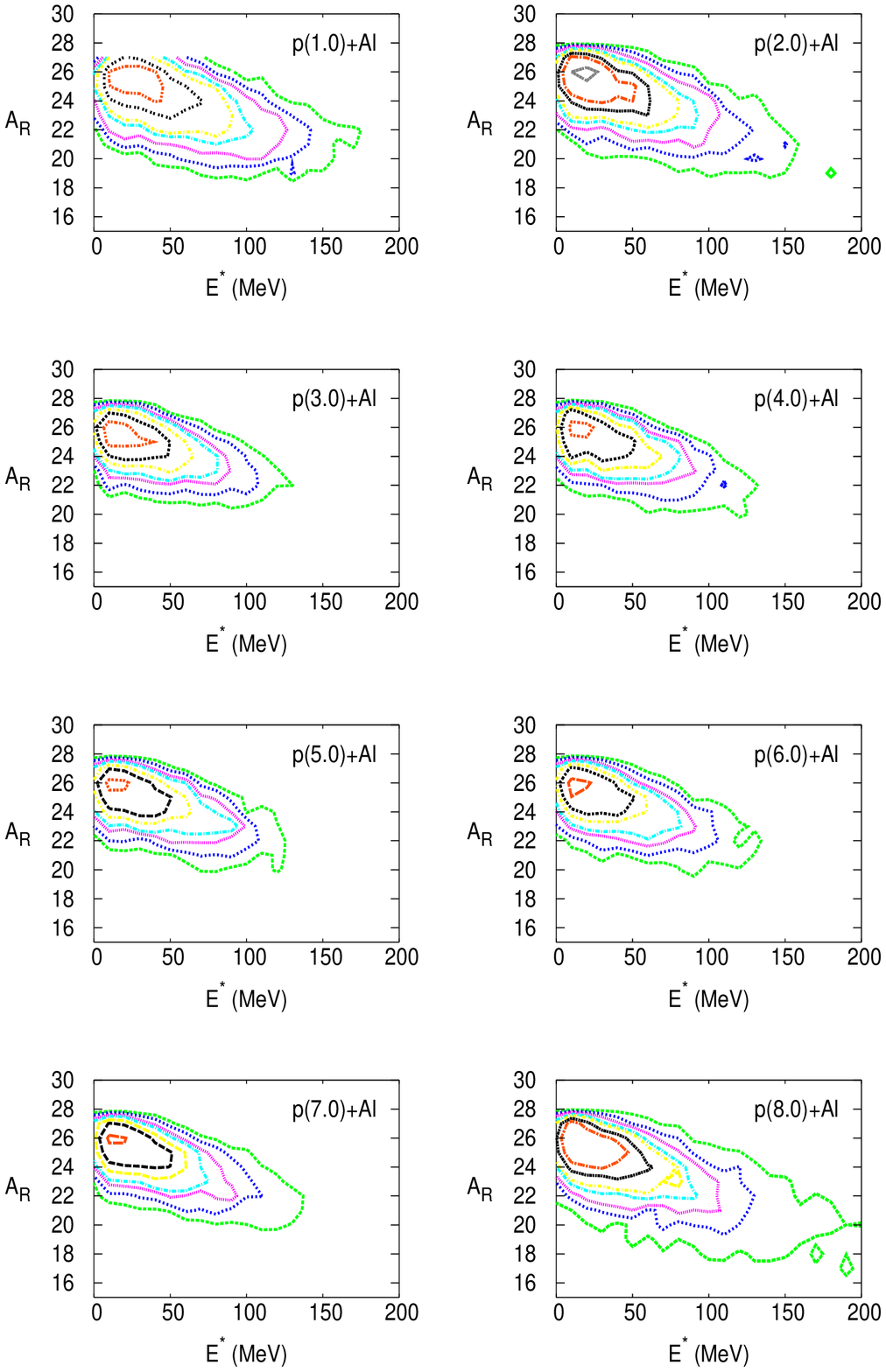}
\caption{{\sl Contour plots of the two-dimensional distributions of excitation 
energy and mass of residual nuclei at the end of first stage of proton induced 
reaction on Al target, at several values of incident energy; results of the HSD
model calculations (the most central contour line corresponds to a maximum of 
the distribution, the more and more outer contours correspond to the decrease 
of a yield)}}
\label{fig:ar_exc_pAl}
\end{figure}

Example two - dimensional distributions of mass and excitation energy of 
residual nuclei formed in p+Al reaction, at several projectile energies are 
displayed in Fig. \ref{fig:ar_exc_pAl}. Again, one observes that the higher 
excitation energy of residual nuclei, the smaller their mass.

\begin{figure}[!htcb]
\vspace{-2cm}
\hspace{-2cm}
\includegraphics[height=16cm, width=15cm, angle=0]{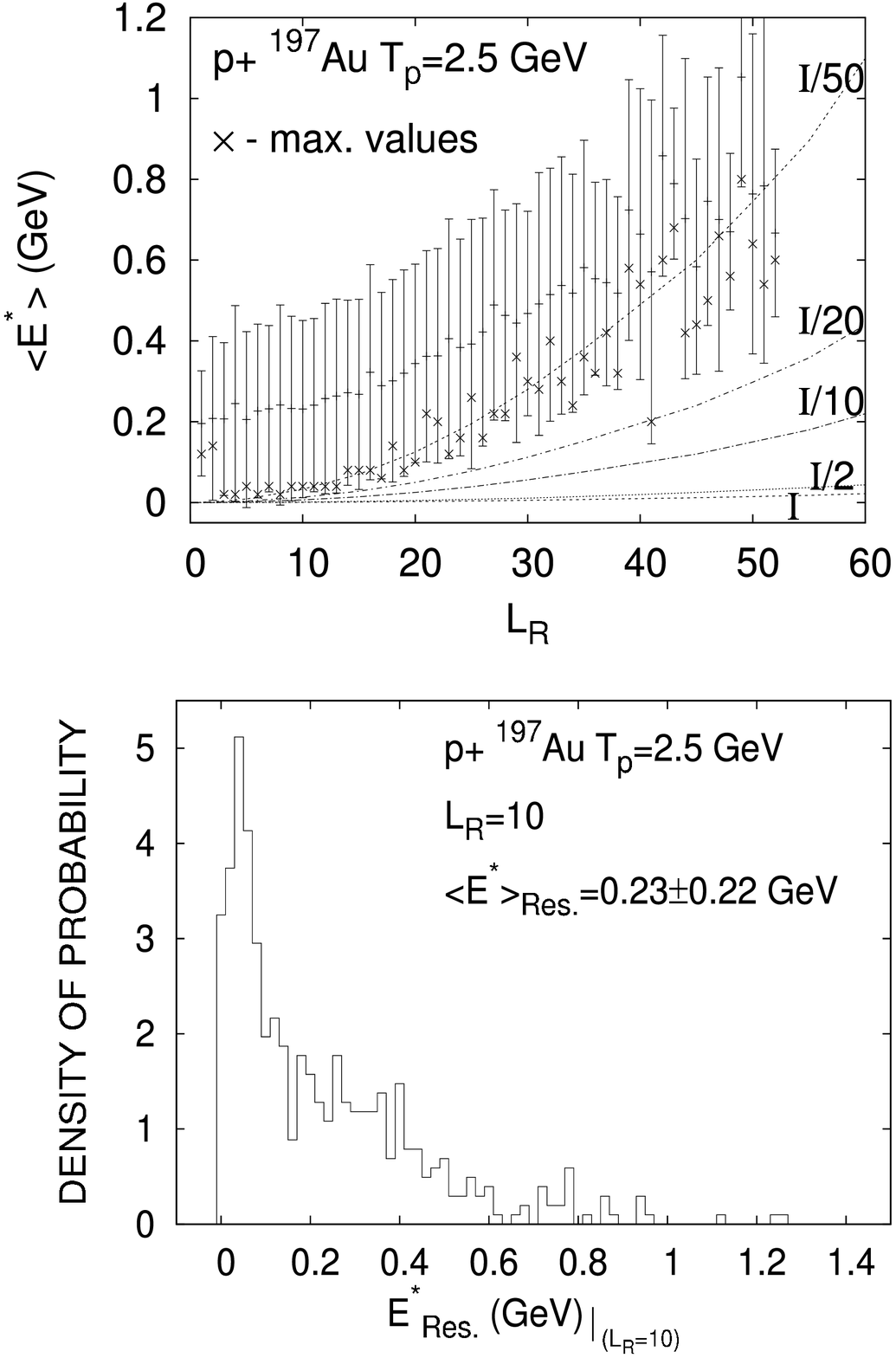}
\caption{{\sl Upper part: regression function of excitation energy on angular
momentum of residual nuclei from an example proton induced reaction on Au 
target, at 2.5 GeV of incident energy; the five lines correspond to the Yrast 
lines calculated with different values of moment of inertia;   
lower part: distribution of the excitation energy for a chosen value 
(10 $\hbar$) of angular momentum; results of the HSD model calculations}}
\label{fig:kr_exc_lr_pAu}
\end{figure}

Dependence of excitation energy on angular momentum of residual nuclei is 
presented in Fig. \ref{fig:kr_exc_lr_pAu}.
The five lines (in upper part of the Fig. \ref{fig:kr_exc_lr_pAu}) 
correspond to the Yrast lines (definition is given in Chapter 
\ref{chapt:statist_emiss}) calculated according to the formula: 
$E^{*}=\frac{\hbar^{2}}{2 I} L_{R}(L_{R}+1)$ with different values of the 
moment of inertia ($I$), where: $I$ is the moment of inertia of 
rigid body ($I=\frac{2}{5} M R^{2}$, where $M$ and $R$ stand for mass and 
radius of nucleus, respectively), $I/50$ corresponds to the hydrodynamical 
limit of nucleus (meaning that only a part of nucleus behaves like the rigid 
body, the other part - in uncorrelated way, see Ref. \cite{Strz79}).   
One observes that angular momentum increases with increase of the average value
 of excitaton energy. It is evident, that the dependence is not inconsistent 
with the Yrast line. 

Physicists, who are studing fragmentation have found that this process occurs 
at some particular value of projectile energy (T$_{p}$) (based on (p,$^{7}$Be) 
reaction, for various targets at different T$_{p}$, see \cite{Buba04}). 
Fragmentation is a phase transition, what means, that a proper amount of energy
 needs to be pumped into a system.  
It follows, that fragmentation occurs at some particular value of 
$(E^{*}/N)_{R}$, which experimentally has been found to be at least equal to 
about 5 MeV/N \cite{Beau00, Viol06}.
\begin{figure}[!htcb]
\vspace{-2cm}
\hspace{-2cm}
\includegraphics[height=16cm, width=15cm, angle=0]{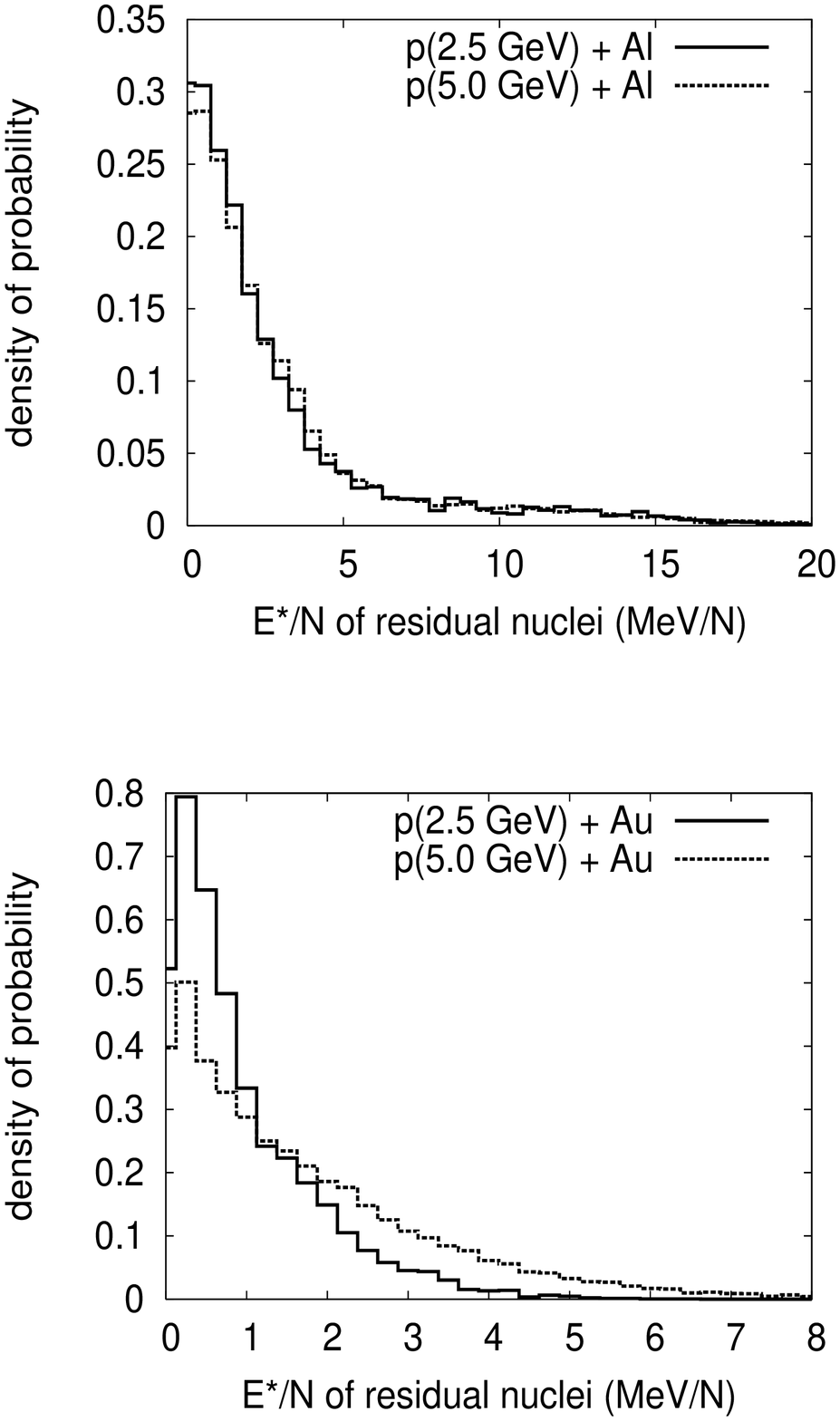}
\caption{{\sl Distributions of excitation energy per nucleon of residual 
nuclei from an example proton induced reaction on Al and Au target, at 2.5 GeV 
and 5.0 GeV of incident energy; results of the HSD model calculations}}
\label{fig:69nn_pAl_pAu_2.5_5.0}
\end{figure}
Using the condition of 5 MeV/N to the HSD model results, presented  
in Fig. \ref{fig:69nn_pAl_pAu_2.5_5.0}, it is seen that fragmentation is 
a very low probable process in proton induced reactions. For example, in p+Al 
reaction at T$_{p}$=5.0 GeV, fragmentation of excited nucleus may occur in 
about 17 $\%$ of cases (i.e. the part of spectrum for $(E^{*}/N)_{R} >$ 5 MeV/N 
is 17 $\%$ of the whole $(E^{*}/N)_{R}$ spectrum), whereas in p+Au 
reaction at the same incident energy, only in about 5
$\%$ of cases. These are cases from the tail of $(E^{*}/N)_{R}$ distribution. 
Based in the Fig. \ref{fig:69nn_pAl_pAu_2.5_5.0} one can conclude that 
fragmentation process is more probable in reactions on light than on heavy 
targets, and at higher incident energies.   

One can suppose that fragmentation of only a part of nucleus could be more 
probable. This requires inhomogeneous excitation of residual nucleus. It means, 
not all of nucleons inside the nucleus are involved in carrying of the 
excitation, i.e. for different target, $(E^{*}/N)_{R}$ should be calculated not
 per each nucleon, but only per some amount of nucleons, which take active part 
in the cascade. 
However based on observations of evolution of nucleon density and distribution 
of kinetic energy inside excited nuclei, it is seen that such partial heating 
up of nucleus is very low probable. It occurs only in about 1 $\%$ of cases, 
as discussed in Chapter \ref{chapt:bulk_prop}.

Additionally, experimentalists, who are interested in fragmentation, look for   
it in central collisions, accompanied with high multiplicity of emitted 
particles, as it is shown qualitatively by results of the HSD model 
calculations, presented in Fig. \ref{fig:1705_50_xyz_pAu5.0}. 
Nevertheless, in the HSD simulations, 
the very low probable partial heating up of nucleus has been found in 
peripheral collisions (see Fig. \ref{fig:1705_50_75_100_b.5.0_xz_pAu2.5}).    

Investigated here reactions are very low invasive processes, involving
only a few nucleons, causing only minor fluctuations of nucleon density. 
So, it is evident that there is a negligible probability for fragmentation   
in proton - nucleus interacting system, in energy range considered in the 
frame of this work. Spallation is evidently dominant process.     

\section{Parametrization}
\label{sec:param}
\markboth{5.1 Parametrization}{Chapter 5. Properties of residual nuclei}

\begin{figure}[!htcb]
\vspace{-1cm}
\hspace{-2cm}
\includegraphics[height=18cm, width=16cm, angle=0]{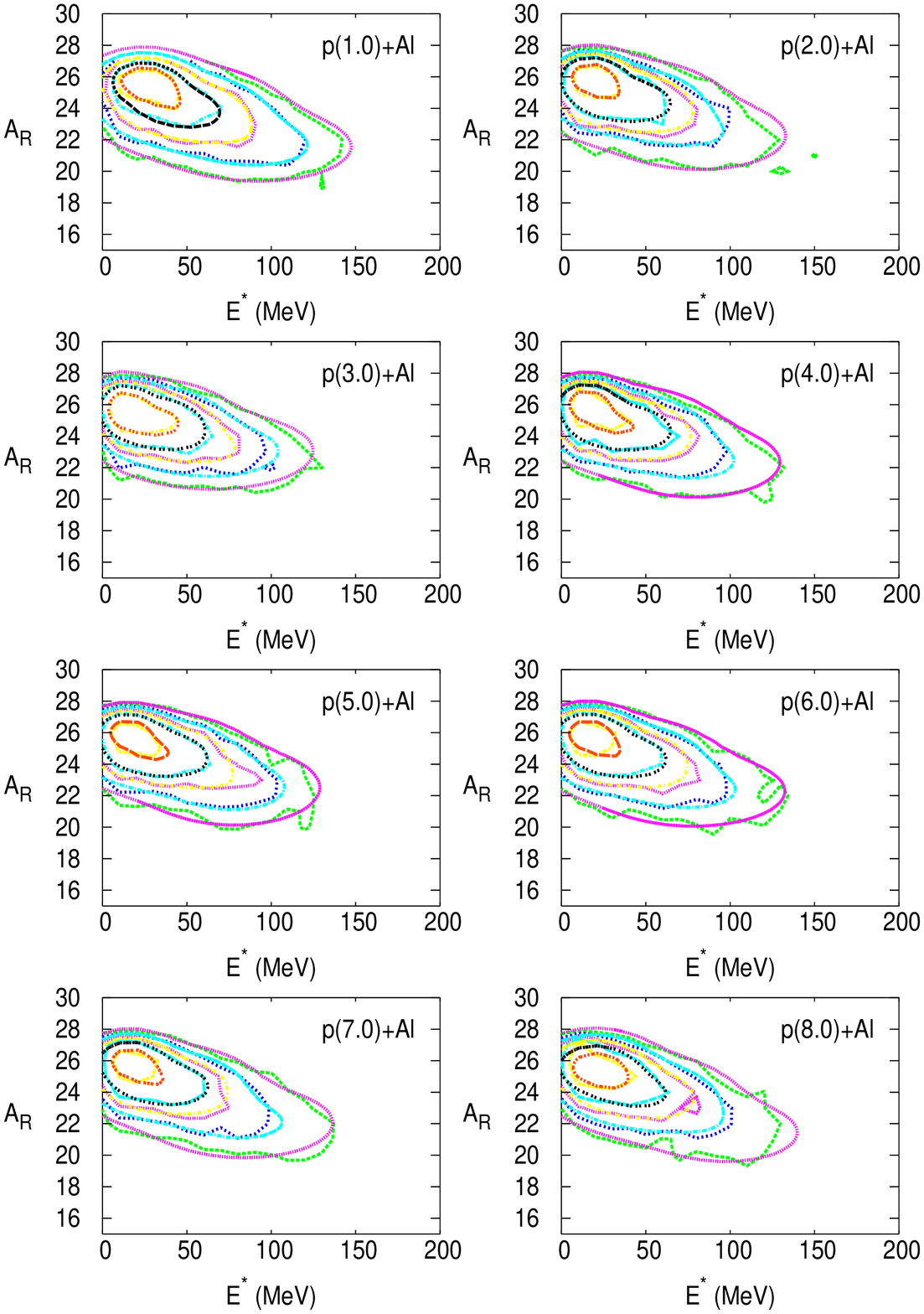}
\caption{{\sl Contour plots of the two-dimensional distributions of excitation
energy and mass of residual nuclei in p+Al reactions, at several values 
of beam energy; jagged lines correspond to results of the HSD model 
calculations, smooth lines define parametrization 
(the most central contour line corresponds to a maximum of the distribution, 
the more and more outer contours correspond to the decrease of a yield)}}
\label{fig:fit_ar_exc_pAl}
\end{figure}

\begin{figure}[!htcb]
\vspace{-2cm}
\hspace{-2cm}
\includegraphics[height=19cm, width=17cm, angle=0]{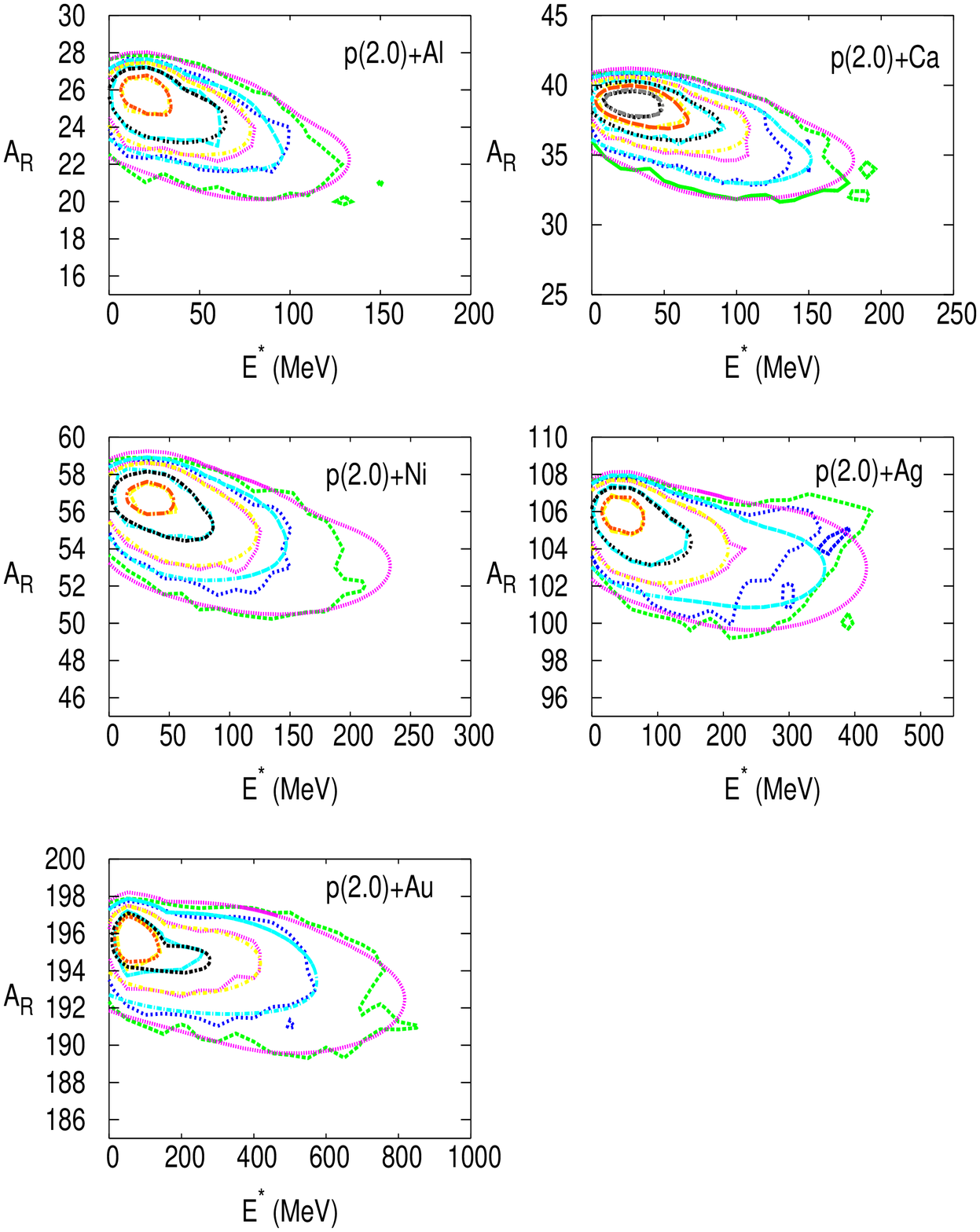}
\caption{{\sl Contour plots of the two-dimensional distributions of excitation
energy and mass of residual nuclei in 2.0 GeV proton induced reaction on
 different targets; jagged lines correspond to results of the HSD model 
calculations, smooth lines define parametrization (the most central contour 
line corresponds to a maximum of the distribution, the more and more outer 
contours correspond to the decrease of a yield)}}
\label{fig:fit_ar_exc_2.0}
\end{figure}

It is impossible to measure or even to calculate the reaction of proton of
all energies on all target nuclei. One comes to idea to find a global
parametrization for distributions of excitation energy versus mass of residual
nuclei for some chosen sample of targets and energies, involving finally
dependency on the target mass and projectile energy, in order to have the
possibility to interpolate for other energies and targets. \\
Two-dimensional distributions of excitation energy and mass of residual
nuclei remaining after first stage of proton induced reaction, on a few targets,
 at several projectile energies have been parametrized by a sum of at least 
three two - dimensional Gaussians:
\begin{equation}
f(A_{R},E^{*})=\sum_{i=0,1,2} a_{i} \cdot exp(-b_{i}(A_{R}-A_{R0i})^{2} - c_{i}
(E^{*}-E^{*}_{0i})^{2})
\label{eq:gauss3_ar_exc}
\end{equation}
where: $a_{i}$ stands for heights of the Gaussians, $b_{i}$ and $c_{i}$ -
define widths in $A_{R}$ and $E^{*}$ directions, likewise $A_{R0i}$ and
$E^{*}_{0i}$ - the mean position of the Gaussians in $A_{R}$ and $E^{*}$
directions, respectively.
Example results of such parametrization are presented in Figures 
\ref{fig:fit_ar_exc_pAl} and \ref{fig:fit_ar_exc_2.0}. It is seen, that the
distributions, calculated with the HSD model, are very well described by a sum
of Gaussians (\ref{eq:gauss3_ar_exc}).

\begin{figure}[!htcb]
\vspace{-2cm}
\hspace{-2cm}
\includegraphics[height=21cm, width=19cm, bbllx=0pt, bblly=80pt, bburx=594pt, bbury=842pt, clip=, angle=0]{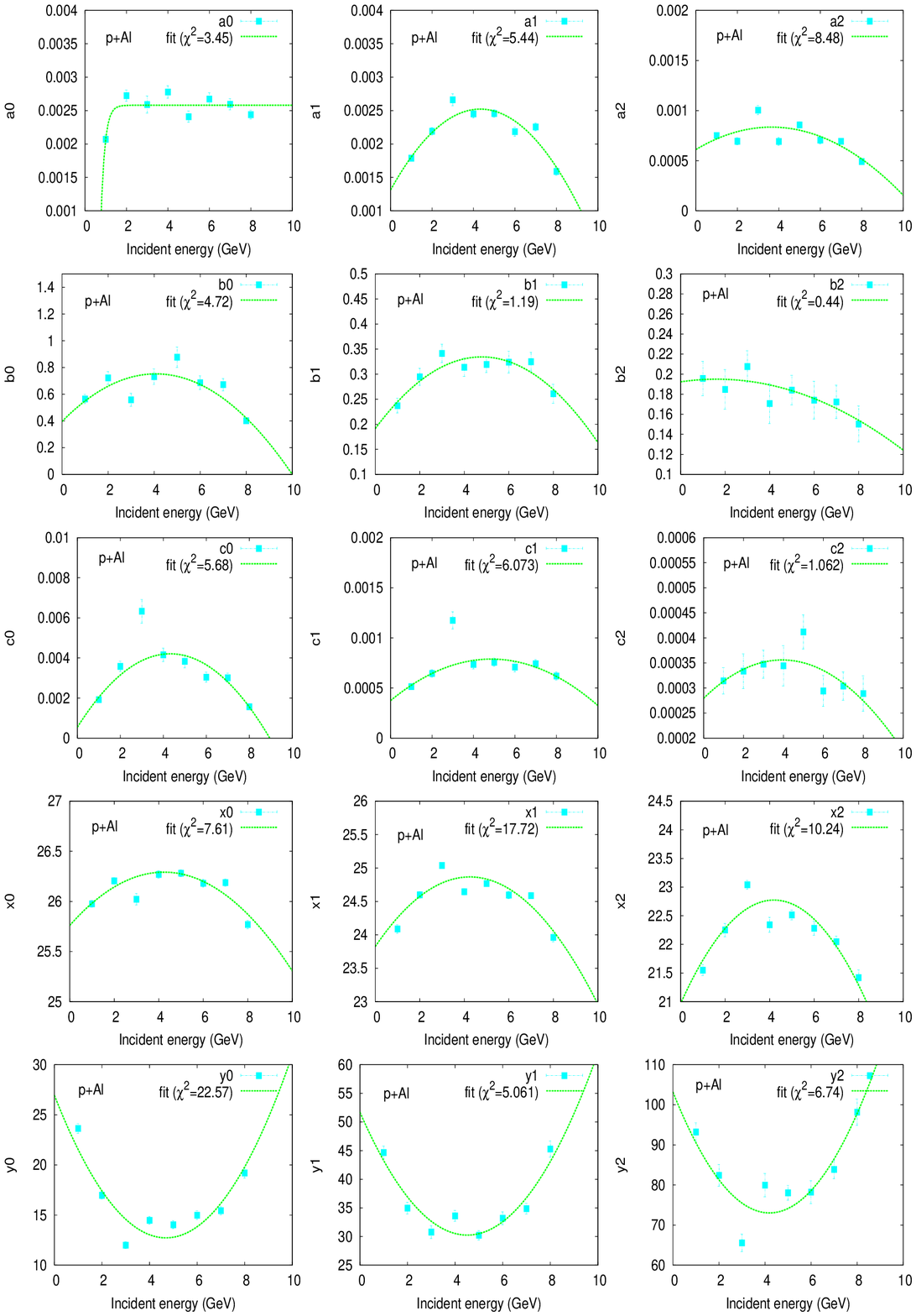}
\caption{{\sl Parametrization of values of the Gaussians parameters as
function of incident energy for the p+Al reaction; e.g. $A_{R01} \equiv$ x1, 
$E^{*}_{01} \equiv$ y1, see formula \ref{eq:gauss3_ar_exc} 
($\chi ^{2}$ means chisquare per degree of freedom)}}
\label{fig:par_Tp_pAl}
\end{figure}

\begin{figure}[!htcb]
\vspace{-2cm}
\hspace{-2cm}
\includegraphics[height=21cm, width=19cm, bbllx=0pt, bblly=80pt, bburx=594pt, bbury=842pt, clip=, angle=0]{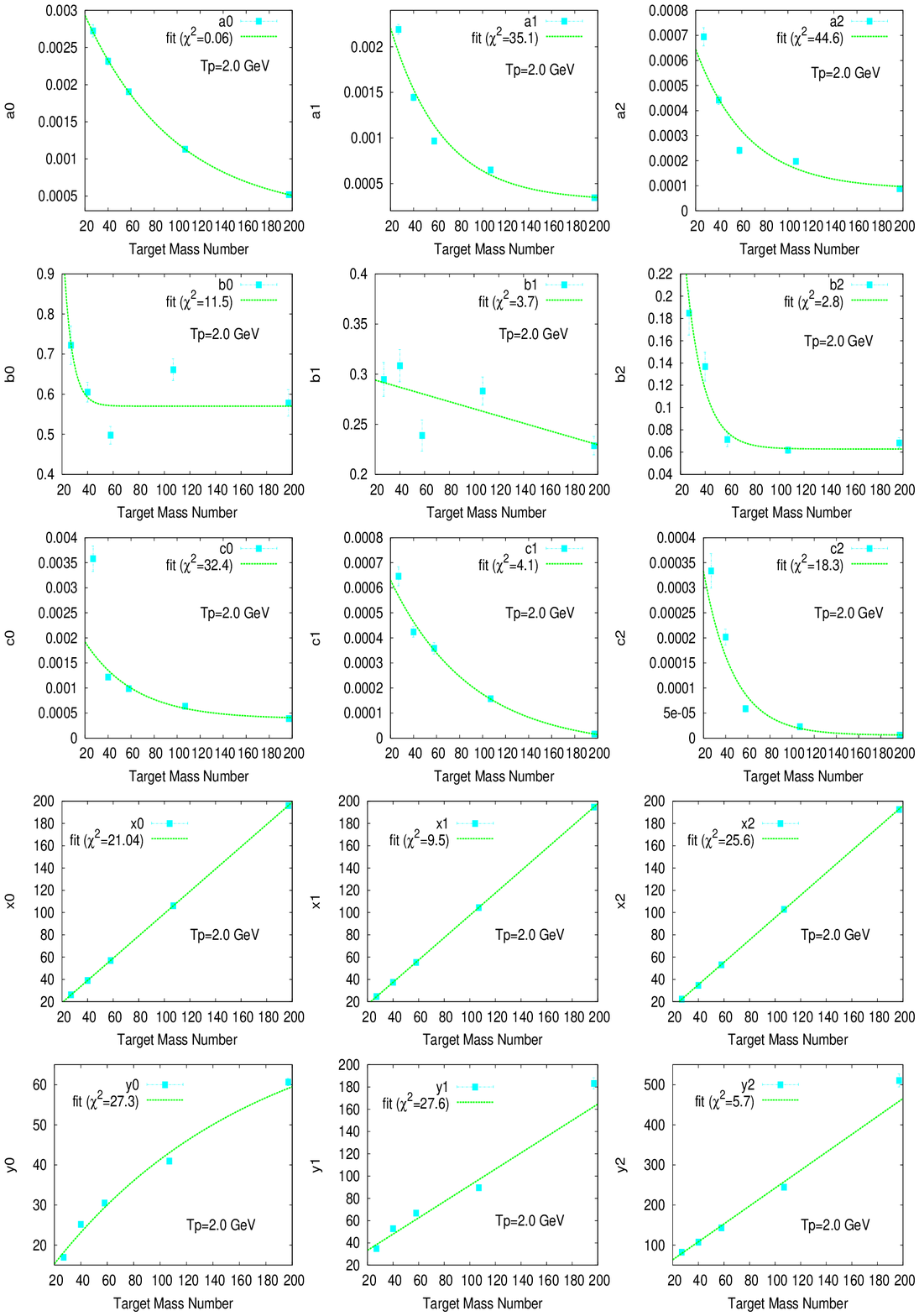}
\caption{{\sl Parametrization of values of the Gaussians parameters as function
 of target mass number for the 2.0 GeV proton induced reaction; e.g. $A_{R01} 
\equiv$ x1, $E^{*}_{01} \equiv$ y1, see formula \ref{eq:gauss3_ar_exc} 
($\chi ^{2}$ means chisquare per degree of freedom)}}
\label{fig:par_Tp_2.0}
\end{figure}

Values of parameters of the Gaussians evolve rather smoothly with incident
energy and mass of target nuclei. Example results of parametrization of such 
dependences are presented in Figures \ref{fig:par_Tp_pAl} and 
\ref{fig:par_Tp_2.0}.

\begin{figure}[!htcb]
\vspace{-2cm}
\hspace{-2cm}
\includegraphics[height=10cm, width=9.5cm, angle=0]{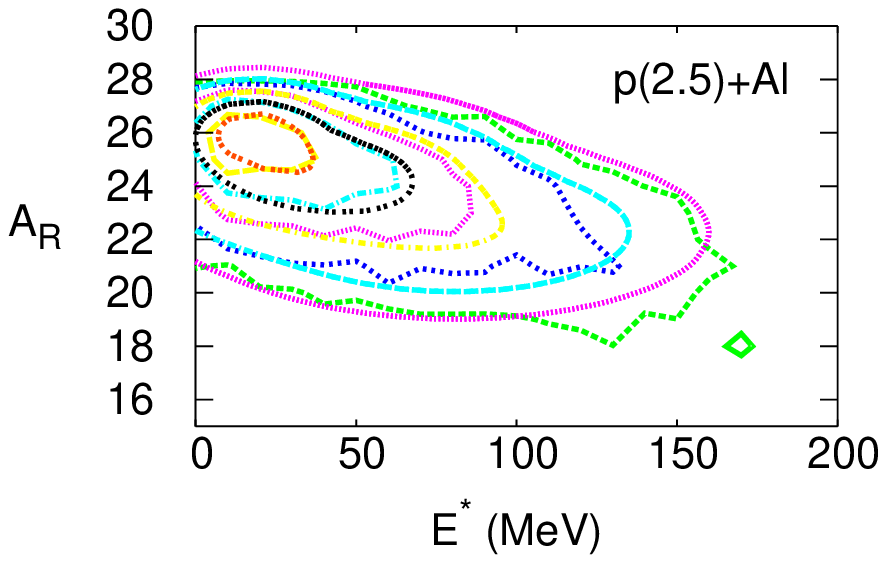}
\hspace{-2.5cm}
\includegraphics[height=10cm, width=9.5cm, angle=0]{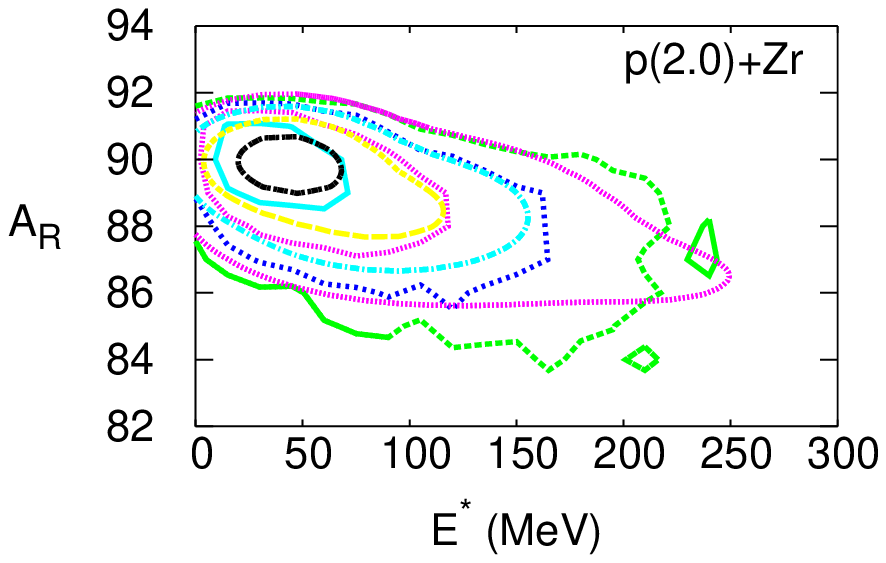}
\caption{{\sl Comparison of interpolated distributions and results of the HSD 
calculations (smooth and jagged lines) for example p+Al reaction at 2.5 GeV 
beam energy and p+Zr reaction at 2.0 GeV beam energy}}
\label{fig:fit_ar_exc_pAl2.5_pZr2.0}
\end{figure}

Using the parametrizations, two-dimensional distributions of excitation 
energy versus mass of residual nuclei could be obtained for any value of 
projectile energy and target nuclei from the chosen ranges, for the considered 
cases.
In order to verify the parametrization, interpolated distributions have been 
compared to results of the HSD calculations. Comparisons of distributions for 
example p+Al reaction at 2.5 GeV beam energy (interpolation of parameters in 
function of projectile energy, see Fig. \ref{fig:par_Tp_pAl}) and p+Zr  
reaction at 2.0 GeV beam energy (interpolation of parameters in function of 
mass of target nuclei, see Fig. \ref{fig:par_Tp_2.0}) are presented in Fig. 
\ref{fig:fit_ar_exc_pAl2.5_pZr2.0}.
Obtained results are satisfactory. This indicates possibility of realization 
of such a parametrization for other target nuclei and other incident energies.


\chapter{Pion spectra}
\markboth{ }{Chapter 6. Pion spectra}

Pions are the most abundantly produced mesons during proton induced reactions on
atomic nuclei. Creation of pions is very important process in the reactions due
 to the fact that they carry away significant part of four-momentum introduced 
by incoming particle during the first stage of the reaction.
This is because pion mass ($\sim$ 140 MeV) is small enough to allow
for copious production of pions in nucleon-nucleon collisions even at low beam
energies, and large enough to contribute significantly to the energy
and momentum transfer.\\
It should be emphasized that in spite of the fact that momentum
carried away by pions is several times smaller than appropriate
momentum of nucleons (see e.g. the energy balance presented in Sec.  
\ref{sec:en_balance}), the contribution of pions to the momentum and
energy transfer may be comparable to that of nucleons. This is because pions, 
due to the low production threshold can be produced and reabsorbed 
several times during the reaction (threshold for charged pion production in 
free nucleon - nucleon collision is equal to about 289 MeV; inside nucleus, it 
can be decreased even up to about 141 MeV \cite{Mart00}, due to the Fermi 
motion). Thus they influence the whole dynamical evolution of the reaction. 
Proper treatment of pion production and absorption
may be crucial for realistic estimation of excitation energy and other 
properties of residual nuclei after the fast cascade of nucleon - nucleon 
collisions.\\ 
Calculations of the first stage of the
reaction are sufficient in order to receive realistic pion spectra,
since all pions are produced only in violent nucleon - nucleon collisions, 
where the locally available amount of four-momentum is large enough.

\begin{figure}[!htcb]
\vspace{-1cm}
\hspace{-2cm}
\includegraphics[height=19cm, width=17cm, angle=0]{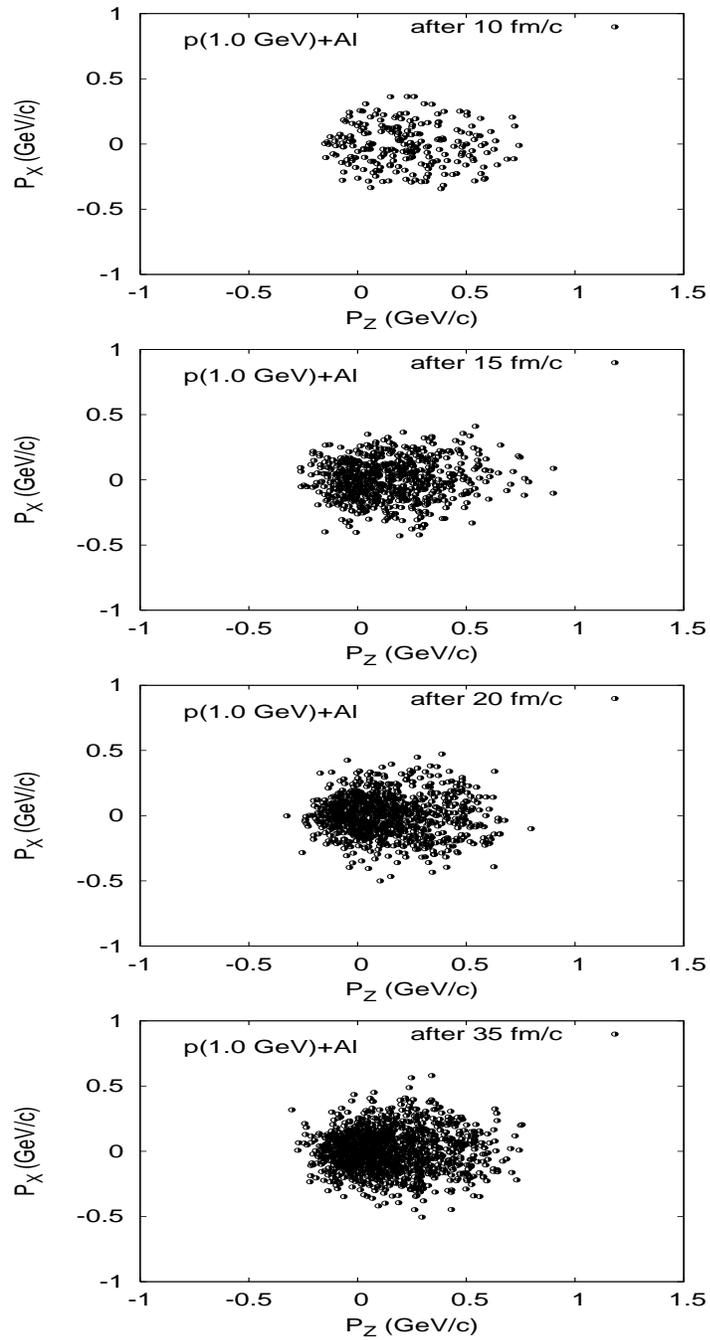}
\caption{{\sl Momentum distributions of pions emitted during p+Al reaction, 
at 1.0 GeV beam energy; $P_{Z}$ corresponds to momentum in beam direction;  
results of the HSD model calculations (scatter plot)}}
\label{fig:PION_pAl1.0_pzpx}
\end{figure}

\begin{figure}[!htcb]
\vspace{-3cm}
\hspace{-1cm}
\includegraphics[height=18cm, width=15cm, bbllx=0pt, bblly=160pt, bburx=594pt, bbury=842pt, clip=, angle=0]{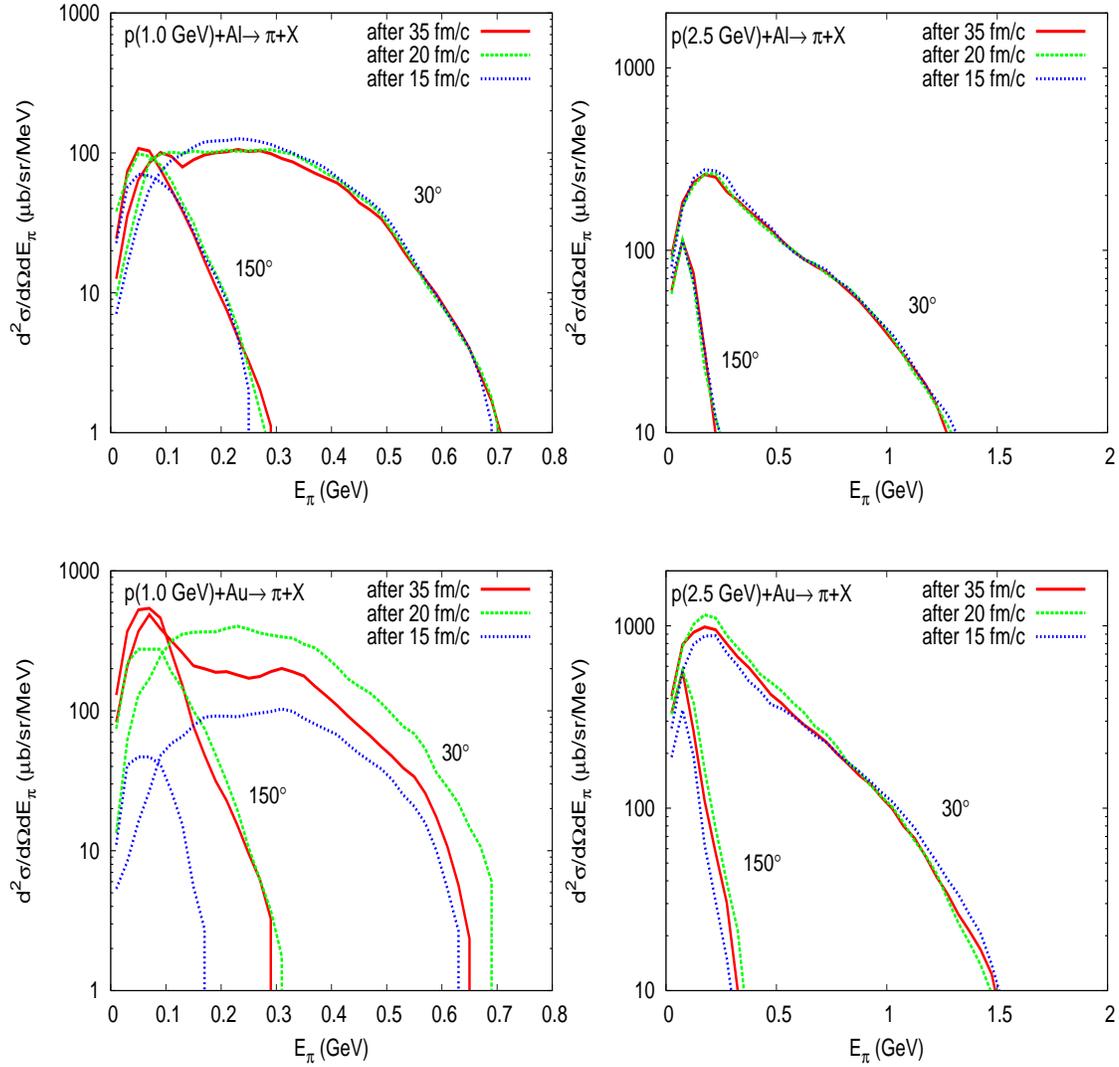}
\caption{{\sl Double differential kinetic energy spectra of all kind pions 
produced during p+Al and p+Au reactions, at 1.0 GeV and 2.5 GeV proton beam 
energy; results of the HSD model calculations}}
\label{fig:cpr_PION_dsig_dom_dek}
\end{figure}

One can distinguish two components in pion angular and energy distributions: 
high energy anisotropic part, i.e. pions emitted dominantly in forward 
direction and isotropic part consisted of low energy thermal pions emitted in 
the whole angular range (0$^{\circ}$ - 180$^{\circ}$). \\
Pions emitted from the first chance NN collisions have high energy, whereas 
those from secondary collisions are less energetic. 
Building up of pion spectra in time is presented in Figures  
\ref{fig:PION_pAl1.0_pzpx} and \ref{fig:cpr_PION_dsig_dom_dek}. 
Typical momentum distributions of pions emitted from 1.0 GeV proton induced 
reaction on Al target are shown; $P_{Z}$ corresponds to the momentum 
in beam direction. The time evolution of pion distributions is depicted 
more precisely in the Fig. \ref{fig:cpr_PION_dsig_dom_dek}, where    
double differential kinetic energy spectra of pions produced during 
the example reactions on light (Al) and heavy (Au) target, at low (1.0 GeV) and 
higher (2.5 GeV) projectile energy are displayed. 
It is seen, that in case of 1.0 GeV proton induced reactions, high energy
anisotropic part of pion distribution is formed first. Then, the isotropic 
part is built up. Formation of the low energy part of pion spectra takes longer 
time, since it is result of several acts of absorption and emission of pions. 
In the Fig. \ref{fig:cpr_PION_dsig_dom_dek}, all produced pions at the 
chosen reaction time are plotted, at further time some of them 
can be absorbed or absorbed and re-emitted in other direction, so the number of 
presented pions can be lower than at ealier time. After 35 fm/c, all presented 
pions are emitted. The number of produced pions is higher in case of 2.5 GeV 
proton induced reactions. Pion distributions are evolving with 
reaction time only slightly. 
At fixed energy of incident protons, pions produced on heavy target are more 
abundant than those produced on light target and low energy pions are dominant. 
Pion on its way can be several times absorbed and emitted (its mean free path 
is equal to 1.0 - 1.5 fm), its initial direction could be changed and its 
energy is suppressed. So, it is much more difficult for pion to be emitted 
in forward direction in case of heavy target, where it has to pass a longer 
distance inside nuclear matter. Therefore, high energy pions emitted in 
forward direction dominate in case of light target. 
Quantitative comparison of yield of the produced pions is displayed in Table 
\ref{table_mult_time}. It is noteworthy, that almost all of the pions have been 
produced due to Delta resonance decay; only a very few pions have been produced
due to N(1440) decay.   
\begin{table}[tbp]
\caption{{\sl  Time evolution of multiplicity of pions from p+Al and p+Au 
reaction at 1.0 GeV and 2.5 GeV proton beam energy, results of the HSD model 
calculations}}
\begin{center}
\begin{tabular}{|c||c|c|c||c|c|c|}
\hline
& \multicolumn{3}{c||}{{\bf p(1.0 GeV)+Al}} & \multicolumn{3}{|c|}{{\bf p(2.5 GeV)+Al}} \\
\hline
&$\pi ^{-}$ &$\pi ^{0}$ &$\pi ^{+}$ &$\pi ^{-}$ &$\pi ^{0}$ &$\pi ^{+}$\\
\hline
15 fm/c: &0.067 &0.18 &0.29 &0.38 &0.51 &0.50\\
\hline
20 fm/c: &0.071 &0.18 &0.29 &0.37 &0.47 &0.47\\
\hline
35 fm/c: &0.069 &0.17 &0.26 &0.37 &0.47 &0.46\\
\hline
\hline
& \multicolumn{3}{c||}{{\bf p(1.0 GeV)+Au}} & \multicolumn{3}{|c|}{{\bf p(2.5 GeV)+Au}} \\
\hline
& $\pi ^{-}$ &$\pi ^{0}$ &$\pi ^{+}$ &$\pi ^{-}$ &$\pi ^{0}$ &$\pi ^{+}$\\
\hline
15 fm/c: &0.010 &0.045 &0.060 &0.35 &0.46 &0.41\\
\hline
20 fm/c: &0.085 &0.19 &0.26 &0.53 &0.60 &0.52\\
\hline
35 fm/c: &0.11 &0.17 &0.23 &0.49 &0.53 &0.46\\
\hline
\end{tabular}
\end{center}
\label{table_mult_time}
\end{table}

\begin{figure}[!htcb]
\vspace{-2cm}
\hspace{-2cm}
\includegraphics[height=20cm, width=17cm, bbllx=0pt, bblly=25pt, bburx=594pt, bbury=842pt, clip=, angle=0]{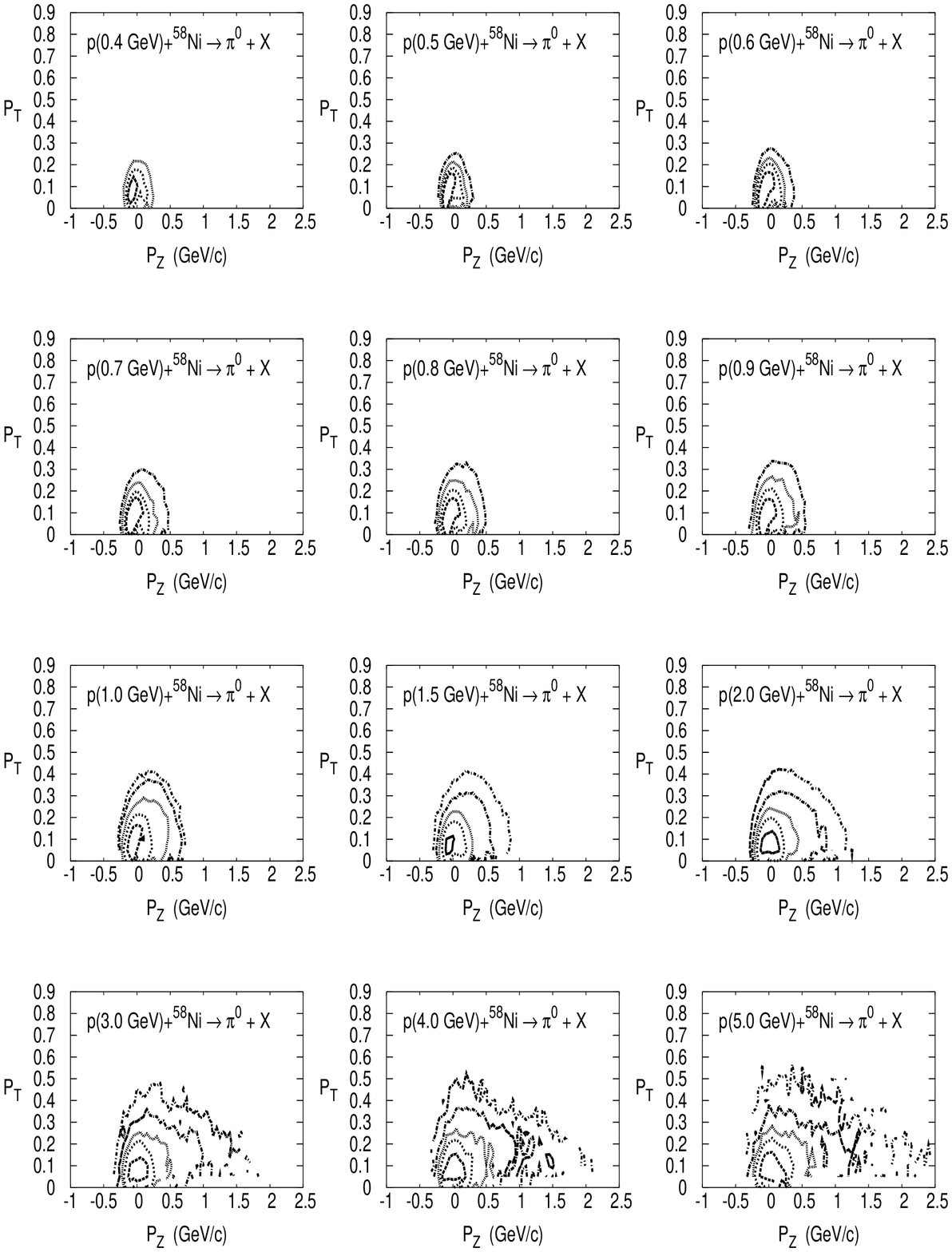}
\caption{{\sl Two-dimensional distributions of transversal versus longitudinal 
momentum of neutral pions emitted during proton induced reaction on Ni target, 
at various incident energies; results of the HSD model calculations (the most 
central contour line corresponds to a maximum of distribution)}}
\label{fig:c_pNi_pizX}
\end{figure}
\begin{figure}[!htcb]
\begin{center}
\vspace{-1.5cm}
\hspace{-2cm}
\includegraphics[height=9cm, width=10cm, angle=0]{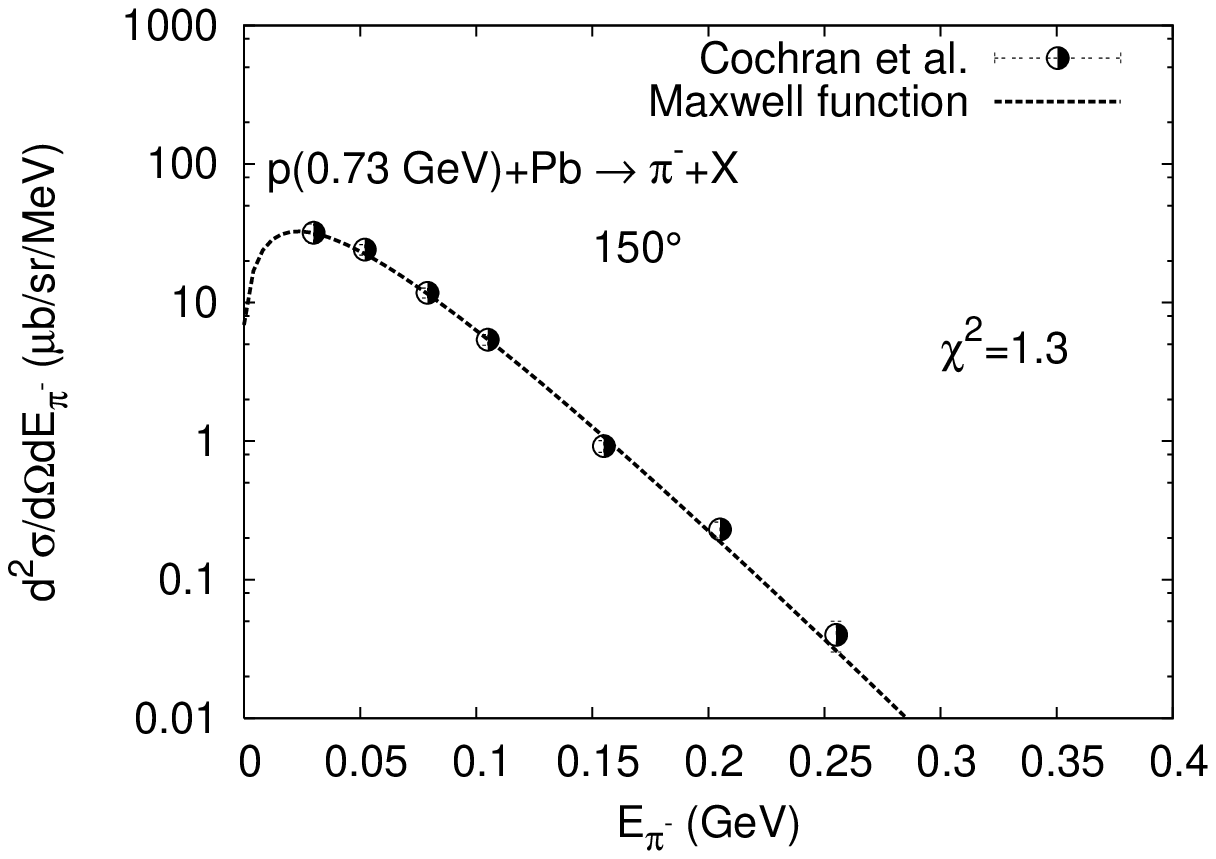}
\caption{{\sl Kinetic energy experimental spectrum of negative pions emitted 
in backward direction in p+Pb reaction, at 0.73 GeV proton energy, measured 
by Cochran et al. \cite{Coch72}, parametrized by Maxwell function}}
\label{fig:par_pPb_pimX_150_0.73}
\end{center}
\end{figure}

Spectra of pions change significantly with projectile energy.
It is seen in Fig. \ref{fig:c_pNi_pizX}, where examplary 
distributions of transversal ($P_{T}$) versus longitudinal ($P_{Z}$) momentum 
of neutral pions emitted during proton induced reaction on Ni target
are plotted for various incident energies. These are results of HSD model 
calculations, where $P_{Z}$ corresponds to the pion momentum in beam 
direction and $P_{T}$ is defined as $P_{T}=\sqrt{P_{X}^{2}+P_{Y}^{2}}$.\\
It can be seen, how the evolution of pion spectra looks like with increase of 
incident energy. At low projectile energies pions have not enough energy to 
pass through a target nucleus, so they cannot be emitted in forward direction. 
In this case, only a small part of target is accessible for pions, the rest of 
nucleus corresponds to the so-called {\sl nuclear shadow}.
Higher pion momenta are associated with increasing incident energy.
With increase of the incident energy, the anisotropic 
part of distribution increases. It means, more and more pions have chance to 
be emitted in forward directions. The isotropic, low - momentum part of pion 
spectra at high projectile energy corresponds to thermal pions, which result 
from sequence of several absorptions and emissions. 
One should also notice that at high projectile energies (in 
the Fig. \ref{fig:c_pNi_pizX} - higher or equal to 2.0 GeV), the isotropic part 
of pion momentum spectra is symmetric around $P_{Z}=0$. 
At low projectile energies 
(in the Fig. \ref{fig:c_pNi_pizX} - less than 2.0 GeV), the isotropic part of 
spectra, i.e. the most central contour line corresponding to a maximum of 
distribution is not symmetric around $P_{Z}=0$, but shifted to negative values 
of $P_{Z}$; to large extent it is due to mentioned nuclear shadow. 
The isotropic part of pion spectra that consists of thermal pions has the 
Maxwell distribution. This is illustrated by Fig. 
\ref{fig:par_pPb_pimX_150_0.73}, where the experimental distribution of 
negative pions, measured by Cochran et al. \cite{Coch72} at backward direction
 (150$^{\circ}$), is presented. 
The following form of Maxwell function has been used for the parametrization: \\
$f(E_{\pi ^{-}}) = A_{0} \cdot 
((E_{\pi ^{-}}+m_{\pi ^{-}})^{2}-m_{\pi ^{-}}^{2})
\cdot e^{-E_{\pi ^{-}}/T_{0}}$ \\
where: $E_{\pi ^{-}}$ and $m_{\pi ^{-}}$ are the kinetic energy of the pions 
and their mass, respectively. $A_{0}$ and $T_{0}$ are fitted parameters. 
The following values of the parameters are obtained: 
$A_{0} = 11476.1 \pm 364.5$ $[$GeV$^{-3}]$ and $T_{0} = 0.024 \pm 0.00039$ 
$[$GeV$]$.
\begin{figure}[!htcb]
\vspace{-2cm}
\hspace{-2cm}
\includegraphics[height=7cm, width=8cm, angle=0]{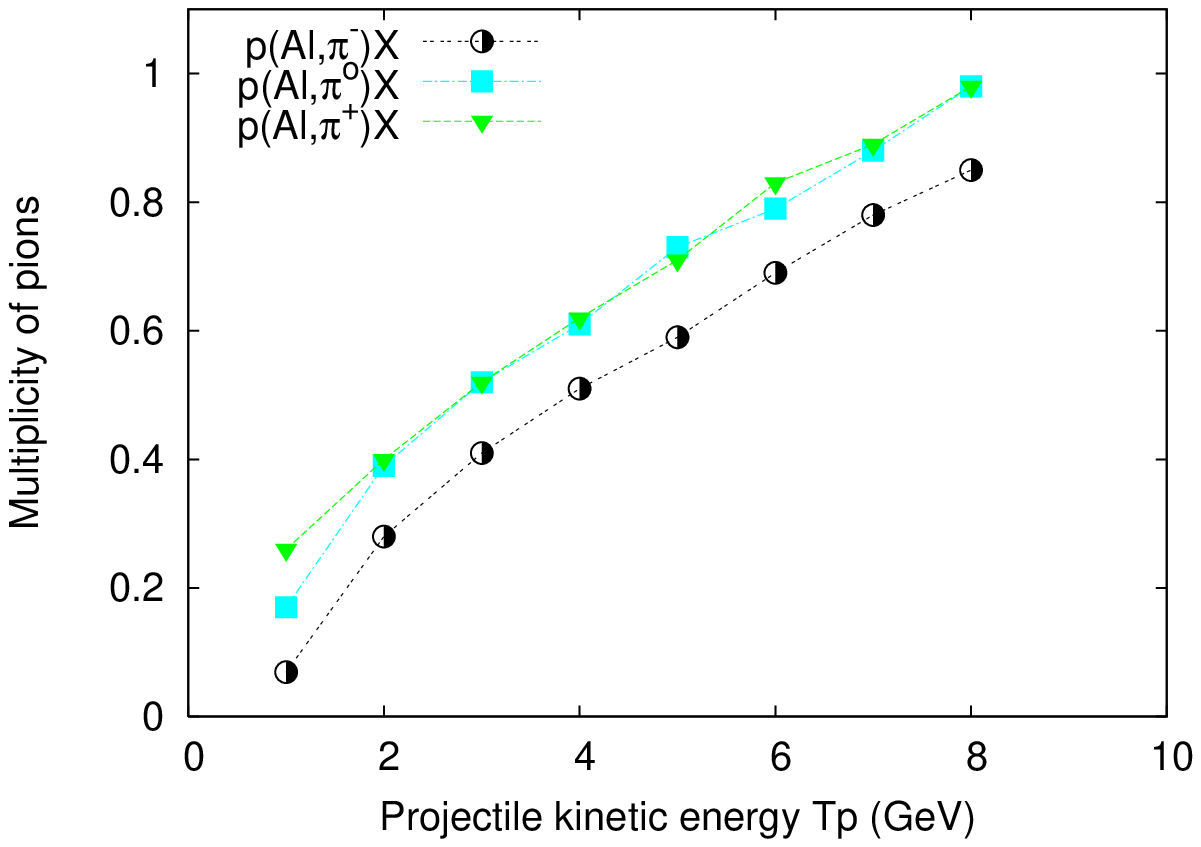}
\hspace{-0.1cm}
\includegraphics[height=7cm, width=8cm, angle=0]{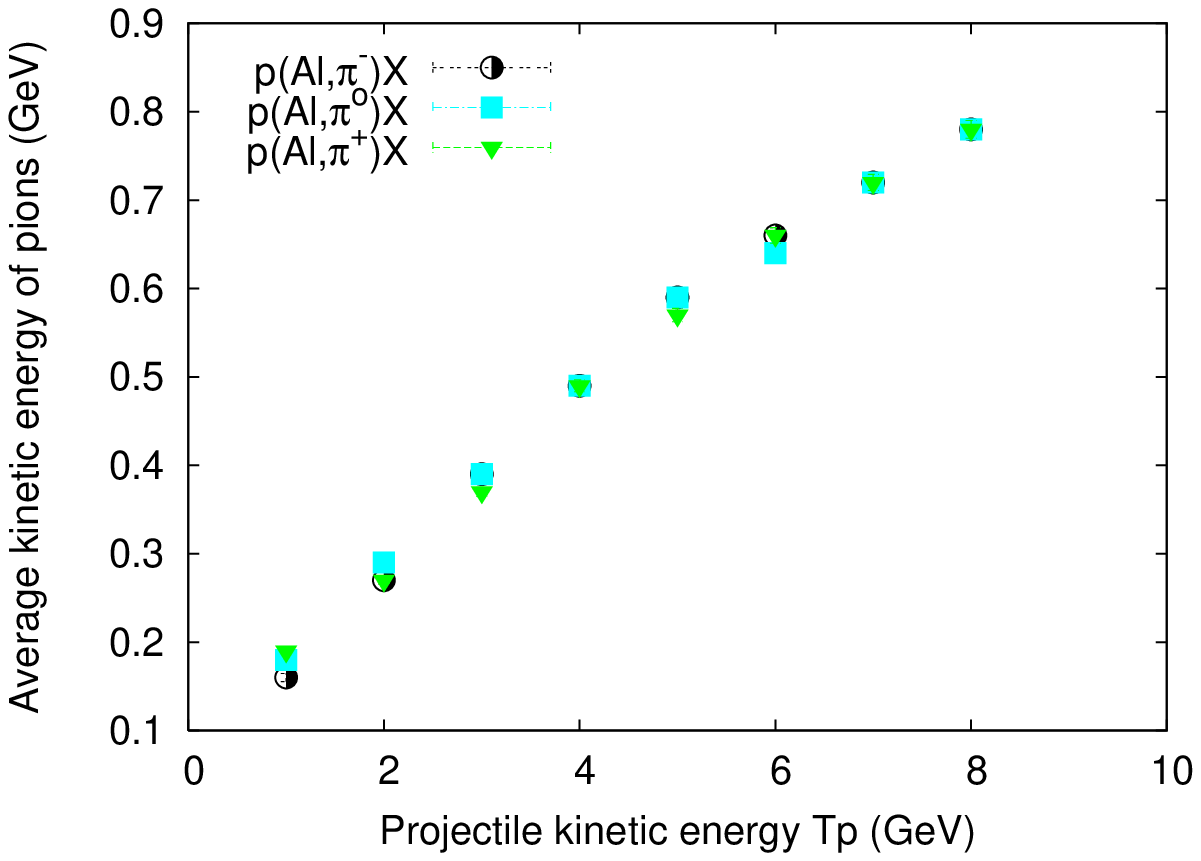}
\caption{{\sl Multiplicity (left) and average values of kinetic energy per 
individual pion (right)
of pions emitted during proton induced reaction on Al target, in function of 
incident energy; results of the HSD model calculations (error bars indicate 
values of standard deviation of the average values, divided by a square root of
 number of events)}}
\label{fig:pAl_mult_ekin_pi}
\end{figure}

\begin{figure}[!htcb]
\vspace{-0.5cm}
\hspace{-2cm}
\includegraphics[height=7cm, width=8cm, angle=0]{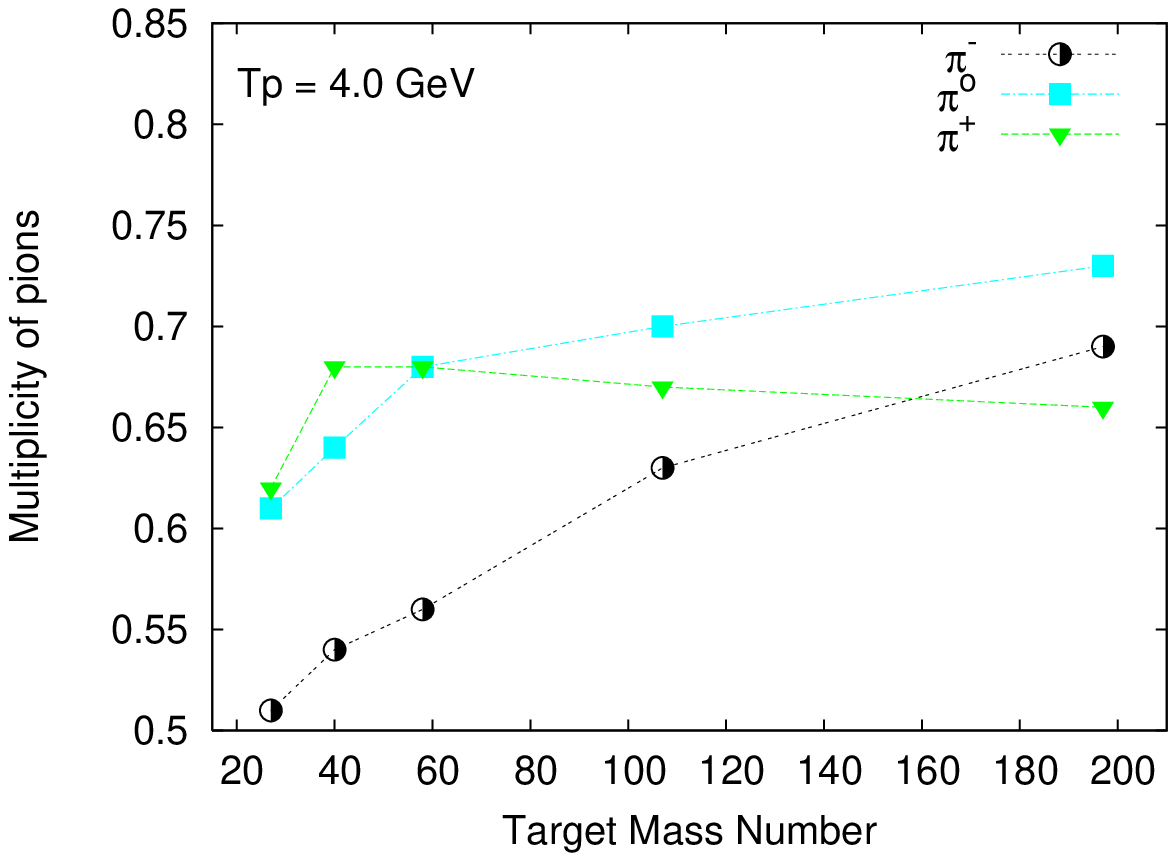}
\hspace{-0.1cm}
\includegraphics[height=7cm, width=8cm, angle=0]{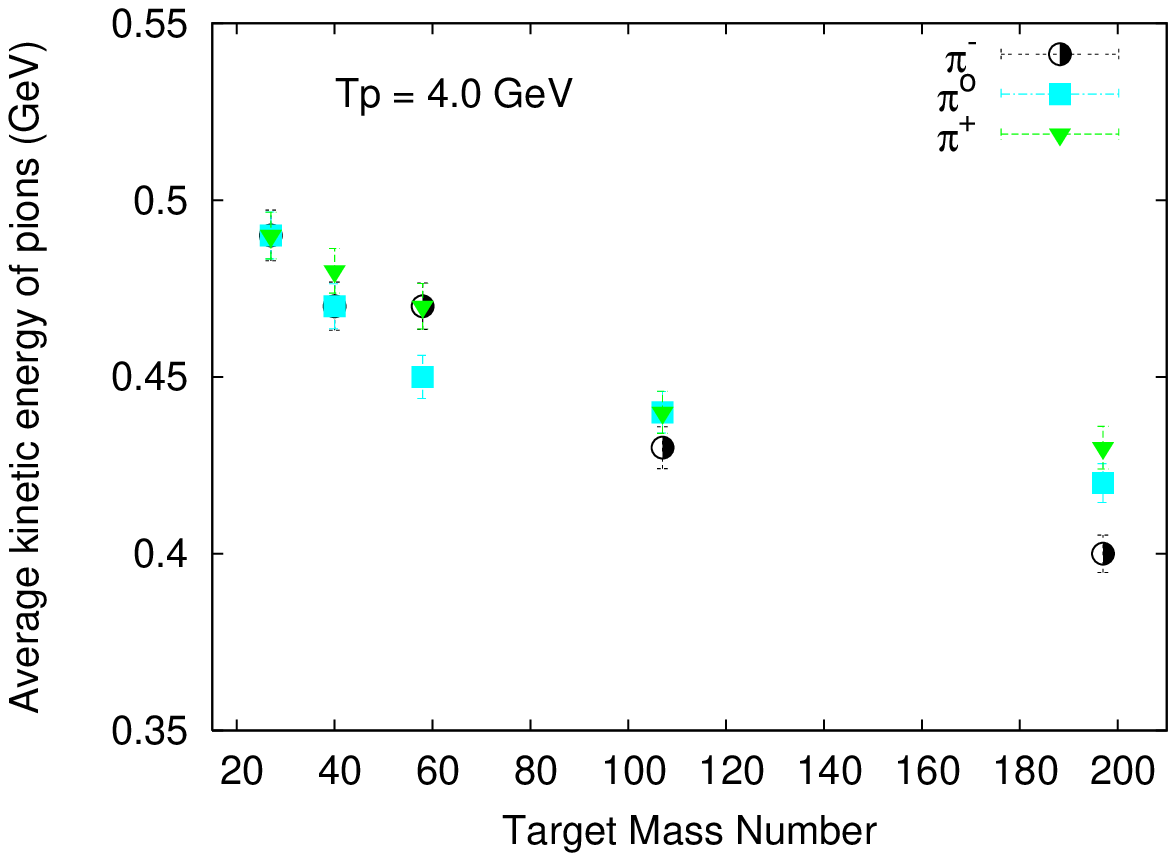}
\caption{{\sl Multiplicity (left) and average values of kinetic energy per 
individual pion (right)
of pions emitted during 4.0 GeV proton induced reactions on a few targets, in 
function of target mass number; results of the HSD model calculations 
(error bars indicate values of standard deviation of the average values, 
divided by a square root of number of events)}}
\label{fig:p4.0_mult_ekin_pi}
\end{figure}
It is interesting to see, how the average yield of pions produced in proton 
induced spallation reactions changes with projectile energy and mass of target.
The HSD model calculations predict that multiplicity of pions, in average, 
increases both with incident energy and mass of target. 
One can expect that due to charge conservation, the relative yield of produced 
pions should fulfill the relation: $\pi ^{+}$ $>$ $\pi ^{0}$ $>$ $\pi ^{-}$.  
The incident energy dependence (presented in Fig. \ref{fig:pAl_mult_ekin_pi}) 
shows, that positive and neutral pion production is favoured, as expected. 
For heavier targets, where there are more collisions, multiplicity of positive 
pions decreases and for the heaviest target it is even lower than multiplicity 
of neutral and negative pions, as displayed in Fig. \ref{fig:p4.0_mult_ekin_pi}.
 The target mass dependence shows, that reabsorptions and secondary collisions 
play an important role, especially for heavy targets. \\
Looking at kinetic energy carried by individual emitted pions, it is seen (Fig. 
\ref{fig:pAl_mult_ekin_pi}, right) that all kinds of pions produced in 
proton induced reactions on a chosen target carry, in average, the same amount 
of kinetic energy. The higher energy of the incident proton, the bigger amount 
of kinetic energy carried out by pions. In reactions on heavier targets (Fig. 
\ref{fig:p4.0_mult_ekin_pi}), emitted pions carry away, in average, less 
kinetic energy. 
In case of the heaviest targets, the negative pions take the least kinetic 
energy of all pions. The suppression of kinetic energy of emitted pions from 
heavier target is an expected result of numerous possible 
reabsorptions and emissions of pions, which become less and less 
energetic.  
   
\begin{figure}[!htcb]
\vspace{-2cm}
\hspace{-1.5cm}
\includegraphics[height=20cm, width=18cm, bbllx=0pt, bblly=40pt, bburx=594pt, bbury=842pt, clip=, angle=0]{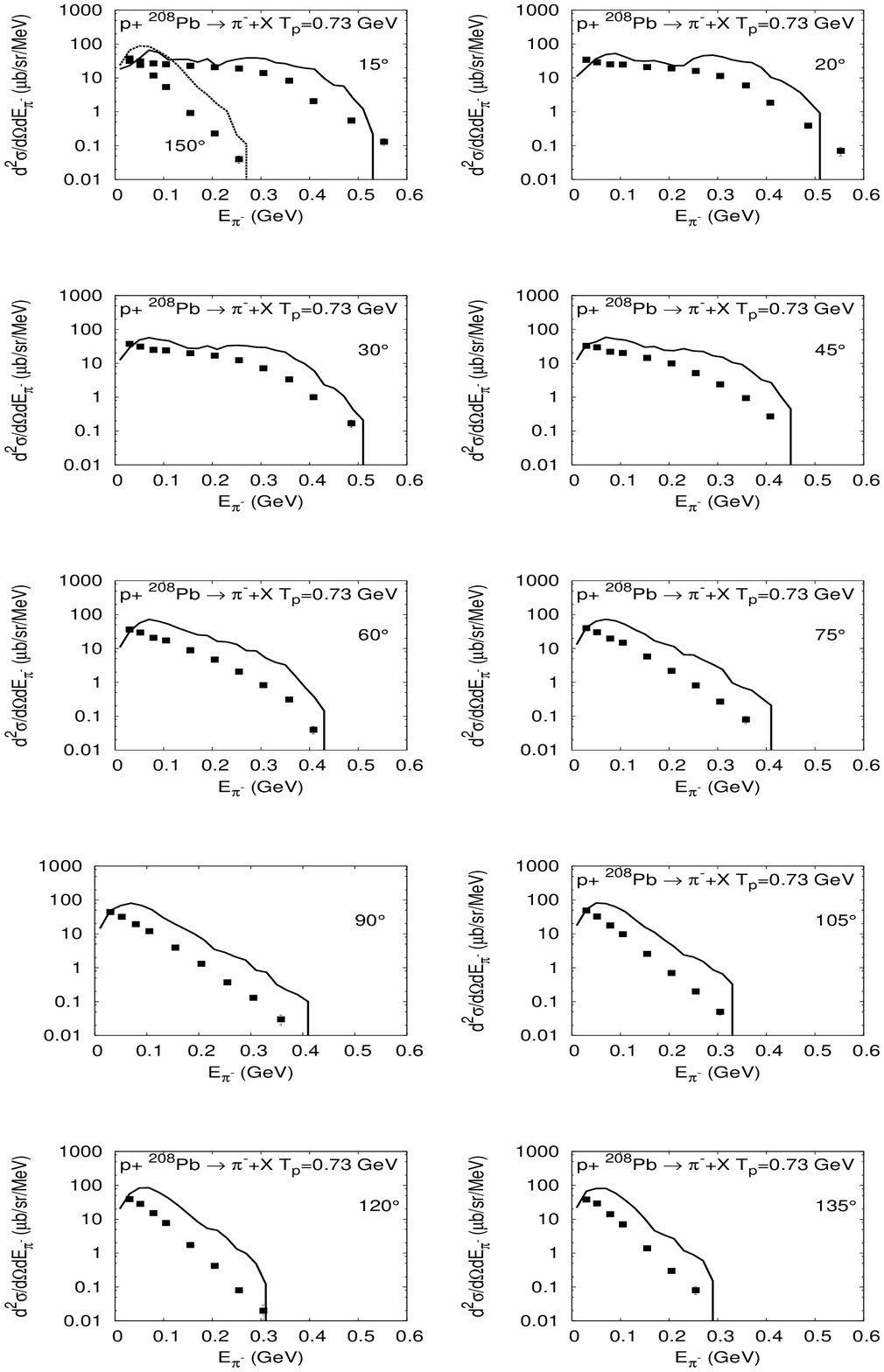}
\caption{{\sl Double differential negative pion spectra from p+Pb reaction at
0.73 GeV proton beam energy; lines show results calculated with HSD model,
symbols indicate the experimental data \cite{Coch72}}}
\label{fig:pPb_pimX_0.73}
\end{figure}

\begin{figure}[!htcb]
\vspace{-2cm}
\hspace{-1.5cm}
\includegraphics[height=20cm, width=18cm, bbllx=0pt, bblly=40pt, bburx=594pt, bbury=842pt, clip=, angle=0]{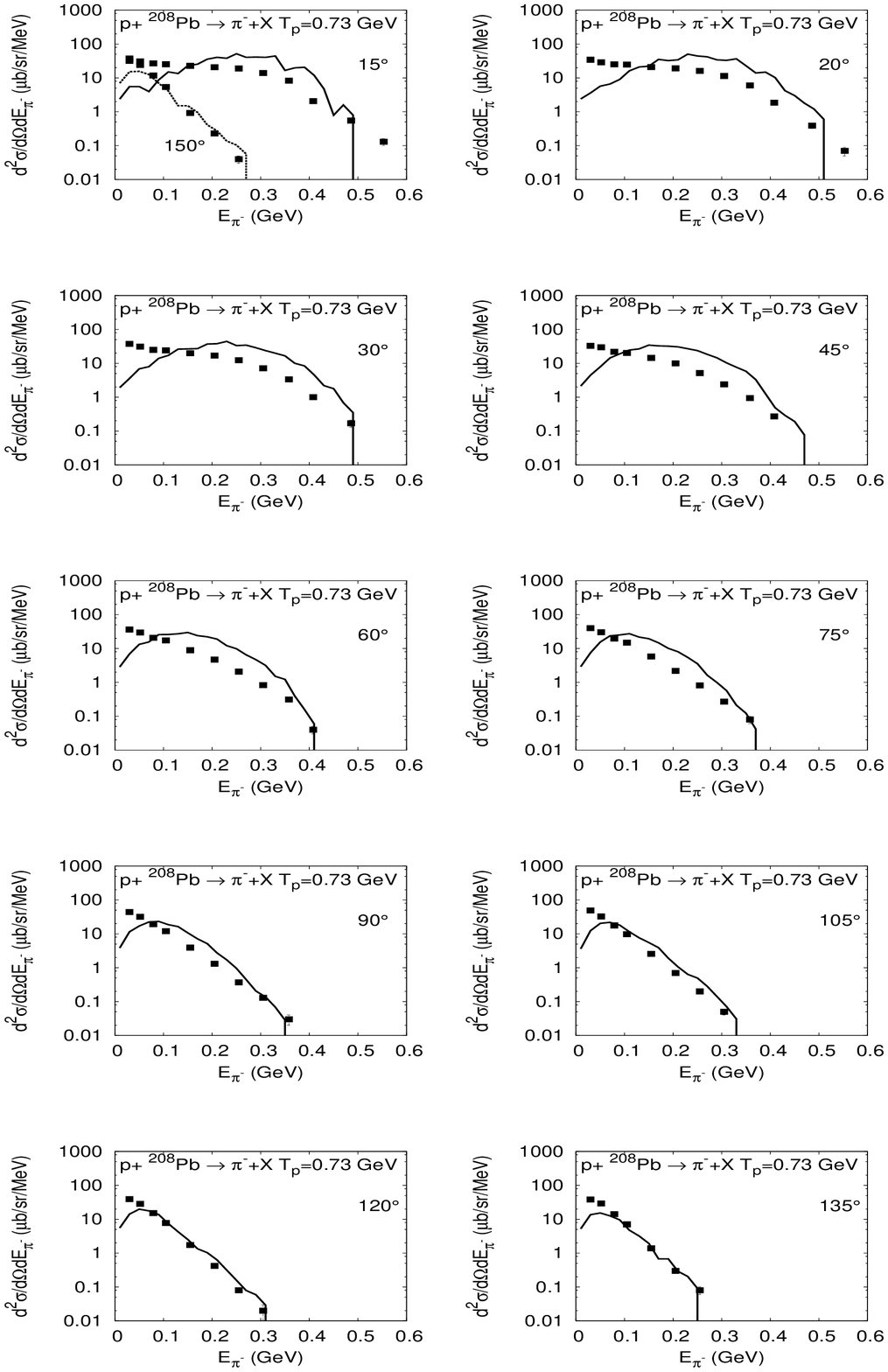}
\caption{{\sl Double differential negative pion spectra from p+Pb reaction at
0.73 GeV proton beam energy; lines show results calculated with HSD model, with
enlarged Delta resonance lifetime, symbols indicate the experimental data
\cite{Coch72}}}
\label{fig:pPb_pimX_0.73_1_d}
\end{figure}

\begin{figure}[!htcb]
\vspace{-2cm}
\hspace{-1.5cm}
\includegraphics[height=20cm, width=18cm, bbllx=0pt, bblly=40pt, bburx=594pt, bbury=842pt, clip=, angle=0]{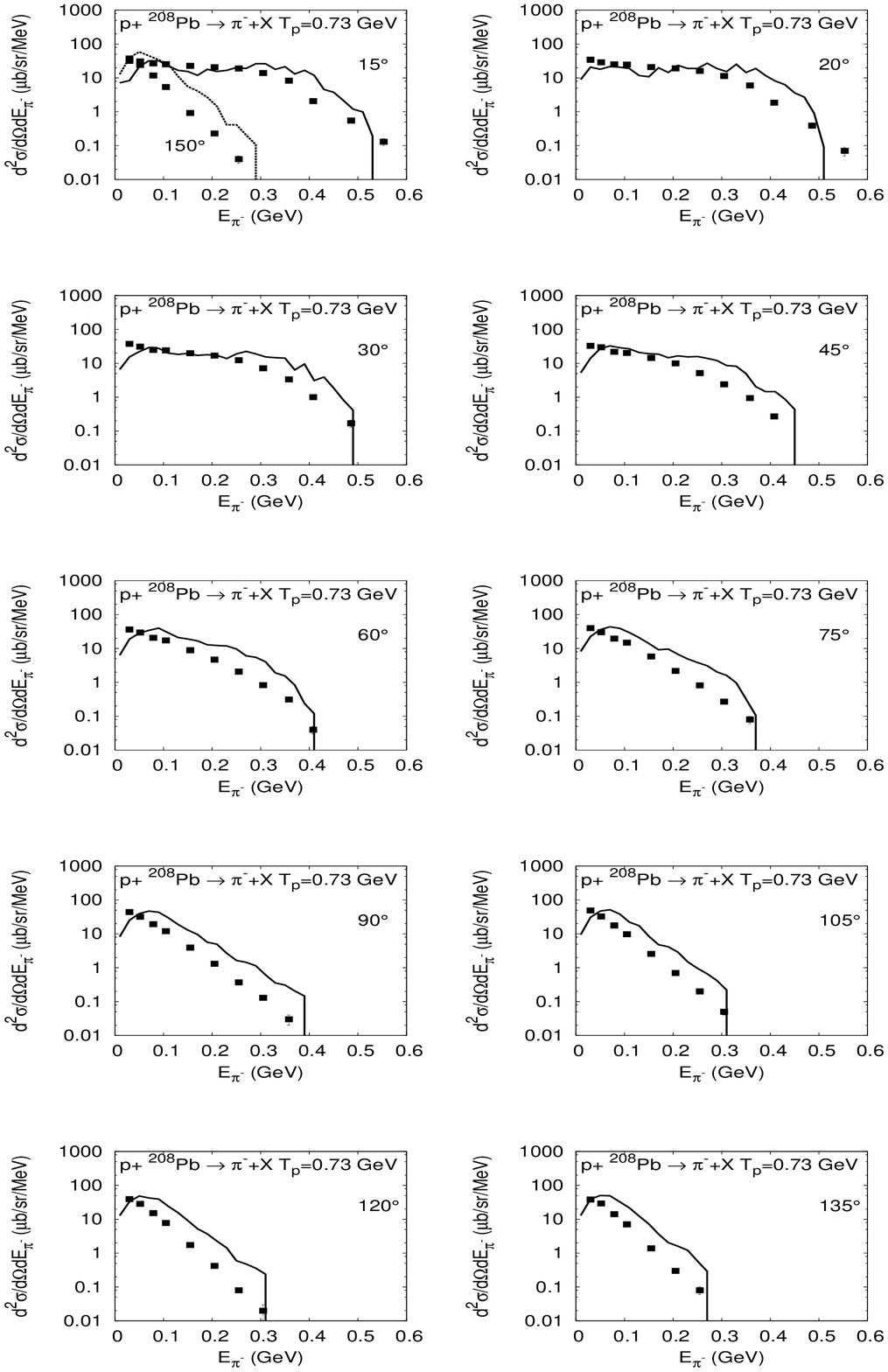}
\caption{{\sl Double differential negative pion spectra from p+Pb reaction at
0.73 GeV proton beam energy; lines show results calculated with HSD model, with
enlarged cross section for the absorption of Delta resonance in reaction
$ N \Delta \rightarrow NN$, symbols indicate the experimental data
\cite{Coch72}}}
\label{fig:pPb_pimX_0.73_1_2}
\end{figure}

\begin{figure}[!htcb]
\vspace{-2cm}
\hspace{-1.5cm}
\includegraphics[height=20cm, width=18cm, bbllx=0pt, bblly=40pt, bburx=594pt, bbury=842pt, clip=, angle=0]{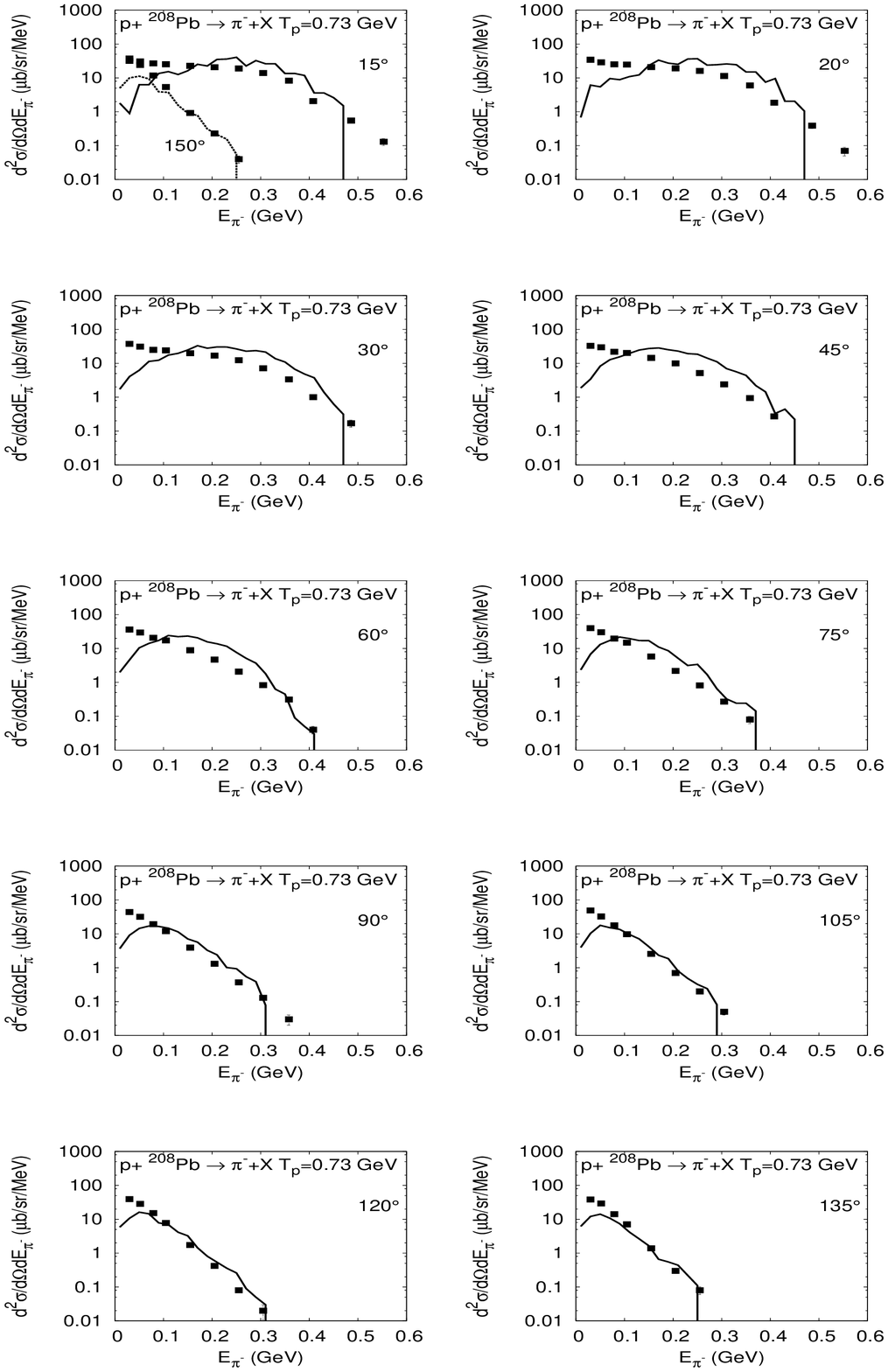}
\caption{{\sl Double differential negative pion spectra from p+Pb reaction at
0.73 GeV proton beam energy; lines show results calculated with HSD model, with
enlarged both cross section for the absorption of Delta resonance in reaction
$ N \Delta \rightarrow NN$ and Delta resonance lifetime, symbols indicate the
experimental data \cite{Coch72}}}
\label{fig:pPb_pimX_0.73_2_d}
\end{figure}

\begin{figure}[!htcb]
\begin{center}
\vspace{-2cm}
\hspace{-1.5cm}
\includegraphics[height=20cm, width=18cm, bbllx=0pt, bblly=40pt, bburx=594pt, bbury=842pt, clip=, angle=0]{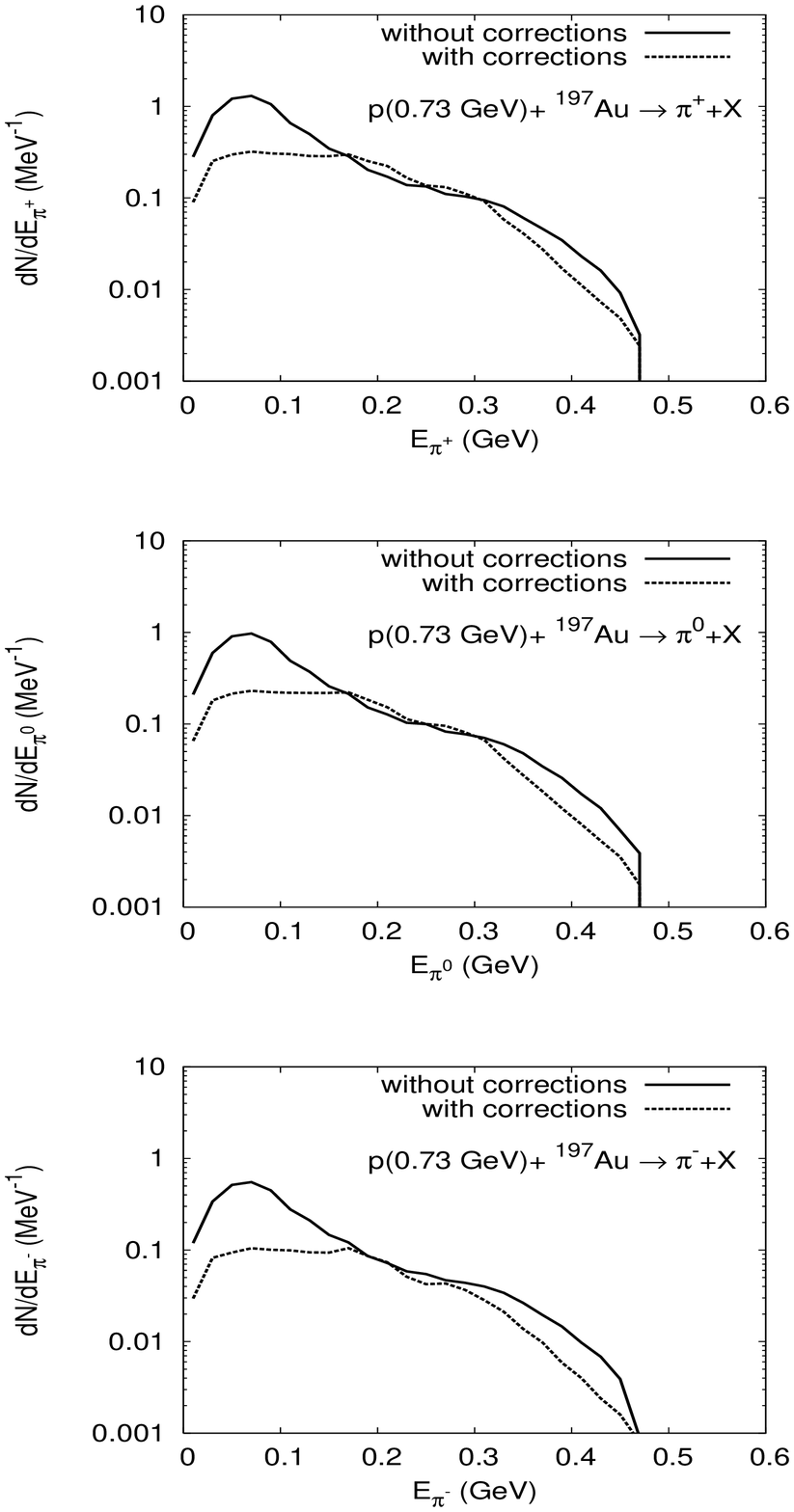}
\caption{{\sl Pions kinetic energy spectra from p+Au reaction at 0.73
GeV proton beam energy; solid lines show results calculated with HSD model,
without any corrections, dashed lines show results calculated with
enlarged cross section for the absorption of Delta resonance in reaction
$ N \Delta \rightarrow NN$ and enlarged Delta resonance lifetime}}
\label{fig:pAu_piX_2.5_0.73}
\end{center}
\end{figure}

\begin{figure}[!htcb]
\begin{center}
\vspace{-2cm}
\hspace{-1.5cm}
\includegraphics[height=20cm, width=18cm, bbllx=0pt, bblly=40pt, bburx=594pt, bbury=842pt, clip=, angle=0]{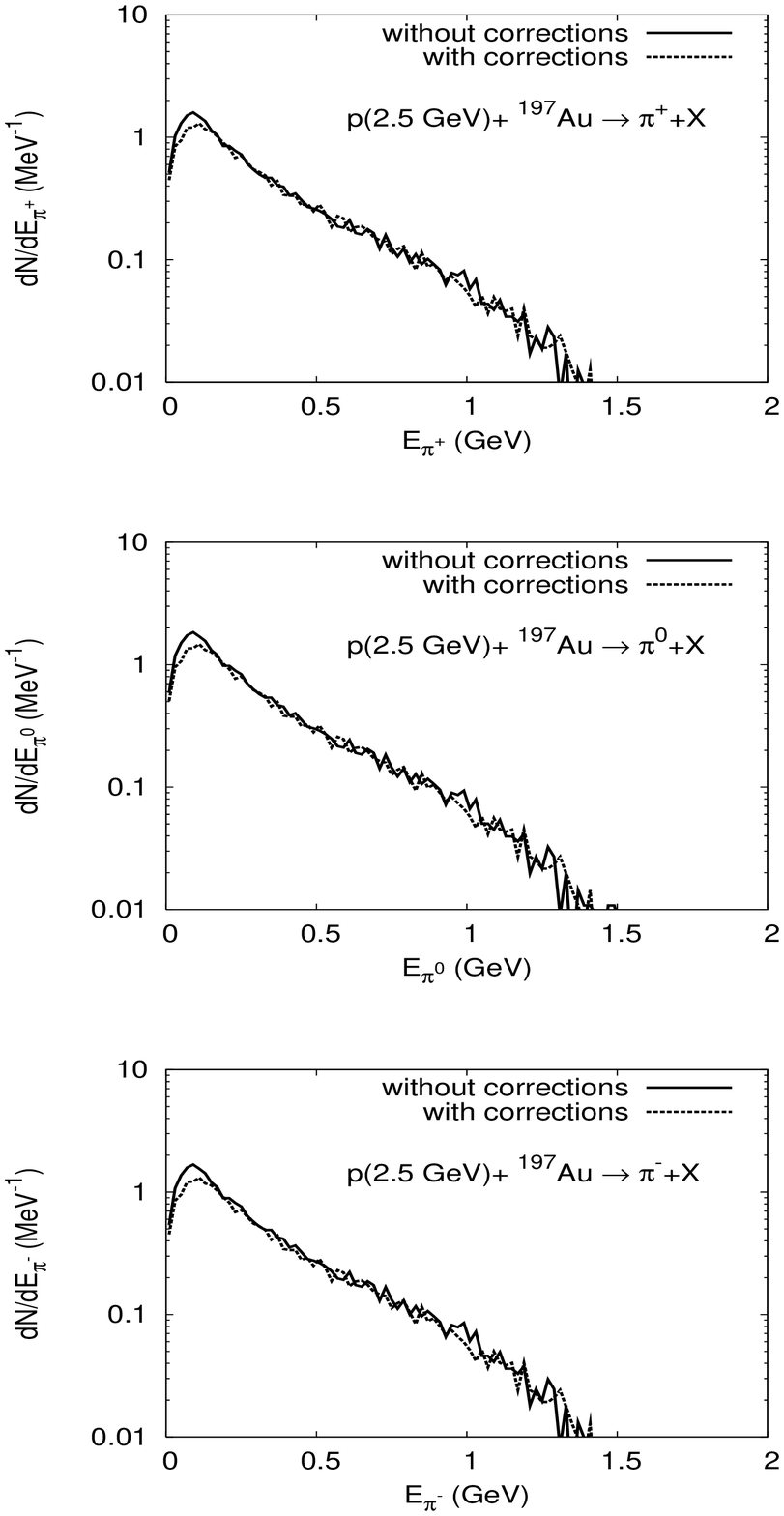}
\caption{{\sl Pions kinetic energy spectra from p+Au reaction at 2.5
GeV proton beam energy; solid lines show results calculated with HSD model,
without any corrections, dashed lines show results calculated with
enlarged cross section for the absorption of Delta resonance in reaction
$ N \Delta \rightarrow NN$ and enlarged Delta resonance lifetime}}
\label{fig:pAu_piX_2.5_lesspi}
\end{center}
\end{figure}

Inclusive differential kinetic energy spectra of pions emitted at different 
angles, calculated with the HSD model, have been confronted with available 
experimental data \cite{Coch72}, as displayed in Fig. \ref{fig:pPb_pimX_0.73}, 
for an example distributions of negative pions from 
p+Pb reaction, at 0.73 GeV of incident energy. 
One finds, that the calculations overestimate the experimental data.
It is evident, that the pion dynamics in the HSD model needs to be improved,
i.e. the number of produced pions should be reduced.
Pions are created mainly due to decay of Delta(1232) resonances. 
It means, pions production and absorption proceeds mainly
through the $NN \leftrightarrow N \Delta$ and $\Delta \leftrightarrow \pi N$
reactions (see description of the model in Chapter \ref{chapt:HSD}).
Therefore, modifications of the cross section for the absorption of Delta
resonance in reaction $ N \Delta \rightarrow NN$ and the Delta lifetime
should influence the amount of produced pions.
In order to estimate the influence, the following corrections have been tested.
First, the Delta resonance lifetime has been enlarged. Comparison of the
pion spectra calculated with HSD model, with enlarged Delta resonance lifetime
and the experimental data \cite{Coch72} is presented in Fig.
\ref{fig:pPb_pimX_0.73_1_d} (the calculations are performed with 30 times
 longer lifetime of Delta resonance). It is seen, that this correction 
reduces number of pions emitted mainly in backward directions.  
This is due to the fact, that longer lifetime of Delta resonance implies 
smaller number of acts of absorption and emission of pions.
If the cross section for the absorption of Delta resonance in reaction
$N \Delta \rightarrow NN$ is enlarged (by factor 20), the number of pions 
emitted in forward directions is reduced, as it is shown in Fig. 
\ref{fig:pPb_pimX_0.73_1_2}.
Implementation of both of the corrections reduces the amount of produced pions 
both in forward and backward directions, as presented in the Fig.
\ref{fig:pPb_pimX_0.73_2_d} (these are results of calculations performed with 
30 times longer lifetime of Delta resonance and the cross section for its 
absorption enlarged by a factor 1.5). 
The improvement of agreement between the data and model results is substantial, 
however, the introduced modifications are rather drastic.
(Improvement of pion dynamics by modifications of Delta resonance lifetime and 
cross section for Delta absorption has been also performed in frame of the 
INCL model in \cite{Boud02}, but obtained results also do not comply well with 
experimental data).
Furthermore, it is tested, that the influence of the joined corrections on the
pion spectra depends on the value of proton beam energy. The higher beam 
energy, the lower reduction of the pions yield by the corrections, as it is 
seen in Figs. \ref{fig:pAu_piX_2.5_0.73} and \ref{fig:pAu_piX_2.5_lesspi}, 
for spectra of pions produced in p+Au reaction, at low (0.73 GeV) and 
higher (2.5 GeV) beam energy, respectively. The quantitative distinction of 
the pions yield from the reactions at two different incident energies are
collected in Table \ref{table_lesspi}. 
It has been also verified that implemented corrections do not modify yield 
and shape of spectra of nucleons.
Moreover, the corrections do not influence the properties of residual nuclei 
after the first stage of reaction. In particular, they do not change excitation 
energy of the remnants, because the average energy carried out by emitted pions 
with and without the corrections is the same; due to the corrections less 
pions are emitted, but they carried, in average, more kinetic energy.   
\begin{table}[tbp]
\caption{{\sl Multiplicity of pions from p+Au reaction at 0.73 GeV and 2.5 GeV
proton beam energy, without any corrections and with enlarged cross section for
 the absorption of Delta resonance in reaction $ N \Delta \rightarrow NN$ and
enlarged Delta resonance lifetime}}
\begin{center}
\begin{tabular}{|c|c|c|c|c|}
\hline
 & \multicolumn{2}{c|}{Tp = 0.73 GeV} & \multicolumn{2}{c|}{Tp = 2.5GeV} \\
\cline{2-5}
 & without corrections & with corrections & without corrections & with corrections\\
\hline
$\pi ^{+}$: &0.15 &0.073 &0.46 &0.43\\
\hline
$\pi ^{0}$: &0.11 &0.053 &0.54 &0.48\\
\hline
$\pi ^{-}$: &0.065 &0.024 &0.49 &0.43\\
\hline
\end{tabular}
\end{center}
\label{table_lesspi}
\end{table}

The not very satisfactory agreement between calculations and experimental data 
is not too surprising, because pion production in proton - nucleus collisions 
is a complicated process. 
Although the elementary cross sections for pions production due to free nucleon
 - nucleon interactions are measured and very well known (see e.g. 
\cite{Bald87}), in-medium problem appears. 
In the HSD model, the in-medium effects are partly included through e.g. 
employment of the Pauli principle and cross section for Delta absorption. 
However, proper handling of modifications due to in-medium effects are hampered
 by a lack of experimental and theoretical knowledge.\\      
Pion production mechanism is highly sensitive to pion 
and $\Delta$ - resonance dynamics. So, apart from tested above modifications of
 the $\Delta$ parameters, also introduction of density dependent width of 
$\Delta$ - resonance (increasing with density), introduction of pion potential, 
change of an angular distribution in order to describe possible 
anisotropic $\Delta$ decay into $\pi$N channel, etc. can influence the pions 
production. 
In fact, till now none of the trials have given satisfactory results, see 
\cite{Aous06, Enge94}.
The propriety of description of pions dynamics and production due to 
proton - nucleus reaction could not be really verified, because experimental 
data are rather scarce, especially in projectile energy range of a few GeV 
order.

There exists also one more reason that influences the calculated pion 
spectra. In the HSD model, a target nucleus, i.e. the nuclear density 
distribution is constructed up to a constant limit $R + R_{CUT}$, where: 
$R=1.12 \cdot A^{1/3} [fm]$ is a radius of nucleus, defined as a distance at 
which the value of nucleus density decreases to a half of the value of 
density in the center of nucleus; 
$A$ is a mass number of the nucleus; $R_{CUT} = 1.5 fm$.   
This is due to difficulties with modelling of the infinite tail of nuclear 
density at nucleus surface, see \cite{mgr02}. Although the truncated part 
corresponds to only a small fraction of the whole nucleus density distribution,
 it can influence the pion spectra. 
This is because only pions produced close to the surface
 have a chance to be emitted, as the mean free path of pions is equal to 
about 1.0 - 1.5 fm, at standard density. 
The introduced cut off of nuclear surface neglects the 
possibility of further pion - nucleon interactions at surface, which are quite 
probable (since pion is a strongly interacting particle) and lead to 
spuriously higher multiplicity of emitted pions.  
The nucleus density cut off does not influence nucleons production, because 
most of nucleons are emitted closer to the center of nucleus, due to the fact 
that their mean free path is equal to about 3.0 fm. 
Therefore, the HSD model description of proton induced 
spallation reaction is not falsified by the used limited distribution of  
density of target nuclei, only outgoing pion distributions are to some extent 
misshapen.


\chapter{Statistical emission of particles in the second stage of 
proton - nucleus collisions}
\label{chapt:statist_emiss}
\markboth{ }{Chapter 7. Statistical emission of particles}

The second stage of proton - nucleus reaction can be theoretically described 
with various statistical models. Historical origin of the models is related to 
Weisskopf's paper published in 1937 \cite{Weis37}. 
Weisskopf considers emission (evaporation) of a neutron with kinetic energy 
$E_{k}$ from nucleus $A$, excited to energy $E_{A}$, transforming it to nucleus 
$B$, with excitation energy $E_{B}=E_{A}-B_{n}-E_{k}$.   
Probability of such emission is calculated using the principle of detailed 
balance. It can be determined if cross section for the inverse process 
$\sigma (E_{A},E_{k})$ is known (i.e. cross section for production of compound 
nucleus by collision of a neutron with nucleus $B$). The probability is 
expressed as:
\begin{equation}
W_{n}(E_{k})= \sigma (E_{A},E_{k}) g m E_{k} \omega _{B}(E_{B})/ \pi ^{2} \hbar ^{3} \omega _{A}(E_{A})
\end{equation} 
where: $m$ stands for neutron mass, $g$ is a spin degeneracy factor, 
$\omega _{A}$ and $\omega _{B}$ are the level densities of the initial and 
final nuclei, respectively. \\ 
In the Weisskopf expression, angular momentum effects are not taken into 
account. In order to consider the effects, one needs to use 
the Hauser-Feshbach formalism \cite{Haus52}. In this case, the decay  
probability is determined by level density $\rho (E_{f},J_{f})$ of the residual
 nucleus, depending both on its excitation energy $E_{f}$ and angular momentum 
$J_{f}$, and transmission coefficient of emitted particle $T_{l}(E_{k})$ 
($l$ is the orbital angular momentum removed by 
the particle, $E_{k}$ is its kinetic energy). \\
Decay width $\Gamma$ for state of energy $E$ and angular momentum $J$ into a 
specific channel $i$ is expressed as: 
\begin{equation}
\Gamma _{i}(E,J)=D(E,J)/2 \pi \sum _{I,l,J_{f}} \delta (I+l+J_{f}-J) \int _{0}^{E-BE_{i}} T_{l}(E_{k}) \rho _{i}(E-BE_{i}-E_{k},J_{f})dE_{k}
\end{equation}     
where: $D(E,J)$ is the level spacing (reciprocal of the level density) of the 
decaying nucleus, $I$ stands for the particle spin. The following energy and 
angular momentum conservation constraints are employed: \\
$J = I+l+J_{f}$, \\
$E = E_{f}+E_{k}+BE_{i}$, where $BE_{i}$ is the binding energy of the emitted 
particle in decay channel $i$. \\ 
The decay probability into channel $i$ is evaluated as: 
$\Gamma _{i}/ \Gamma _{tot}$, where: $ \Gamma _{tot}= \sum _{i} \Gamma _{i}$, 
$i$ runs over all open channels. \\ 
\begin{figure}[!ht]
\begin{center}
\includegraphics[height=8.5cm, width=8.5cm, angle=0]{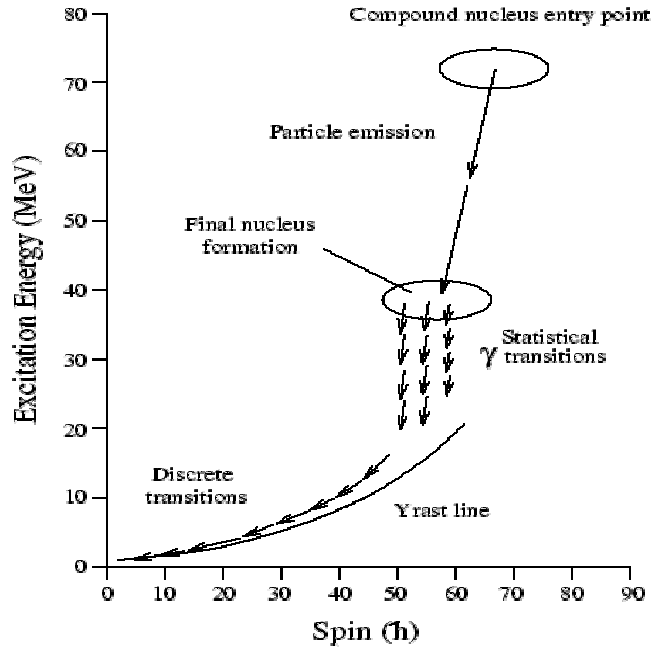}
\caption{{\sl The Statistical Evaporation Process}}
\label{fig:yrast_rys}
\end{center}
\end{figure}

Important step to comprehensive understanding of the dependence of the 
level density on angular momentum of compound nucleus was done by J. R. Grover
 \cite{Grov62}. He invented the "Yrast level" term to describe the 
lowest lying level at some given angular momentum. It means, that for any 
angular momentum $J$ the nucleus has to have an excitation energy at least  
equals to the rotational energy $E_{rot}(J)$. Therefore, there cannot be any 
states with excitation energy below $E_{rot}(J)$, i.e. below "Yrast level". 
Particle decay is inhibited close to the "Yrast line" (i.e. the level density 
is close to zero for the final states allowed by energy and angular momentum 
conservation). Deexcitation proceeds mainly by gamma emission along the 
"Yrast line", as illustrated in Fig. \ref{fig:yrast_rys}.\\   

In the frame of this work, calculations of the second stage of proton - 
nucleus reactions have been done with use of PACE2 \cite{Gavr93} and GEM 
\cite{Furi00} codes.  
Below, basic features and results of the models are presented.   

\section{Evaporation model - PACE 2}

The statistical model code PACE2 ({\bf P}rojection {\bf A}ngular-Momentum 
{\bf C}oupled {\bf E}vaporation, version 2) \cite{Gavr93} is based on 
the Hauser-Feshbach formalism. The decay sequence of an excited nucleus 
is calculated with use of a Monte Carlo procedure, which is followed until the 
nucleus reaches its ground state. At each stage of deexcitation, the decay 
probability is determined by calculating decay width for neutron, proton, 
alpha and gamma. Final states differ in excitation energy, angular momentum, 
nucleon number $A$ and proton number $Z$. \\
The model contains several parameters, which has to be fixed. 
The main ingredients of the code are the level density and transmission 
coefficients.
The level density $\rho(E,J)$, at a specific value $J$, is given by: 
\begin{equation}
 \rho (E,J) = (2J + 1) \omega (E - E_{rot}(J))
\end{equation}
where: $\omega (E)$ is the total level density at excitation energy $E$, 
evaluated according the following formula (given by Weisskopf for Fermi gas, 
see Ref. \cite{Weis37}, \cite{Bohr75}):   
\begin{equation}
\omega (E) \approx exp[2(a E)^{1/2}]
\label{eq:nucl_lev_dens}
\end{equation}
with level density parameter $a=const$ (its dimension is MeV$^{-1}$). \\
For selecting the rotational energy contribution, $E_{rot}(J)$, model of A. J. 
Sierk \cite{Sier86} is used. \\
The transmission coefficients $T_{l}$ are calculated with use of default 
potentials; for protons and neutrons taken from Perey and Perey \cite{Pere76}, 
for alpha particles from Igo and Huizenga \cite{Huiz62}. 
As the nucleus decays, it is assumed that transmission coefficients of charged 
particles are shifted in their kinetic energy dependence: $E_{k}-V_{c}(A,Z)$, 
where $A$ and $Z$ are the nucleon and proton numbers of the emitting nucleus, 
and $V_{c}$ is the Coulomb barrier for the emitted particle.\\ 
Fission of the excited nuclei is also taken into account. 
The fission barrier of Sierk is used \cite{Sier86}.  

Gamma emission is considered statistically using $E1$, $E2$, $M1$ and $M2$
transition intensities.


\section{Generalized Evaporation Model}
\markboth{7.2 Generalized Evaporation Model}{Chapter 7. Statistical emission of particles}

\begin{figure}[!htcb]
\vspace{-0.75cm}
\hspace{-2cm}
\includegraphics[height=19cm, width=17cm, bbllx=0pt, bblly=25pt, bburx=594pt, bbury=842pt, clip=, angle=0]{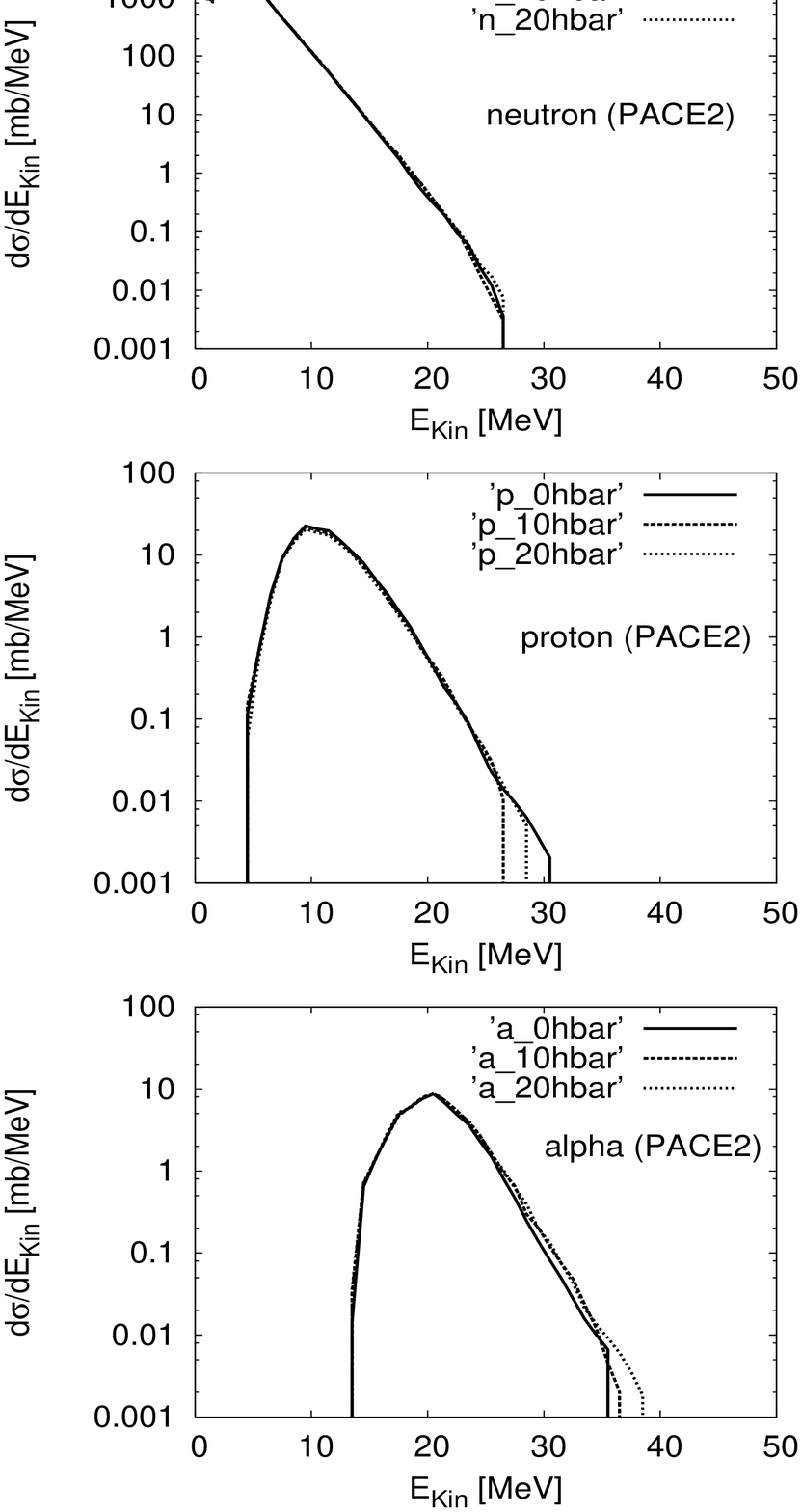}
\caption{{\sl Spectra of particles evaporated from excited Au nucleus, with 
the excitation energy equal to 100 MeV and different values of 
angular momentum; results of the PACE2 model (solid lines - zero value of 
angular momentum of the excited nucleus, dashed lines - angular momentum 
equal to 10 $\hbar$, dotted lines - 20 $\hbar$)}}
\label{fig:npa_0h_10h_20h}
\end{figure}

Another model used for the Monte Carlo simulations of the second stage of 
the spallation reaction is the Generalized Evaporation Model (GEM), 
developed by S. Furihata \cite{Furi00}, based on the Weisskopf-Ewing approach 
\cite{Weis37, Weis40}. 
In the GEM model, 66 nuclides up to Mg are included as ejectiles, not only
 the dominant particles emitted from an excited nucleus (i.e. nucleons and 
helium isotopes), as it is in the PACE2.
The accurate level density function is used for deriving the decay 
width of particle emission, instead of an approximate form, as used in the 
PACE2 model (see the previous Section). 
Additionally, depending on the excitation energy and mass of the decaying 
nucleus, different parametrizations of the level density $\rho (E)$ are 
applied, see Ref. \cite{Furi00}.\\
In the GEM model, contrary to the PACE2, the dependence of the level 
density on the angular momentum is neglected. Nevertheless, negligible  
difference between spectra of evaporated particles, calculated with the PACE2 
model, with zero and non-zero values of angular momentum of emitting nuclei,
has been noticed, as displayed in Fig. \ref{fig:npa_0h_10h_20h}. 
Adequate multiplicities of evaporated particles, calculated with the PACE2 
model, with different values of angular momentum of emitting nucleus are 
collected in Table \ref{table_npa_0h_10h_20h}.

\begin{figure}[!htcb]
\vspace{-0.75cm}
\hspace{-2cm}
\includegraphics[height=19cm, width=17cm, bbllx=0pt, bblly=25pt, bburx=594pt, bbury=842pt, clip=, angle=0]{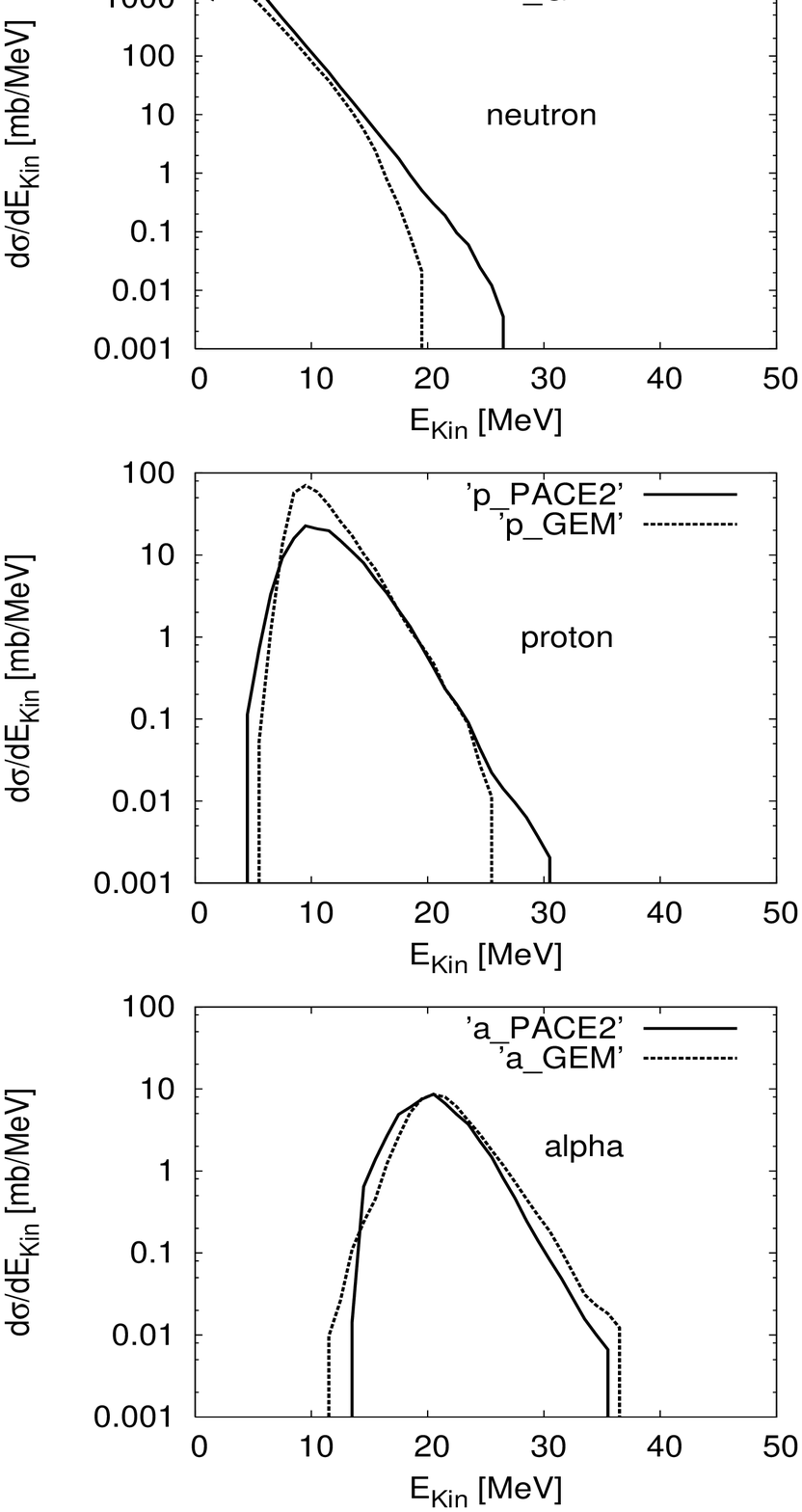}
\caption{{\sl Spectra of particles evaporated from excited Au 
nucleus, with the excitation energy equal to 100 MeV and zero value of angular
momentum; solid lines - results of the PACE2 model, dashed lines - results 
of the GEM model}}
\label{fig:npa_gem_pace2}
\end{figure}

\begin{table}[tbp]
\caption{{\sl Multiplicity of particles evaporated from excited Au nucleus,
with excitation energy equal to 100 MeV and different values of angular 
momentum; results of the PACE2 model calculations}}
\begin{center}
\begin{tabular}{|l| |l| |l| |l|}
\hline
 & 0 $\hbar$ & 10 $\hbar$ & 20 $\hbar$ \\
\hline
n: & 7.80 & 7.80 & 7.70\\
\hline
p: & 0.076 & 0.080 & 0.070\\
\hline
$^{4}$He: & 0.031 & 0.031 & 0.031\\
\hline
\end{tabular}
\end{center}
\label{table_npa_0h_10h_20h}
\end{table}
\noindent Spectra of evaporated particles (nucleons and $^{4}$He)  
calculated with the PACE2 and GEM models are in good agreement, as shown 
in Fig. \ref{fig:npa_gem_pace2} for an example Au nucleus, with the 
excitation energy set to 100 MeV and zero value of angular momentum.  
Adequate multiplicities of the emitted particles are collected in 
Table \ref{table_npa_g_p}.

\begin{table}[tbp]
\caption{{\sl Multiplicity of particles evaporated from excited Au nucleus,
with excitation energy equal to 100 MeV; results of the GEM and the PACE2 model 
calculations}}
\begin{center}
\begin{tabular}{|l| |l| |l|}
\hline
 & GEM & PACE2 \\
\hline
n: & 8.86 & 7.80\\
\hline
p: & 0.17 & 0.076\\
\hline
$^{4}$He: & 0.029 & 0.031\\
\hline
\end{tabular}
\end{center}
\label{table_npa_g_p}
\end{table}

\begin{figure}[!htcb]
\vspace{-0.75cm}
\hspace{-2cm}
\includegraphics[height=19cm, width=17cm, bbllx=0pt, bblly=15pt, bburx=594pt, bbury=842pt, clip=, angle=0]{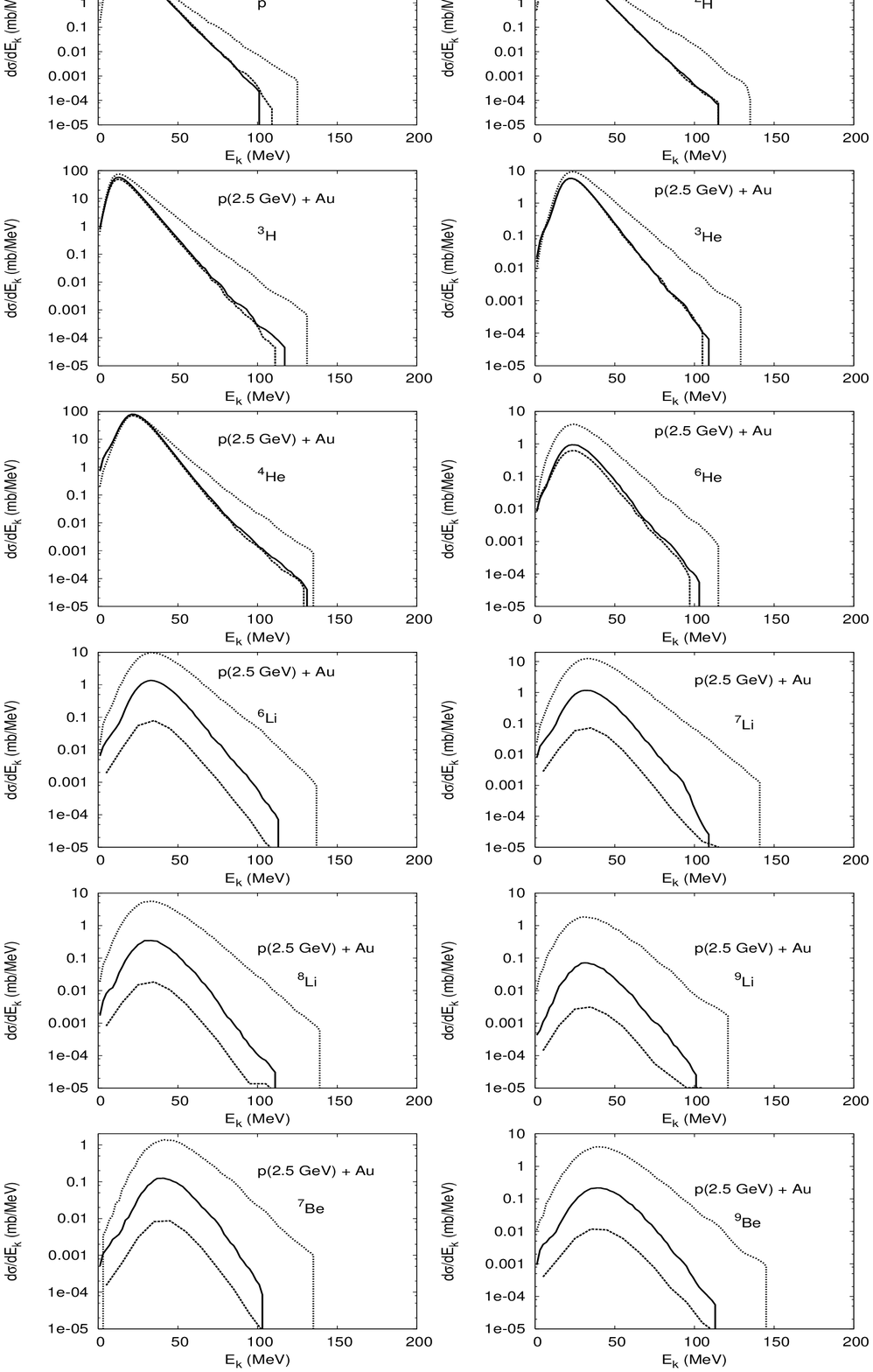}
\caption{{\sl Spectra of isotopes evaporated during p+Au reaction, at 2.5 GeV 
beam energy; results of the HSD+GEM model calculations (solid lines - 
calculations with the GCCI level density parameter, dashed lines - 
calculations with $a=A/8$, dotted lines - $a=A/20$)}}
\label{fig:pAu_10a_2.5}
\end{figure}

\begin{figure}[!htcb]
\vspace{-0.75cm}
\hspace{-2cm}
\includegraphics[height=19cm, width=17cm, bbllx=0pt, bblly=130pt, bburx=594pt, bbury=842pt, clip=, angle=0]{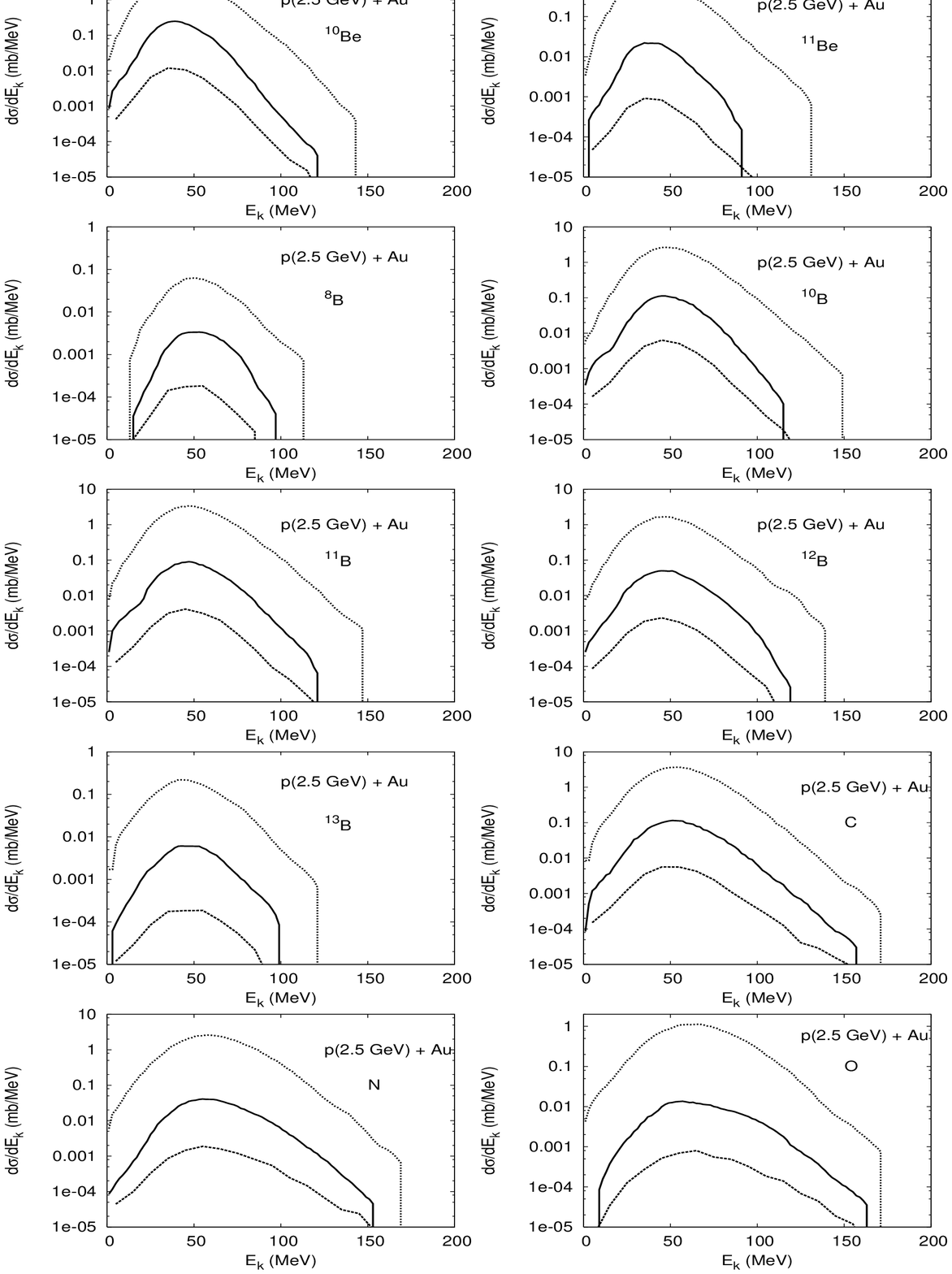}
\caption{{\sl Spectra of isotopes evaporated during p+Au reaction, at 2.5 GeV 
beam energy; results of the HSD+GEM model calculations (solid lines - 
calculations with the GCCI level density parameter, dashed lines - 
calculations with $a=A/8$, dotted lines - $a=A/20$)}}
\label{fig:pAu_10b_2.5}
\end{figure}

Results of the GEM model calculations are very sensitive on the used level 
density parameter $a$ \cite{Furi00}. There is a possibility to use $a$ 
parameter equal to a quotient of mass of emitting nucleus and a 
constant number $a=A/const$ MeV$^{-1}$ (see formula \ref{eq:nucl_lev_dens}), 
where $const \approx 5 \div 20$ (see Ref.\cite{Dostr58, Dostr59}) or 
one can use the precise Gilbert - Cameron - Cook - Ignatyuk (GCCI) level 
density parametrization with the energy 
dependence taken into account, see Ref.\cite{Furi00, Gilb65}.
Dependence on the used method of level density parametrization is noticeable 
e.g. in multiplicities and spectra of evaporated particles.  
Comparison of spectra of various isotopes evaporated during the second 
stage of an example p+Au reaction, at 2.5 GeV beam energy, 
results of the GEM model calculations, with the different level density 
parameters are presented in Figures \ref{fig:pAu_10a_2.5} and 
\ref{fig:pAu_10b_2.5}.
It is seen from the Figures, that distributions of light particles 
(isotopes of hydrogen and helium) calculated with different level density 
parameters are very similar. Contrary, distributions of heavier isotopes 
(lithium, beryllium, etc.) differ significantly. 
Use of the GCCI level density parameter gives much higher abundance 
 of the heavier isotopes, than use of the simple $a=A/8$ parameter. 
Moreover, calculations with the $a=A/20$ parameter result with even higher 
abundance than with the GCCI level density parameter. 
Adequate multiplicities of the evaporated particles are collected in Table 
\ref{table_gcci8}. 
It is seen from the Figures, that choice of level density parameter affects not 
only heights of the spectra, but influences also their slopes. 
In the frame of the work, the GCCI level density parameter is used 
(confrontation with experimental data leads to conclusion, that use of the 
GCCI level density parameter gives the most reliable results).  \\
\begin{table}[tbp]
\caption{{\sl Multiplicity of particles evaporated during p+Au reaction, at 2.5
 GeV proton beam energy, calculated with the GCCI, $a=A/8$ and $a=A/20$ level 
density parameter}}
\begin{center}
\begin{tabular}{|l| |l| |l| |l|}
\hline
 & GCCI & $a=A/8$ & $a=A/20$ 
\\
\hline
n: & 14.78 & 14.90 & 9.016
\\
\hline
p: & 1.76 & 1.97 & 0.85
\\
\hline
$^{2}$H: & 1.11 & 1.11 & 1.17
\\
\hline
$^{3}$H: & 0.51 & 0.44 & 0.78
\\
\hline
$^{3}$He: & 0.063 & 0.062 & 0.12
\\
\hline
$^{4}$He: & 0.84 & 0.76 & 0.85
\\
\hline
$^{6}$He: & 0.012 & 0.0083 & 0.059
\\
\hline
$^{6}$Li: & 0.020 & 0.013 & 0.17
\\
\hline
$^{7}$Li: & 0.018 & 0.012 & 0.22
\\
\hline
$^{8}$Li: & 0.0055 & 0.0032 & 0.104
\\
\hline
$^{9}$Li: & 0.0011 & 0.00057 & 0.035
\\
\hline
$^{7}$Be: & 0.0021 & 0.0016 & 0.027
\\
\hline
$^{9}$Be: & 0.0039 & 0.0024 & 0.081
\\
\hline
$^{10}$Be: & 0.0045 & 0.0024 & 0.11
\\
\hline
$^{11}$Be: & 0.00041 & 0.00019 & 0.021
\\
\hline
$^{8}$B: & 0.00006 & 0.00004 & 0.0013
\\
\hline
$^{10}$B: & 0.0022 & 0.0013 & 0.060
\\
\hline
$^{11}$B: & 0.0018 & 0.00091 & 0.076
\\
\hline
$^{12}$B: & 0.0011 & 0.00053 & 0.039
\\
\hline
$^{13}$B: & 0.00013 & 0.00005 & 0.0049
\\
\hline
C: & 0.0027 & 0.0015 & 0.093
\\
\hline
N: & 0.0011 & 0.00055 & 0.071
\\
\hline
O: & 0.00041 & 0.00025 & 0.034
\\
\hline
\end{tabular} 
\end{center}
\label{table_gcci8}
\end{table}


\chapter{Bulk models predictions for the proton induced nuclear reactions}
\markboth{ }{Chapter 8. Bulk models predictions}

\section{Illustration of energy balance of the reaction}
\label{sec:en_balance}

In order to show, how large fraction of energy is carried out 
during the first and the second stage of proton - nucleus reaction, 
respectively, a sample case of the HSD and statistical evaporation model 
calculations, for p + Au collision at 2.5 GeV proton beam energy is analysed 
in this Section. \\
The total initial kinetic energy introduced into nucleus by incoming proton is 
equal to 2.5 GeV. During the first stage of the reaction, about 1.7 GeV of the 
energy is carried away by nucleons and about 0.4 GeV - by pions. 
The multiplicity of nucleons is near 4.43, whereas the pion multiplicity, 
summed over all pion charges, is equal to 1.48. 
The average kinetic energy of one nucleon is about 0.4 GeV, whereas the average
 kinetic energy of one pion is equal to 0.29 GeV. 
Non-negligible amount of energy is taken away from the excitation energy and 
converted into rest masses of emitted pions, i.e. 0.2 GeV.
Remaining energy of the excited nucleus after the first stage 
of the reaction, is equal to only about 200 MeV. 
This amount of energy is then evaporated in the second stage of the reaction by
 emission of mainly neutrons, protons, alphas and gammas. 
Emitted neutrons accumulate about 20 MeV of the energy, protons carried 
about 3 MeV, alphas - about 1 MeV, and gammas - about 6 MeV. 
The multiplicity of neutrons is equal to about 10 and the average kinetic 
energy of one neutron is equal to about 2 MeV, multiplicity of protons is equal
 to about 1 and the average kinetic energy of one proton is equal to about 3 
MeV. Multiplicity of alphas is equal to about 0.3 and the average kinetic 
energy of one alpha is equal to about 4 MeV and finally multiplicity of gammas 
is equal to about 6 and the average kinetic energy of one gamma is equal to 
1 MeV. 
After taking into account the binding energy of nucleons inside nucleus, i.e. 
about 8 MeV per one nucleon, the whole amount of excitation energy of the 
residual nucleus (about 200 MeV) is carried out during the second stage of the 
reaction. \\
The presented balance shows, that the biggest amount of initial energy 
deposited in nucleus by incoming proton is carried out during the first stage 
of the reaction, i.e. about 85 $\%$, and only about 15 $\%$ of the initial 
energy is carried away during the second stage of the reaction.

\section{Participation of fission processes in spallation reaction}
\label{sec:fission}
\markboth{8.2 Participation of fission processes}{Chapter 8. Bulk models predictions}

During the second stage of the proton - nucleus spallation reaction, 
deexcitation of the hot residual nucleus by evaporation of various isotopes 
takes place, as it is mentioned above in this work. 
However, there is a possibility, that during this process the evaporating 
nucleus undergoes fission. In such case, the particles emission is continued 
from the fission fragments.
Probability of fission of particular nuclei depends on the ratio of Coulomb 
energy and surface energy of individual nucleus. The probability is strongly 
dependent on fissility parameter ($f$) defined as:
\begin{equation}
f=Z^{2} / A
\label{eq:fiss_def}
\end{equation}    
where: $Z$ is the charge of a nucleus and $A$ is the mass number of a nucleus.

\begin{figure}[!ht]
\begin{center}
\includegraphics[height=8cm, width=10cm, angle=0]{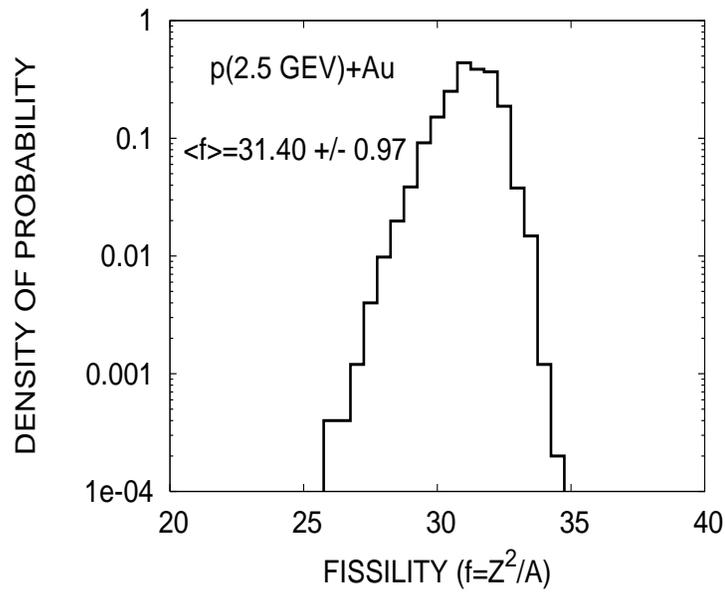}
\caption{{\sl Distribution of the fissility parameter for residual nuclei 
emerged after the first stage of p+Au reaction, at 2.5 GeV beam energy;  
results of the HSD model calculations}}
\label{fig:fiss_hist}
\end{center}
\end{figure}

For example, the fissility parameter for Au nucleus, calculated 
according to the definition (\ref{eq:fiss_def}) is equal to 31.7.
The experimental data collected in Ref. \cite{Carv64} indicate, that Au nucleus 
has rather low probability for fission, in comparison of Th and U nuclei.    
Distribution of the fissility parameter for 
residual nuclei emerged after the first stage of an example p+Au reaction, 
at 2.5 GeV beam energy (result of the HSD model calculations) is presented in 
Fig. \ref{fig:fiss_hist}. 
The average value of the fissility parameter distribution is equal to 
$31.40 \pm 0.97$.

\begin{figure}[!ht]
\begin{center}
\includegraphics[height=6cm, width=8cm, angle=0]{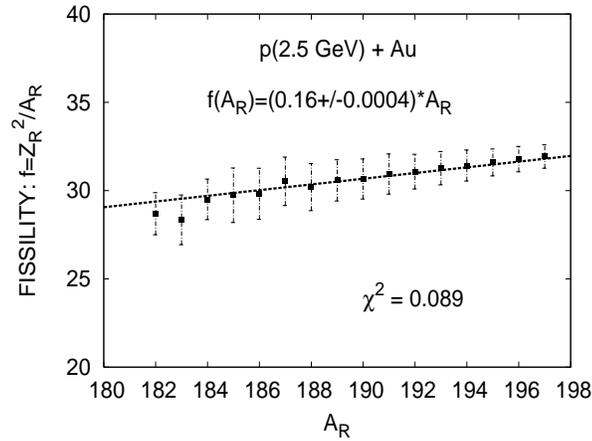}
\caption{{\sl The fissility parameter in function of mass of residual nuclei 
for the example p+Au reaction, at 2.5 GeV beam energy; 
results of the HSD model calculations 
($\chi ^{2}$ means chisquare per degree of freedom)}}
\label{fig:kr_A_fiss}
\end{center}
\end{figure}

\begin{figure}[!ht]
\begin{center}
\includegraphics[height=6cm, width=8cm, angle=0]{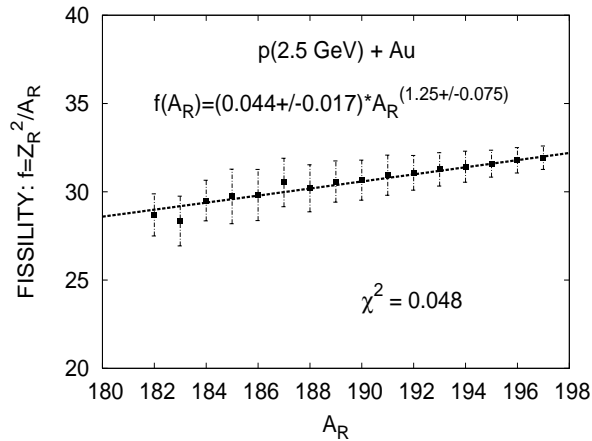}
\caption{{\sl The fissility parameter in function of mass of residual nuclei for
the example p+Au reaction, at 2.5 GeV beam energy; 
results of the HSD model calculations 
($\chi ^{2}$ means chisquare per degree of freedom)}}
\label{fig:kr1_A_fiss}
\end{center}
\end{figure}

On the base of a behavior of the fissility parameter in function of 
mass of residual nuclei, it can be tested if a ratio of charge and mass 
number is constant during the reaction and equal to the ratio of charge and 
mass number of the initial target nucleus. The fissility parameter in function 
of mass of residual nuclei for the example p+Au reaction, at 2.5 GeV proton 
beam energy is presented in Fig. \ref{fig:kr_A_fiss}.    
The ratio of charge and mass number of the Au initial target nucleus is equal 
to: Z(Au)/A(Au) = 79/197 = 0.40. Using the definition of the fissility 
parameter (\ref{eq:fiss_def}) and assuming that: Z=0.40$\cdot$A, 
it is obtained that: f=0.16$\cdot$A.
Parametrization of the dependence presented in the Fig. \ref{fig:kr_A_fiss}, 
by a function proportional to the mass number give exactly the same result as 
received from the definition. This indicates, that in a first approximation, 
a ratio of charge and mass number is constant during the reaction and equal to 
the ratio of charge and mass number of the initial target nucleus.  
However, parametrization of the dependence by a function proportional to some 
power of the mass number (correcting a slope) gives a bit better 
accuracy, as presented in Fig. \ref{fig:kr1_A_fiss}.
It means, that the ratio of charge and mass number is not exactly equal to 
the ratio for the initial target nucleus. This is confirmed by multiplicities 
of protons and neutrons emitted during first stage of the p+Au reaction, at 
2.5 GeV proton beam energy. Namely, due to interactions inside the target 
nucleus, multiplicity of emitted protons (equal to 1.66) 
is a bit higher than multiplicity of emitted neutrons (equal to 1.63).
Contrary, in case of n+Au reaction, at 2.5 GeV neutron beam energy, 
multiplicity of emitted neutrons during the first stage of the reaction 
(equal to 1.89) is a bit higher than multiplicity of emitted protons 
(equal to 1.38). This indicates, that information about type of projectile is 
kept during the first stage of reaction. 

\begin{figure}[!ht]
\begin{center}
\includegraphics[height=6cm, width=8cm, angle=0]{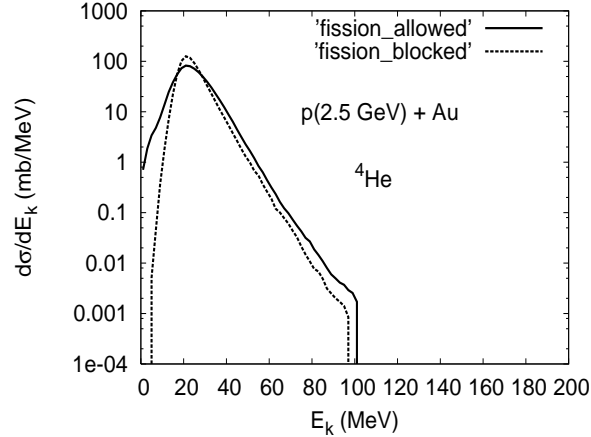}
\caption{{\sl The kinetic energy distributions of $^{4}$He evaporated 
during the second stage of p+Au reaction, at 2.5 GeV beam energy; 
results of the HSD+GEM model calculations
(solid line - fission allowed, dashed line - fission blocked)}}
\label{fig:alfy_fiscomp}
\end{center}
\end{figure}

If looking at the kinetic energy distributions of isotopes emitted during the 
second stage of p+Au reaction, calculated with blocked and allowed fission, 
differences are visible, because Au nucleus possesses non-zero, although rather
 small fissility. The example spectra (result of the GEM model 
calculations) are presented in Fig. \ref{fig:alfy_fiscomp}.      
It is clearly seen from the Figure, that fission shifts spectra of emitted 
particles down in their maximum and causes them wider in energetical range, 
in direction of both lower and higher energies (in Fig. \ref{fig:alfy_fiscomp} 
the effect is shown for alpha particles). This is because of  
evaporation from fission fragments, which move in opposite directions. 
As a consequence there is a bit less particles with intermediate energies, 
emitted from the unfissioned residuum, but more particles with a bit lower and 
higher energies, emitted after fission, from excited fragments moving in 
backward and forward directions, respectively.      

\begin{figure}[!htcb]
\vspace{-2cm}
\hspace{-2cm}
\includegraphics[height=20cm, width=18cm, bbllx=0pt, bblly=50pt, bburx=594pt, bbury=842pt, clip=, angle=0]{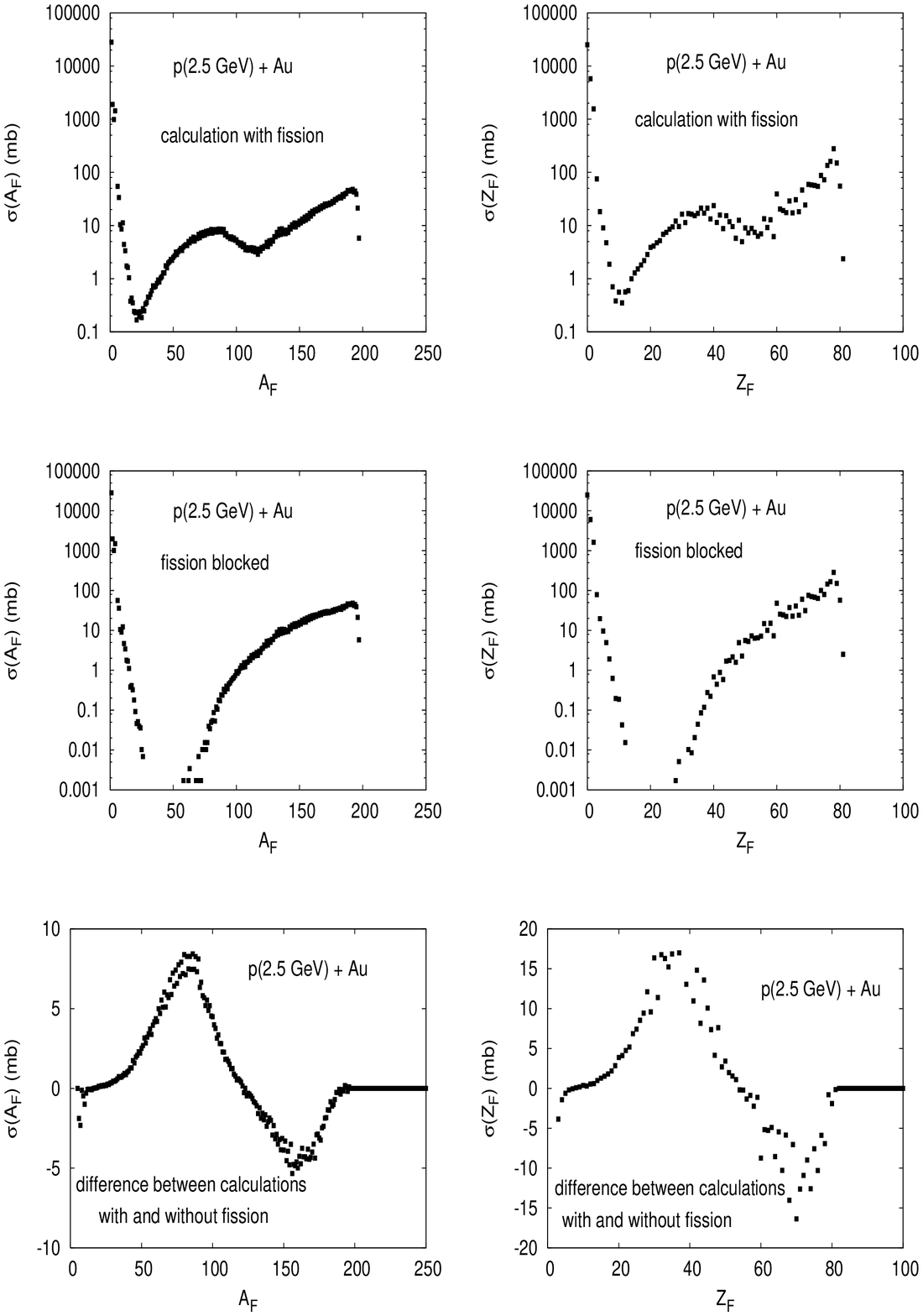}
\caption{{\sl Cross section for different fragments production in function of 
fragment mass and charge from p+Au reaction, at 2.5 GeV beam energy,
with (a) allowed and (b) blocked fission; results of the HSD+GEM model 
calculations; (c) differences between calculations with and without fission}}
\label{fig:af_zf_subtr}
\end{figure}

Cross section for the different fragments production in function of the 
fragments masses and charges (result of the HSD+GEM model calculations), for 
the example p+Au reaction, at 2.5 GeV proton beam energy, calculated with 
blocked and allowed fission, are presented in Fig. \ref{fig:af_zf_subtr}.   
Interesting result is obtained, when one subtracts the cross sections obtained 
with blocked and allowed fission.
One can see in the Figures that fission diminishes amount of nuclei of masses 
close to 150 - 160 and increases amount of nuclei of masses in the range 60 - 
100.
One can conclude, that fission is much slower process than evaporation. First 
evaporation from a hot residual nucleus takes place. 
Due to evaporation both the mass and excitation energy of residuum decrease. 
Also the Coulomb energy and the surface energy of the nucleus is 
changing. If the emission of particles is slow, a possibility of fission 
occurs. Fission is a collective slow process. This is confirmed by 
experimental data collected in Ref. \cite{Carv64}, which show, that fission 
cross section decreases with increasing energy provided into nuclei. 
In case of very fast evaporation, fission has no possibility to occur.
The second stage of spallation reaction is a competition between 
evaporation and fission.
 


\chapter{Comparison of results of the HSD plus evaporation model calculations 
with experimental data}
\markboth{ }{Chapter 9. Comparison of calculations with experimental data} 

Result of the HSD plus statistical evaporation model calculations have been 
examined by confrontation with existing experimental data. 
Effects are discussed in this Chapter. 

\section{Neutron spectra}
\markboth{9.1 Neutron spectra}{Chapter 9. Comparison of calculations with experimental data}

\begin{figure}[!htcb]
\vspace{-1.5cm}
\hspace{-2cm}
\includegraphics[height=20cm, width=18cm, angle=0]{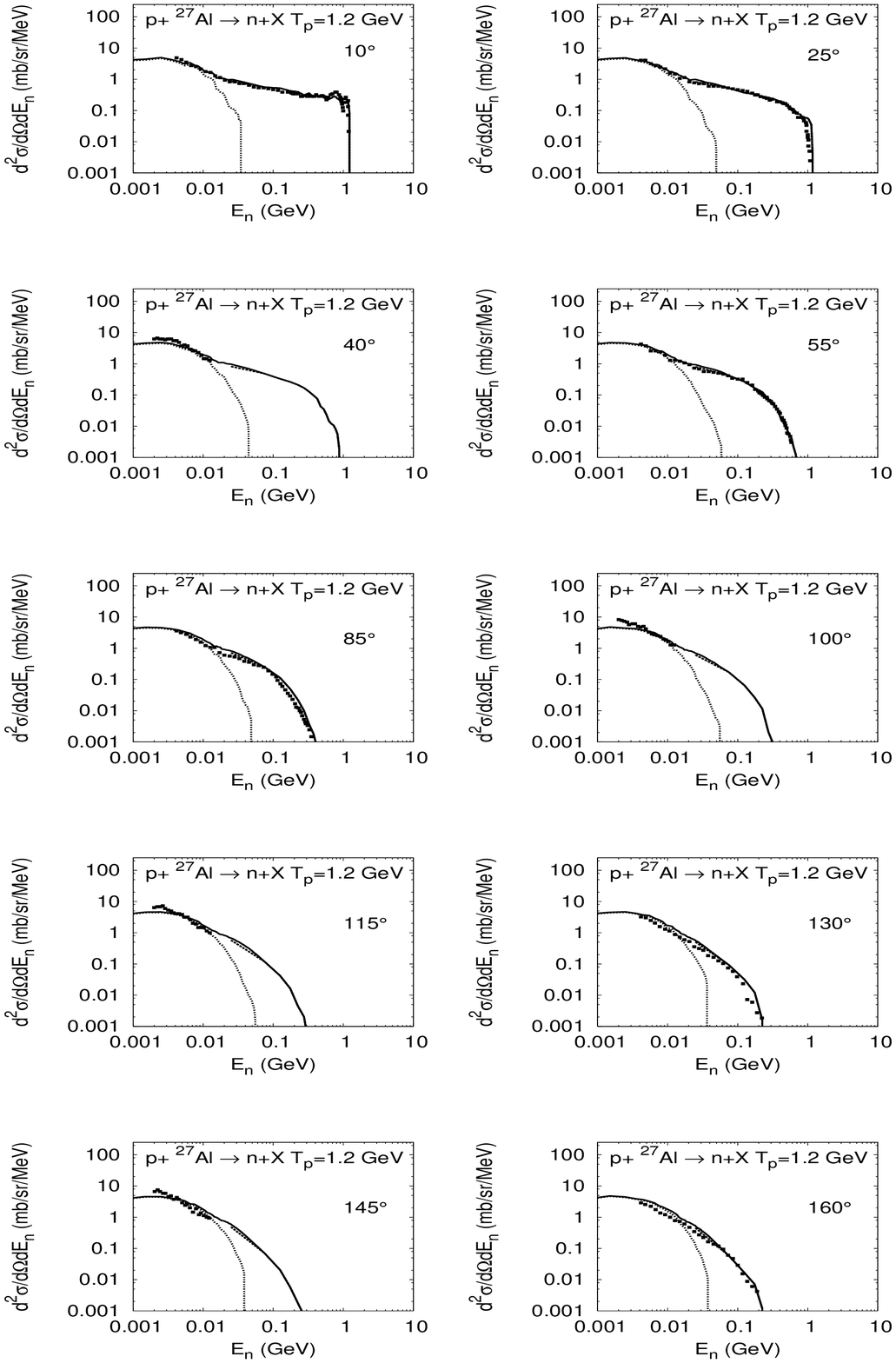}
\caption{{\sl Double differential neutron production cross section for p+Al
reaction, at 1.2 GeV proton beam energy; lines show results of the HSD+GEM 
model calculations 
(dashed and dotted lines are contributions of first and second stage of 
reaction, solid line is their sum), 
symbols indicate the experimental data \cite{Lera02}}}
\label{fig:pAl_nX_1.2}
\end{figure}

\begin{figure}[!htcb]
\vspace{-2cm}
\hspace{-2cm}
\includegraphics[height=20cm, width=18cm, angle=0]{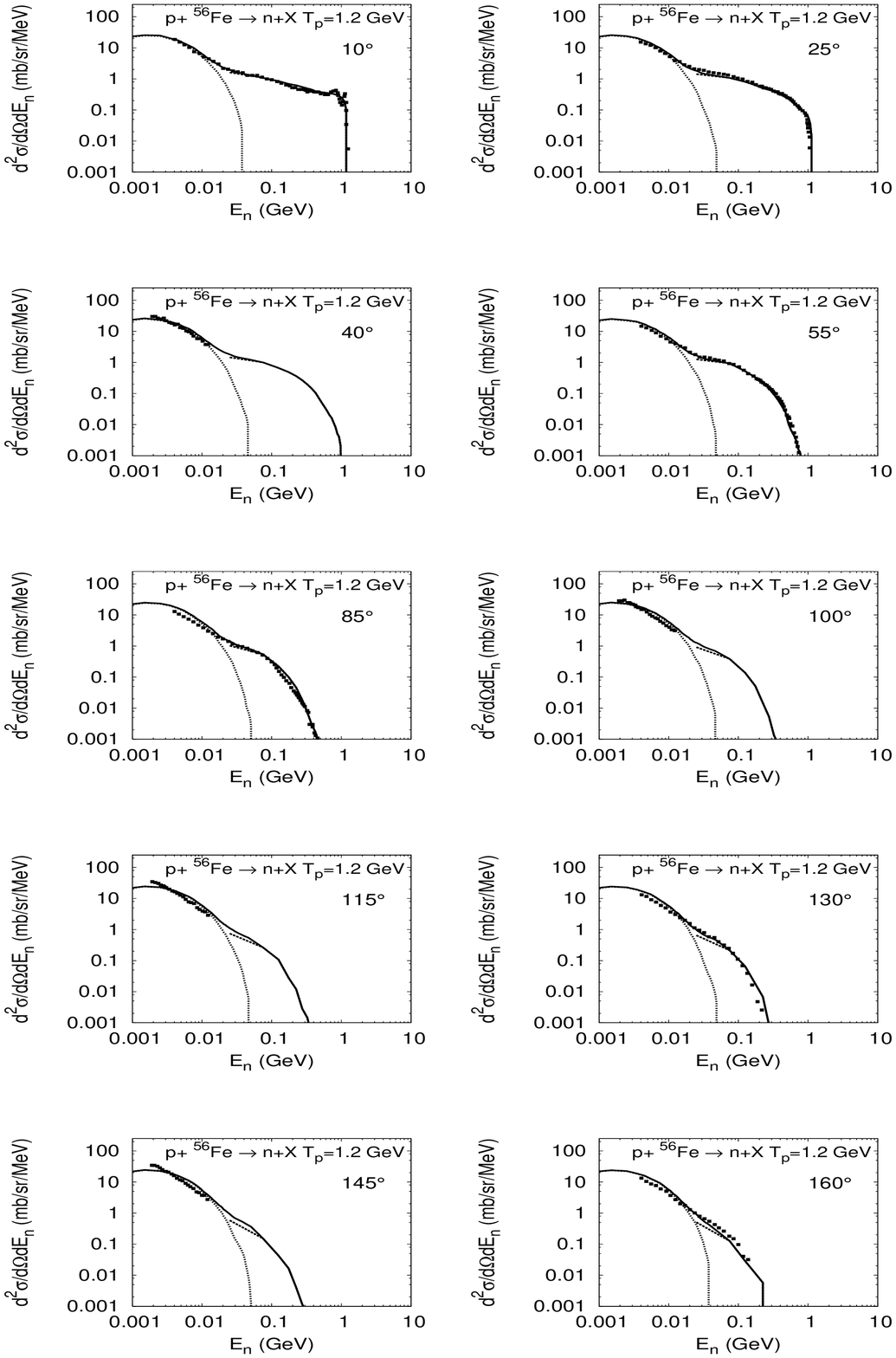}
\caption{{\sl Double differential neutron production cross section for p+Fe
reaction, at 1.2 GeV proton beam energy; lines show results of the HSD+GEM 
model calculations
(dashed and dotted lines are contributions of first and second stage of 
reaction, solid line is their sum), 
symbols indicate the experimental data \cite{Lera02}}}
\label{fig:pFe_nX_1.2}
\end{figure}

\begin{figure}[!htcb]
\vspace{-2cm}
\hspace{-2cm}
\includegraphics[height=20cm, width=18cm, angle=0]{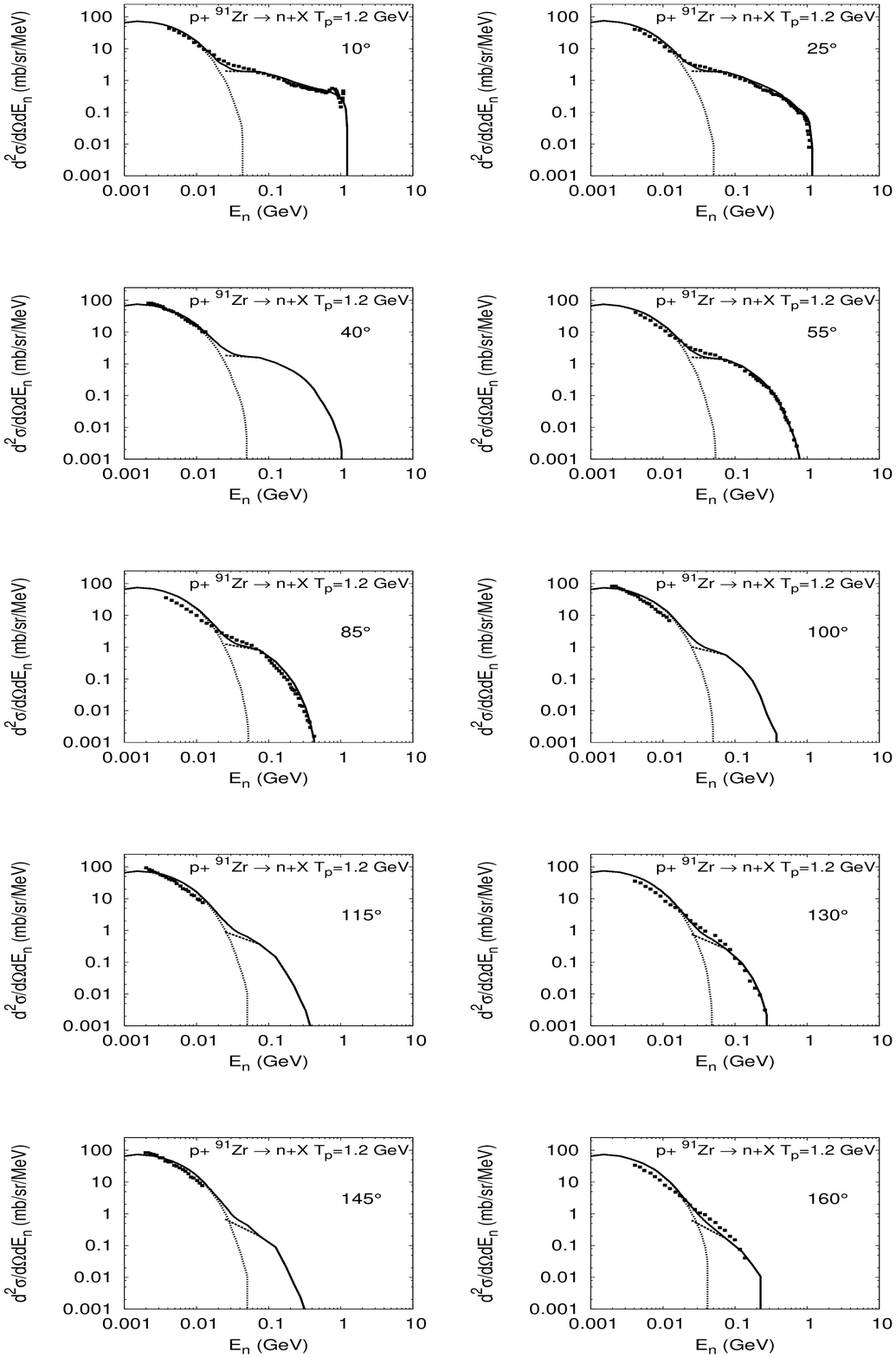}
\caption{{\sl Double differential neutron production cross section for p+Zr
reaction, at 1.2 GeV proton beam energy; lines show results of the HSD+GEM 
model calculations
(dashed and dotted lines are contributions of first and second stage of 
reaction, solid line is their sum), 
symbols indicate the experimental data \cite{Lera02}}}
\label{fig:pZr_nX_1.2}
\end{figure}

\begin{figure}[!htcb]
\vspace{-2cm}
\hspace{-2cm}
\includegraphics[height=20cm, width=18cm, angle=0]{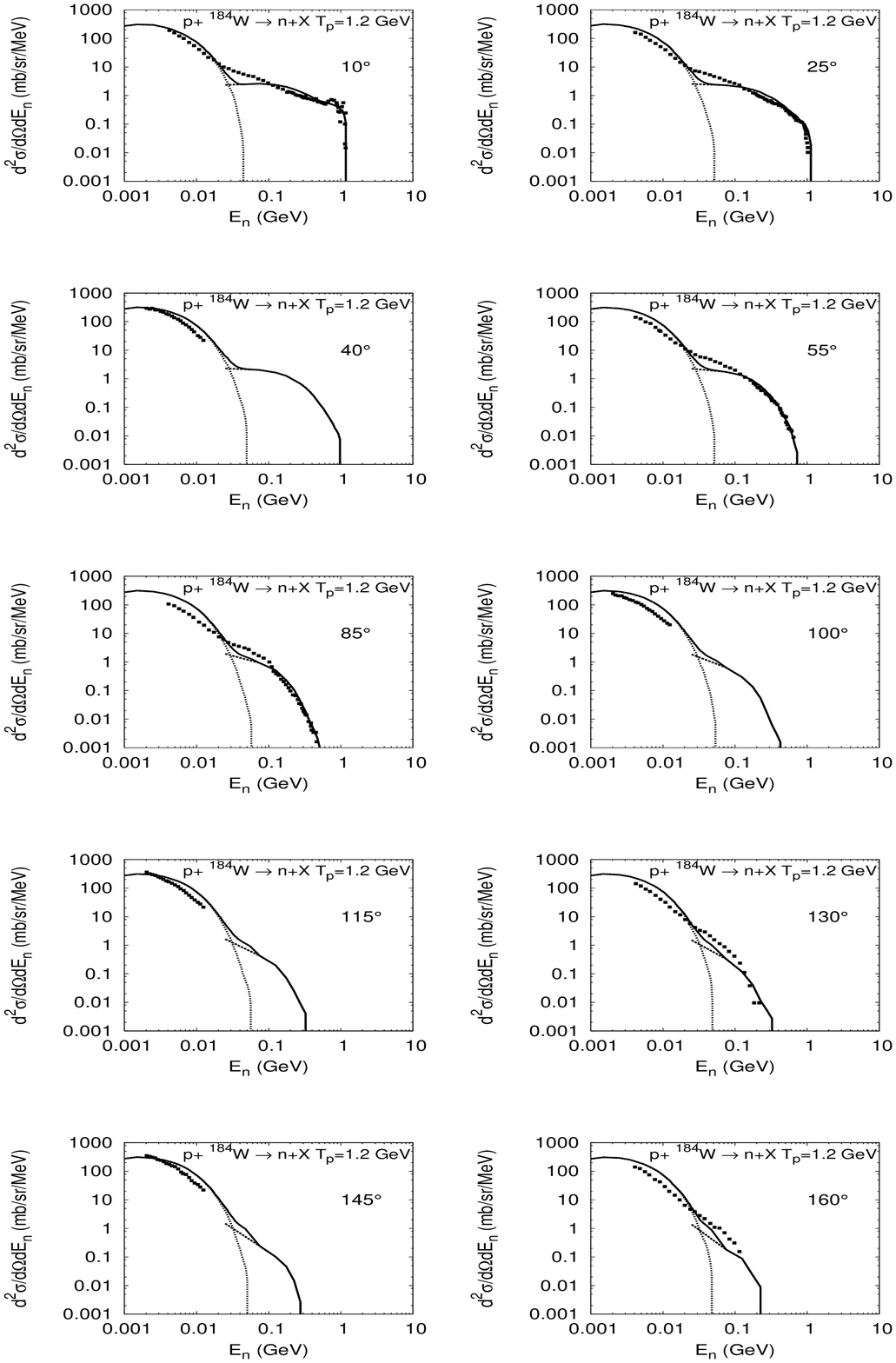}
\caption{{\sl Double differential neutron production cross section for p+W
reaction, at 1.2 GeV proton beam energy; lines show results of the HSD+GEM 
model calculations
(dashed and dotted lines are contributions of first and second stage of 
reaction, solid line is their sum), 
symbols indicate the experimental data \cite{Lera02}}}
\label{fig:pW_nX_1.2}
\end{figure}

\begin{figure}[!htcb]
\vspace{-2cm}
\hspace{-2cm}
\includegraphics[height=20cm, width=18cm, angle=0]{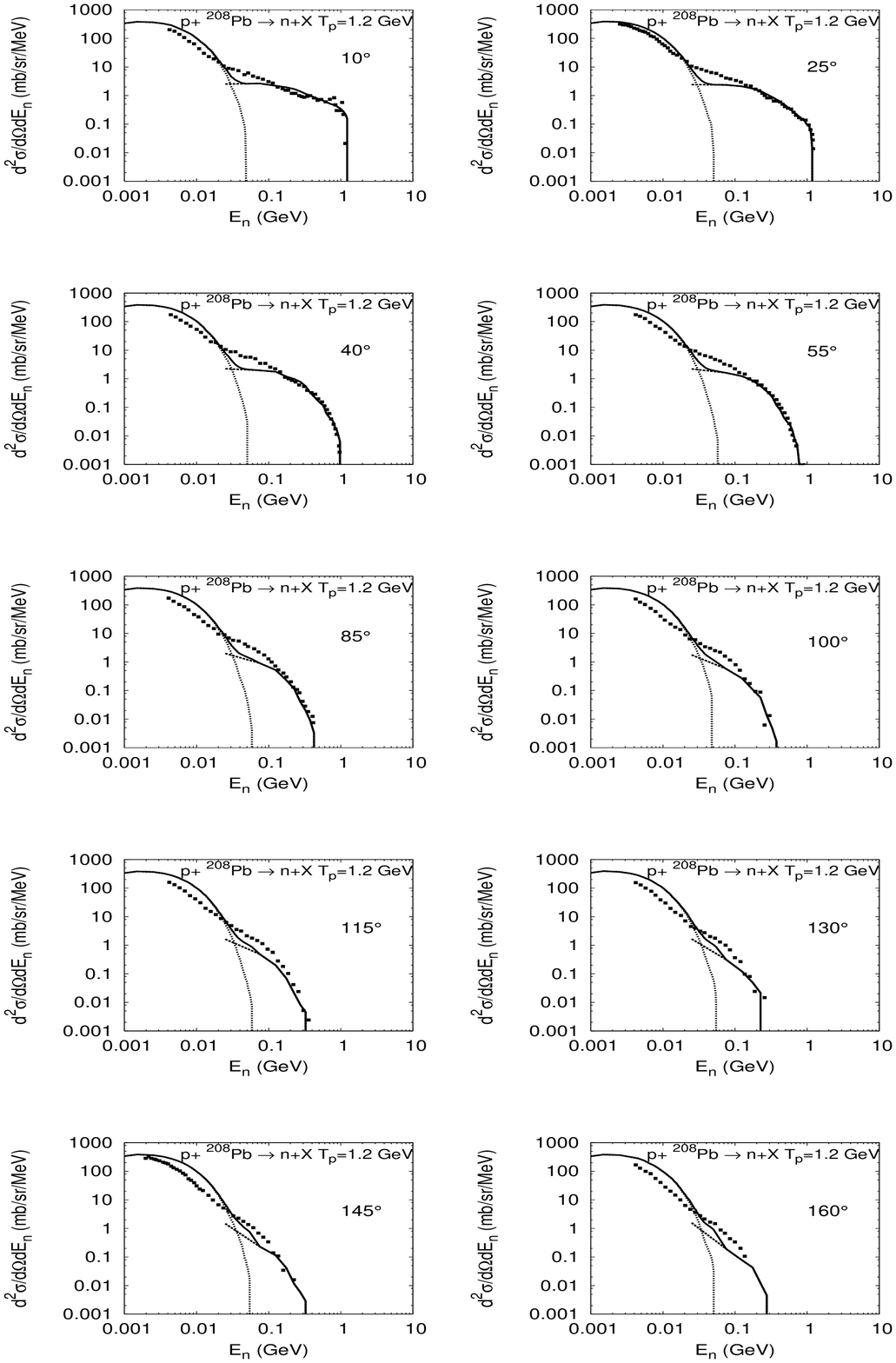}
\caption{{\sl Double differential neutron production cross section for p+Pb
reaction, at 1.2 GeV proton beam energy; lines show results of the HSD+GEM 
model calculations
(dashed and dotted lines are contributions of first and second stage of 
reaction, solid line is their sum), 
symbols indicate the experimental data \cite{Lera02}}}
\label{fig:pPb_nX_1.2}
\end{figure}

\begin{figure}[!htcb]
\vspace{-2cm}
\hspace{-2cm}
\includegraphics[height=20cm, width=18cm, angle=0]{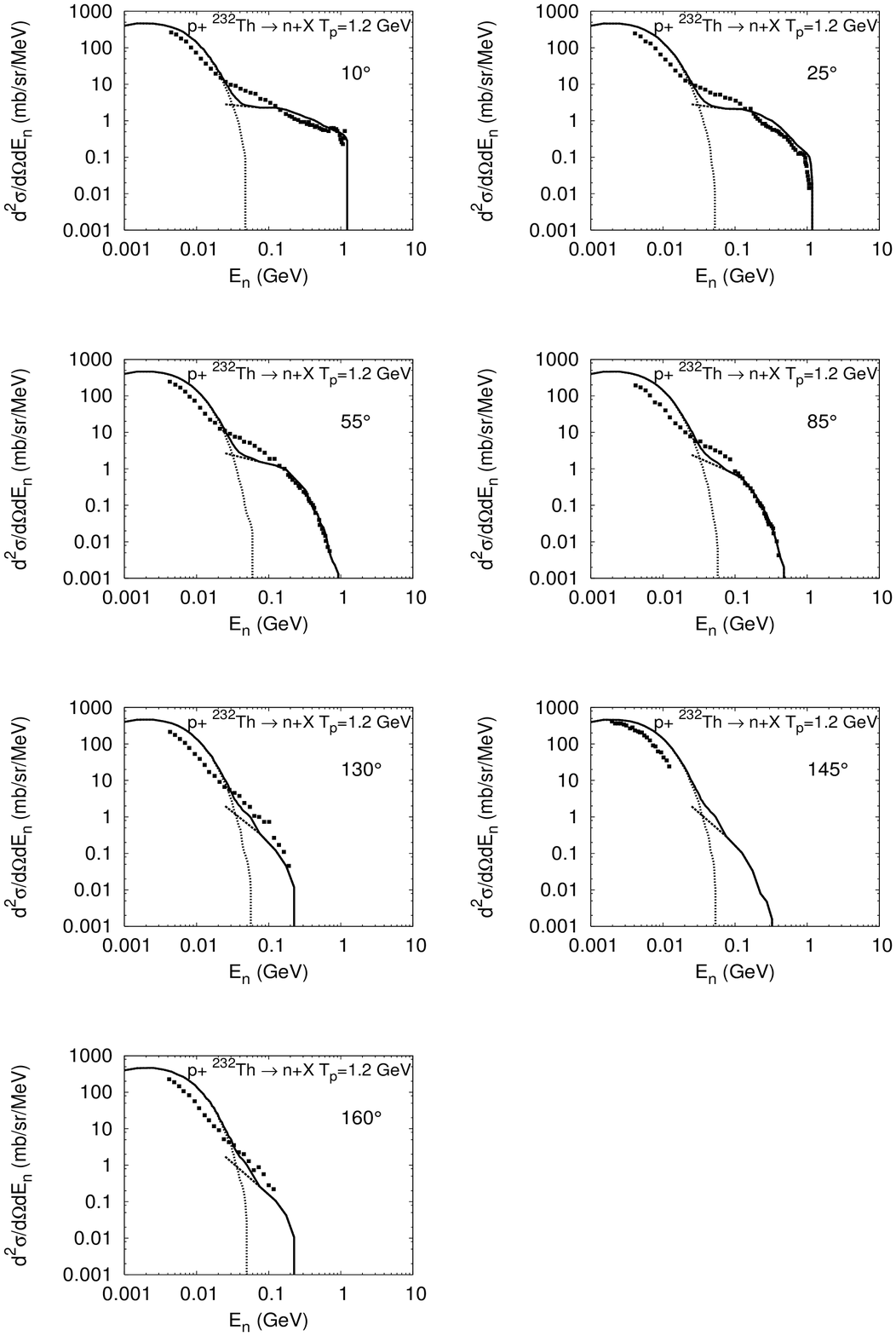}
\caption{{\sl Double differential neutron production cross section for p+Th
reaction, at 1.2 GeV proton beam energy; lines show results of the HSD+GEM 
model calculations
(dashed and dotted lines are contributions of first and second stage of 
reaction, solid line is their sum), 
symbols indicate the experimental data \cite{Lera02}}}
\label{fig:pTh_nX_1.2}
\end{figure}

\begin{figure}[!htcb]
\vspace{-2cm}
\hspace{-2cm}
\includegraphics[height=20cm, width=18cm, angle=0]{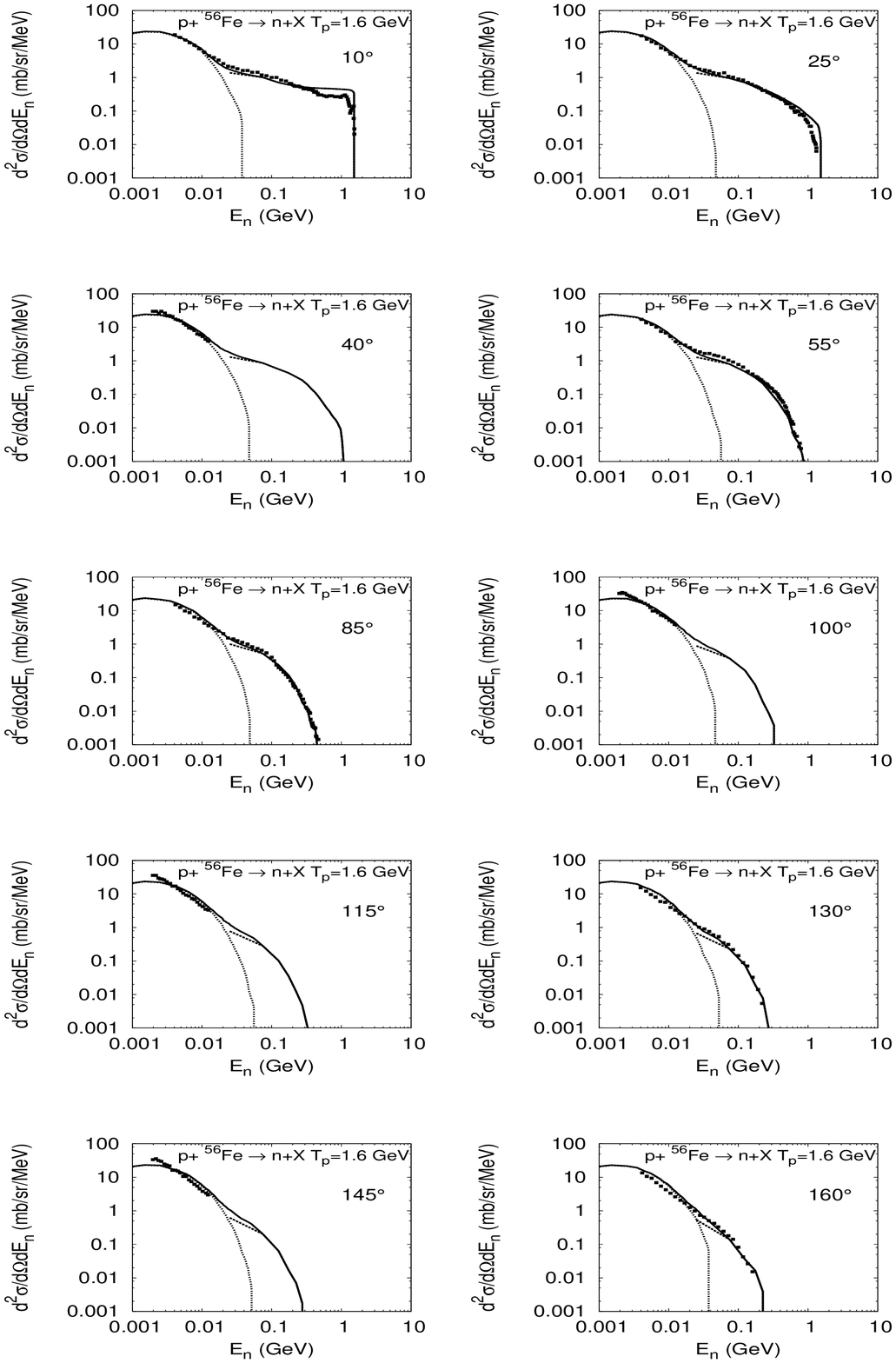}
\caption{{\sl Double differential neutron production cross section for p+Fe
reaction, at 1.6 GeV proton beam energy; lines show results of the HSD+GEM 
model calculations
(dashed and dotted lines are contributions of first and second stage of 
reaction, solid line is their sum), 
symbols indicate the experimental data \cite{Lera02}}}
\label{fig:pFe_nX_1.6}
\end{figure}

\begin{figure}[!htcb]
\vspace{-2cm}
\hspace{-2cm}
\includegraphics[height=20cm, width=18cm, angle=0]{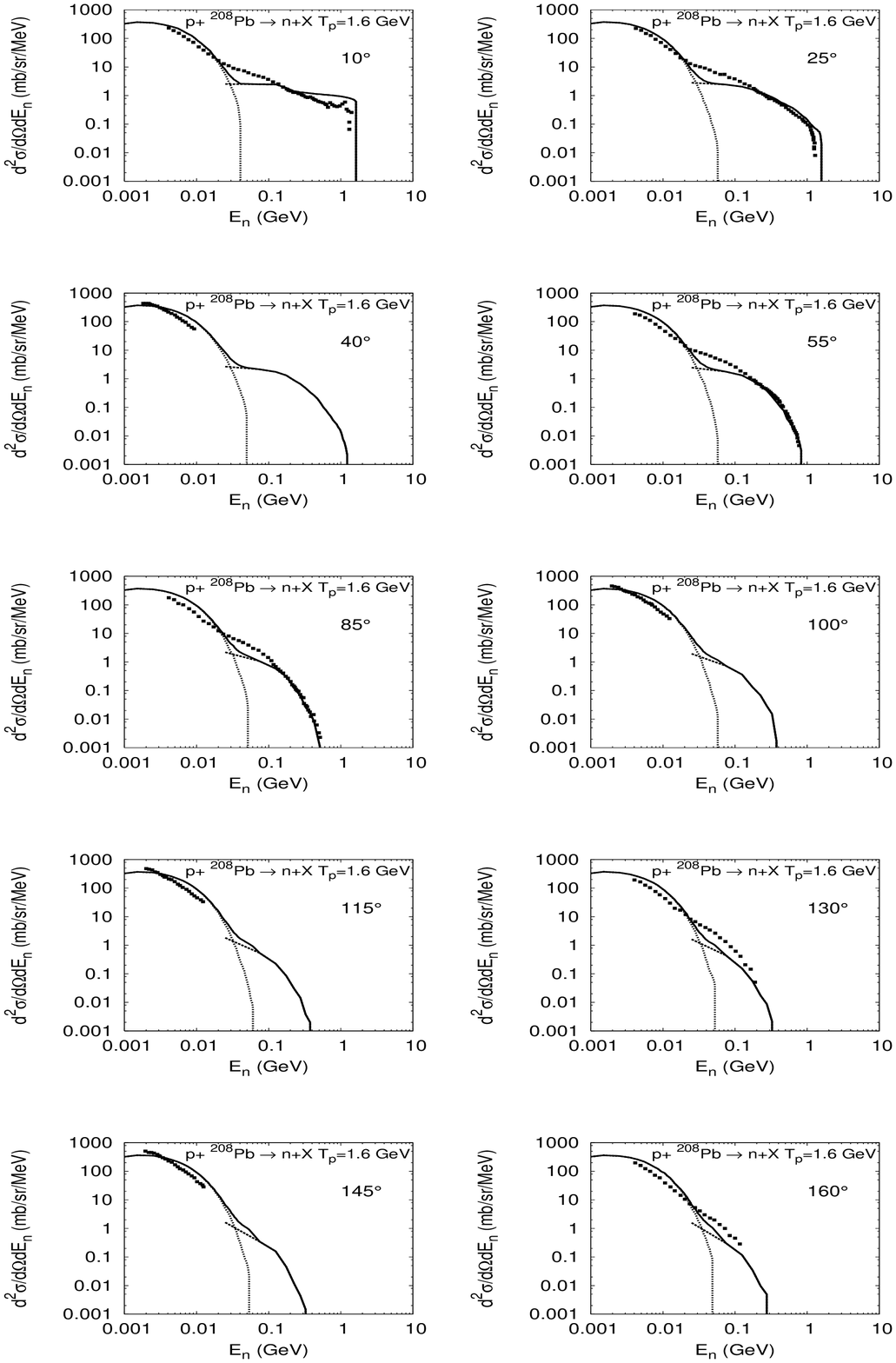}
\caption{{\sl Double differential neutron production cross section for p+Pb
reaction, at 1.6 GeV proton beam energy; lines show results of the HSD+GEM 
model calculations
(dashed and dotted lines are contributions of first and second stage of 
reaction, solid line is their sum), 
symbols indicate the experimental data \cite{Lera02}}}
\label{fig:pPb_nX_1.6}
\end{figure}

\begin{figure}[!htcb]
\vspace{-2cm}
\hspace{-2cm}
\includegraphics[height=20cm, width=18cm, angle=0]{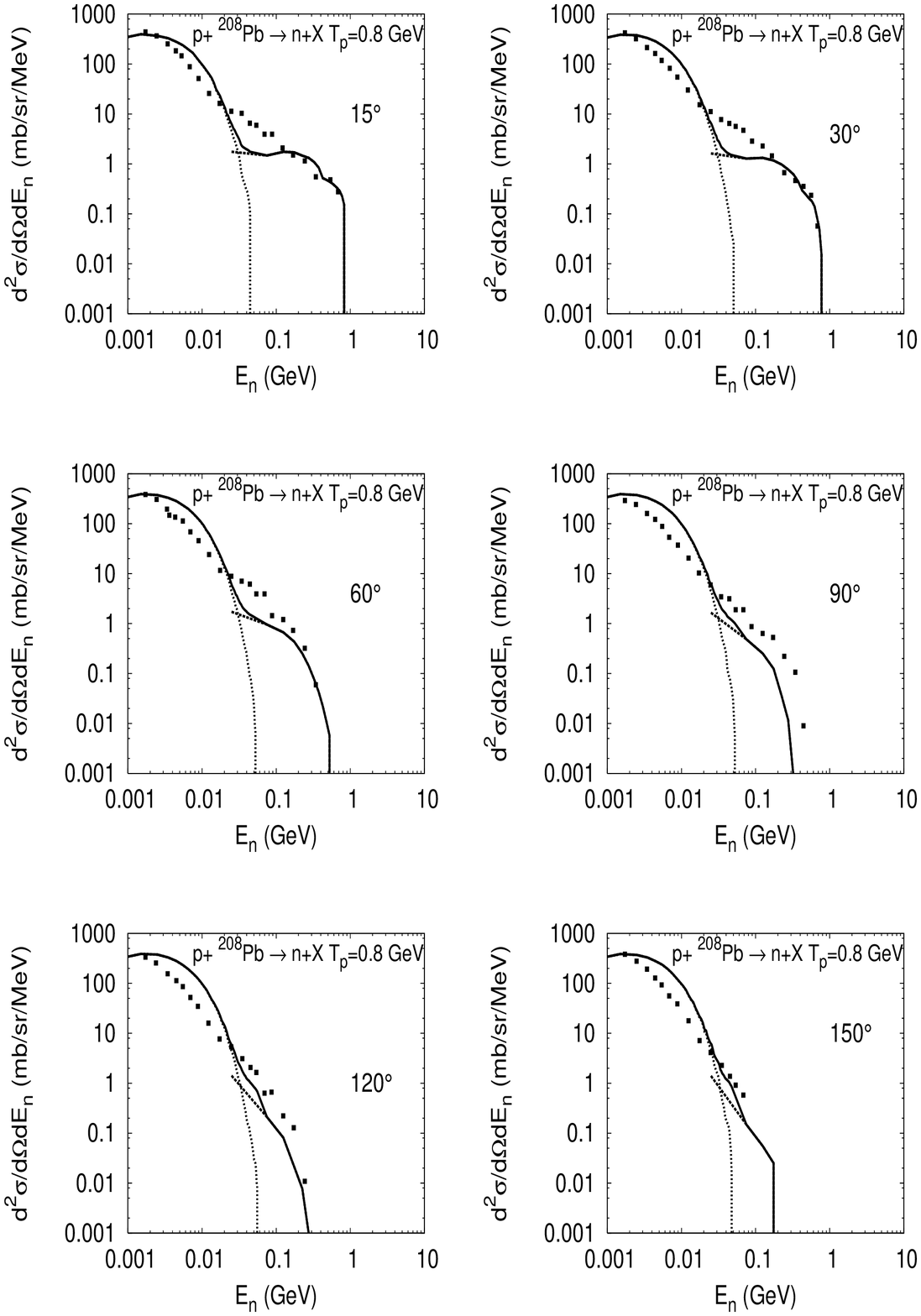}
\caption{{\sl Double differential neutron production cross section for p+Pb
reaction, at 0.8 GeV proton beam energy; lines show results of the HSD+GEM 
model calculations
(dashed and dotted lines are contributions of first and second stage of 
reaction, solid line is their sum), 
symbols indicate the experimental data \cite{Meig99}}}
\label{fig:pPb_nX_0.8}
\end{figure}

\begin{figure}[!htcb]
\vspace{-2cm}
\hspace{-2cm}
\includegraphics[height=20cm, width=18cm, angle=0]{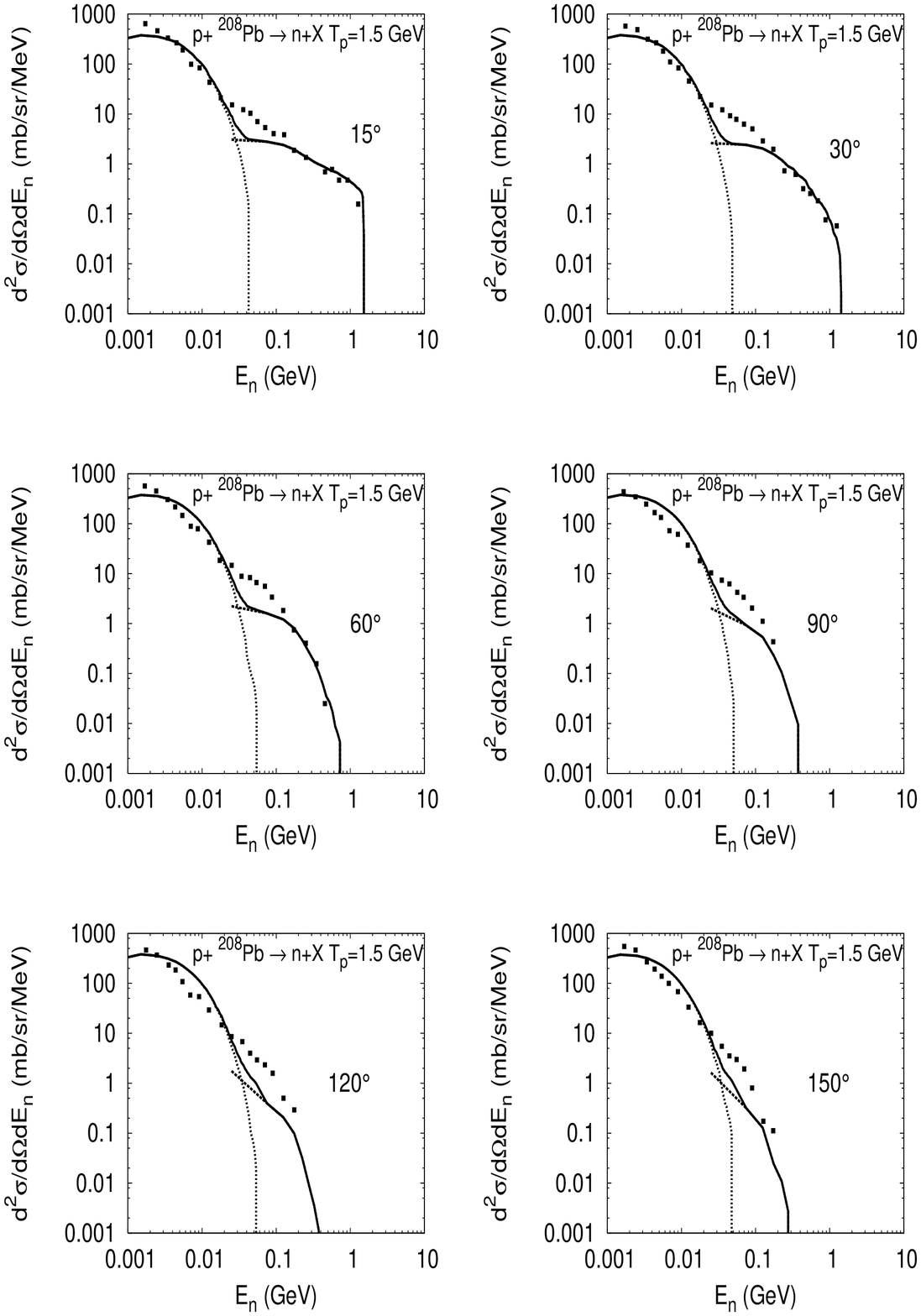}
\caption{{\sl Double differential neutron production cross section for p+Pb
reaction, at 1.5 GeV proton beam energy; lines show results of the HSD+GEM 
model calculations
(dashed and dotted lines are contributions of first and second stage of 
reaction, solid line is their sum), 
symbols indicate the experimental data \cite{Meig99}}}
\label{fig:pPb_nX_1.5}
\end{figure}

Inclusive differential kinetic energy spectra of neutrons, emitted at different
angles during proton - nucleus reactions on various target nuclei and at
various impact energies, calculated with the HSD plus evaporation model 
and compared with available experimental data are analysed in this Section.\\
High energy part of neutron distribution, composed of so-called fast
neutrons, produced during the first stage of the reaction is a result
of the HSD model calculations. 
Such neutrons are emitted mainly in forward direction. \\
Low energy part of neutron distribution, composed of slow neutrons
produced during the second stage of the reaction is a
result of evaporation model calculations. 
Such neutrons are emitted both in forward and backward directions. \\
Neutron spectra obtained as a result of summing up these
two parts are compared with the following experimental data.
Distributions of spallation neutrons produced in proton induced
reactions on Al, Fe, Zr, W, Pb and Th targets at 1.2 GeV beam energy and on
Fe and Pb targets at 1.6 GeV proton beam energy, emitted at angles:
10$^{\circ}$, 25$^{\circ}$, 40$^{\circ}$, 55$^{\circ}$, 85$^{\circ}$,
100$^{\circ}$, 115$^{\circ}$, 130$^{\circ}$, 145$^{\circ}$ and 160$^{\circ}$
are compared with data measured at the SATURNE accelerator in
Saclay \cite{Lera02}. Similarly, calculated spectra of neutrons produced in
proton induced reactions on Pb target at 0.8 and 1.5 GeV beam energies,
emitted at angles: 15$^{\circ}$, 30$^{\circ}$, 60$^{\circ}$, 90$^{\circ}$,
120$^{\circ}$ and 150$^{\circ}$ are compared with experimental data
measured by Ishibashi et al. \cite{Meig99}. 
The comparisons are presented in Figures:
 \ref{fig:pAl_nX_1.2}, \ref{fig:pFe_nX_1.2}, \ref{fig:pZr_nX_1.2},
\ref{fig:pW_nX_1.2}, \ref{fig:pPb_nX_1.2}, \ref{fig:pTh_nX_1.2},
\ref{fig:pFe_nX_1.6}, \ref{fig:pPb_nX_1.6}, \ref{fig:pPb_nX_0.8} and
\ref{fig:pPb_nX_1.5}, respectively. 

One can see, that independently on the value of projectile energy, the 
calculated neutron spectra produced in proton induced
reactions on light targets (i.e. Al, Fe and Zr, presented in the Figures 
\ref{fig:pAl_nX_1.2}, \ref{fig:pFe_nX_1.2}, \ref{fig:pZr_nX_1.2} and
\ref{fig:pFe_nX_1.6}, respectively) are in perfect agreement with the
adequate experimental data.
Contrary, in case of spectra of neutrons produced in proton induced reactions
on heavy targets (i.e. W, Pb and Th, presented in the Figures 
\ref{fig:pW_nX_1.2}, \ref{fig:pPb_nX_1.2}, \ref{fig:pTh_nX_1.2},
\ref{fig:pPb_nX_1.6}, \ref{fig:pPb_nX_0.8} and \ref{fig:pPb_nX_1.5},
respectively), the good agreement between calculations and proper experimental
data is seen only in backward angles. In forward angles, calculations
underestimate experimental data in the central parts of the distributions.
Furthermore, based on the presented results of the several reactions, 
on the six target nuclei, the regular growth of the 
missing part of the calculated spectra is observed.
The effect is continuous and increase monotonically with mass of target.
 The missing parts, in case of reactions on heavy targets indicate, that 
apart from the two stages of spallation reaction (fast and slow stage), there 
must be an additional intermediate stage, probably preequilibrium stage. 
It seems, that neutron production during this intermediate stage is dominant 
in case of reactions with heavy targets and negligible for reactions with 
light targets.


Multiplicities of neutrons emitted during first and second stage of the 
reactions are collected in Table \ref{table: multn_f12}.
It is clearly seen, that neutrons are emitted mainly in the second stage of 
spallation reaction.
\newpage
\begin{center}
\begin{table*}[!h]
\caption{{\sl Multiplicities of neutrons emitted during fast and slow stage
of reaction}}
\vspace{0.5cm}
\hspace{2.0cm}
\begin{tabular}{|c||c|c|}
\hline
reaction & first stage & second stage \\
\hline \hline
p(1.2 GeV)+Al & 1.65 & 1.20 \\
\hline
p(1.2 GeV)+Fe & 2.16 & 2.84 \\
\hline
p(1.6 GeV)+Fe & 2.18 & 2.84 \\
\hline
p(1.2 GeV)+Zr & 2.24 & 6.15 \\
\hline
p(1.2 GeV)+W & 2.10 & 17.48 \\
\hline
p(0.8 GeV)+Pb & 1.31 & 21.02 \\
\hline
p(1.2 GeV)+Pb & 1.96 & 21.31 \\
\hline
p(1.5 GeV)+Pb & 2.19 & 20.57 \\
\hline
p(1.6 GeV)+Pb & 2.23 & 19.85 \\
\hline
p(1.2 GeV)+Th & 2.00 & 25.80 \\
\hline
\end{tabular}
\label{table: multn_f12}
\end{table*}
\end{center}

\section{Proton spectra}
\markboth{9.2 Proton spectra}{Chapter 9. Comparison of calculations with experimental data}

\begin{figure}[!htcb]
\vspace{-2cm}
\hspace{-2cm}
\includegraphics[height=20cm, width=18cm, bbllx=0pt, bblly=25pt, bburx=594pt, bbury=842pt, clip=, angle=0]{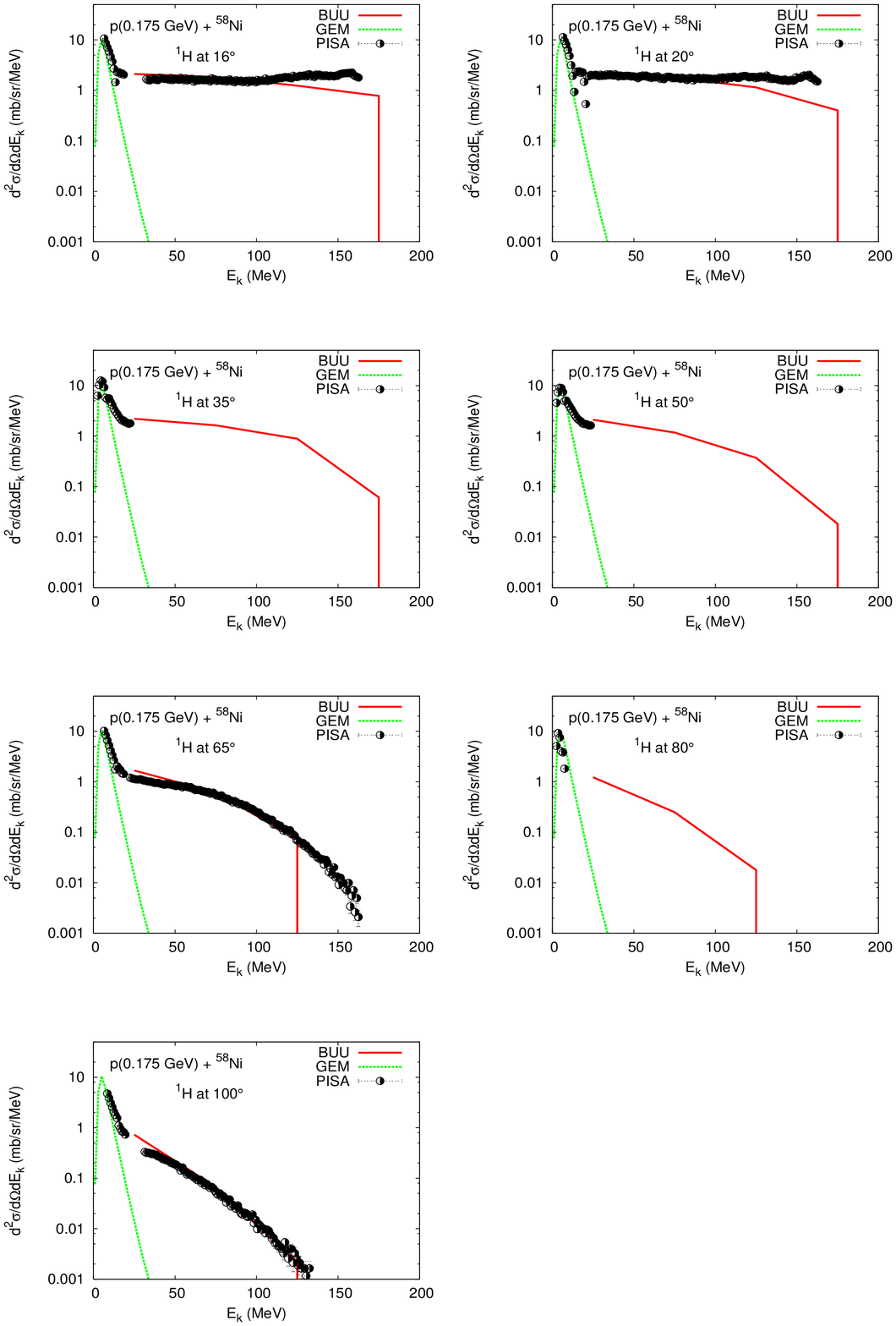}
\caption{{\sl Double differential proton spectra from p+Ni reaction at 0.175 
GeV incident energy; lines show results of HSD and GEM model calculations 
(solid line - first stage, dashed line - second stage),  
symbols indicate the experimental data measured by experiment PISA 
\cite{Buba07}}}
\label{fig:pNi_pX_0.175}
\end{figure}

\begin{figure}[!htcb]
\vspace{-2cm}
\hspace{-2cm}
\includegraphics[height=20cm, width=18cm, bbllx=0pt, bblly=40pt, bburx=594pt, bbury=842pt, clip=, angle=0]{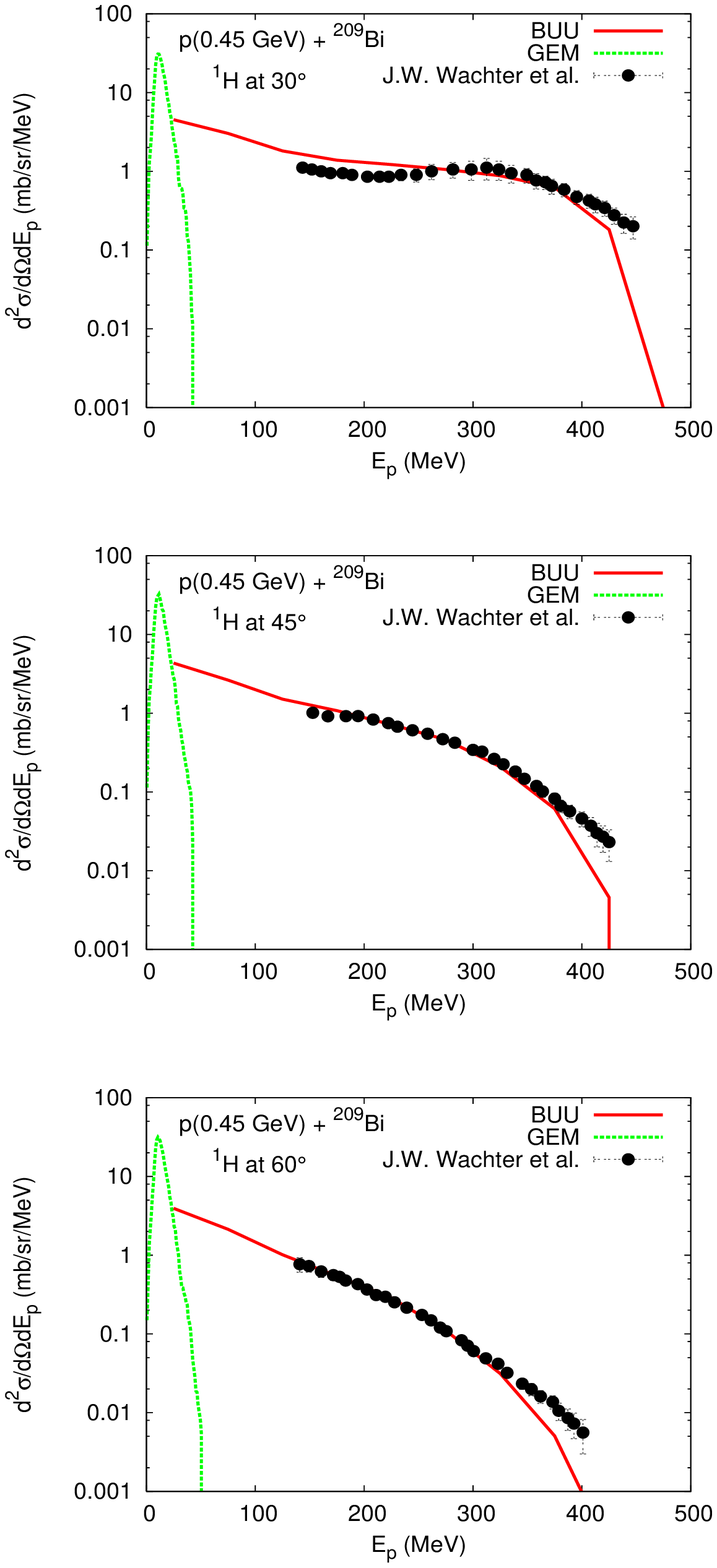}
\caption{{\sl Double differential proton spectra from p+Bi reaction at 0.45 GeV 
proton beam energy; lines show results of HSD and GEM model calculations
(solid line - first stage, dashed line - second stage),  
symbols indicate the experimental data \cite{Wach72}}}
\label{fig:pBi_pX_0.45}
\end{figure}

Inclusive, double differential kinetic energy spectra of protons emitted at 
different angles in proton induced reactions on various targets, at broad range 
 of incident energy, calculated with the HSD+GEM model, are compared 
with available experimental data. 
Model results of low energy reactions: p+Ni at 0.175 GeV beam energy (in this
 case, time of first stage calculations is equal to 55 fm/c), compared with 
experimental data measured by PISA collaboration \cite{Buba07} and p+Bi at 
0.45 GeV beam energy (with time of first stage calculations equals to 65 fm/c),
 compared with data published in \cite{Wach72}, are 
presented in Figures \ref{fig:pNi_pX_0.175} and \ref{fig:pBi_pX_0.45}, 
respectively. 
Very good agreement between calculations and data is clearly seen.   

\begin{figure}[!htcb]
\vspace{-2cm}
\hspace{-2cm}
\includegraphics[height=20cm, width=18cm, bbllx=0pt, bblly=25pt, bburx=594pt, bbury=842pt, clip=, angle=0]{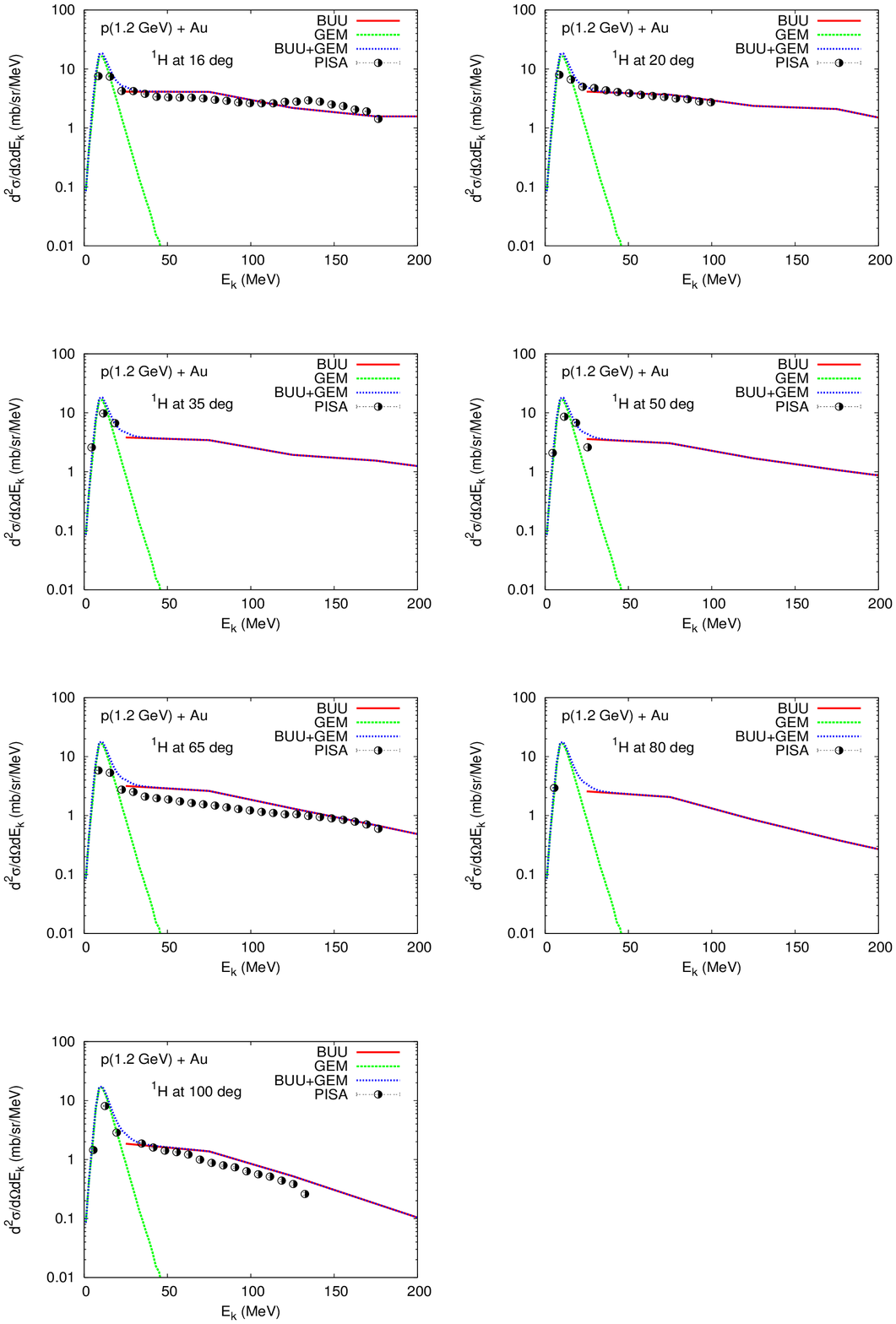}
\caption{{\sl Double differential proton spectra from p+Au reaction at 1.2
GeV proton beam energy; lines show results of HSD and GEM model calculations 
(solid line - first stage, dashed line - second stage),  
symbols indicate data measured by experiment PISA \cite{Buba07}}}
\label{fig:pAu_pX_1.2}
\end{figure}

\begin{figure}[!htcb]
\vspace{-2cm}
\hspace{-2cm}
\includegraphics[height=20cm, width=18cm, bbllx=0pt, bblly=25pt, bburx=594pt, bbury=842pt, clip=, angle=0]{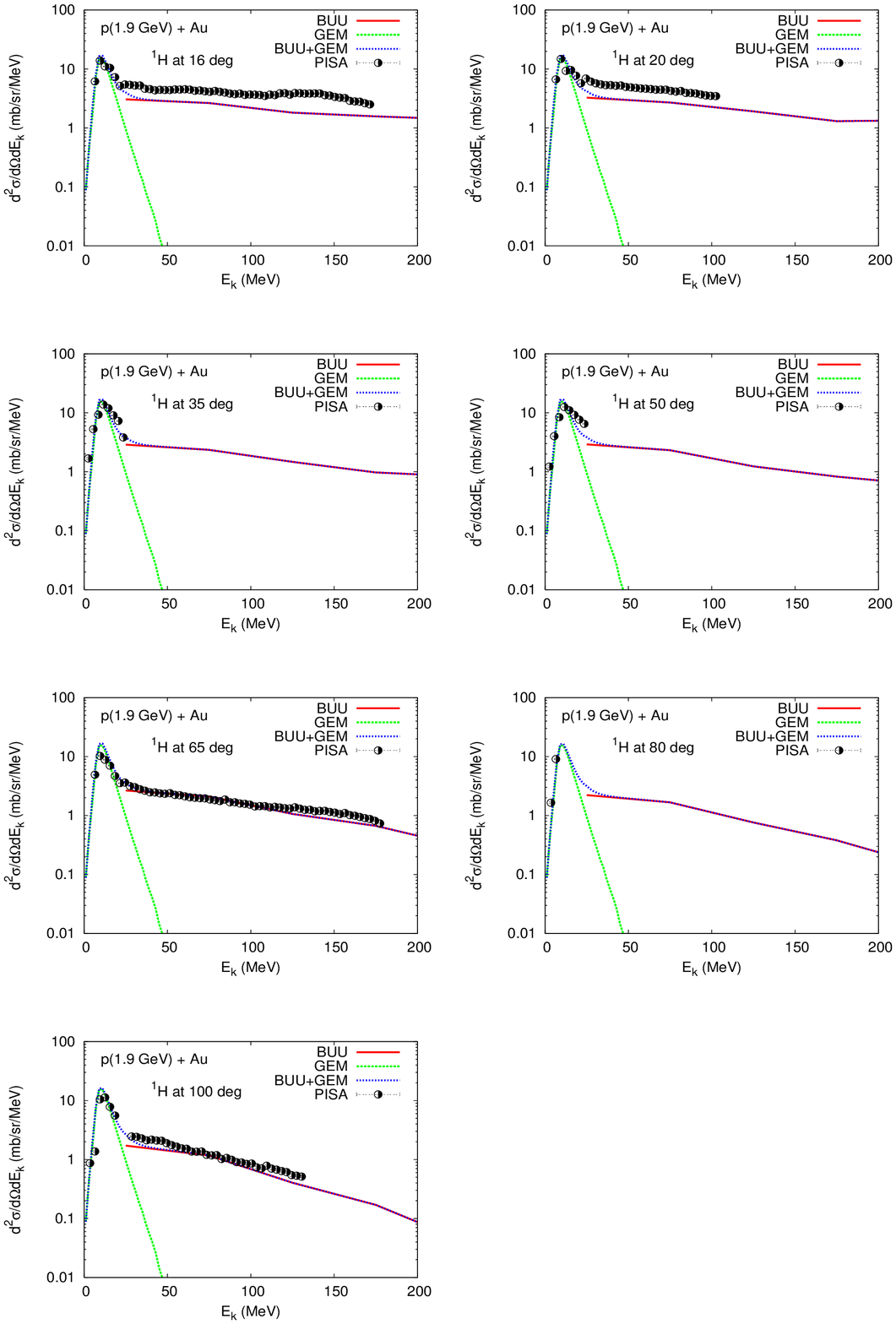}
\caption{{\sl Double differential proton spectra from p+Au reaction at 1.9
GeV proton beam energy; lines show results of HSD and GEM model calculations
(solid line - first stage, dashed line - second stage),
symbols indicate data measured by experiment PISA \cite{Buba07}}}
\label{fig:pAu_pX_1.9}
\end{figure}

\begin{figure}[!htcb]
\vspace{-2cm}
\hspace{-2cm}
\includegraphics[height=20cm, width=18cm, bbllx=0pt, bblly=25pt, bburx=594pt, bbury=842pt, clip=, angle=0]{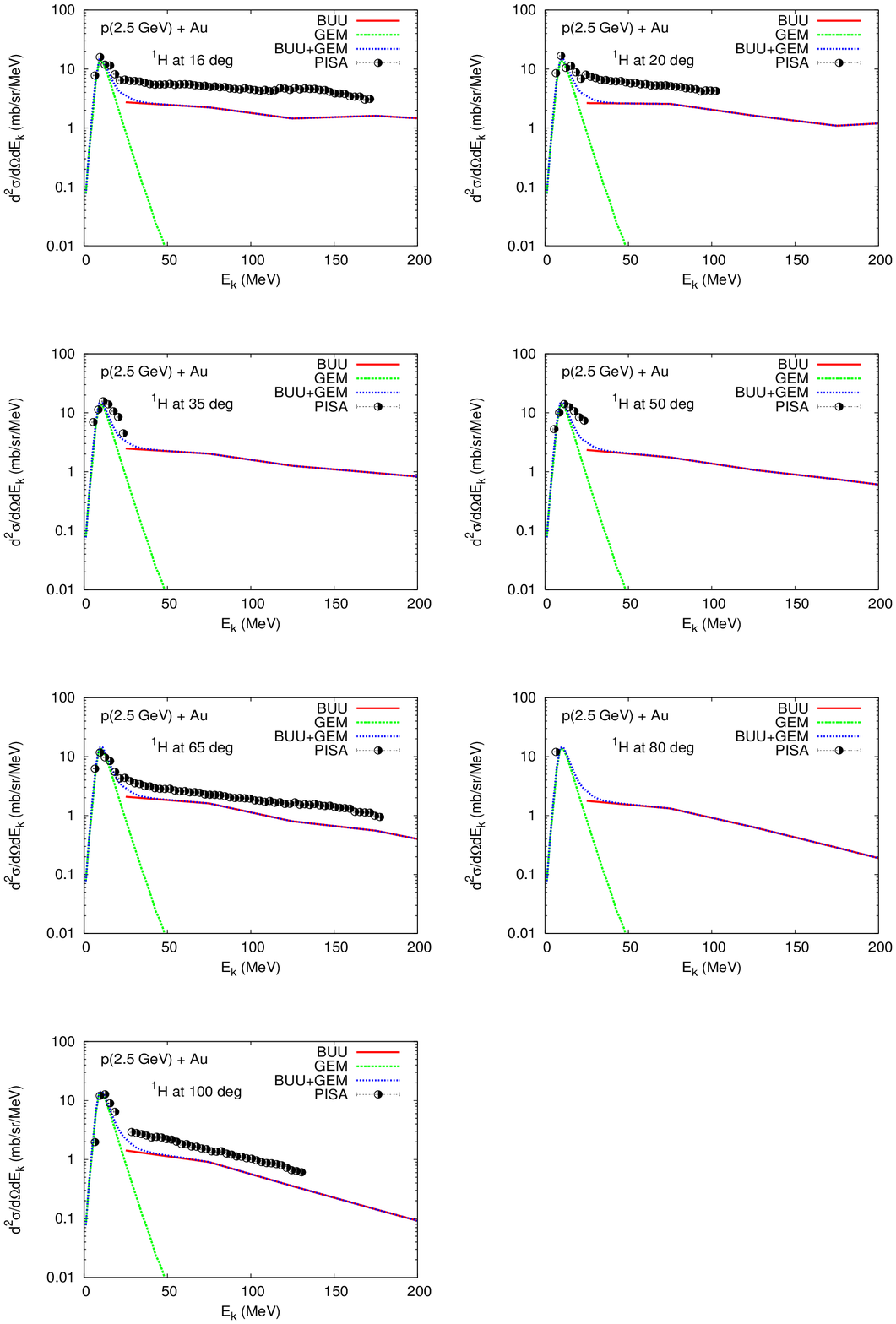}
\caption{{\sl Double differential proton spectra from p+Au reaction at 2.5
GeV proton beam energy; lines show results of HSD and GEM model calculations
(solid line - first stage, dashed line - second stage),
symbols indicate data measured by experiment PISA \cite{Buba07}}}
\label{fig:pAu_pX_2.5}
\end{figure}

Comparisons of calculated proton spectra produced in reactions induced by higher
incident energies, e.g. p+Au at 1.2 GeV, 1.9 GeV and 2.5 GeV projectile energy 
(with time of first stage calculations equal to 45 fm/c), confronted with data 
measured by PISA collaboration \cite{Buba07}, are shown in 
Figures \ref{fig:pAu_pX_1.2}, \ref{fig:pAu_pX_1.9} and \ref{fig:pAu_pX_2.5}. 
In order to make the comparison easier, the high energy parts of 
the calculated distributions are plotted only up to 200 MeV of proton kinetic 
energy. It is seen, that in case of 1.2 GeV of incident energy 
(Fig. \ref{fig:pAu_pX_1.2}), the high energy parts of experimental 
distributions are described well, the low energy parts are slightly 
overestimated by calculations. 
In the case of 1.9 GeV beam energy (Fig. \ref{fig:pAu_pX_1.9}), the 
whole spectra, especially protons emitted in backward directions are described 
very well. Low energy part of experimental distributions of protons from 
reaction at 2.5 GeV (Fig. \ref{fig:pAu_pX_2.5}) are also very well described by
 calculations. The high energy parts are slightly underestimated. 
Nevertheless, the average agreement between the HSD+GEM model results 
concerning proton spectra and adequate experimental data is satisfactory.

In the presented Figures, contributions from intra-nuclear cascade 
(high energy part, result of the HSD model calculations) and evaporation 
(low energy part, result of the GEM model calculations) are visible. 
The cascade contribution evidently dominates. 
This indicates, that protons are emitted mainly in the first stage of 
spallation reaction. This is additionally confirmed by values of multiplicities
 of protons emitted during first and second stage of reaction, compared in 
Table \ref{table: multp_f12} (quite contrary to the neutron case, consider 
Table \ref{table: multn_f12}).  

\begin{figure}[!htcb]
\vspace{-2cm}
\hspace{-2cm}
\includegraphics[height=20cm, width=16cm, bbllx=0pt, bblly=130pt, bburx=594pt, bbury=842pt, clip=, angle=0]{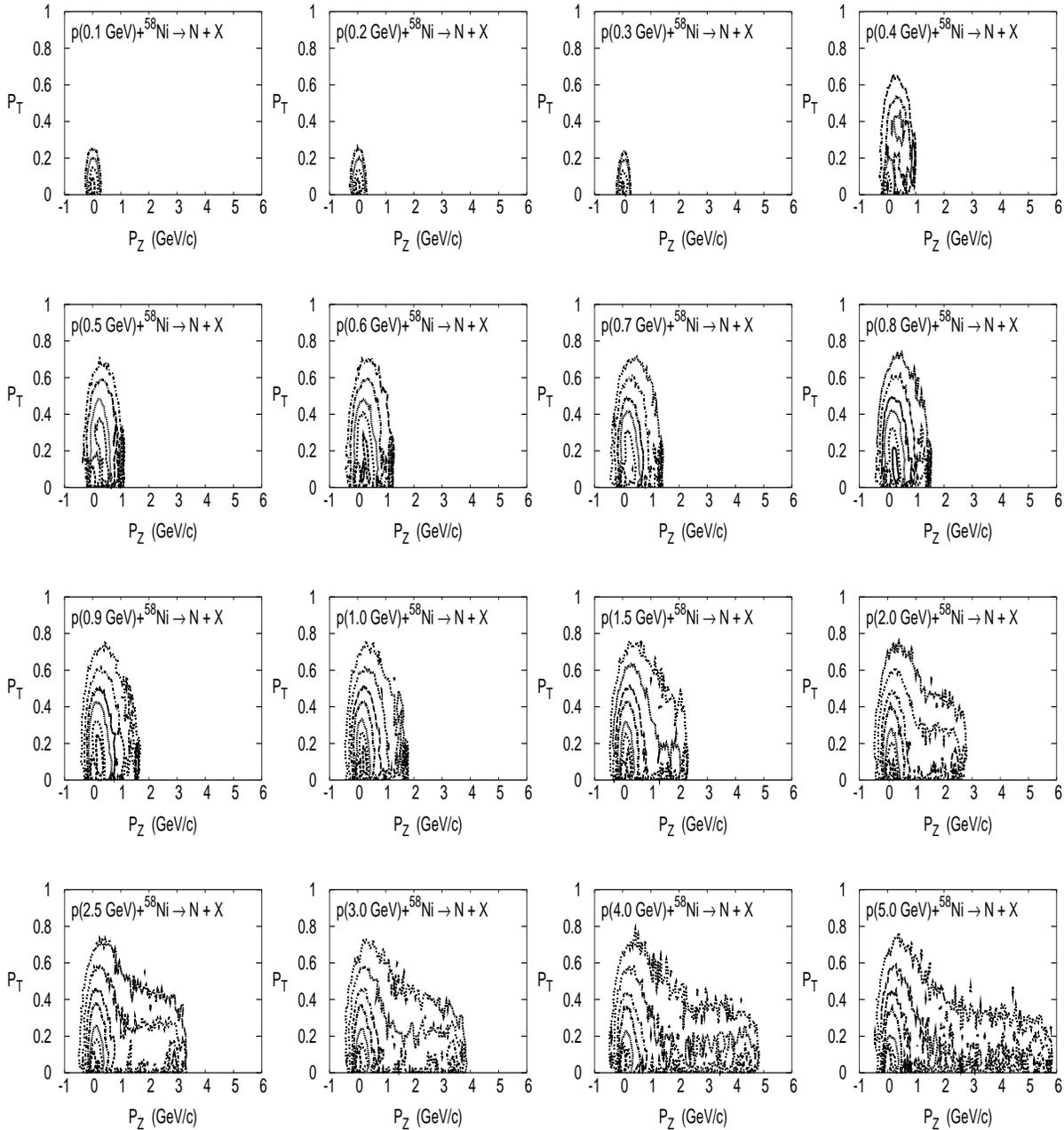}
\caption{{\sl Two-dimensional distributions of transversal versus longitudinal
momentum of nucleons emitted during the first stage of proton induced reaction 
on Ni target, at various incident energies; results of the HSD model 
calculations (the most central contour line, symmetrical around zero value,  
corresponds to a maximum of distribution)}}
\label{fig:c_pNi_NX}
\end{figure}

Nucleon distributions depend strongly on projectile energy. The example 
two-dimensional distributions of transversal ($P_{T}$) versus longitudinal 
($P_{Z}$) momentum of nucleons emitted during the first stage of proton 
induced reactions on Ni target, at various incident energies are presented in 
Fig. \ref{fig:c_pNi_NX}. These are results of HSD model calculations, 
where $P_{Z}$ corresponds to the nucleons momentum in beam direction and 
$P_{T}$ is defined as $P_{T}=\sqrt{P_{X}^{2}+P_{Y}^{2}}$.     
The two-dimensional distributions depict, how the nucleon spectra evolve with 
incident energy. 
In all the cases, low energy part is symmetric around zero value 
(it is not so in the case of pion production, see Fig. \ref{fig:c_pNi_pizX}).
 With increase of projectile energy, the distribution is building up mainly in 
longitudinal (forward) direction. 
It spreads always up to a value corresponding to the value of beam momentum. 
That is because, it is possible that incoming proton will 
scatter only slightly on a surface of target nucleus and will appear in
forward direction with almost unchanged momentum.
In the Fig. \ref{fig:c_pNi_NX}, momentum spectra of all nucleons are 
presented, since distributions of protons and neutrons emitted during first 
stage of the reaction are very similar. The only exception concerns momentum 
corresponding to the beam momentum.
\begin{center}
\begin{table*}[tbp]
\caption{{\sl Multiplicities of protons emitted during first and second stage 
of reaction}}
\vspace{0.5cm} 
\hspace{2.0cm} 
\begin{tabular}{|c||c|c|}
\hline
reaction & first stage & second stage \\
\hline \hline
p(0.175 GeV)+Ni & 1.61 & 0.91 \\ 
\hline
p(0.45 GeV)+Bi & 2.16 & 1.00 \\
\hline
p(1.2 GeV)+Au &2.67 &1.13 \\
\hline
p(1.9 GeV)+Au &2.42 &1.099 \\
\hline
p(2.5 GeV)+Au &2.20 &0.96 \\
\hline
\end{tabular}
\label{table: multp_f12}
\end{table*}
\end{center}

\section{Deuteron spectra}
\label{sec:deut}
\markboth{9.3 Deuteron spectra}{Chapter 9. Comparison of calculations with experimental data}

\begin{figure}[!htcb]
\vspace{-2cm}
\hspace{-1cm}
\includegraphics[height=15cm, width=15cm, bbllx=0pt, bblly=300pt, bburx=594pt, bbury=842pt, clip=, angle=0]{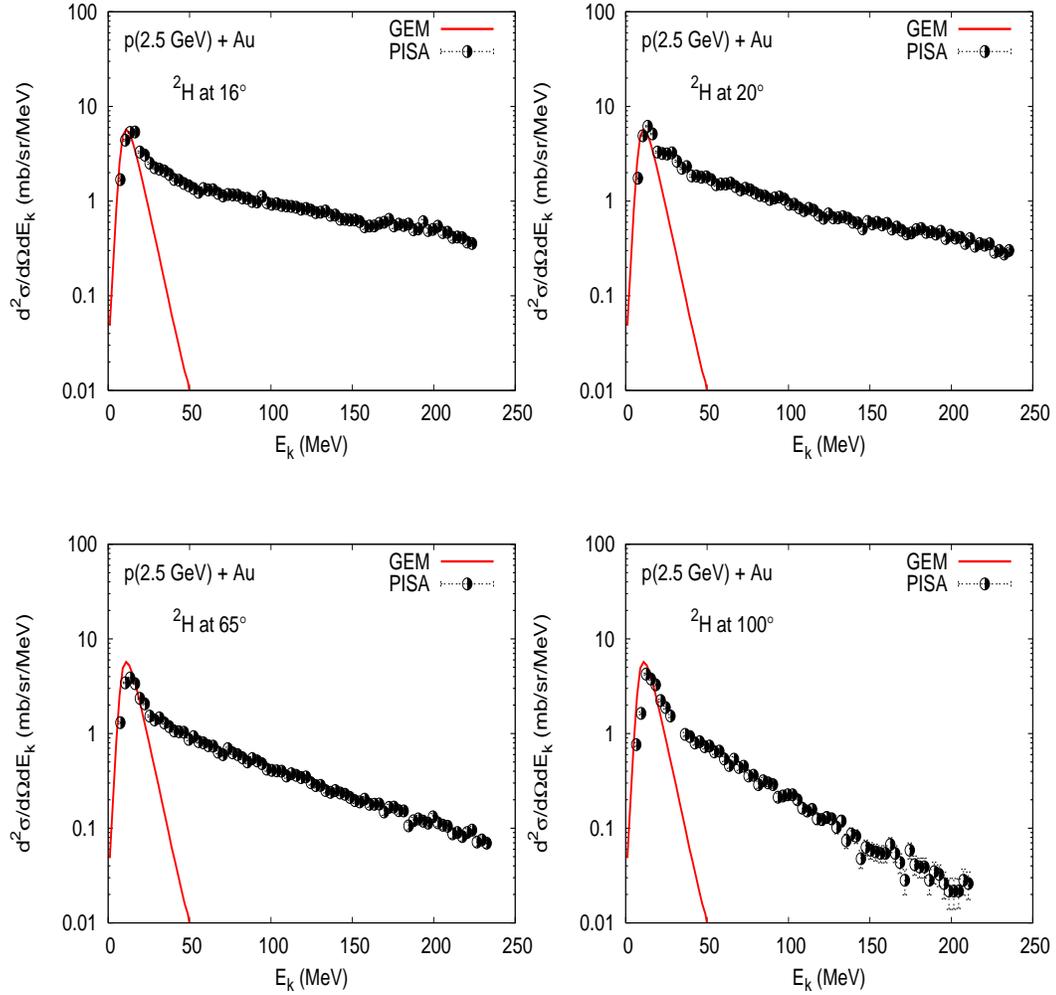}
\caption{{\sl Double differential deuteron spectra from p+Au reaction, at 2.5
GeV proton beam energy; lines show results of the HSD+GEM model calculations,
symbols indicate data measured by PISA experiment \cite{Buba07}}}
\label{fig:pAu_dX_2.5}
\end{figure}

Comparison of inclusive differential energy spectra of deuterons
calculated with the HSD+GEM model, with example experimental data is presented
in Fig. \ref{fig:pAu_dX_2.5}.\\

\begin{figure}[!htcb]
\vspace{-2cm}
\hspace{-2cm}
\includegraphics[height=20cm, width=18cm, bbllx=0pt, bblly=245pt, bburx=594pt, bbury=842pt, clip=, angle=0]{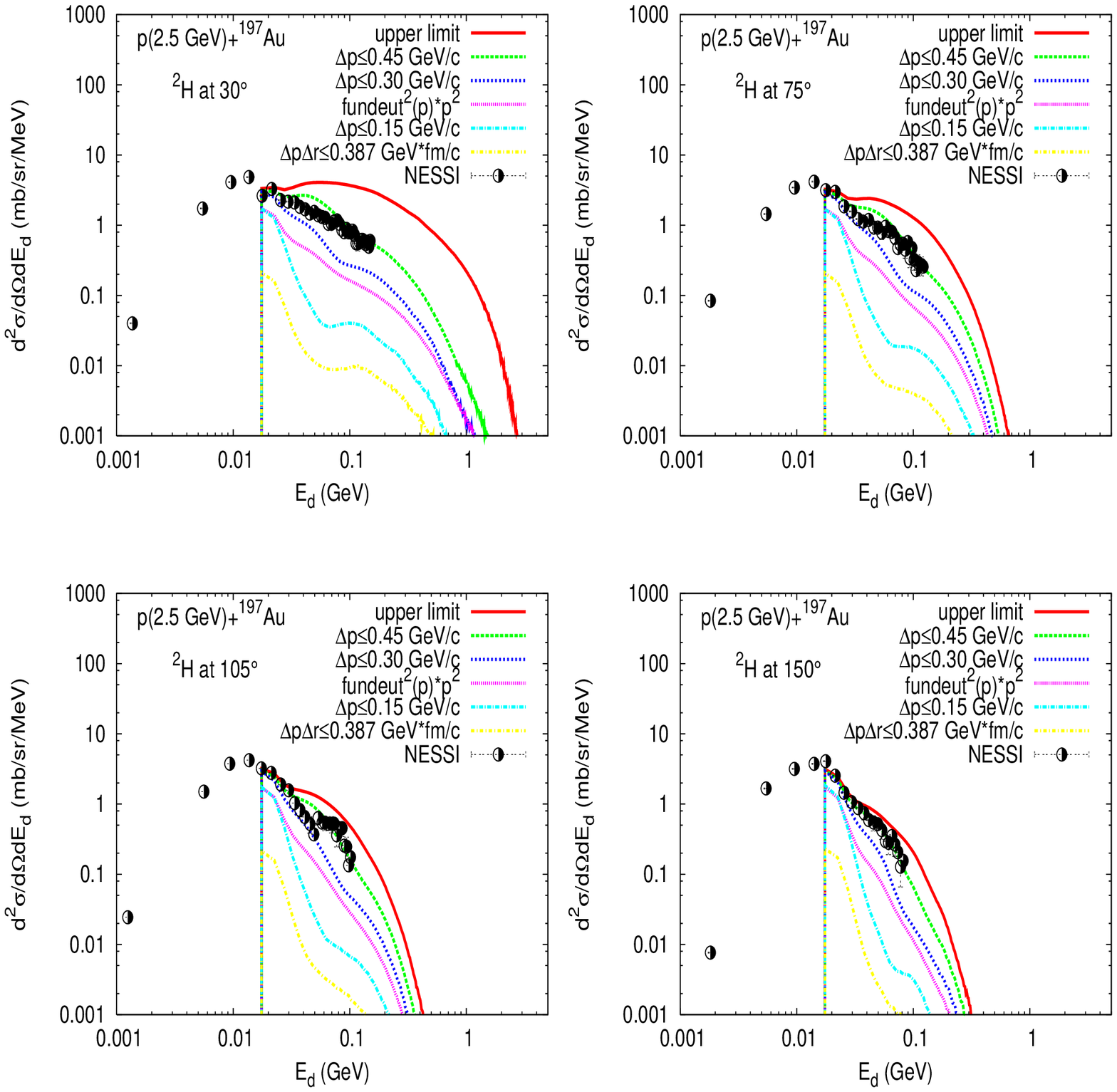}
\caption{{\sl Differential spectra of deuterons of coalescence of proton and
neutron emitted during fast stage of p+Au reaction, at 2.5 GeV proton beam
energy, compared with experimental data measured by NESSI \cite{Leto02} 
("upper limit" means all proton - neutron pairs convert into deuterons)}}
\label{fig:pAu2.5_deut_fast}
\end{figure}

\begin{figure}[!htcb]
\vspace{-2cm}
\hspace{-3cm}
\includegraphics[height=20cm, width=18cm, bbllx=0pt, bblly=245pt, bburx=594pt, bbury=842pt, clip=, angle=0]{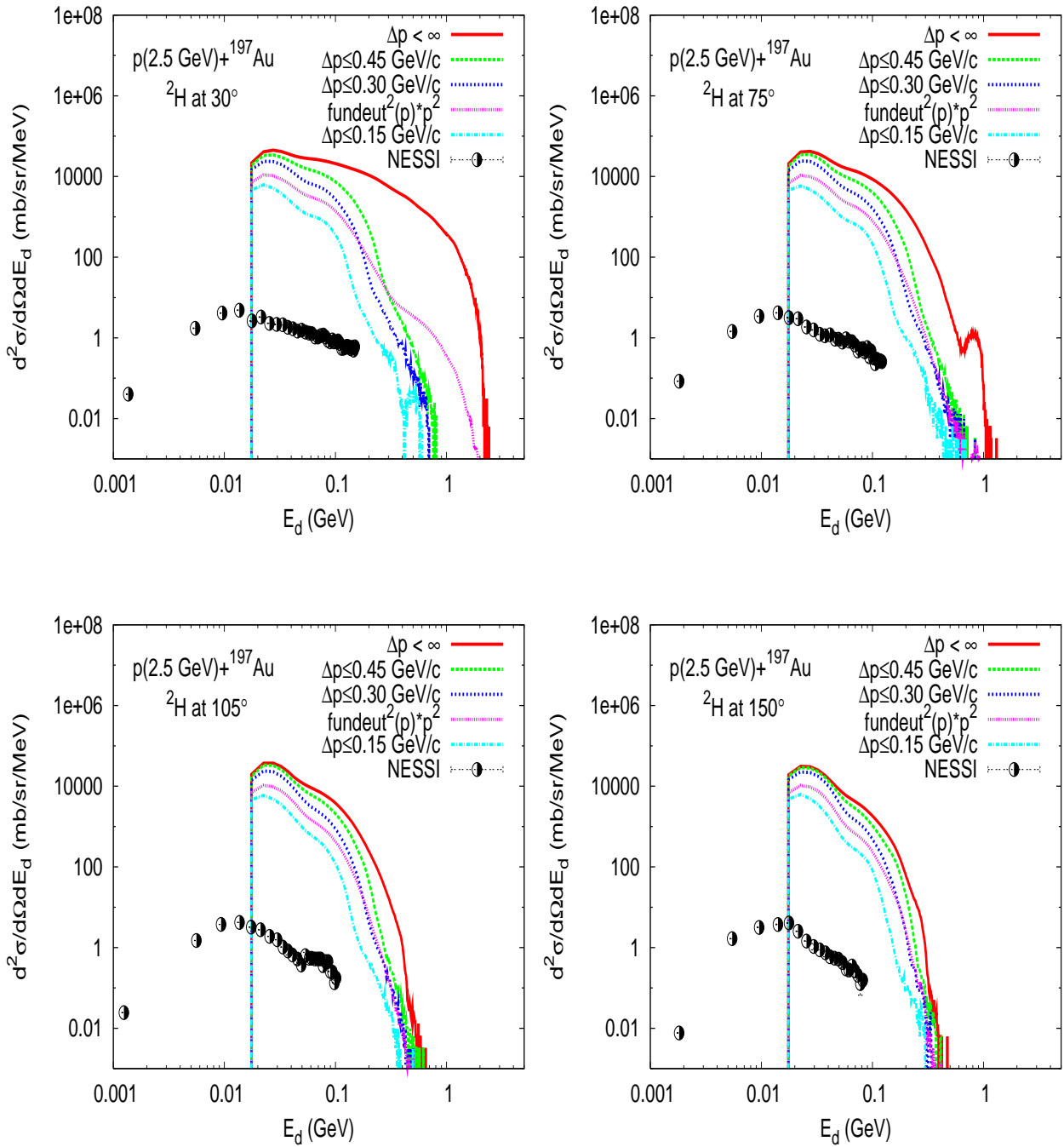}
\caption{{\sl Differential spectra of deuterons of coalescence of fast emitted
proton and a neutron from residual nucleus remained after fast stage of p+Au
reaction, at 2.5 GeV proton beam energy, compared to NESSI data \cite{Leto02}}}
\label{fig:pAu2.5_deut_slow}
\end{figure}

One can see, that only low energy part (i.e. deuterons
 evaporated from the residual nucleus during second stage of spallation
reaction) is described by the model calculations. The absence of high energy
part of modelled distributions indicates a lack of deuterons produced during
first stage of the reaction. Certainly, description of the whole distributions
necessitates implementation of some additional mechanisms into the first stage
model. Since deuteron spectra behave very similar as proton distributions
(compare e.g. Figures \ref{fig:pAu_pX_2.5} and \ref{fig:pAu_dX_2.5}), one can 
suppose that the high energy part of deuteron spectra are produced due to
a coalescence of proton and neutron during first stage of reaction.
Production of clusters (i.e. light ions) cannot be followed with the HSD 
approach, since it is a one body theory.
Nevertheless, a trial of construction of deuterons from the fast
stage of the spallation reaction, with use of informations provided by the HSD 
approach, have been undertaken in the frame of this work.
The idea of generation of such deuterons, called here {\sl deuterons of
coalescence}, is based on the information of positions and momenta of
nucleons belonging to the residual nucleus and nucleons being emitted during
the first stage. The information is taken at a certain point of time, i.e. in 
the end of the first stage calculations. Deuterons can be built up in two 
manners: by fast proton and proper fast neutron, which are already emitted, 
or by
fast proton emitted during fast stage of the reaction and
proper neutron which belongs actually to the excited residual nucleus, but
could have been glued with and taken away by the escaping proton during the
first stage of reaction. In both cases, it is assumed that probability of
creation a deuteron by defined above nucleon pair, depends on their relative
momentum ($\Delta p$).
It means, two nucleons are considering as a deuteron if their relative momentum
 fulfills assumed condition. The following conditions have been tested:
 each pair of neutron and proton can create a deuteron with the same
probability (this gives the highest border of the absolute normalization),
next, deuterons can be created by proton and neutron only if their relative
momentum is lower than or equal to, respectively: $\Delta p \leq$ 0.45 GeV/c,
$\Delta p \leq$ 0.30 GeV/c and $\Delta p \leq$ 0.15 GeV/c.\\
All formed deuterons are additionally repulsed by Coulomb force.
The Coulomb potential is approximated by potential of homogeneously
charged sphere. This is because an incident proton does not cause any
significant changes in the nucleon density distribution inside target nucleus, 
as shown in Chapter \ref{chapt:bulk_prop}.
Therefore, the kinetic energy of each deuteron is enlarged by value:
\begin{equation}
E_{Coul}=\frac{0.00144*Z_{T}*Z_{d}}{R_{T}} \; [GeV]
\label{eq:v_coul}
\end{equation}
where: $Z_{T}$ and $Z_{d}$ are charges of target nucleus and deuteron,
respectively, $R_{T}$ is a radius of target.\\
Obtained differential energy spectra of deuterons produced in p+Au 
reaction at 2.5 GeV proton beam energy are presented in Figures 
\ref{fig:pAu2.5_deut_fast} and \ref{fig:pAu2.5_deut_slow}. The calculated 
distributions are compared with experimental data measured by NESSI 
collaboration \cite{Leto02}.

\begin{figure}[!htcb]
\vspace{-2cm}
\hspace{-1cm}
\includegraphics[height=16cm, width=15cm, bbllx=0pt, bblly=245pt, bburx=594pt, bbury=842pt, clip=, angle=0]{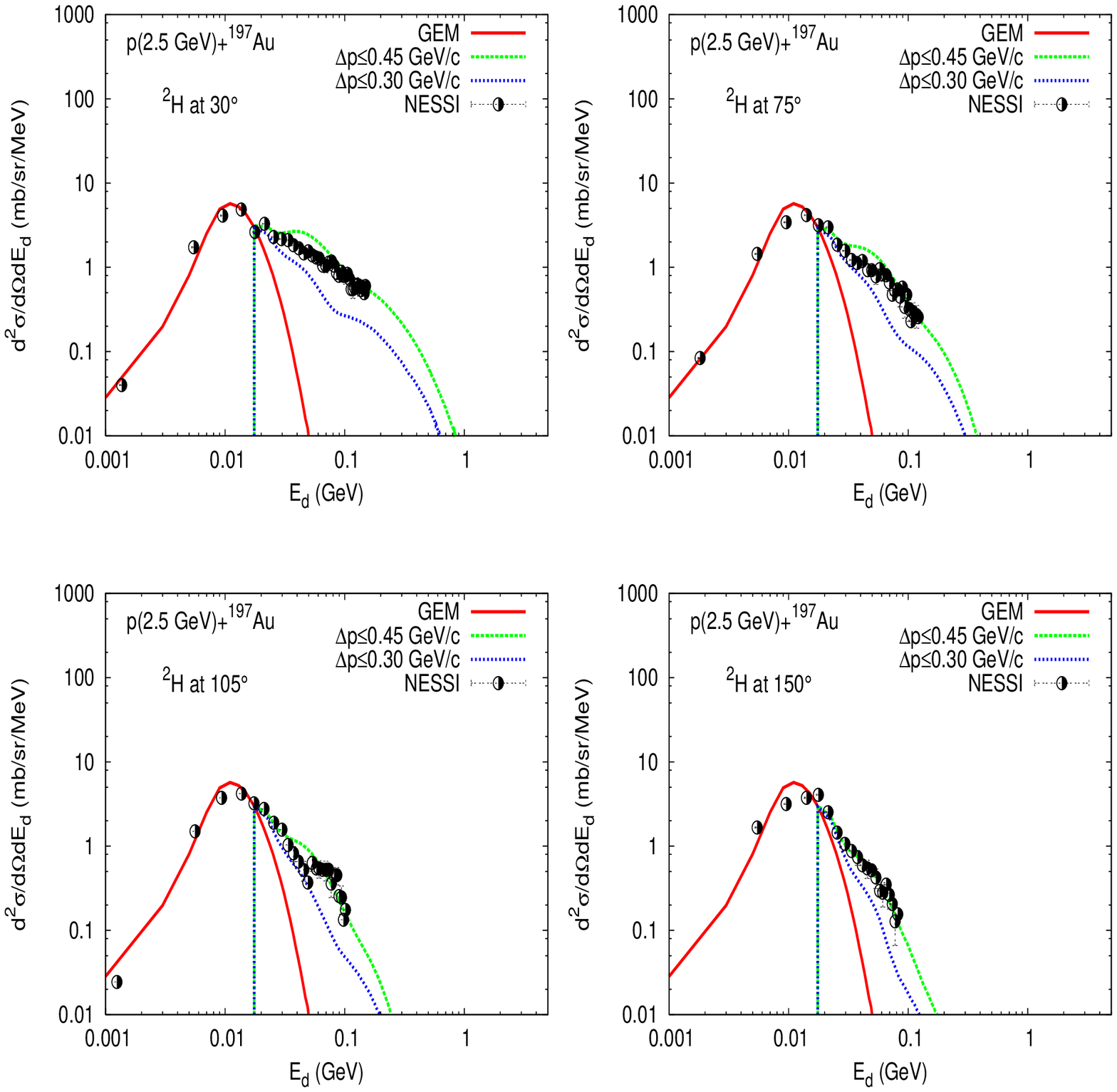}
\caption{{\sl Double differential deuteron spectra from p+Au reaction, at 2.5
GeV proton beam energy; lines show results of the HSD+GEM model calculations,
symbols indicate the experimental data measured by NESSI \cite{Leto02}
("GEM" means "statistical evaporation only")}}
\label{fig:f2_pAu_dX_2.5_fast}
\end{figure}
In more precise definition, deuteron is consider as a set of nucleons, which
are sufficiently close to each other in phase space.
Therefore, distributions of deuterons constructed of pairs of fast nucleons,
with condition set not only on their relative momentum ($\Delta p$), but also
their relative position ($\Delta r$) have been tested, too.
The following unit volume of phase space have been selected: $\Delta p \Delta r
 \le$ 0.387 GeV$\cdot$fm/c (= the Fermi momentum $\times$ 1.4 fm) \cite{Boud04}.
The result underestimates experimental data, as shown in 
Fig. \ref{fig:pAu2.5_deut_fast} and \ref{fig:pAu2.5_deut_slow}. That is because
 of lost of information about the time evolution of relative positions of the
nucleons during the reaction. The condition of relative positions of
nucleons, imposed on the situation at the end of first stage of reaction 
rejects nucleon pairs, that it has no record whether in some earlier period of 
time nucleons were close enough to create deuterons. 
Therefore, it determines, at most, the lowest border of coalescence effect. \\
There is also possible to define the probability (in fact the density of
probability) of creation a deuteron by a pair of nucleons depending on their
relative momentum, as equal to the modulus of the deuteron
wave function squared: $|\psi _{0} (p)|^{2} \cdot p^{2}$.
The wave function can be parametrized itself as:
\begin{equation}
\psi _{0} (p) = \sum _{i} A_{i} exp{(- \alpha _{i} p^{2})}
\label{eq:ffal_deut}
\end{equation}
Values of the coefficients $A_{i}$ and $\alpha _{i}$ are taken from
\cite{Albe75}. The used parametrization of the deuteron wave function is
accurate within a few percent, up to momenta of 1 GeV/c,
see Ref. \cite{Albe75}. \\
Obtained in this way spectra of deuterons underestimate the experimental data,
see Figures \ref{fig:pAu2.5_deut_fast} and \ref{fig:pAu2.5_deut_slow}.\\
The calculated spectra of deuterons constructed by two fast nucleons
(Fig. \ref{fig:pAu2.5_deut_fast} and \ref{fig:f2_pAu_dX_2.5_fast}) are 
normalized absolutely. The spectra of deuterons constructed by fast protons and
 neutrons from the residual nucleus (Fig. \ref{fig:pAu2.5_deut_slow}) cannot be 
normalized absolutely. 
It is because the information, whether a fast proton before leaving target 
nucleus has been near a neutron long enough to
create a deuteron is lost, what gives in this case evident overestimation of
experimental data.

It is seen from the Fig. \ref{fig:f2_pAu_dX_2.5_fast}, that the high energy
part of the experimental
spectra are well described by deuterons of coalescence created by a pair of
fast nucleons with relative momentum lower or equal to about 0.40 GeV/c.
It is also seen that distributions of deuterons of coalescence are perfect 
continuation of the low energy part.
As result, it gives full description of experimental data.
However, taking into account shapes of the distributions from Fig. 
\ref{fig:pAu2.5_deut_slow}, one cannot reject possibility that some of the 
deuterons are produced as coalescence of e.g. fast proton and neutron from 
residual nucleus.



\section{Other Ejectiles}
\markboth{9.4 Other Ejectiles}{Chapter 9. Comparison of calculations with experimental data}

\begin{figure}[!htcb]
\vspace{-0.75cm}
\hspace{-2cm}
\includegraphics[height=19cm, width=17cm, bbllx=0pt, bblly=155pt, bburx=594pt, bbury=842pt, clip=, angle=0]{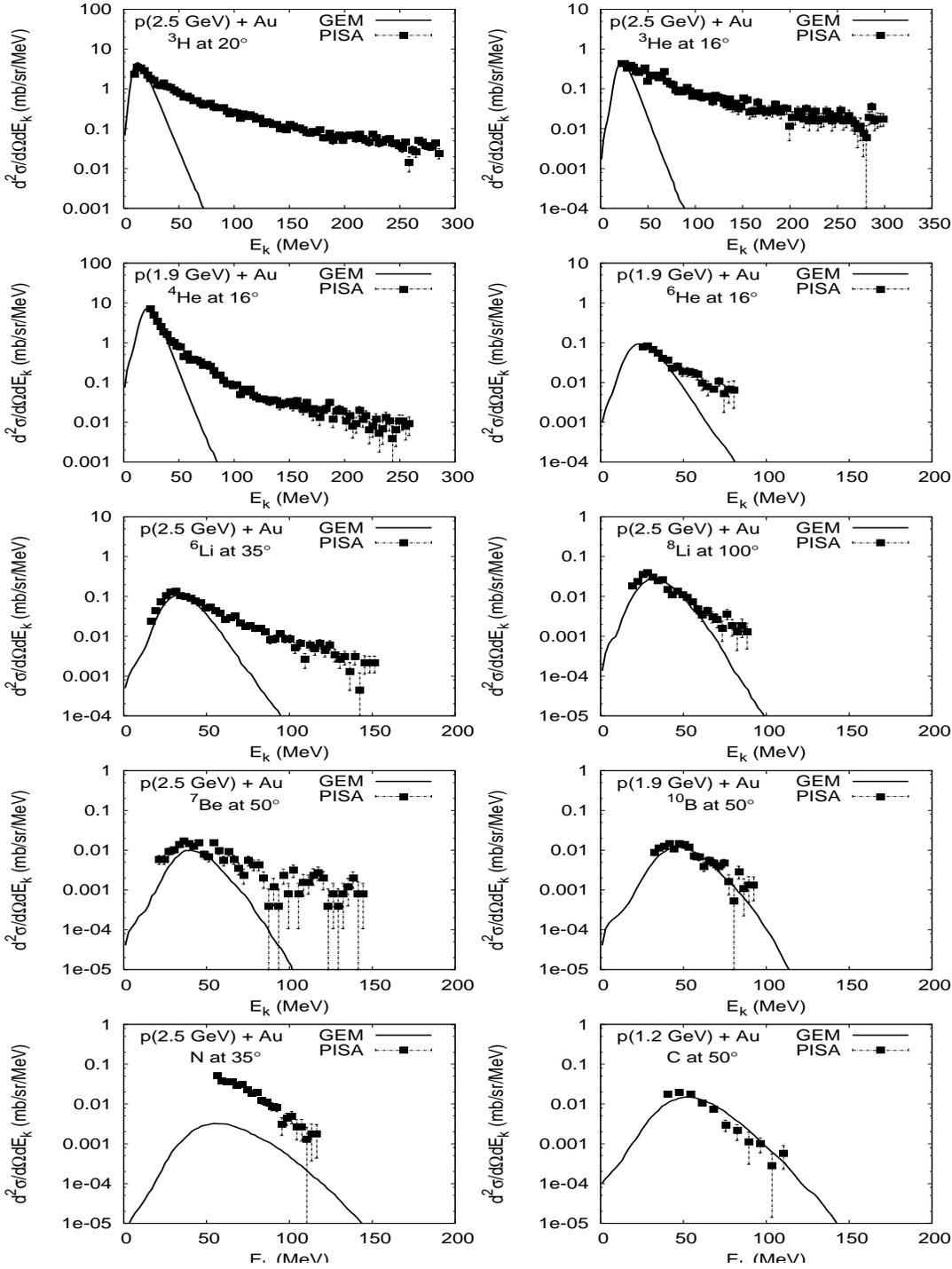}
\caption{{\sl Example double differential spectra of various ejectiles emitted
during an example reaction p+Au, at few incident energies;
results of the HSD+GEM model calculations (solid lines) compared with
experimental data measured by PISA at COSY \cite{Buba07} (squares)}}
\label{fig:pAu_oth_ej}
\end{figure}

Looking at comparison of distributions of heavier particles emitted in 
proton - nucleus reaction, calculated with the HSD+GEM model with  
experimental data, displayed in Fig. \ref{fig:pAu_oth_ej}, it is seen 
that only low energy part is described. 
That is because, in the frame of the model, the light composites are produced 
only in evaporation stage, as discussed in Sec. \ref{sec:deut}. 
 
\begin{figure}[!htcb]
\vspace{-0.75cm}
\hspace{-2cm}
\includegraphics[height=19cm, width=17cm, bbllx=0pt, bblly=15pt, bburx=594pt, bbury=842pt, clip=, angle=0]{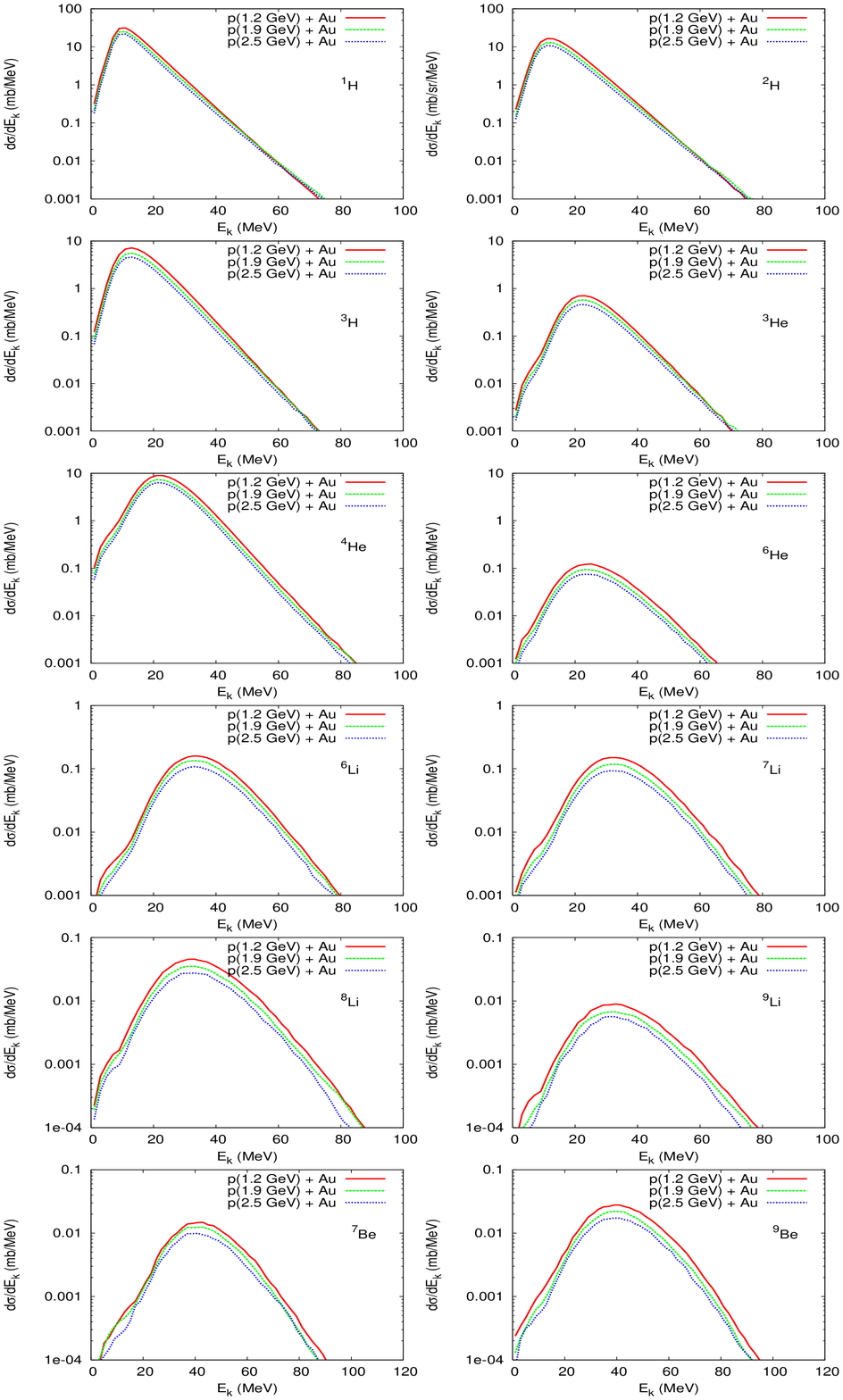}
\caption{{\sl Kinetic energy spectra of different ejectiles emitted in proton 
induced reaction on Au target, at several proton beam energies; results of the 
HSD+GEM model calculations}}
\label{fig:isot_Au_a}
\end{figure}

\begin{figure}[!htcb]
\vspace{-0.75cm}
\hspace{-2cm}
\includegraphics[height=19cm, width=17cm, bbllx=0pt, bblly=155pt, bburx=594pt, bbury=842pt, clip=, angle=0]{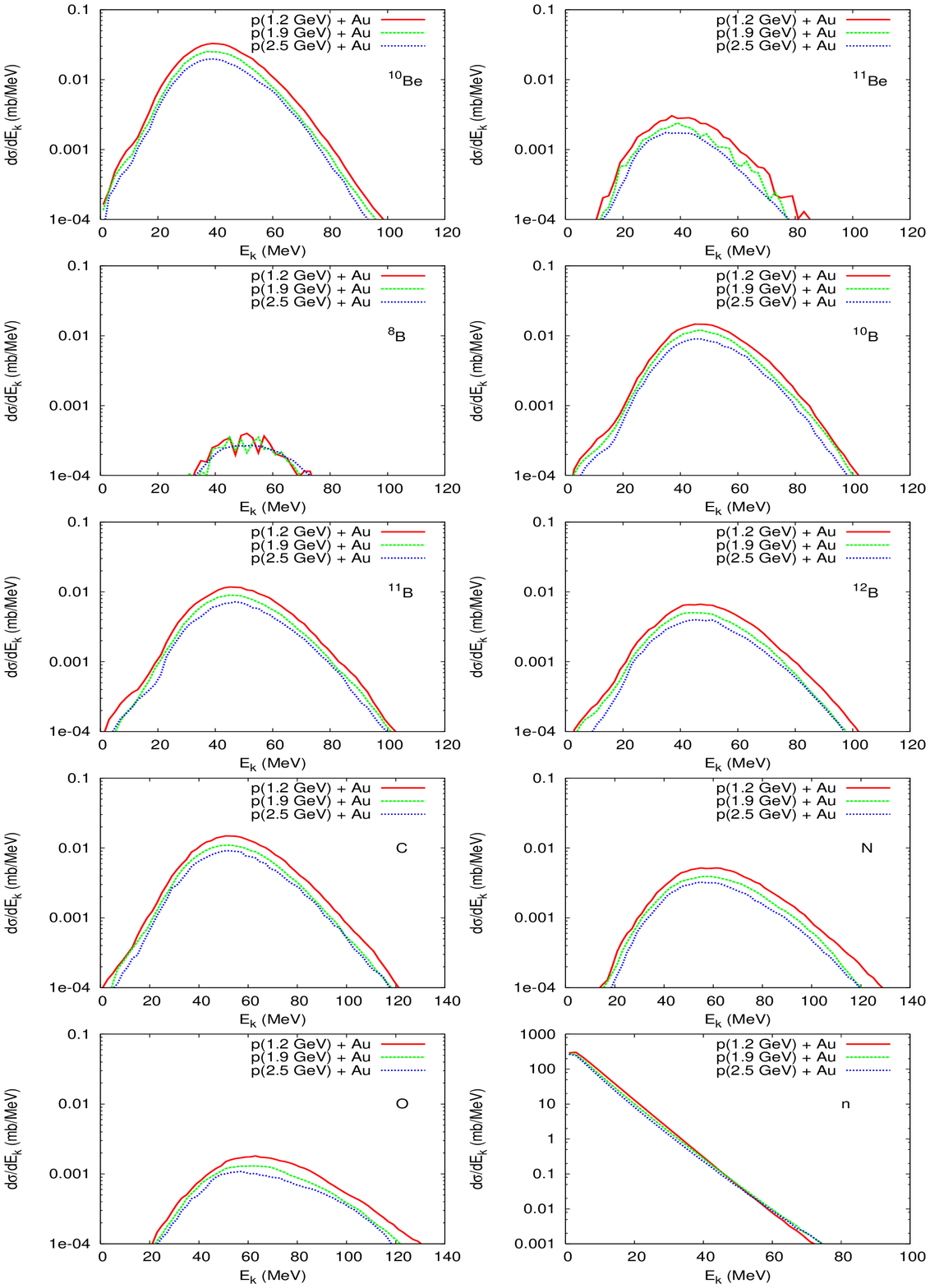}
\caption{{\sl Kinetic energy spectra of different ejectiles emitted in proton 
induced reaction on Au target, at several proton beam energies; results of the 
HSD+GEM model calculations}}
\label{fig:isot_Au_b}
\end{figure}

\begin{figure}[!htcb]
\vspace{-0.75cm}
\hspace{-2cm}
\includegraphics[height=19cm, width=17cm, bbllx=0pt, bblly=15pt, bburx=594pt, bbury=842pt, clip=, angle=0]{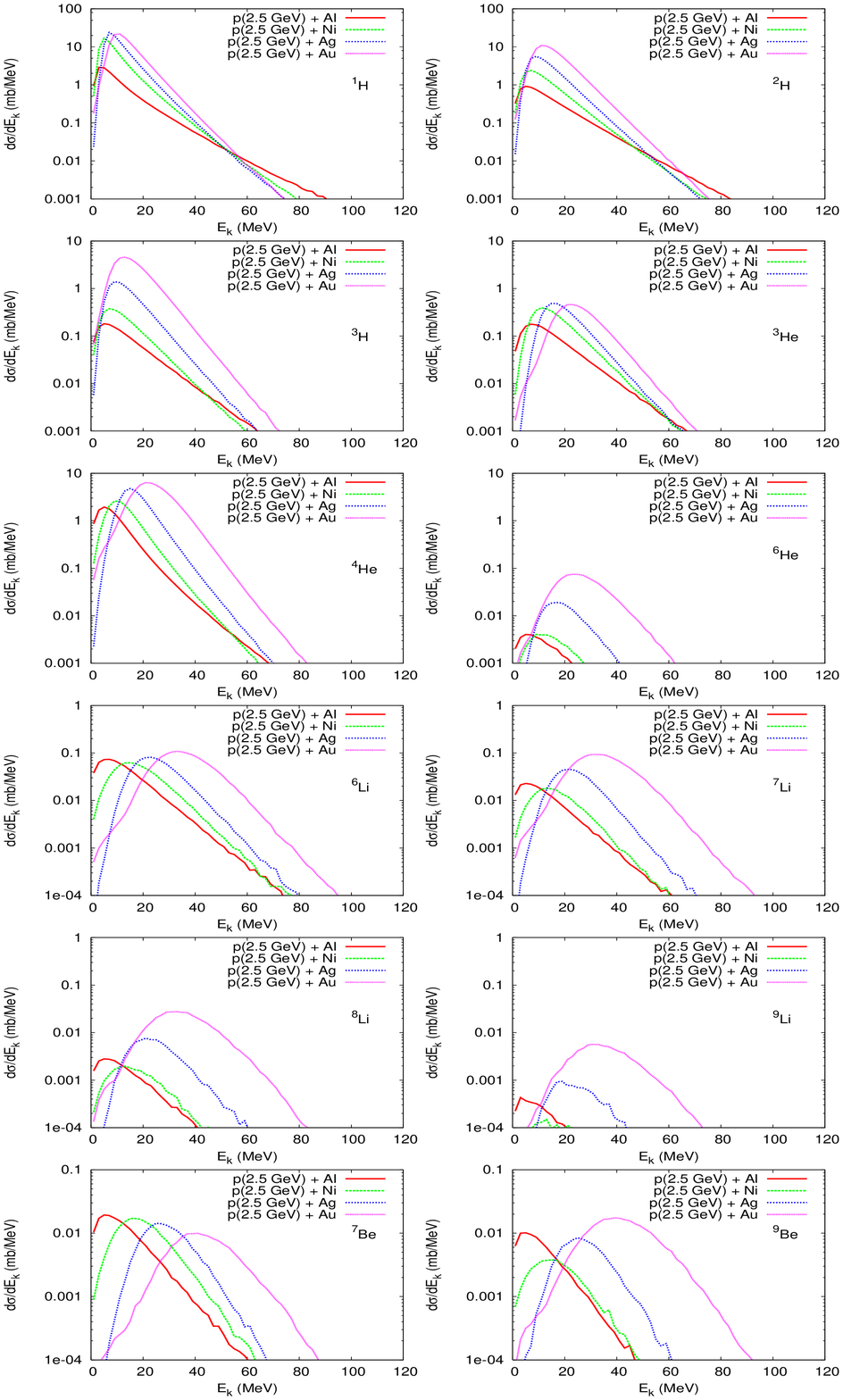}
\caption{{\sl Kinetic energy spectra of different ejectiles emitted in 2.5 GeV 
proton induced reaction on several targets; results of the HSD+GEM model 
calculations}}
\label{fig:isot_2.5_a}
\end{figure}

\begin{figure}[!htcb]
\vspace{-0.75cm}
\hspace{-2cm}
\includegraphics[height=19cm, width=17cm, bbllx=0pt, bblly=155pt, bburx=594pt, bbury=842pt, clip=, angle=0]{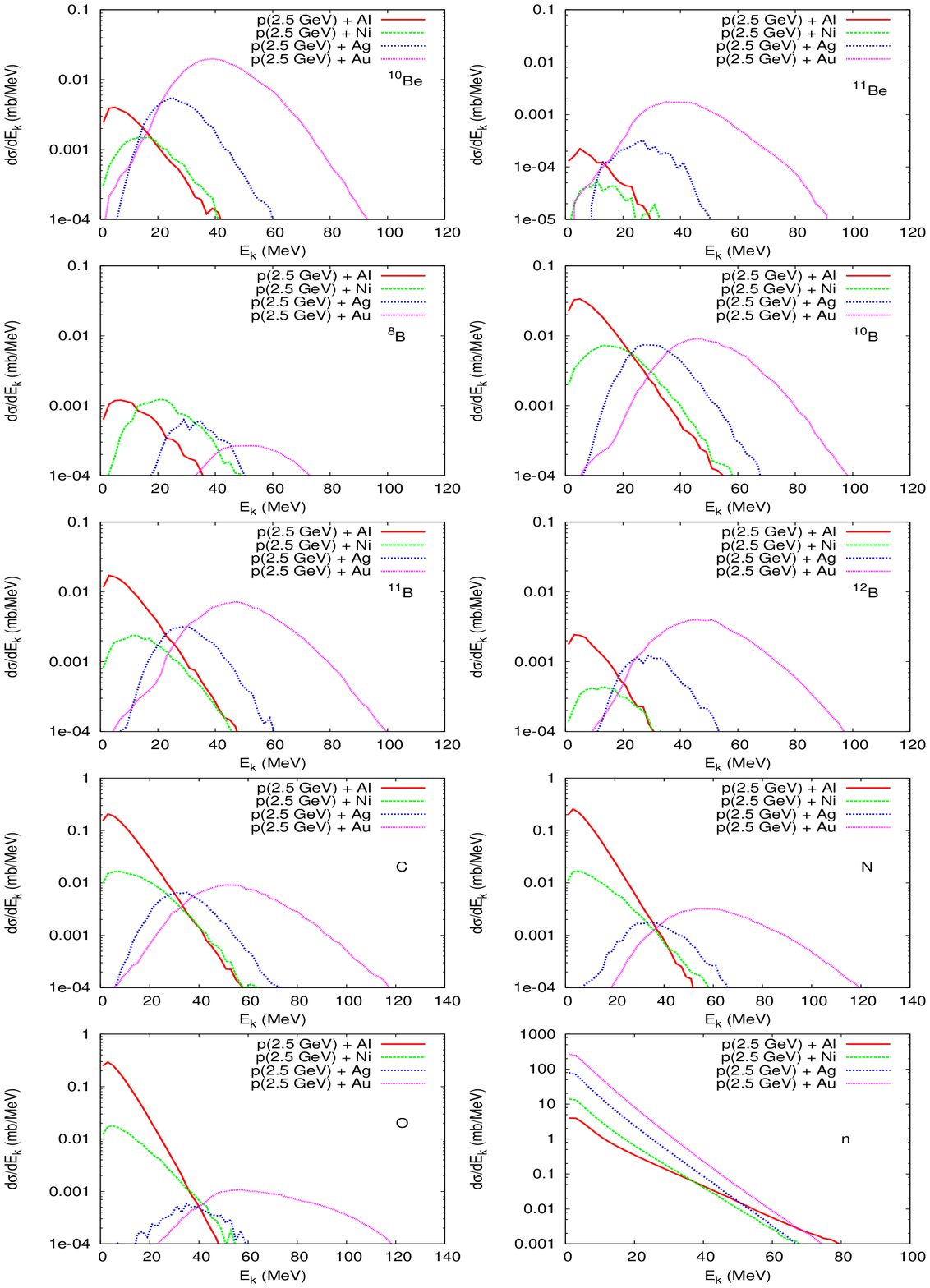}
\caption{{\sl Kinetic energy spectra of different ejectiles emitted in 2.5 GeV 
proton induced reaction on several targets; results of the HSD+GEM model 
calculations}}
\label{fig:isot_2.5_b}
\end{figure}

It is interesting to know, what is the predictive power of the model, how 
distributions of the evaporated composites vary with projectile energy and 
mass of target.  
Example energy spectra of different ejectiles (isotopes of H, He, Li, Be, 
B, ...) emitted in proton induced reaction on Au target, at a few 
chosen beam energies are presented in Figures \ref{fig:isot_Au_a}, 
and \ref{fig:isot_Au_b}. Spectra of particles ejected 
during 2.5 GeV proton induced reactions on few example targets are shown in 
Figures \ref{fig:isot_2.5_a} and \ref{fig:isot_2.5_b}. 
Looking at the distributions of each ejectile, it is clearly seen, that  
shape of the particular distributions almost does not change with projectile 
energy, but strongly depends on used target. In all presented cases, the 
heavier target, the broader the distributions.    
Average kinetic energy of emitted particle increases with increase of mass of 
used target, but in this case neutrons are an exception.   
This indicates, that reason for such behavior is connected 
with charge of the particles, i.e. Coulomb repulsion force between the 
ejectiles and an emitting source. The higher charge of the emitting source, the
 higher average kinetic energy gained by charged ejectiles due to the Coulomb 
repulsion.   
Similarly, the particles yield almost does not change with incident energy, but 
varies significantly with mass of target. 
Multiplicities of the particles, produced due to discussed here reactions are 
compared in Tables \ref{table: isot_Au_abc} and \ref{table: isot_2.5_abc}. 
The yield of neutrons increases 
linearly with increase of the target mass. Multiplicity of each individual 
charged ejectile varies differently with the target mass. In some cases, it
first decreases with the target mass number (for example presented in the 
Table \ref{table: isot_2.5_abc}, the minimal yield corresponds usually to Ni 
target) and then increases with further increase of mass of target (e.g. 
proton yield behaves in opposite way). In many cases, it is also observed 
roughly linear decrease of the multiplicity with increase of the target mass 
(as e.g. in case of $^{6}Li$).  
The trends for the particles yield can be to some extent qualitatively 
understood, since it is known, that the particles emission is influenced by 
the Coulomb barrier, which is higher for heavier nuclei and by the binding 
energy, which in average first increases with mass number (A) of target 
(up to A equal to about 56) and then decreases with A. 
The relative yield of different ejectiles emitted in one chosen reaction (
Tables \ref{table: isot_Au_abc} and \ref{table: isot_2.5_abc}) is also  
determined by separation energy of individual ejectile, which is not linear 
with the mass number of ejectile.

\newpage
\begin{center}
\begin{table*}[tbp]
\caption{{\sl Multiplicities of various ejectiles emitted during second stage
of proton induced reactions on Au target, at a few impact energies}}
\vspace{0.5cm}
\hspace{2.0cm}
\begin{tabular}{|c||c|c|c|}
\hline
impact energy & 1.2 GeV & 1.9 GeV & 2.5 GeV \\
\hline \hline
n: &14.17  &12.79  &11.88  \\
\hline
$^{1}H$: &1.13  &1.099  &0.96  \\
$^{2}H$: &0.59  &0.59  &0.51 \\
$^{3}H$: &0.28  &0.28  &0.24 \\
\hline
$^{3}He$: &0.020  &0.023  &0.021 \\
$^{4}He$: &0.59  &0.56  &0.48 \\
$^{6}He$: &0.0054  &0.0058  &0.0052 \\
\hline
$^{6}Li$: &0.0045  &0.0058  &0.0056 \\
$^{7}Li$: &0.0067  &0.0076  &0.0067 \\
$^{8}Li$: &0.0018  &0.0022  &0.0019 \\
$^{9}Li$: &0.00039  &0.00045  &0.00039 \\
\hline
$^{7}Be$: &0.00045  &0.00060  &0.00055 \\
$^{9}Be$: &0.0015  &0.0017  &0.0015 \\
$^{10}Be$: &0.0018  &0.0020  &0.0017 \\
$^{11}Be$: &0.00017  &0.00017  &0.00016 \\
\hline
$^{8}B$: &0.000010  &0.000018  &0.000011 \\
$^{10}B$: &0.00065  &0.00081  &0.00069 \\
$^{11}B$: &0.00068  &0.00078  &0.00067 \\
$^{12}B$: &0.00047  &0.000497  &0.00043 \\
$^{13}B$: &0.000061  &0.000051  &0.000048 \\
\hline
\hline
C: &0.0011  &0.0012  &0.0012 \\
\hline
N: &0.00044  &0.00045  &0.00043 \\
\hline
O: &0.0002  &0.00021  &0.00017 \\
\hline
\end{tabular}
\label{table: isot_Au_abc}
\end{table*}
\end{center}

\newpage
\begin{center}
\begin{table*}[tbp]
\caption{{\sl Multiplicities of various ejectiles emitted during second stage
of 2.5 GeV proton induced reactions on several targets}}
\vspace{0.5cm}
\hspace{2.0cm}
\begin{tabular}{|c||c|c|c|c|}
\hline
reaction &p+Al &p+Ni &p+Ag &p+Au \\
\hline \hline
n: &0.97  &1.76  &6.11  &11.88  \\
\hline
$^{1}H$: &0.82  &2.20  &2.19  &0.96  \\
$^{2}H$: &0.36  &0.51  &0.75  &0.51 \\
$^{3}H$: &0.076  &0.089  &0.20  &0.24 \\
\hline
$^{3}He$: &0.082  &0.10  &0.082  &0.021 \\
$^{4}He$: &0.58  &0.53  &0.64  &0.48 \\
$^{6}He$: &0.0017  &0.0012  &0.0036  &0.0052 \\
\hline
$^{6}Li$: &0.033  &0.021  &0.017  &0.0056 \\
$^{7}Li$: &0.0097  &0.0057  &0.0092  &0.0067 \\
$^{8}Li$: &0.0012  &0.00065  &0.0016  &0.0019 \\
$^{9}Li$: &0.00016  &0.000046  &0.00020  &0.00039 \\
\hline
$^{7}Be$: &0.0087  &0.0059  &0.0032  &0.00055 \\
$^{9}Be$: &0.0040  &0.0013  &0.0019  &0.0015 \\
$^{10}Be$: &0.0016  &0.00055  &0.0013  &0.0017 \\
$^{11}Be$: &0.000075  &0.000015  &0.000068  &0.00016 \\
\hline
$^{8}B$: &0.00059  &0.00046  &0.00015  &0.000011 \\
$^{10}B$: &0.013  &0.0029  &0.0019  &0.00069 \\
$^{11}B$: &0.0061  &0.00088  &0.00079  &0.00067 \\
$^{12}B$: &0.00087  &0.00016   &0.00030  &0.00043 \\
$^{13}B$: &0.000092 &0.0000069   &0.000018  &0.000048 \\
\hline
C: &0.064  &0.0062  &0.0018  &0.0012 \\
\hline
N: &0.070  &0.0049  &0.00053  &0.00043 \\
\hline
O: &0.077  &0.0048  &0.00018  &0.00017 \\
\hline
\end{tabular}
\label{table: isot_2.5_abc}
\end{table*}
\end{center}


\begin{center}
\begin{table*}[tbp]
\caption{{\sl Average values and corresponding standard deviations of mass of 
heavy nuclei remained after first and second stage of p+Pb reaction, at 
different incident energies (Tp)}}
\vspace{0.5cm}
\hspace{2.0cm}
\begin{tabular}{|c||c|c|}
\hline
 incident energy & after first stage & after second stage \\
\hline \hline
Tp = 0.5 GeV &208.36$\pm$1.14 &188.21$\pm$7.17  \\
\hline
Tp = 1.0 GeV &206.73$\pm$1.48 &186.65$\pm$7.31  \\
\hline
Tp = 2.0 GeV &205.79$\pm$2.24 &188.33$\pm$9.70 \\
\hline
Tp = 3.0 GeV &205.70$\pm$2.39 &189.53$\pm$9.98 \\
\hline
Tp = 4.0 GeV &205.26$\pm$2.82 &188.68$\pm$10.24 \\
\hline
Tp = 5.0 GeV &204.57$\pm$3.25 &187.70$\pm$10.58 \\
\hline
Tp = 7.0 GeV &203.86$\pm$3.92 &186.89$\pm$11.28 \\
\hline
Tp = 9.0 GeV &202.78$\pm$4.73 &185.04$\pm$11.83 \\
\hline
\end{tabular}
\label{table: A_Res_pPb}
\end{table*}
\end{center}

\section{Massive Fragments}
\markboth{9.5 Massive Fragments}{Chapter 9. Comparison of calculations with experimental data}

\begin{figure}[!htcb]
\vspace{-1.5cm}
\hspace{-2cm}
\includegraphics[height=20cm, width=18cm, bbllx=0pt, bblly=25pt, bburx=594pt, bbury=842pt, clip=, angle=0]{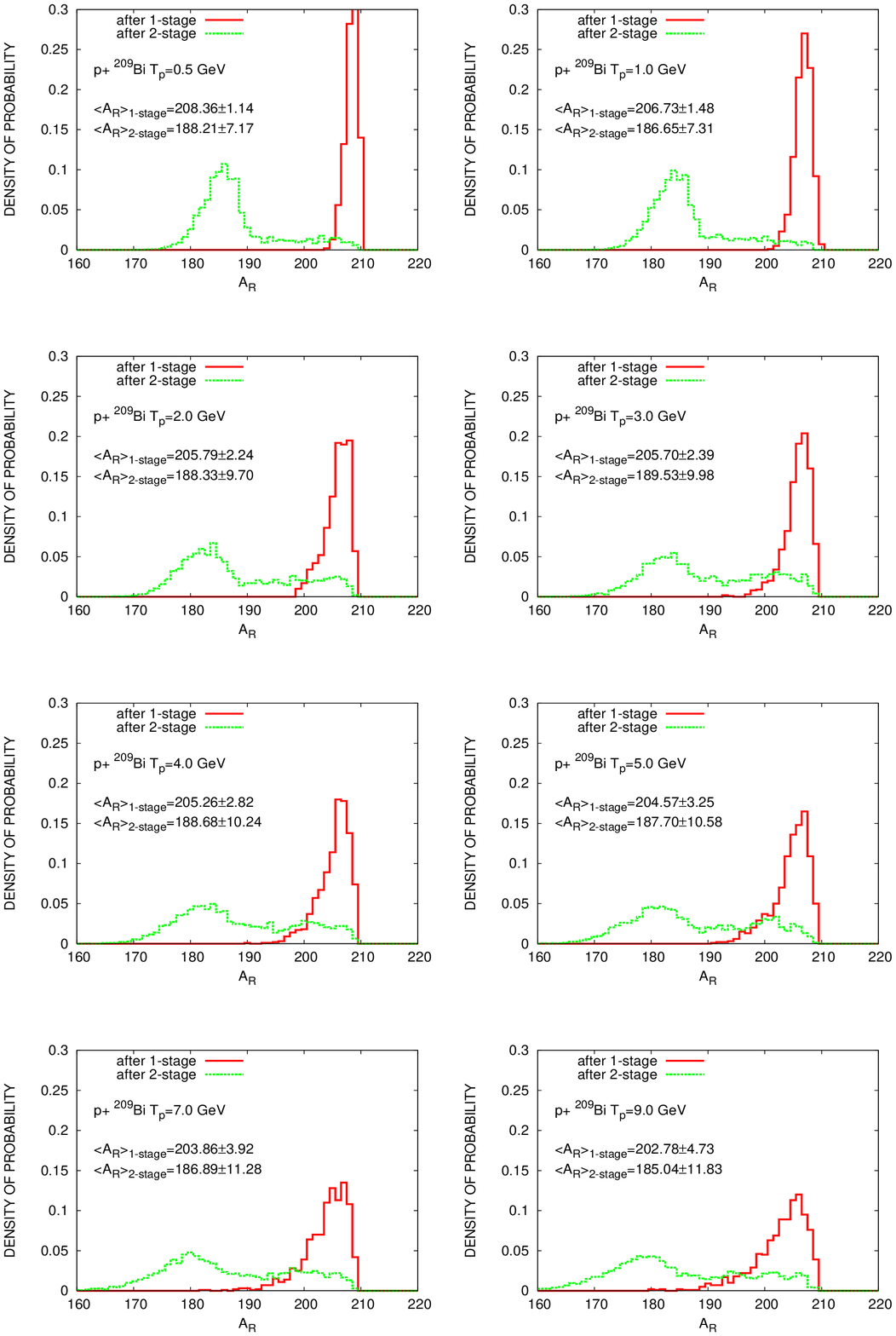}
\caption{{\sl Mass distributions of heavy fragments remained after first and 
second stage of p+Pb reaction at few proton beam energies; results of the HSD 
(first stage) and PACE2 (second stage) model calculations (see also Table 
\ref{table: A_Res_pPb})}}
\label{fig:A_Res_pPb}
\end{figure}

As it is mentioned above, spallation is characterized by observations of one 
heavy (in respect to the mass of initial target) nucleus, a small number of 
light fragments and several individual nucleons.  
Mass distributions of the heavy fragments, produced as result of proton induced 
spallation reaction on a chosen target vary with incident energy.
In Figure \ref{fig:A_Res_pPb}, mass distributions of heavy fragments formed in 
example p+Pb reactions, at several incident energies (calculated with the 
HSD + statistical evaporation model), set together with mass distributions of 
corresponding residual nuclei remaining after first stage of the reactions 
(results of the HSD model only), are shown.
It is seen, that both of the distributions (after first stage and 
after whole reaction) broaden with increase of incident energy.
Additionally, the higher projectile energy, the lower average mass of nuclei 
after the first stage of the reaction. Distribution of mass of nuclei 
produced after the second stage of the reaction is much broader than mass 
distribution of the remnants of the first stage. 
The following changes of the distribution of mass of nuclei produced after the 
second stage, as function of projectile energy are observed. At the lowest 
presented projectile energy, a maximum at low masses is seen and very small 
amount of nuclei with masses close to masses of nuclei after first stage of 
reaction (which are close to mass of the initial target).
With increase of incident energy, the distribution broadens.
The higher the incident energy, the smaller is the maximum at lower masses.  
The lighter and lighter nuclei appears.
Also more nuclei with masses close to mass of the initial nucleus are produced.
Quantitative differences are presented in Table \ref{table: A_Res_pPb}. 

\begin{figure}[!htcb]
\vspace{-1.5cm}
\hspace{-2cm}
\includegraphics[height=20cm, width=17cm, bbllx=0pt, bblly=25pt, bburx=594pt, bbury=842pt, clip=, angle=0]{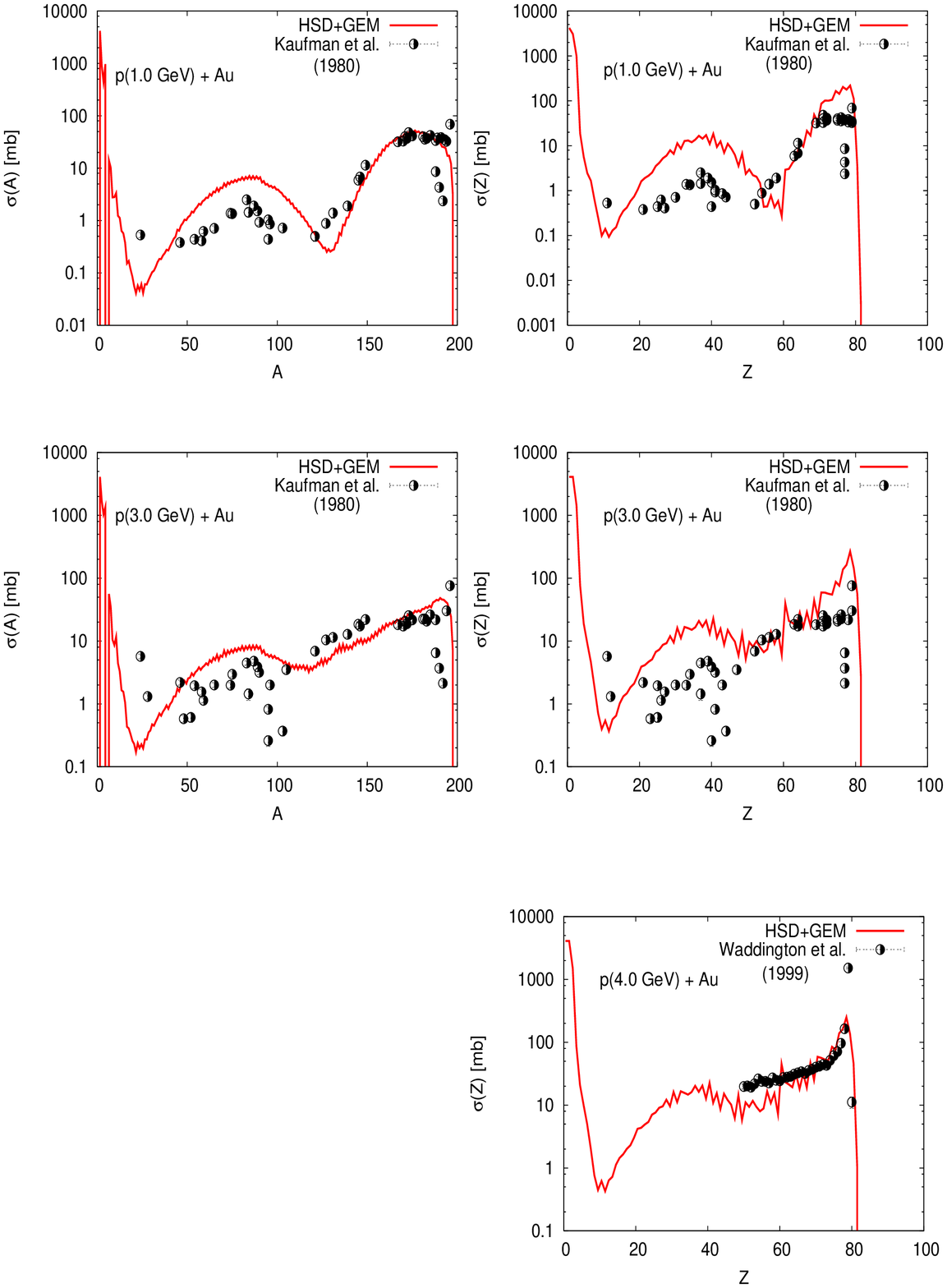}
\caption{{\sl Mass (left) and charge (right) distributions of products of the 
second stage of p+Au reaction, at a few proton beam energies; results of the 
HSD+GEM model calculations (solid lines), compared to experimental data 
\cite{Kauf80} and \cite{Wadd99}}}
\label{fig:pAu_sigaz_iaz}
\end{figure}

\begin{figure}[!htcb]
\begin{center}
\vspace{-2cm}
\hspace{-2cm}
\includegraphics[height=17cm, width=14cm, bbllx=0pt, bblly=80pt, bburx=594pt, bbury=842pt, clip=, angle=0]{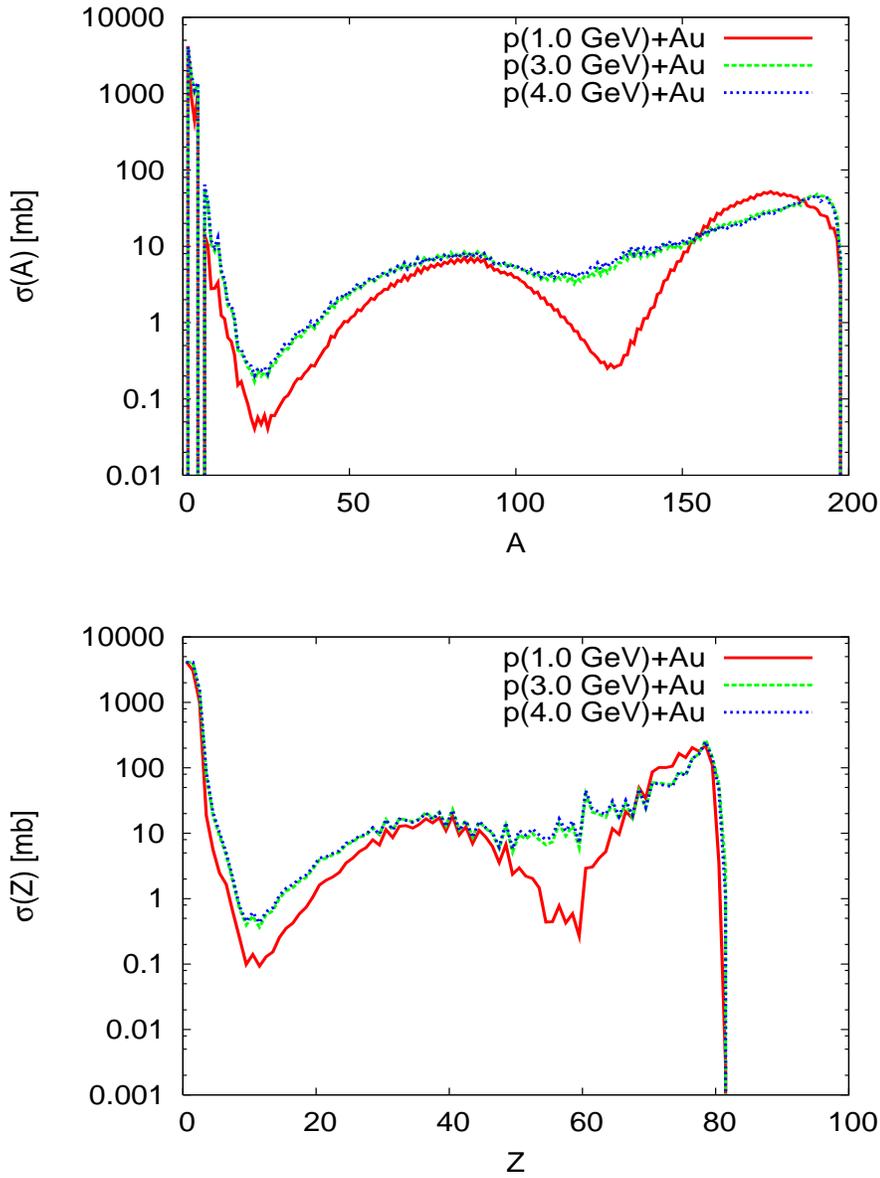}
\caption{{\sl Comparison of mass and charge distributions of products of the 
second stage of p+Au reaction emitted at different proton beam energies; 
results of the HSD+GEM model calculations 
(one observes that for higher incident energies of proton, the dips in the 
distributions become shallower)}}
\label{fig:obl_pAu_sigaz_iaz}
\end{center}
\end{figure}

In Figure \ref{fig:pAu_sigaz_iaz}, comparison of calculated mass and charge 
distributions of products of the proton induced reaction on Au target, at a 
few incident energies, with experimental data \cite{Kauf80} and \cite{Wadd99}, 
is presented. Agreement between the calculations and measurements is quite 
good. \\
If one compares mass and charge distributions of products of proton induced 
reactions on a chosen target, at different projectile energies, significant 
difference is seen in mass range $A > 25$ and charge range 
$Z > 10$, as shown in Fig. \ref{fig:obl_pAu_sigaz_iaz}. 
At the lower projectile energy, two almost separate distributions are visible: 
one ($25 < A < 130$, $10 < Z < 60$, lower yield) corresponds to fission 
products, the second ($130 < A < 200$, $60 < Z < 80$, higher yield) indicates 
spallation products (fission process is discussed in Sec. \ref{sec:fission}).  
At higher incident energy, the split of the distributions vanishes.

\chapter{Comparison of different models predictions on the first stage 
of proton induced reactions}
\markboth{ }{Chapter 10. Comparison of different models predictions}

Apart from the HSD model, a few other microscopic models have been 
developed in order to describe the fast stage of proton - nucleus reaction. 
All the models are based on different assumptions or simplifications 
(as described in Chapters \ref{chapt:theo_models} and \ref{chapt:HSD}). 
It is interesting to compare their predictions.
 In this chapter, results of the HSD, INCL4 (\cite{Cugn97}) and QMD 
(\cite{Aich91}) models are examined.

The most important issue for comparison of results of various models 
of the first stage of proton - nucleus reaction are: particles production 
during the first stage of the reaction and properties of residual 
nuclei remaining after the stage.
Predictions on pion production are particularly interesting for comparison.
This is due to the fact, that calculations of the first stage of reaction
are sufficient in order to receive realistic pion spectra, as all pions are
produced only in the fast cascade of nucleon - nucleon collisions, where the 
available amount of four-momentum is large enough. One of the important 
features which can influence the magnitude of pion production and absorption 
is treatment of the nucleon-nucleus mean field (that corresponds to the main 
difference between the models). To check whether assumptions
concerning the mean field in the model calculations affect
significantly multiplicity of produced pions, results of the HSD model are 
compared with the INCL4 and QMD models calculations and presented in 
Tables \ref{table_piNi} and \ref{table_piAu}, for 2.5 GeV proton induced 
reactions on Ni and Au target, respectively.
\begin{table}[tbp]
\caption{{\sl Multiplicity of pions produced during p+$^{58}$Ni reaction at 
2.5 GeV proton beam energy; results of the HSD, INCL4 \cite{MKistr} and QMD 
 \cite{MPuch} model calculations}}
\begin{center}
\begin{tabular}{|c|c|c|c|}
\hline
 &HSD &INCL4 &QMD \\
\hline
$\pi ^{+}$ &0.51 &0.35 &0.42 \\
\hline
$\pi ^{0}$ &0.46 &0.34 &0.36 \\
\hline
$\pi ^{-}$ &0.38 &0.18 &0.28 \\
\hline
\end{tabular}
\end{center}
\label{table_piNi}
\end{table}

\begin{table}[tbp]
\caption{{\sl Multiplicity of pions produced during p+$^{197}$Au reaction
at 2.5 GeV proton beam energy; results of the HSD, INCL4 \cite{MKistr} and 
QMD \cite{MPuch} model calculations}}
\begin{center}
\begin{tabular}{|c|c|c|c|}
\hline
 &HSD &INCL4 &QMD \\
\hline
$\pi ^{+}$ &0.49 &0.29 &0.34 \\
\hline
$\pi ^{0}$ &0.53 &0.35 &0.34 \\
\hline
$\pi ^{-}$ &0.46 &0.25 &0.32 \\
\hline
\end{tabular}
\end{center}
\label{table_piAu}
\end{table}
It is seen that the relative behavior of
the pion multiplicities obtained from the different models is
similar: for lighter target (Ni) the $\pi(+)$ : $\pi(0)$ : $\pi(-)$
ratio is equal to 1 : 0.86 : 0.67 for QMD \cite{MPuch}, 1 : 0.97 : 0.51 for 
INCL4 \cite{MKistr} and 1 : 0.90 : 0.75 for the HSD, while for the heavy target
 (Au) 1 : 1 : 0.94 (QMD \cite{MPuch}), 1 : 1.21 : 0.86 (INCL4 \cite{MKistr}) 
and 1 : 1.08 : 0.94 (HSD). However, absolute values 
of pion multiplicities obtained from the HSD model are always higher
by factor $\sim$1.5 than the values received from the INCL4 and the QMD models,
although results of the HSD model are closer to that of the QMD model. 

\begin{figure}[!htcb]
\vspace{-2cm}
\hspace{-1cm}
\includegraphics[height=15cm, width=15cm, bbllx=0pt, bblly=180pt, bburx=594pt, bbury=842pt, clip=, angle=0]{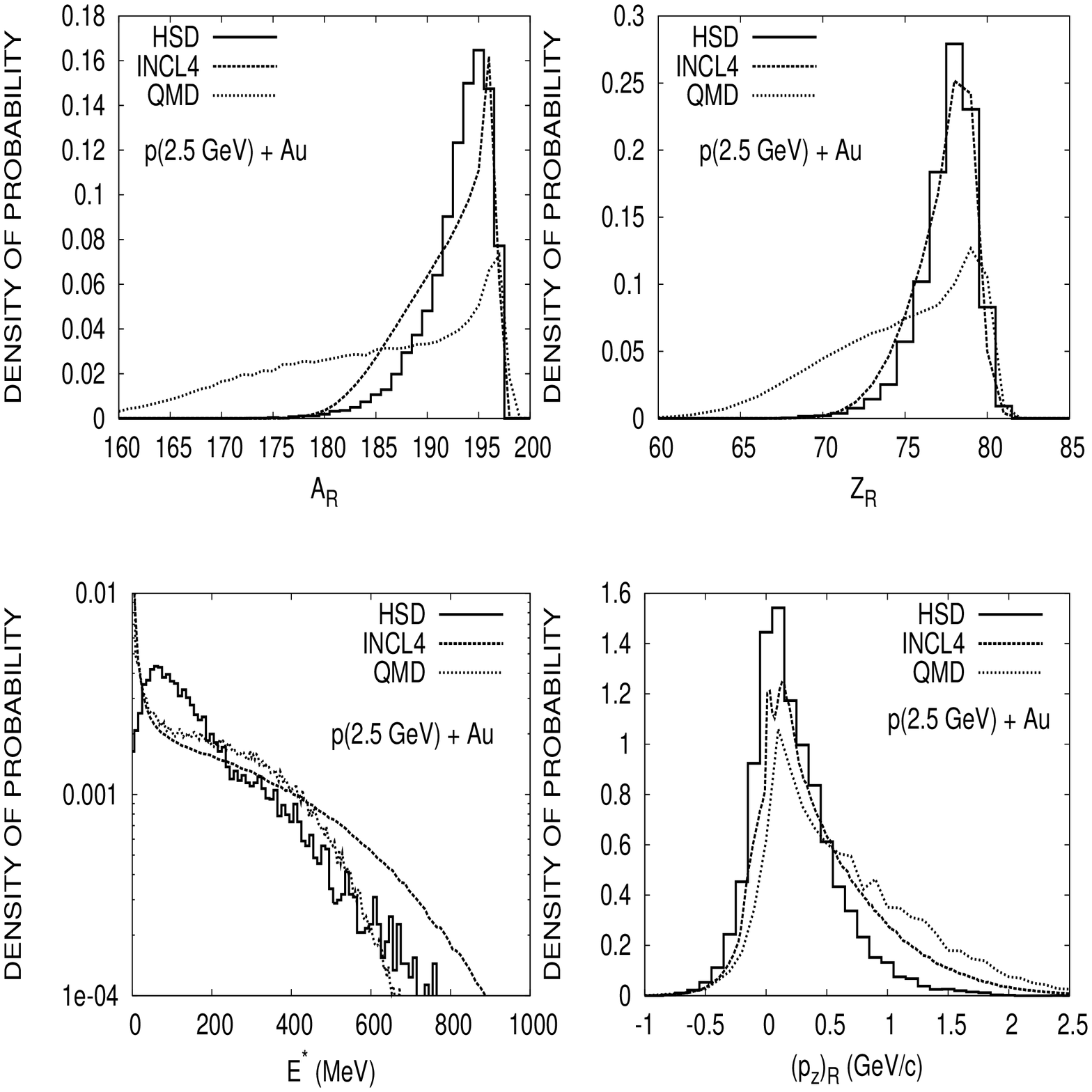}
\caption{{\sl Distributions of properties of the residual nuclei 
after first stage of p + Au reaction, at 2.5 GeV proton beam energy; 
comparison of results of the HSD (solid lines), INCL4 \cite{MKistr} 
and QMD \cite{MPuch} (scatter lines) model calculations}}
\label{fig:q_inc_buu_pAu2.5_res}
\end{figure}

Let's compare now also an amount of kinetic energy carried out by the most 
abundantly produced particles during the first stage of the reaction, i.e. 
nucleons and pions. 
The average multiplicity and average kinetic energy of nucleons and pions 
emitted during the first stage of example 2.5 GeV proton induced reaction on 
Au target are displayed in Table \ref{table_mult_ekin_nucl}. 
\begin{table}[tbp]
\caption{{\sl Multiplicity, average values and standard deviations of total  
kinetic energy of nucleons and pions emitted during first stage of p+Au 
reaction, at 2.5 GeV proton beam energy; comparison of results of the HSD, 
INCL4 \cite{MKistr} and QMD \cite{MPuch} model calculations}}
\begin{center}
\begin{tabular}{|c||c|c||c|c||c|c|}
\hline
&\multicolumn{2}{c||}{{\bf HSD}} &\multicolumn{2}{|c||}{{\bf INCL4}} &\multicolumn{2}{|c|}{{\bf QMD}}\\
\hline
&$M$ &$<E_{k}>[GeV]$ &$M$ &$<E_{k}>[GeV]$ &$M$ &$<E_{k}>[GeV]$\\
\hline
n &2.23 &0.78$\pm$1.27 &3.48 &0.80$\pm$1.46 &7.63 &0.76$\pm$1.83 \\
\hline
p &2.20 &0.90$\pm$1.36 &2.68 &0.96$\pm$1.58 &5.63 &0.96$\pm$2.31 \\
\hline
$\pi ^{-}$ &0.49 &0.14$\pm$0.13 &0.25 &0.07$\pm$0.08 &0.32 &0.09$\pm$0.11 \\
\hline
$\pi ^{0}$ &0.53 &0.15$\pm$0.15 &0.35 &0.13$\pm$0.13 &0.34 &0.10$\pm$0.12 \\
\hline
$\pi ^{+}$ &0.46 &0.13$\pm$0.13 &0.29 &0.11$\pm$0.11 &0.34 &0.11$\pm$0.12 \\
\hline
\end{tabular}
\end{center}
\label{table_mult_ekin_nucl}
\end{table}
The values collected in Table \ref{table_mult_ekin_nucl} indicate, that 
although the average multiplicity of emitted particles provided by the  
models are slightly different, the average amount of kinetic energy carried out
 by the particles is almost identical in all of the models. 
Consequently, properties of residual nuclei evaluated by the different models 
should be similar.
In Figure \ref{fig:q_inc_buu_pAu2.5_res}, comparison of the properties of 
residual nuclei, i.e. distributions of mass, charge, excitation energy and 
momentum in beam direction of the residual nucleus created in example p+Au 
reaction, at 2.5 GeV beam energy, are shown.
The adequate average values are listed in Table \ref{table_inc_buu_pAu2.5_res}.
One observes that the HSD and INCL4 predictions are quite similar. 

\begin{figure}[!htcb]
\vspace{-0.5cm}
\hspace{-1cm}
\includegraphics[height=6cm, width=8cm, angle=0]{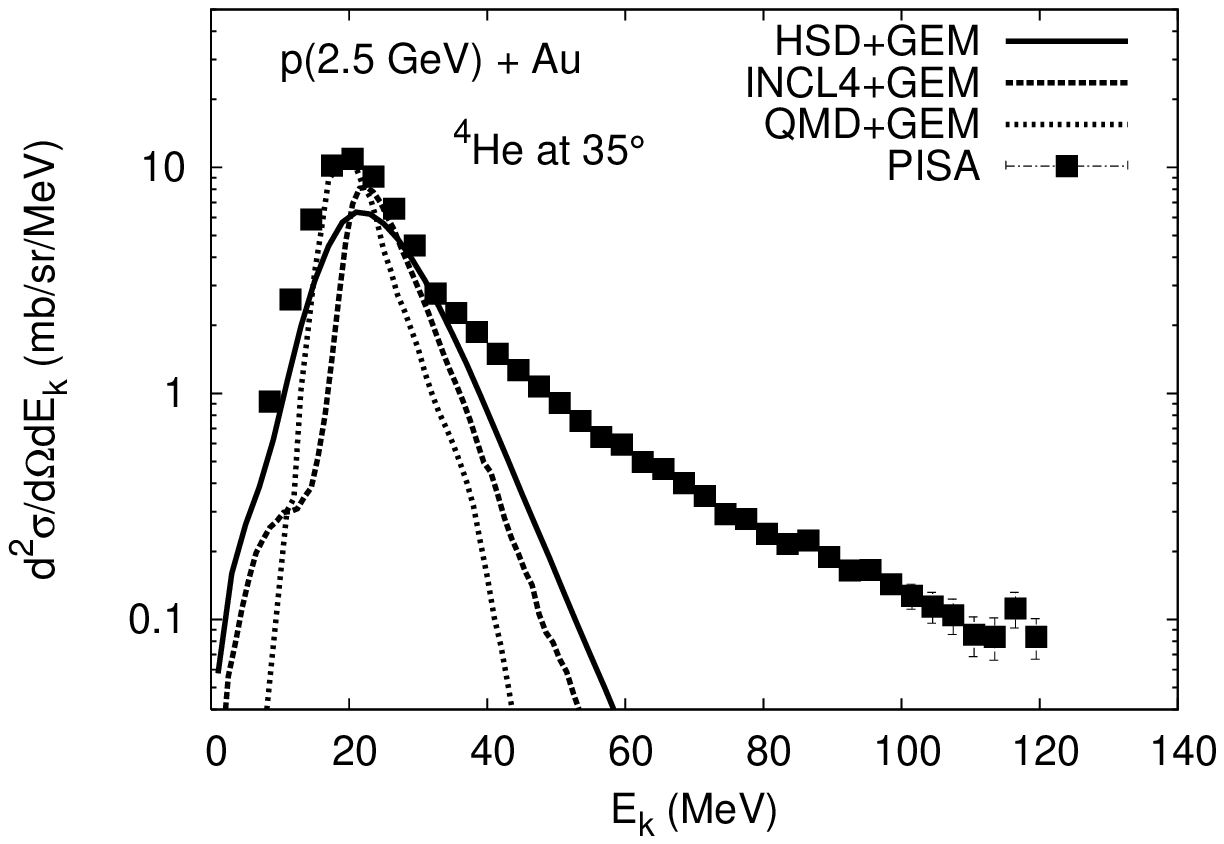}
\hspace{-0.6cm}
\includegraphics[height=6cm, width=8cm, angle=0]{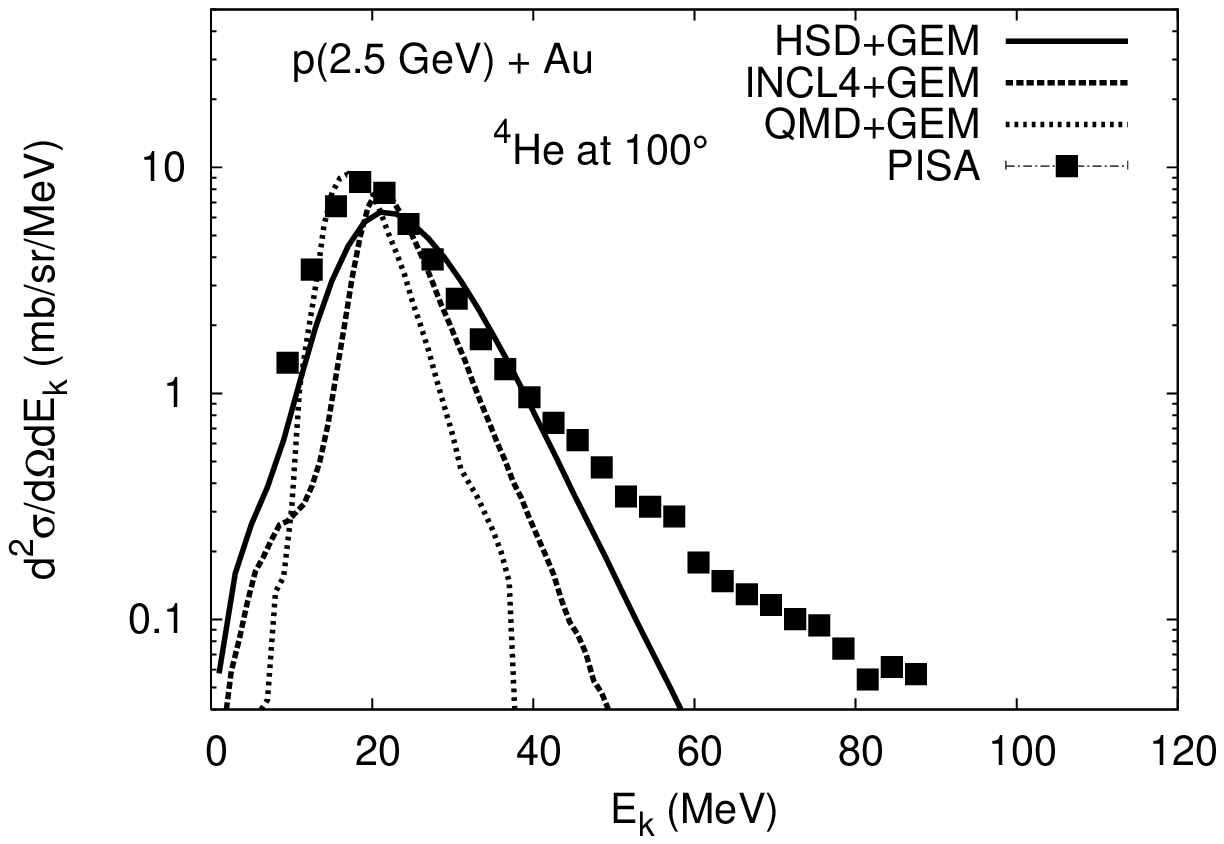}
\caption{{\sl Double differential distributions of $^{4}$He
emitted during p+Au reaction, at 2.5 GeV incident energy; results of the
HSD+GEM, INCL4+GEM (\cite{MKistr}) and QMD+GEM (\cite{MPuch})
model calculations compared to experimental data measured by PISA collaboration
\cite{Buba07}}}
\label{fig:q_inc_buu_pAu2.5_he4}
\end{figure}

\begin{figure}[!hcb]
\vspace{-1.5cm}
\hspace{-2cm}
\includegraphics[height=20cm, width=17cm, bbllx=0pt, bblly=25pt, bburx=594pt, bbury=842pt, clip=, angle=0]{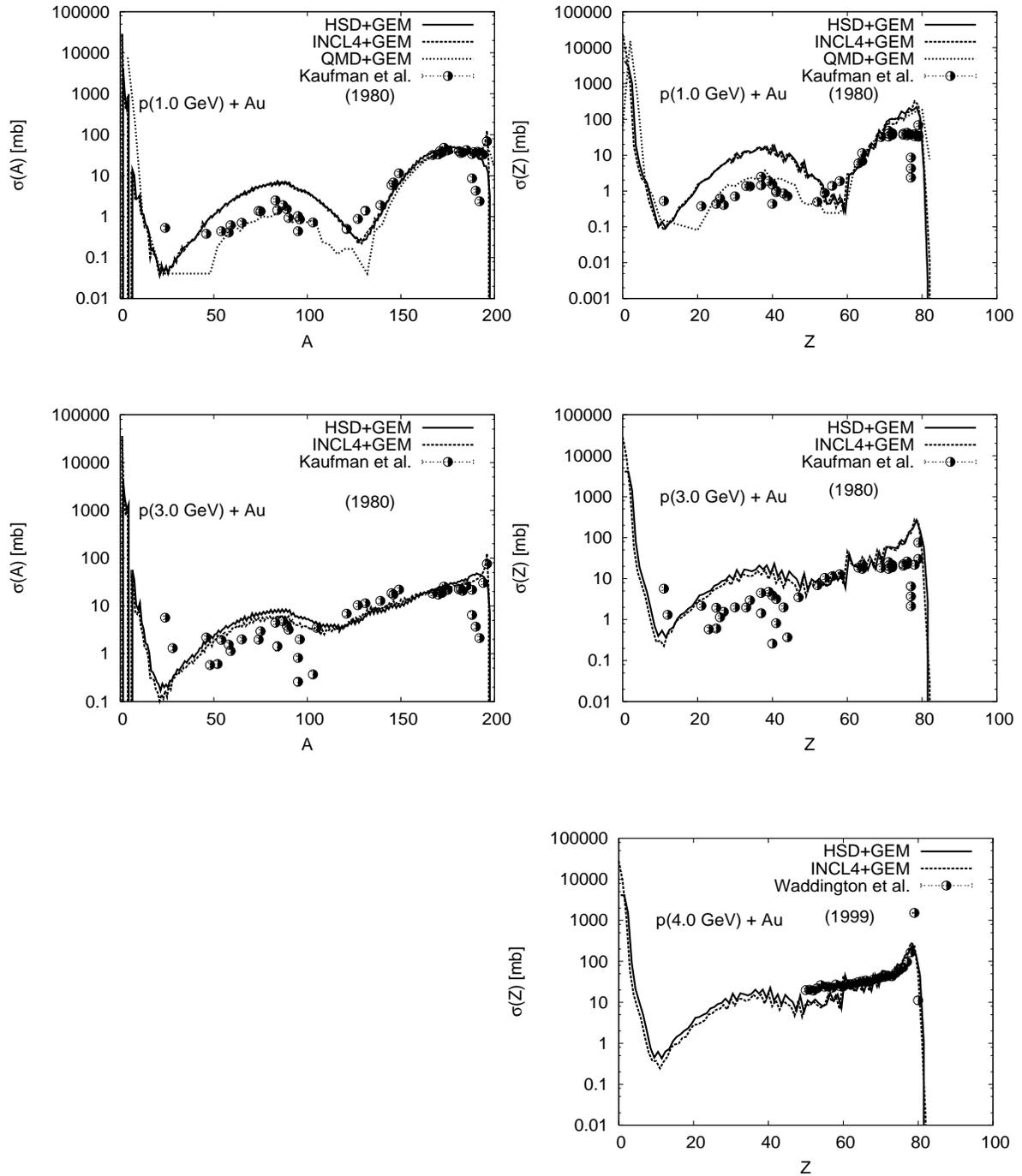}
\caption{{\sl Comparison of mass (left) and charge (right) distributions of 
products of the second stage of p+Au reaction emitted at different proton beam 
energies; results of the HSD+GEM, INCL4+GEM (\cite{MKistr}) and QMD+GEM 
(for T$_{p}$ = 1.0 GeV only, \cite{MPuch}) model calculations compared to 
experimental data \cite{Kauf80} and \cite{Wadd99}}}
\label{fig:q_inc_buu_pAu_sigaz_iaz}
\end{figure}

\begin{table}[tbp]
\caption{{\sl Average values and standard deviations of properties of residual 
nuclei after first stage of p+Au reaction, at 2.5 GeV proton beam energy; 
comparison of results of the HSD, INCL4 \cite{MKistr} and QMD \cite{MPuch} 
model calculations}}
\begin{center}
\begin{tabular}{|c|c|c|c|}
\hline
 &HSD &INCL4 &QMD \\
\hline
$<A_{R}>$ &193.0$\pm$3.3 &191.8$\pm$4.03 &183.5$\pm$10.1 \\
\hline
$<Z_{R}>$ &77.6$\pm$1.8 &77.3$\pm$1.9 &73.8$\pm$4.3 \\
\hline
$<E^{*}>[MeV]$ &201.7$\pm$171.3 &253.6$\pm$226.3 &195.5$\pm$165.8 \\
\hline
$<(p_{z})_{R}>[GeV/c]$ &0.28$\pm$0.35 &0.44$\pm$0.51 &0.58$\pm$0.62 \\
\hline
\end{tabular}
\end{center}
\label{table_inc_buu_pAu2.5_res}
\end{table}
Results of the models are sensitive to used parameters  
(in particular time duration of the first stage calculations, which is 
determined for each model individually, see \cite{Cugn97, Aich91}). 
Therefore it is worthy to compare results obtained from each of the model and 
estimate their predictive power by confrontation with experimental data.
In order to describe the nuclear reaction in unified way to verify predictions 
of the different models with experimental observations, the first 
stage models have been incorporated with one chosen statistical evaporation 
model applied for the second stage of reaction, i.e. GEM \cite{Furi00}.
Output of the HSD, INCL4 and QMD model, respectively (i.e. mass, charge, 
momentum and excitation energy of the residual nuclei) defines input for the 
GEM model. \\
In order to verify which of the distributions of the properties displayed in 
the Fig. \ref{fig:q_inc_buu_pAu2.5_res} is the most reliable, calculated 
double differential kinetic energy spectra of $^{4}$He emitted at two different
 angles during p+Au reaction at 2.5 GeV incident energy are compared with 
experimental data \cite{Buba07}, and shown in Fig. 
\ref{fig:q_inc_buu_pAu2.5_he4}. Distributions of $^{4}$He 
 are the best choice for such a comparison due to the fact, that in the 
$^{4}$He channel evaporation is dominant. As it is shown in Ref.\cite{Buba07}, 
about 90$\%$ of the $^{4}$He spectrum consists of the particles produced 
during the evaporation stage. One can see, that the models give similar 
results, although the HSD and the QMD model results are somehow closer to 
experimental data.\\
In Figure \ref{fig:q_inc_buu_pAu_sigaz_iaz}, comparison of calculated example 
fragment mass and charge distributions compared with experimental data 
\cite{Kauf80} and \cite{Wadd99} are displayed.
It is seen, that also in this case the models give similar results.  
In particular, results of the HSD and the INCL4 models are almost identical. 
If looking at the QMD model calculations, plotted for reaction p+Au at 1.0 
GeV beam energy, in the emitted fragment mass number and charge range 
corresponding to the fission peak (i.e. $50<A<100$ and $20<Z<50$), they are 
lower than calculations of the other models, but closest to the 
experimental data. However, outside the range (i.e. $A<50$, $A>100$ and $Z<20$, 
$Z>50$) results of all considered here models are very similar.

Based on the comparisons, one can conclude that results 
of the HSD model do not differ significantly from results of the other models. 
The additional verification by confrontation of the calculations with 
experimental data indicates, that results of the HSD and the QMD model are 
particularly close to the experimental data. 
As results of the HSD model are not distinctly different from 
that of the QMD model (which is actually the most advanced one), a big 
advantage of the HSD model, connected with respectively short time of 
calculations should be mentioned. The average time needed to calculate a 
reaction with the QMD model is of the order of a few days, while time needed to
 calculate the adequate reaction with the HSD model - only few hours.    

\chapter{Summary and Conclusions}
\markboth{ }{Chapter 11. Summary and Conclusions}

Global information of properties of proton - nucleus reactions 
(i.e. particles production, properties of residual nuclei after 
first stage of the reactions) has been presented in wide range of projectile 
energy and mass of target nuclei, with the Hadron String Dynamics (HSD) 
\cite{WCass, Geis98} model for the first stage of the reaction, 
incorporated with statistical evaporation model (GEM \cite{Furi00} and PACE2 
\cite{Gavr93}) for the second stage. It has been confronted with experimental 
data and results of other models. \\ 
The version of the HSD code \cite{WCass} has been developed in order to 
describe proton induced spallation reactions in wide energy range 
 in the following aspects. First of all, to calculate properties of residual
nuclei after first stage of the reaction, defining an input for
models of the second stage calculations.
The code has been developed for description of pion production 
in proton - nucleus reactions.
Additionally, a version of HSD code that allows for calculations of pion 
induced reactions has been prepared.\\
Stopping time for the first stage model calculations has been established 
based on behaviors of excitation energy, longitudinal momentum and angular 
momentum as a function of time. These are decisive for determination of the 
duration time of the first stage of proton - nucleus collisions, due to the 
fact that in the dependences two phases can be distinguished: a phase of rapid 
variations, followed by a phase of much slower variations, indicating 
thermodynamical equilibrium of the nucleus. Properties of residual nuclei are 
evaluated by exploring conservation of energy, mass, charge, momentum and 
angular momentum.

It has been shown, that distributions of properties of residual nuclei 
(i.e. mass, charge, excitation energy, momentum, angular momentum) differ 
significantly with mass of target, but behave similarly for varied projectile 
energies.
Multiplicity of emitted pions increases with incident energy and with target 
mass number. Contrary, e.g. average kinetic energy of other individual emitted 
particles (i.e. nucleons, pions) increases with kinetic 
energy of incoming proton, but decreases with target mass number.
Higher nucleon and pion momenta and their increasingly forward peaked 
distributions are associated with increasing incident energy. That is 
connected with increasingly forward focused nature of nucleon - nucleon 
collisions with increase of incident energy; the total nucleon - nucleon 
scattering cross section remains approximately constant, but ratio of 
inelastic to elastic collisions increases rapidly with impact energy 
\cite{PDG94}.

It has been shown, that the biggest amount of initial energy deposited in 
nucleus by incoming proton is carried out during the first stage of the 
reaction, i.e. about 85 $\%$, and only about 15 $\%$ of the initial energy is 
carried away during the second stage of the reaction.
Most of the initial energy is carried out by nucleons and pions.
Neutrons are emitted mainly in the second stage of reaction, 
whereas protons - mainly during the fast cascade stage. 
Pions are emitted only in the first stage of reaction, where the locally 
available amount of four-momentum is large enough.  
In case of light composites, only low energy part of 
distribution, corresponding to a result of evaporation stage, is described by 
the HSD plus evaporation model. 
E.g. for deuterons, the absence of high energy part of modelled distributions 
indicates a lack of deuterons produced during first stage of the reaction. 
Description of the whole distributions necessitates future implementation of 
some additional mechanisms into the first stage model, 
e.g. coalescence processes. 
It is observed already in neutron spectra that these additional mechanisms have 
to depend strongly on size of the system: for heavier targets the neutron 
spectra in some region are visibly underestimated by model calculations, see 
Fig. \ref{fig:pPb_nX_1.5}.

Very interesting conclusion of the work is a prevision for a negligible 
probability for fragmentation in proton induced reactions, at considered here 
range of incident energy. This statement is based on results of the HSD model 
calculations of excitation energy per nucleon of residual nuclei and time 
evolution of nucleon density. 
Results of the HSD model show, that a value of 
excitation energy greater than 5 MeV/N occurs very rarely, in only few percent 
of cases (studies of fragmentation have found that it occurs at a 
value of excitation energy per nucleon of a residual nuclei at least equal to 
about 5 MeV/N \cite{Beau00, Viol06}). 
Fragmentation could be more probable if a part of nuclei takes part in 
excitation, i.e. the excitation energy should be calculated only per some 
amount of the {\sl active} nucleons; only this part of nucleus would undergo 
fragmentation. However modelling of 
nucleon density evolution with the reaction time shows, that proton 
induced reactions are very low - invasive processes. Incoming 
proton causes insignificant modifications of the nuclear density. Energy is 
rather uniformly distributed among the nucleons of residual nuclei. A situation,
 when two differently excited parts of nucleus appear, occurs only in about 
$1\%$ of proton - nucleus reaction cases and 
rather in peripheral collisions, whereas experimentalists associate 
fragmentation with central collisions. 
 
It has been shown that in proton induced reaction, in the considered here 
incident energy range, spallation is evidently dominant process. Moreover, the 
second stage of spallation reaction is a competition between evaporation and 
fission.

Results of the HSD model calculations have been compared with available 
experimental data and with adequate results of other microscopic models 
(i.e. INCL4 \cite{Cugn97} and QMD \cite{Aich91}). The comparisons model - 
model show that results of 
the HSD model do not differ significantly from results of the other models 
calculations. Results of the HSD model are also consistent with various 
experimental data. 
Qualitatively, the data are well described by the calculations. 
Quantitatively, the maximal noticed discrepancy is of factor 2. It concerns 
pion spectra, which are the worst described cases, i.e. overestimated by a 
factor of 2. That is because, despite  
description of NN interactions is well established, details of pions 
propagation in nuclear matter are very delicate. 
Legth of free path for pions in nuclear matter of standard density 
is 1 - 1.5 fm. It means, pions that arrive to detectors stem from surface of 
nucleus. In models like HSD it is just the surface that is described only 
approximately. 
Pions production due to proton - nucleus 
collisions is a complicated process, since they are not emitted directly, but 
result from a succession of productions and absorptions inside a nucleus, 
mediated by Delta resonance. 
 
The comparisons indicates that assumptions and 
simplifications employed in the HSD model are correct.

In order to obtain the possible best description of proton induced reactions, 
trials of various feasible modifications of the HSD model parameters has been 
investigated. However, the possible modifications, as in case of each of 
semi-classical models are limited. 
A necessity of the exact fully quantum model appears. 
Unfortunately, construction of such model of many body strongly interacting 
system is at present practically impossible. 

Summarizing, the global properties of proton induced reactions have been 
presented in wide range of mass of target and incident energy. 
It has been shown, that these are quite non-invasive reactions, where 
spallation is the dominant process. 
Properties of residual nuclei after the first stage of the reaction are 
weakly dependent on incident energy, but strongly dependent on mass of target. 
High momenta of emitted particles are associated with high incident energies.
The second stage of the reaction is a competition of evaporation and fission. 
The agreement of results of the HSD calculations with experimental data and 
with results of the other microscopic models calculations indicates that 
proper assumptions have been employed in the HSD model.
As discussed in Sec. \ref{sec:param}, results obtained from the HSD model vary 
smoothly as function of proton incident energy and mass of target, so 
interpolation of results is quite feasible. 

\newpage
\section*{Acknowledgments}
\markboth{ }{ }
\thispagestyle{empty}

I would like to thank all the people who helped me during the years of \\ 
Ph. D. studies.  

First of all I would like to express my enormous gratitude to Dr hab. 
Zbigniew Rudy, for devoting his time, many ideas that helped to bring this 
dissertation into existence, for his guidance, patience, support, encouragement
 and help whenever I need it. 

I am grateful to Prof. Bogus\l aw Kamys for allowing me to prepare 
this dissertation in the Faculty of Physics, Astronomy and Applied Computer 
Science of the Jagiellonian University. 

I am extremely grateful to Prof. Bogus\l aw Kamys and Prof. Lucjan 
Jarczyk for giving me an opportunity to work in PISA collaboration.\\ 
I am indebted to them for their helpful ideas, suggestions, their kindness,  
support and encouragement during all the years of my studies. 

A big "thank you" to all colleagues from the PISA collaboration, \\
especially to:
Prof. D.Filges, Dr F.Goldenbaum, Prof. L.Jarczyk, Prof. B.Kamys, Dr M.Kistryn, 
Dr hab. St.Kistryn, Dr E.Kozik, Prof. A.Magiera, B.Piskor-Ignatowicz, 
M.Pucha\l a, Dr K.Pysz, Dr hab. Z.Rudy \\and M.Wojciechowski.

I'd like to express my gratitude to Dr Ma\l gorzata Kistryn and Ma\l gorzata 
Pucha\l a for communication concerning results of INCL and QMD model 
calculations.
 
I would like to acknowledge Prof. Ulrich Mosel for allowing me to work 
in his group, at the Institut of Theoretical Physics of Justus-Liebig 
University in Giessen.

I am extremely grateful to Prof. Wolfgang Cassing for giving me an 
opportunity to work under his guidance at the Institut of Theoretical Physics 
in Giessen and ability to use his HSD code. I am indebted to him for help in 
developing of the code, for sharing his experience, devoting his time and for 
his patience. 

I would like to acknowledge Prof. Jim Ritman for allowing me to work in 
Research Center Juelich.

I am grateful to Dr Frank Goldenbaum, for the invitation to the Research Center 
Juelich, for his kindness and help.

I am very grateful to Dr hab. Zbigniew Rudy for careful reading and correcting 
this dissertation.\\
I am also very grateful to Prof. Bogus\l aw Kamys, Prof. Lucjan Jarczyk and 
Dr Frank Goldenbaum for correcting some parts of this thesis.

Thanks to all my colleagues from "the room 03A", from Giessen and from IKP 
for the pleasant atmosphere of the daily work.


\end{document}